\documentclass[12pt,twoside,english]{article}
\usepackage[T1]{fontenc}
\usepackage{units}
\usepackage{mathtools}
\usepackage{amsmath}
\usepackage{amssymb}
\usepackage{graphicx}

\makeatletter

\providecommand{\tabularnewline}{\\}

\newcommand{\lyxaddress}[1]{
	\par {\raggedright #1
	\vspace{1.4em}
	\noindent\par}
}

\usepackage{a4wide}
\usepackage{graphicx}
\usepackage{fancyhdr}
\usepackage{amsmath}
\usepackage{subfigure}
\usepackage{setspace}
\usepackage{xcolor}
\usepackage{lmodern} 
\usepackage{verbatim} 
\numberwithin{equation}{section}

\newcommand{\vek}[1]{\mathchoice{\displaystyle\boldsymbol#1}
{\textstyle\boldsymbol#1}{\scriptstyle\boldsymbol#1}
{\scriptscriptstyle\boldsymbol#1}}
\newcommand{\mat}[1]{\mathchoice{\displaystyle\mathbf#1}
{\textstyle\mathbf#1}{\scriptstyle\mathbf#1}
{\scriptscriptstyle\mathbf#1}}

\newcommand\ManUndef[1][]{\Gamma_{\!\!\vek X}^{c}}
\newcommand\ManDef[1][]{\Gamma_{\!\!\vek x}^{c}}
\newcommand\ManUndefD[1][]{\Gamma_{\!\!\vek X,\mathrm{D}}^{c}}
\newcommand\ManUndefN[1][]{\Gamma_{\!\!\vek X,\mathrm{N}}^{c}}
\newcommand\ManDefD[1][]{\Gamma_{\!\!\vek x,\mathrm{D}}^{c}}
\newcommand\ManDefN[1][]{\Gamma_{\!\!\vek x,\mathrm{N}}^{c}}
\newcommand\ShapeFcts[1][]{B_{i}}

\makeatother

\usepackage{babel}
\begin{document}
\title{On the Simultaneous Solution of Structural Membranes on all Level
Sets within a Bulk Domain}
\author{T.P. Fries, M.W. Kaiser}
\maketitle

\lyxaddress{\begin{center}
Institute of Structural Analysis\\
Graz University of Technology\\
Lessingstr. 25/II, 8010 Graz, Austria\\
\texttt{www.ifb.tugraz.at}\\
\texttt{fries@tugraz.at}
\end{center}}
\begin{abstract}
A mechanical model and numerical method for structural membranes implied
by \emph{all }isosurfaces of a level-set function in a three-dimensional
bulk domain are proposed. The mechanical model covers large displacements
in the context of the finite strain theory and is formulated based
on the tangential differential calculus. Alongside curved two-dimensional
membranes embedded in three dimensions, also the simpler case of curved
ropes (cables) in two-dimensional bulk domains is covered. The implicit
geometries (shapes) are implied by the level sets and the boundaries
of the structures are given by the intersection of the level sets
with the boundary of the bulk domain. For the numerical analysis,
the bulk domain is discretized using a background mesh composed by
(higher-order) elements with the dimensionality of the embedding space.
The elements are by no means aligned to the level sets, i.e., the
geometries of the structures, which resembles a fictitious domain
method, most importantly the Trace FEM. The proposed numerical method
is a hybrid of the classical FEM and fictitious domain methods which
may be labeled as ``Bulk Trace FEM''. Numerical studies confirm
higher-order convergence rates and the potential for new material
models with continuously embedded sub-structures in bulk domains.

\end{abstract}
\newpage\tableofcontents{}\newpage{}

\section{Introduction\label{X_Introduction}}

One may classify models in structural mechanics based on their dimensionality
which is mostly related to their geometric properties. For example,
thin structures with one prominent direction subjected to bending
may often be modeled as one-dimensional beams, and flat structures
with small thickness as plates. For straight beams and plates, one
may thereby reduce the dimensionality of the naturally three-dimensional
structure to 1 or 2, respectively, often resulting in largely simplified
boundary value problems (BVPs) compared to the original, three-dimensional
case. However, for \emph{curved} structures such as curved beams,
membranes and shells \cite{Bischoff_2017a,Calladine_1983a,Ciarlet_1997a,Chapelle_2011a,Ibrahimbegovic_1993a},
the situation is often more complex as they are embedded in some exterior
space with higher dimensionality. As the deformation takes place in
this background space, the complexity is still related to the dimensionality
of the background space. Furthermore, due to the curvature of the
lower-dimensional structures, differential geometry becomes an inherent
part of the mechanical models, i.e., the resulting BVPs. In this work,
the focus is on structural membranes being curved two-dimensional
structures in three dimensions undergoing large displacements; the
simpler reduction of this scenario are curved one-dimensional ropes
(cables) in two dimensions.

Curved lower-dimensional manifolds such as curved lines in 2D and
curved surfaces in 3D may be defined \emph{explicitly} through parametrizations
\cite{Carmo_1976a,Walker_2015a,Delfour_2011a,Farin_1999a} or \emph{implicitly}
through the level-set method \cite{Osher_2003a,Osher_2001a,Sethian_1999b}.
In the latter case, often the zero-level set of some scalar function
implies the domain of interest (e.g., the geometry of one membrane).
However, it is important to note that a level-set function, in fact,
implies infinitely many level sets featuring different constant values
each. One may then ask for a successive analysis of different curved
structures implied by (some) selected level sets. It turns out that
the displacement fields in the vicinity of a selected level set (typically)
vary smoothly, suggesting that it is possible to formulate a mechanical
model for the displacement field of \emph{all} structures implied
by all level sets within some bulk domain \emph{at once}; such a model
and the related numerical analysis are in the focus of this work.
Innovative applications of the proposed mechanical model may be found
in new material models where some (elastic) bulk material is equipped
with embedded sub-structures (i.e., continuously distributed fibres
and membranes), introducing a new concept of anisotropy. The simultaneous
analysis of structures also features advantages in design and optimization
as minimal and maximal values of displacements or stress quantities
are immediately obtained for a whole family of geometries (i.e., all
level sets).

The basic idea is outlined as follows: Let there be some undeformed
bulk domain with the dimensionality of the embedding space and a level-set
function. The level sets inside the bulk domain imply infinitely many
bounded, curved manifolds of codimension 1, which herein represent
a set of curved, undeformed membranes in 3D or ropes in 2D, respectively.
We now seek a mechanical model which determines the displacement field
of the bulk domain, such that the level sets interpreted in this deformed
bulk domain are the deformed structures (membranes). Therefore, a
finite strain theory is formulated in the frame of the tangential
differential calculus (TDC). With TDC, we only refer to the modern
differential geometric point of view of formulating BVPs on curved
manifolds based on surface differential operators rather than on local
curvilinear coordinate systems and Christoffel symbols. The advantages
of the TDC-framework have already been demonstrated by the authors
in \cite{Fries_2020a}, with focus on generalizing the geometrical,
mechanical and numerical description and treatment of single membranes,
see also \cite{Schoellhammer_2019b,Schoellhammer_2019a} for shells.
The potential for simultaneous multi-membrane studies on all level
sets was not foreseen at that point, yet it may now be seen as another
important advantage of proposing mechanical models for curved structures
in the frame of the TDC. 

For the numerical analysis, the (undeformed) bulk domain is discretized
using higher-order elements of the same dimensionality as the embedding
space. The elements conform to the boundary of the bulk domain, however,
they are by no means aligned to the level sets which imply the actual
shape of the structures. This may be interpreted as a hybrid between
the classical FEM and fictitious or embedding domain methods. Here,
classical FEM refers to a single-membrane study employing a surface
mesh which conforms to the shape and boundary of the membrane; this
may also be called the Surface FEM. A fictitious domain approach for
a single-membrane study would be to use, e.g., the Cut FEM or Trace
FEM as presented in \cite{Fries_2020a} where a background mesh is
employed that neither conforms to the shape nor the boundary of the
membrane of interest, see also \cite{Burman_2014a,Burman_2010a,Burman_2012a}
for general references on the Cut FEM and \cite{Olshanskii_2017a,Reusken_2015a,Grande_2018a}
for the Trace FEM. Just as most fictitious domain methods, the Trace
FEM comes at the price of dealing with cut elements in terms of integration
and stabilization and increased efforts in considering boundary conditions.
The method presented here does also use a background mesh in the bulk
domain, however, without any need for cut elements and boundary conditions
are prescribed as usual in the classical FEM. In this sense, it resembles
most features of the classical FEM, yet it is emphasized that $d$-dimensional
background elements are used to simultaneously analyze a set of curved,
$\left(d-1\right)$-dimensional structures.

It seems natural to label the resulting approach, which herein is
applied in the context of structural membranes and ropes (in large
displacement theory), the \textquotedblleft Bulk Trace FEM\textquotedblright .
The most important fact that the domains of interest are imposed by
level sets justifies the use of the term \textquotedblleft Trace FEM\textquotedblright{}
and the addition of \textquotedblleft bulk\textquotedblright{} refers
to the fact that rather than solving a BVP on one level set, it is
solved simultaneously on all level sets in some bulk domain.

To the best of the author\textquoteright s knowledge, we are not aware
of similar approaches for the simultaneous modeling and analysis of
curved, lower-dimensional \emph{structures}, even less so in the context
of large displacements. However, for flow and transport problems,
models for the simultaneous solution on all level sets are found,
e.g., in \cite{Adalsteinsson_2003a,Dziuk_2008a,Dziuk_2008b} and for
some elliptic partial differential equations in \cite{Burger_2009a}.
Often, low-order meshes are employed in the bulk domains and the transport
takes place on closed level sets so that boundary conditions play
a minor role. Then, the emphasis is often on transient problems and
possibly moving bulk domains \cite{Adalsteinsson_2003a,Dziuk_2008b}.
For an overview of finite element methods for PDEs on surfaces in
general and an introduction on the simultaneous analysis on all level
sets within a bulk domain for some elliptic and parabolic PDEs, see
\cite{Dziuk_2013a}. Minimizing the bulk domain to a \emph{narrow
band} around some selected level set of interest leads to the narrow-band
method proposed in \cite{Deckelnick_2010a,Deckelnick_2014a}. Narrow-band
methods based on finite differences on a standard Cartesian grid for
transport problems are presented in \cite{Bertalmio_2001a,Greer_2006a,Greer_2006b}.
As mentioned before, when the solution on a \emph{single} level set
is sought, fictitious domain methods such as the Trace FEM and Cut
FEM may be used as presented, e.g., in \cite{Olshanskii_2014a,Olshanskii_2017a,Chernyshenko_2015a,Grande_2016a,Burman_2015a,Burman_2018a,Burman_2019a}.
In structural mechanics, Cut and Trace FEM approaches are published
in \cite{Cenanovic_2016a} for linear membranes, \cite{Schoellhammer_2020a}
for Reissner-Mindlin shells, \cite{Gfrerer_2018a} for Kirchhoff-Love
shells and in \cite{Fries_2020a} for non-linear ropes and membranes
in large displacement theory.

The paper is organized as follows: The general concept of using level
sets over bulk domains for the implicit and simultaneous geometry
definition of curved structures is outlined in Section \ref{X_Preliminaries},
including the definition of tangential differential operators on the
level sets. The mechanical setup in the context of large displacement
theory is outlined in Section \ref{X_MechanicalModel}, where reference
and spatial configurations are distinguished and equilibrium is enforced
in the latter. The complete BVP for the simultaneous mechanical modeling
on all level sets in the bulk domain is given, including boundary
conditions. When the bulk domain is optionally equipped with mechanical
properties (e.g., an elastic bulk material), the concept of continuously
embedded sub-structure models is outlined in Section \ref{XX_ElasticBulkDomains}.
The Bulk Trace FEM for the numerical approximation of the BVP is defined
in Section \ref{X_BulkTraceFEM}. Numerical results are presented
in Section \ref{X_NumericalResults} in two- and three-dimensional
bulk domains and demonstrate higher-order convergence rates provided
that the solutions of the BVP are sufficiently smooth. The paper ends
in Section \ref{X_Conclusions} with a summary and conclusions.

\section{Level sets in bulk domains: Geometric setup and differential operators\label{X_Preliminaries}}

Before dealing with the mechanical setup in terms of deformed and
undeformed configurations in finite strain theory, the situation is
first outlined generically with focus on geometric quantities as implied
by the level sets within some bulk domain and resulting differential
operators.

\subsection{Bulk domains and level-set functions\label{XX_BulkDomainsAndLevelSets}}

\begin{figure}
\centering

\subfigure[$\Omega$ and $\phi$]{\includegraphics[width=0.24\textwidth]{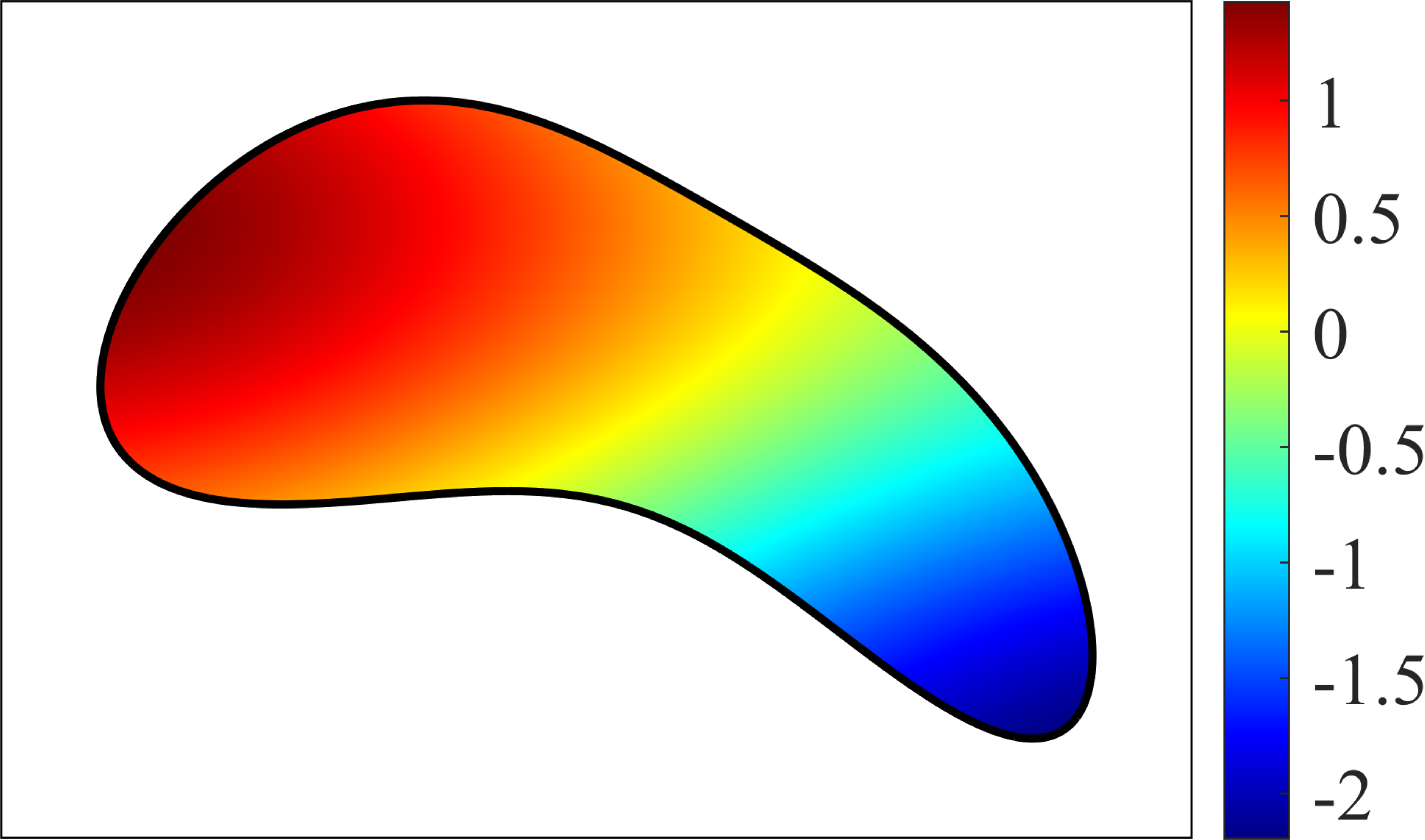}}\hfill\subfigure[$\Omega$ and level sets of $\phi$]{\includegraphics[width=0.24\textwidth]{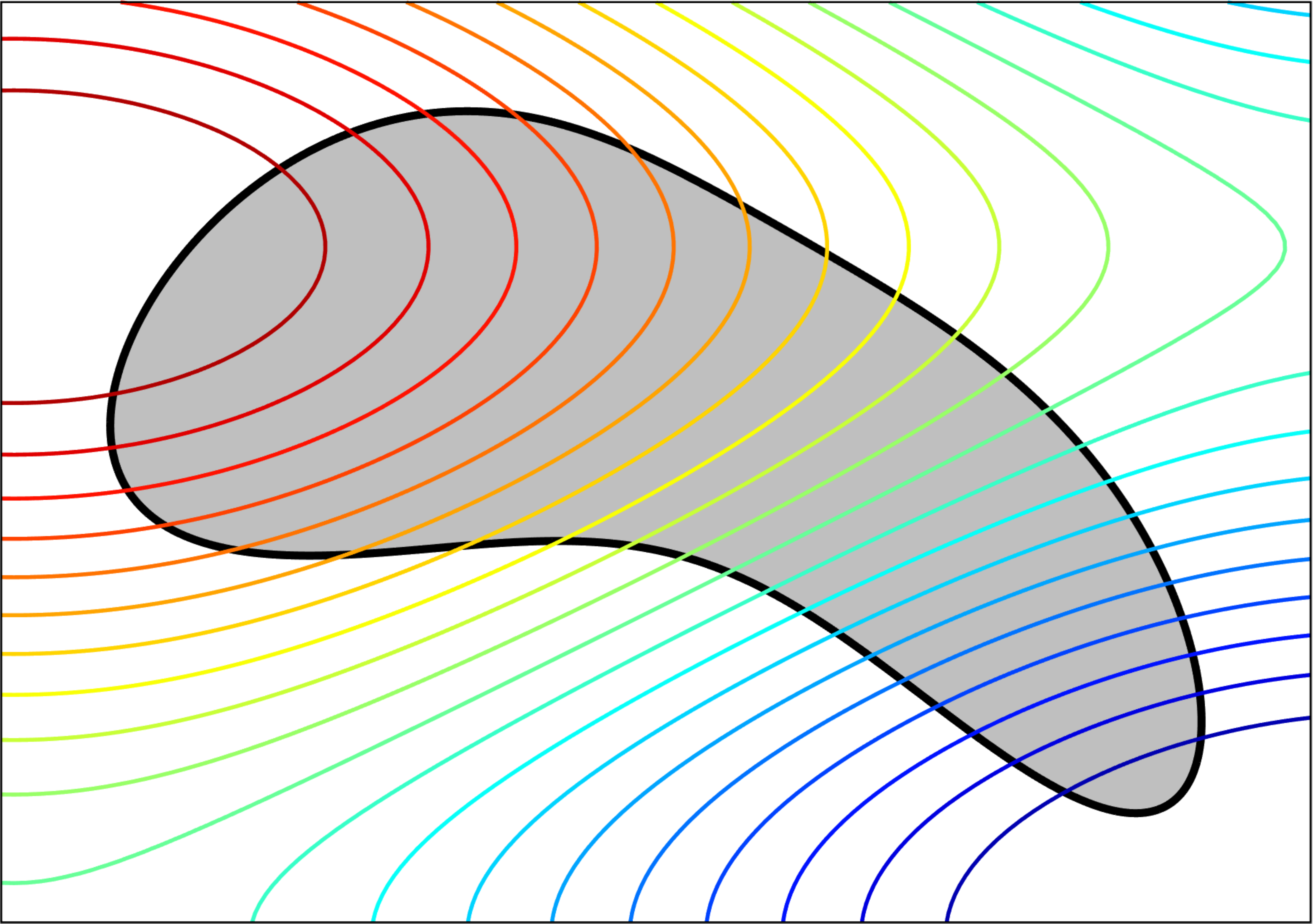}}\hfill\subfigure[$\Omega$ and level sets $\Gamma^{c}$]{\includegraphics[width=0.24\textwidth]{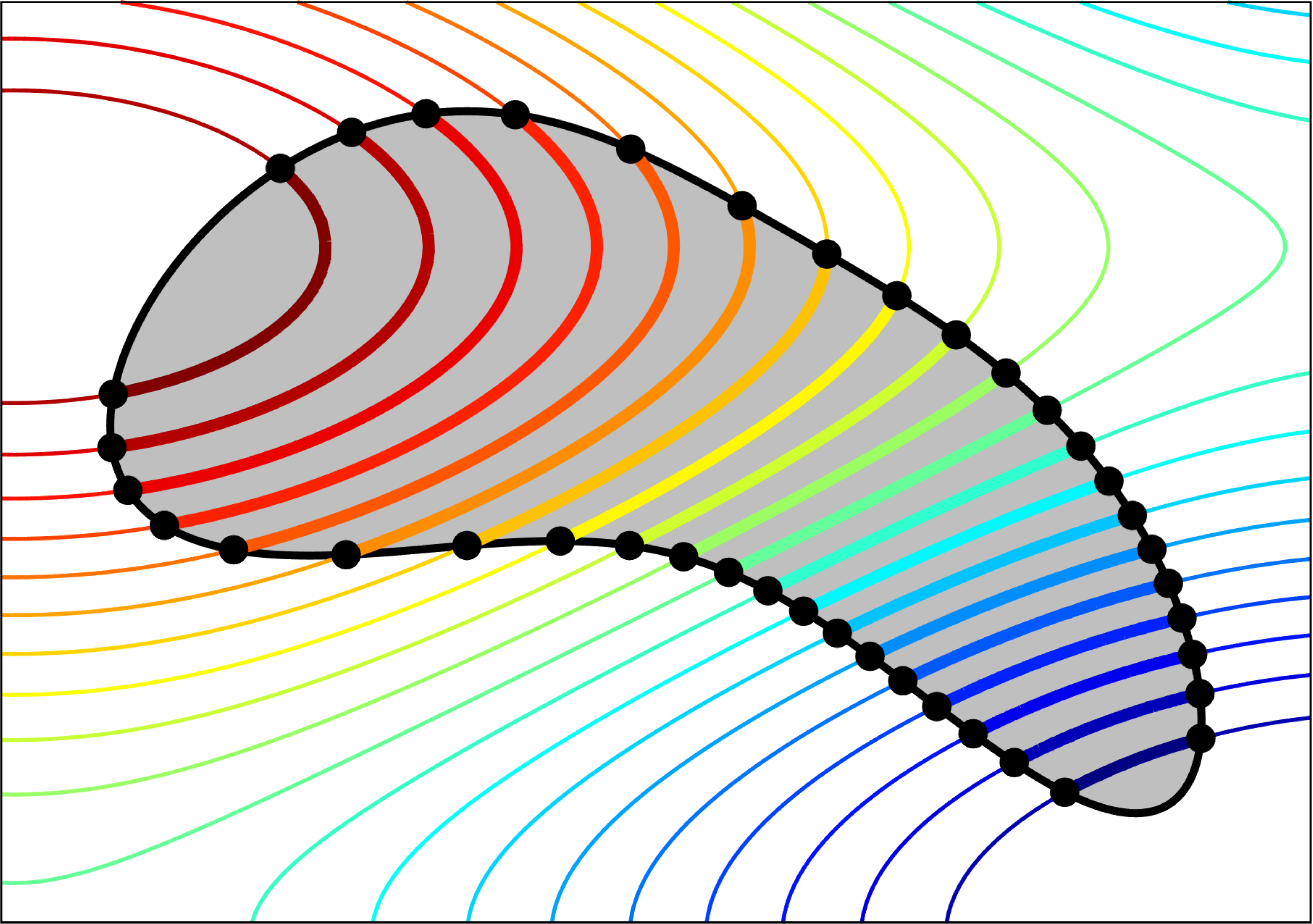}}\hfill\subfigure[level sets $\Gamma^{c}$]{\includegraphics[width=0.24\textwidth]{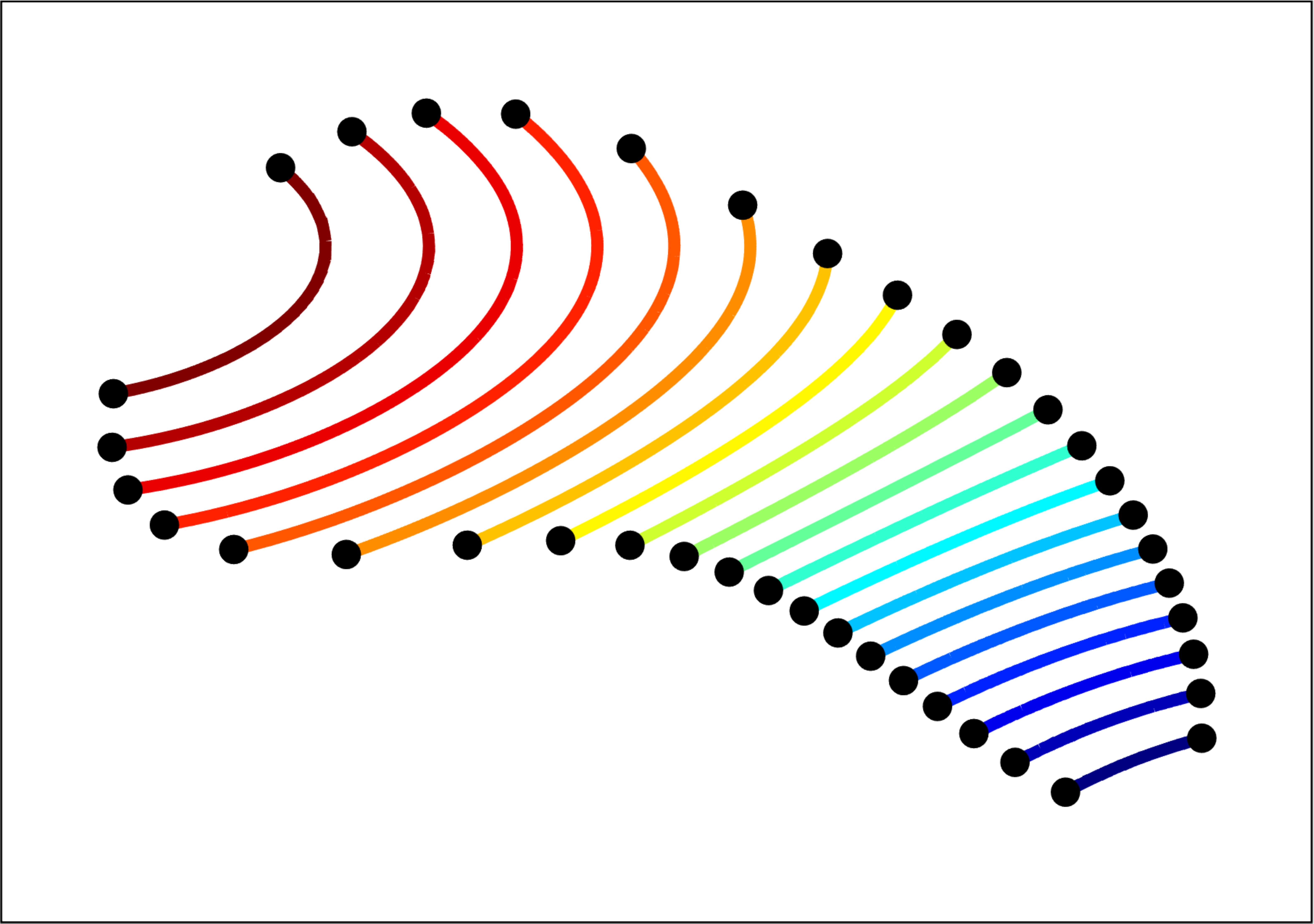}}

\subfigure[$\Omega$ and $\phi$]{\includegraphics[width=0.24\textwidth]{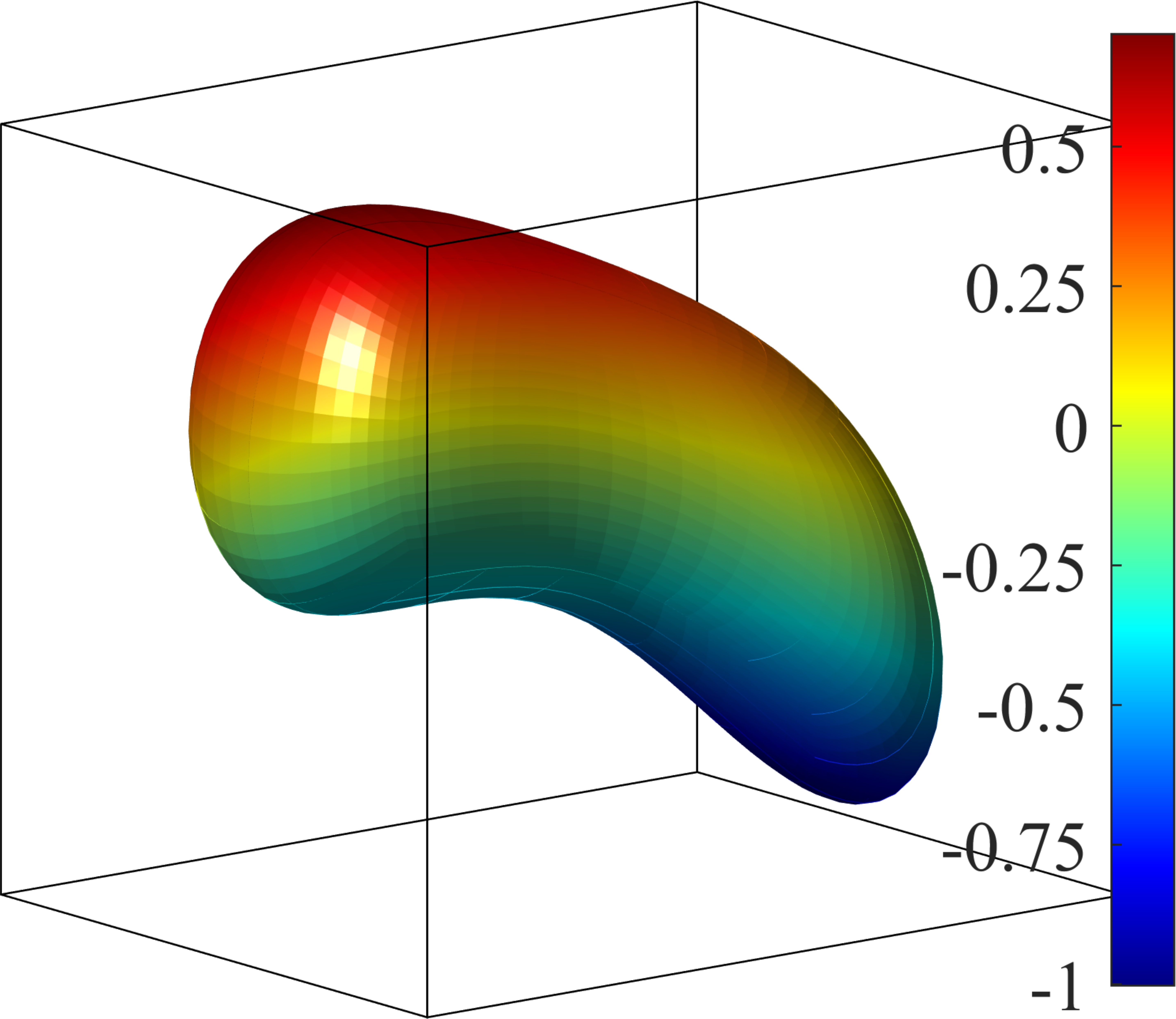}}\hfill\subfigure[$\Omega$ and level sets of $\phi$]{\includegraphics[width=0.24\textwidth]{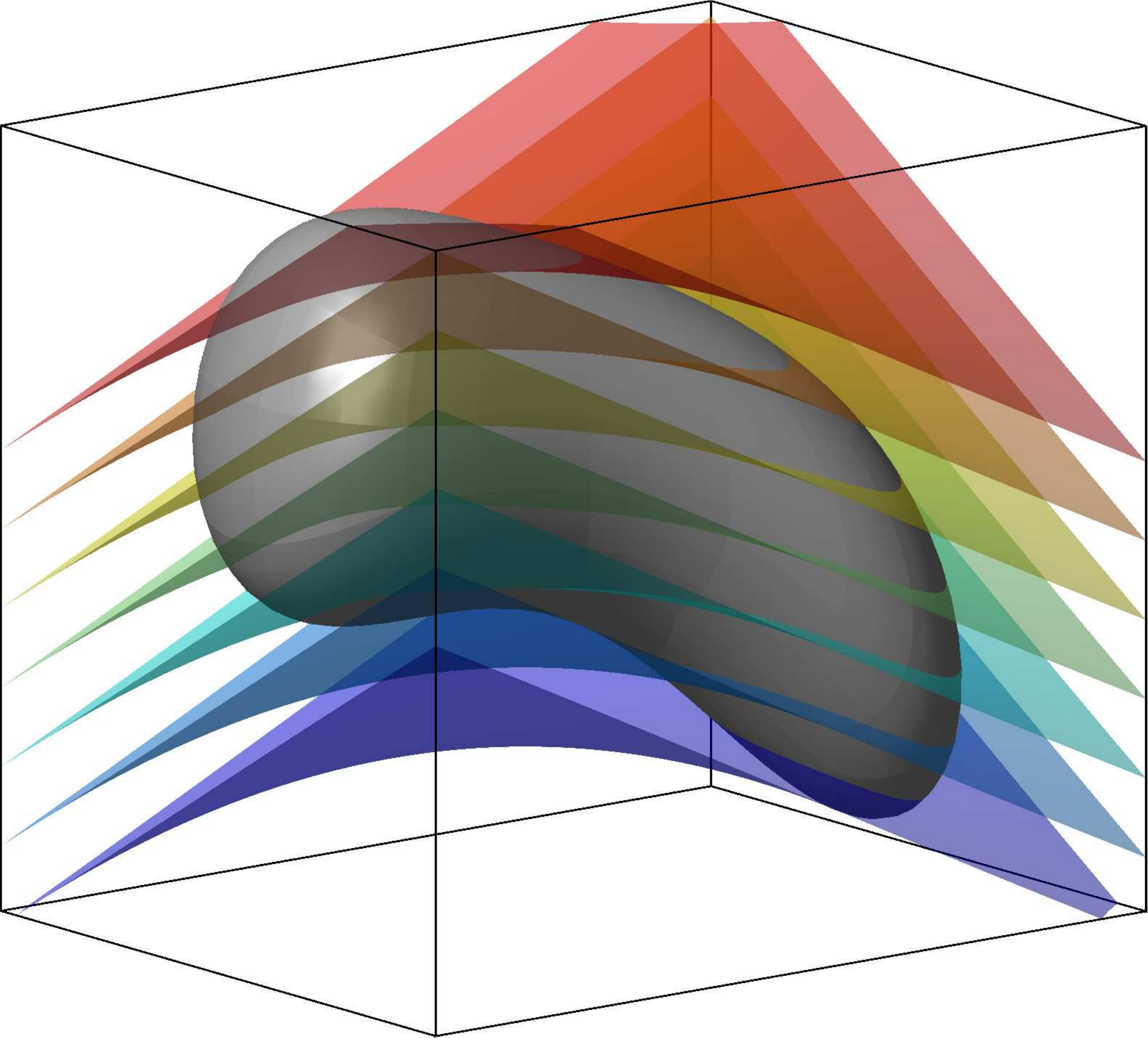}}\hfill\subfigure[$\Omega$ and level sets $\Gamma^{c}$]{\includegraphics[width=0.24\textwidth]{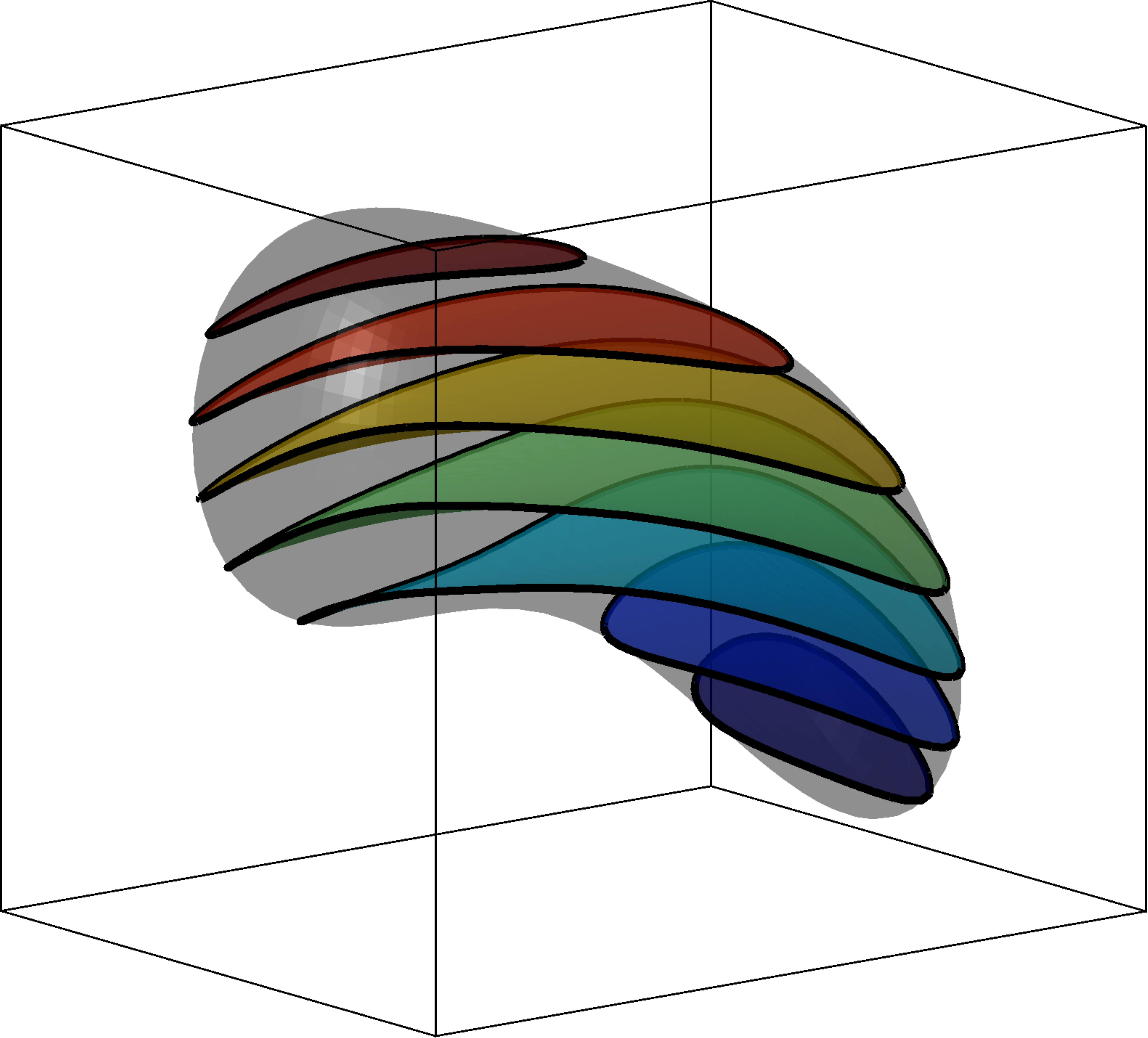}}\hfill\subfigure[level sets $\Gamma^{c}$]{\includegraphics[width=0.24\textwidth]{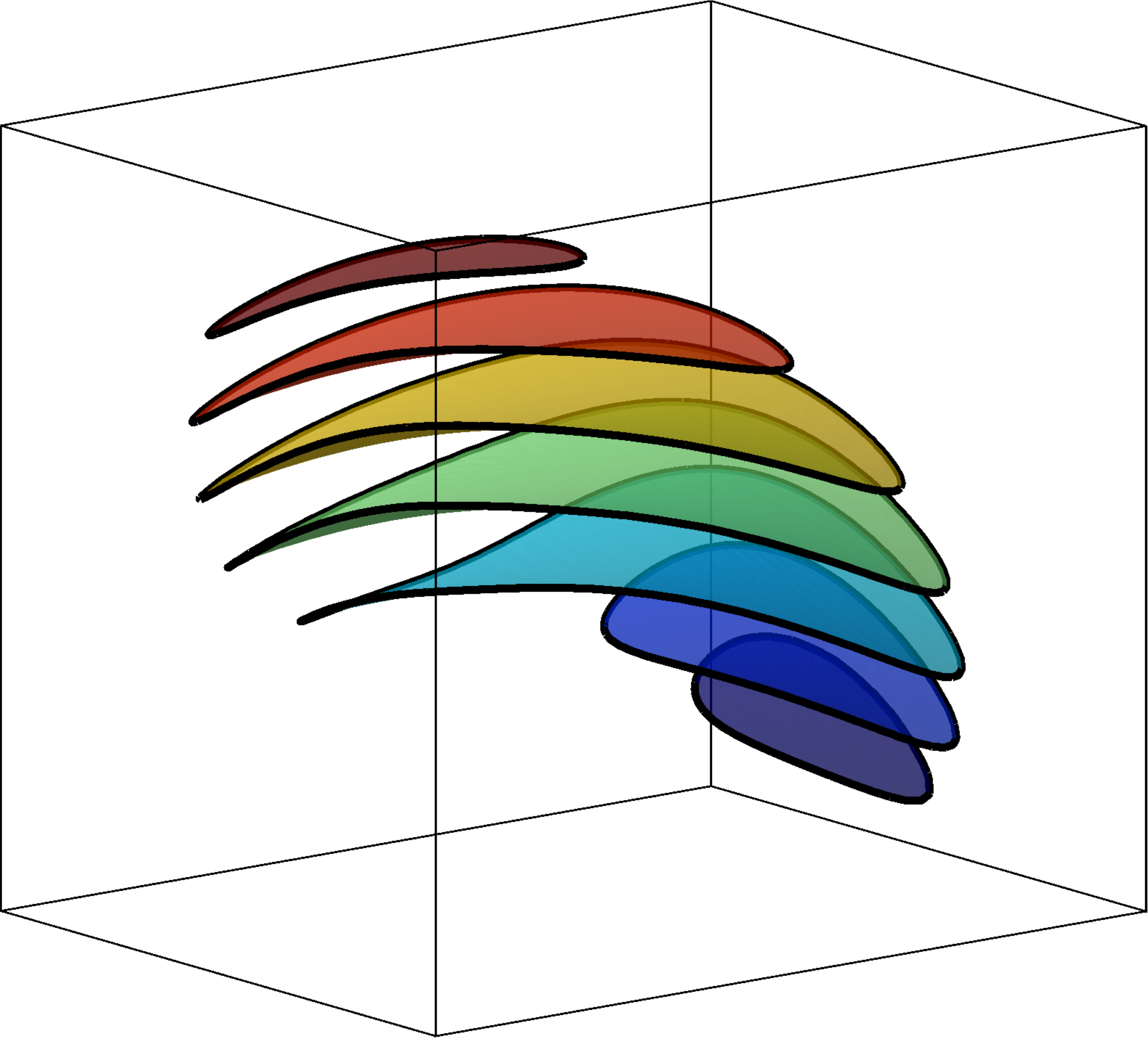}}

\caption{\label{fig:BulkDomainWithLevelSets}Some bulk domain $\Omega$ and
level-set function $\phi\left(\vek x\right)$ in two and three dimensions
and the implied level sets $\Gamma^{c}$.}
\end{figure}

Let there be a $d$-dimensional bulk domain $\Omega\subset\mathbb{R}^{d}$
and a level-set function $\phi\left(\vek x\right):\Omega\rightarrow\mathbb{R}$.
We call the smallest and largest value of $\phi$ in the bulk domain
$\phi^{\min}=\inf\phi\left(\vek x\right)$ and $\phi^{\max}=\sup\phi\left(\vek x\right)$.
Then, the individual level sets $\Gamma^{c}$ related to constant
level-set values $c\in\mathbb{R}$, 
\begin{equation}
\Gamma^{c}=\left\{ \vek x\in\Omega:\,\phi(\vek x)=c\in\mathbb{R}\right\} ,\,\phi^{\min}<c<\phi^{\max},\label{eq:LevelSets}
\end{equation}
are bounded, typically curved, $(d-1)$-dimensional manifolds (i.e.,
they have codimension $1$), see Fig.~\ref{fig:BulkDomainWithLevelSets}.
Consequently, the set of all bounded level sets $\Gamma^{c}$ coincides
with $\Omega$. The boundary of the bulk domain $\Omega$ is called
$\partial\Omega$. The boundary of some level set $\Gamma^{c}$ is
labeled $\partial\Gamma^{c}$ and is the intersection point or line
of the level set with constant value $c$ and the boundary of the
background domain $\partial\Omega$, see again Fig.~\ref{fig:BulkDomainWithLevelSets}.
As such, the set of all $\partial\Gamma^{c}$ coincides with $\partial\Omega$.

\begin{figure}
\centering

\subfigure[$\Omega$ and $\phi$]{\includegraphics[width=0.24\textwidth]{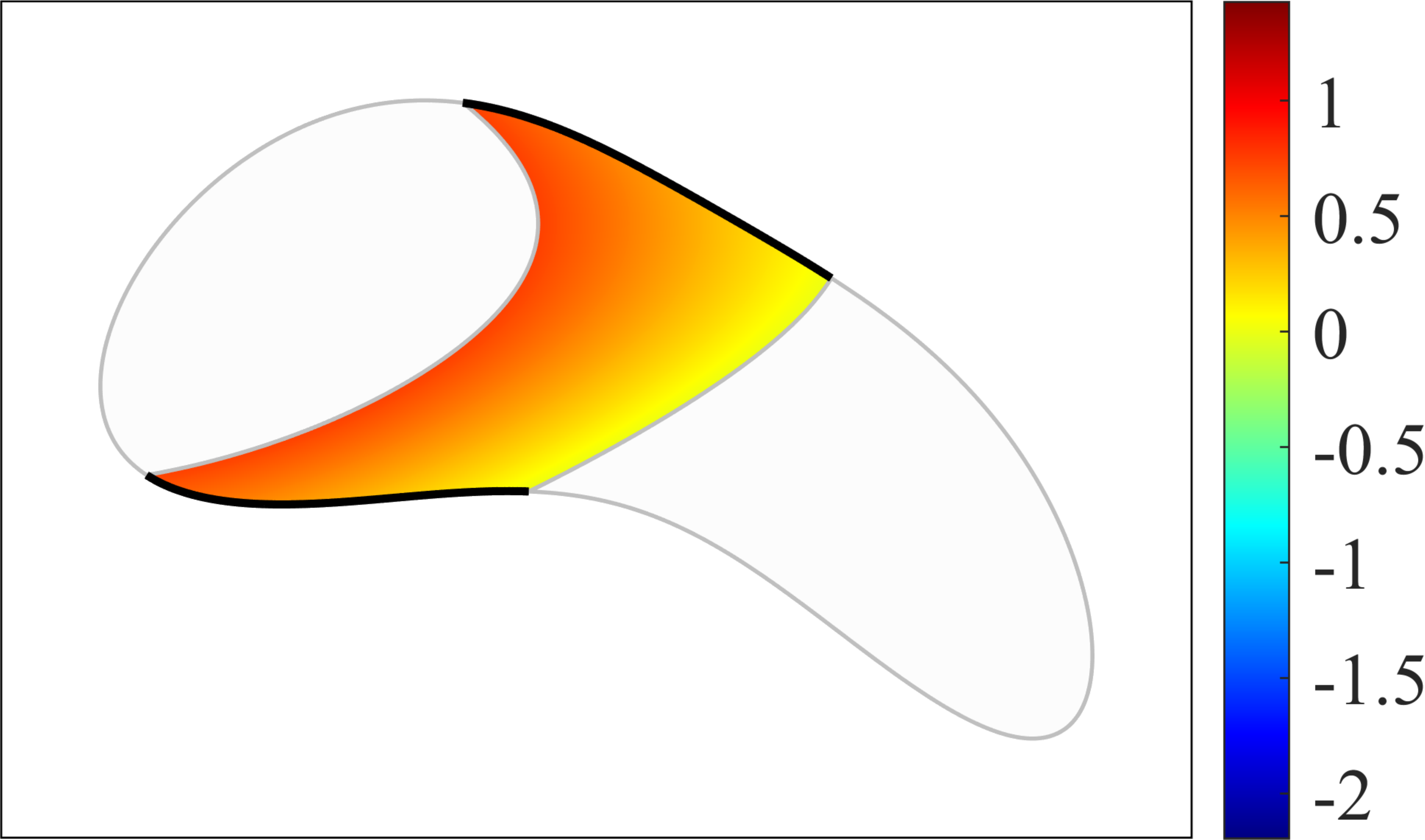}}\hfill\subfigure[$\Omega$ and level sets of $\phi$]{\includegraphics[width=0.24\textwidth]{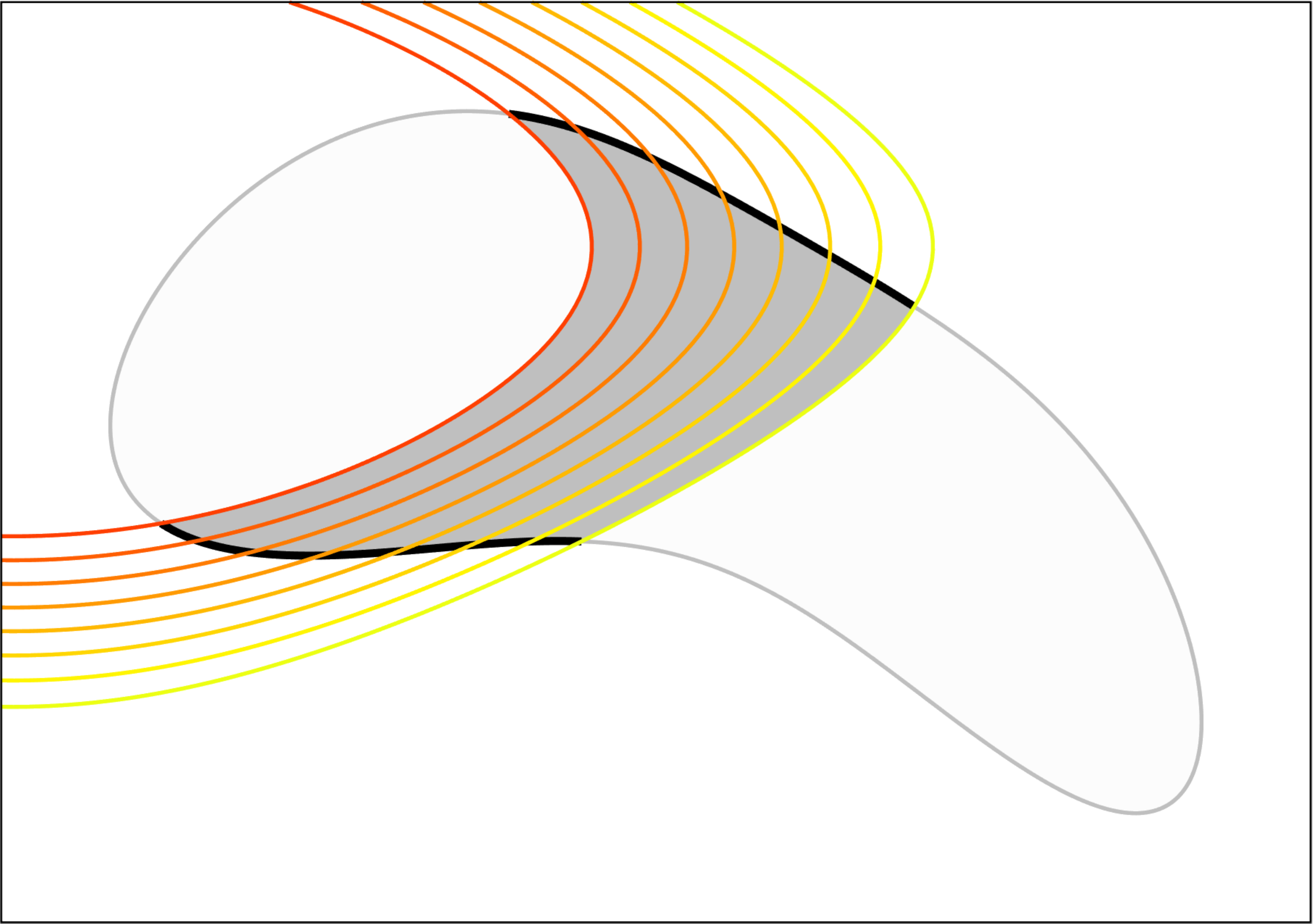}}\hfill\subfigure[$\Omega$ and level sets $\Gamma^{c}$]{\includegraphics[width=0.24\textwidth]{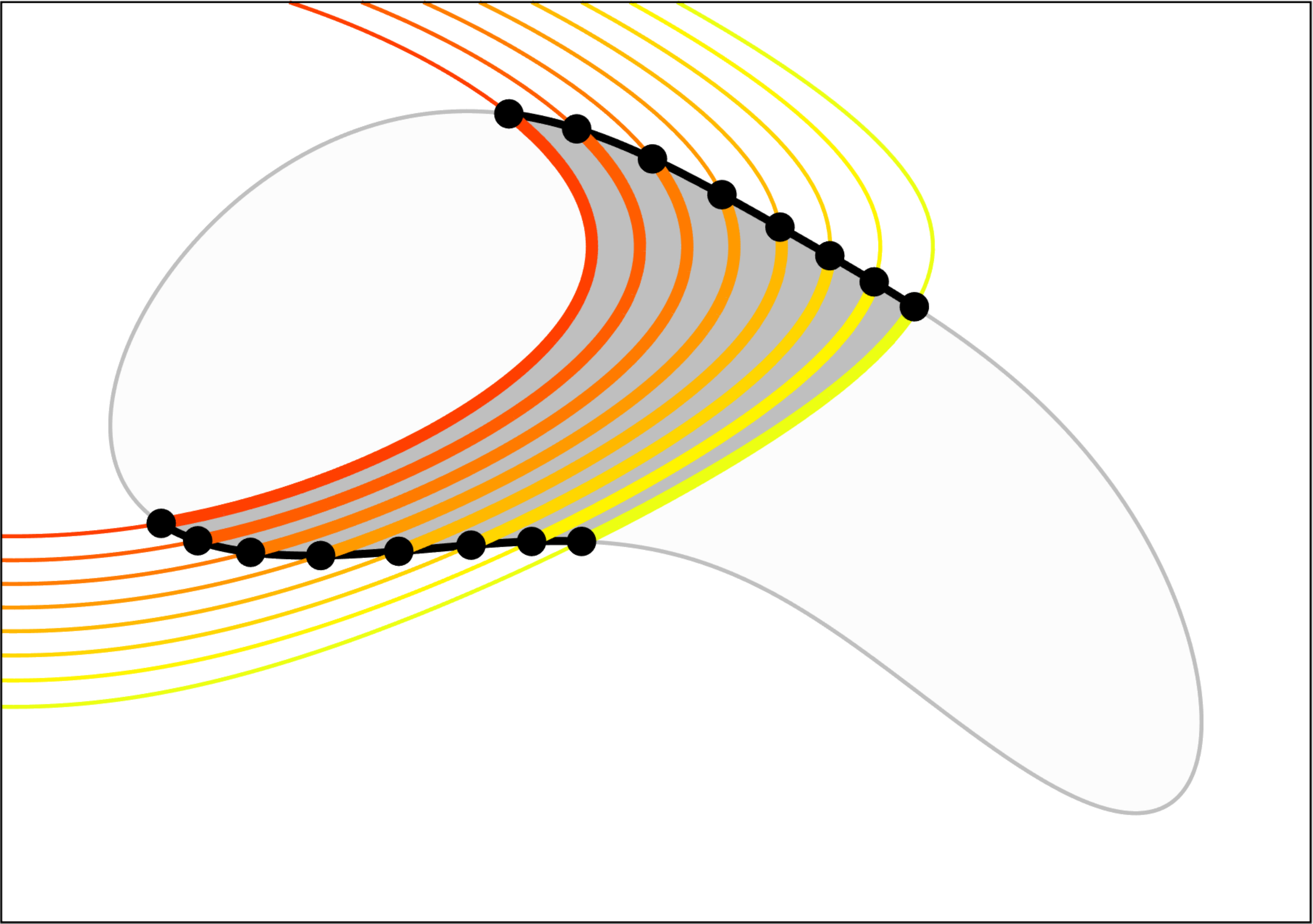}}\hfill\subfigure[level sets $\Gamma^{c}$]{\includegraphics[width=0.24\textwidth]{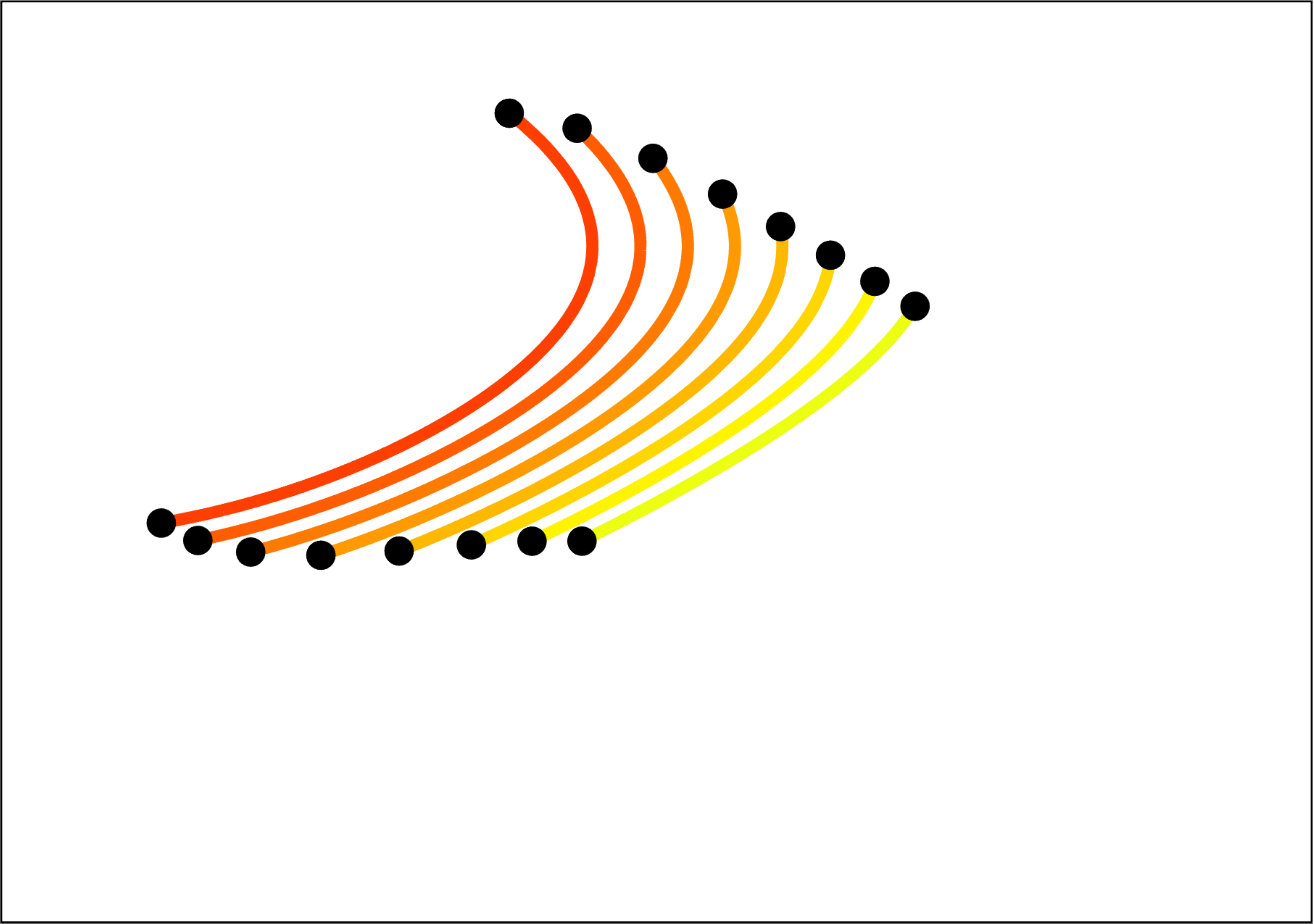}}

\subfigure[$\Omega$ and $\phi$]{\includegraphics[width=0.24\textwidth]{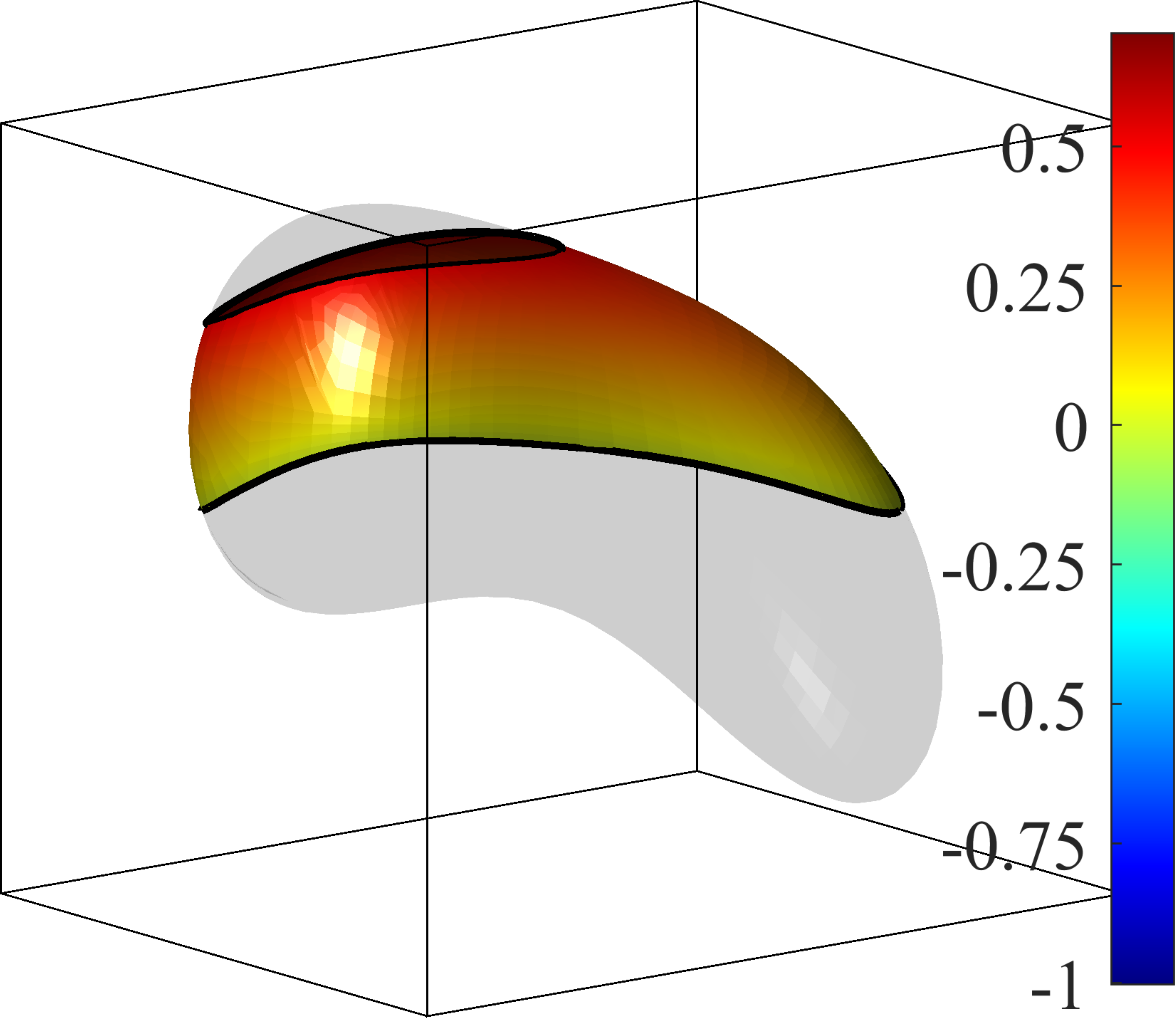}}\hfill\subfigure[$\Omega$ and level sets of $\phi$]{\includegraphics[width=0.24\textwidth]{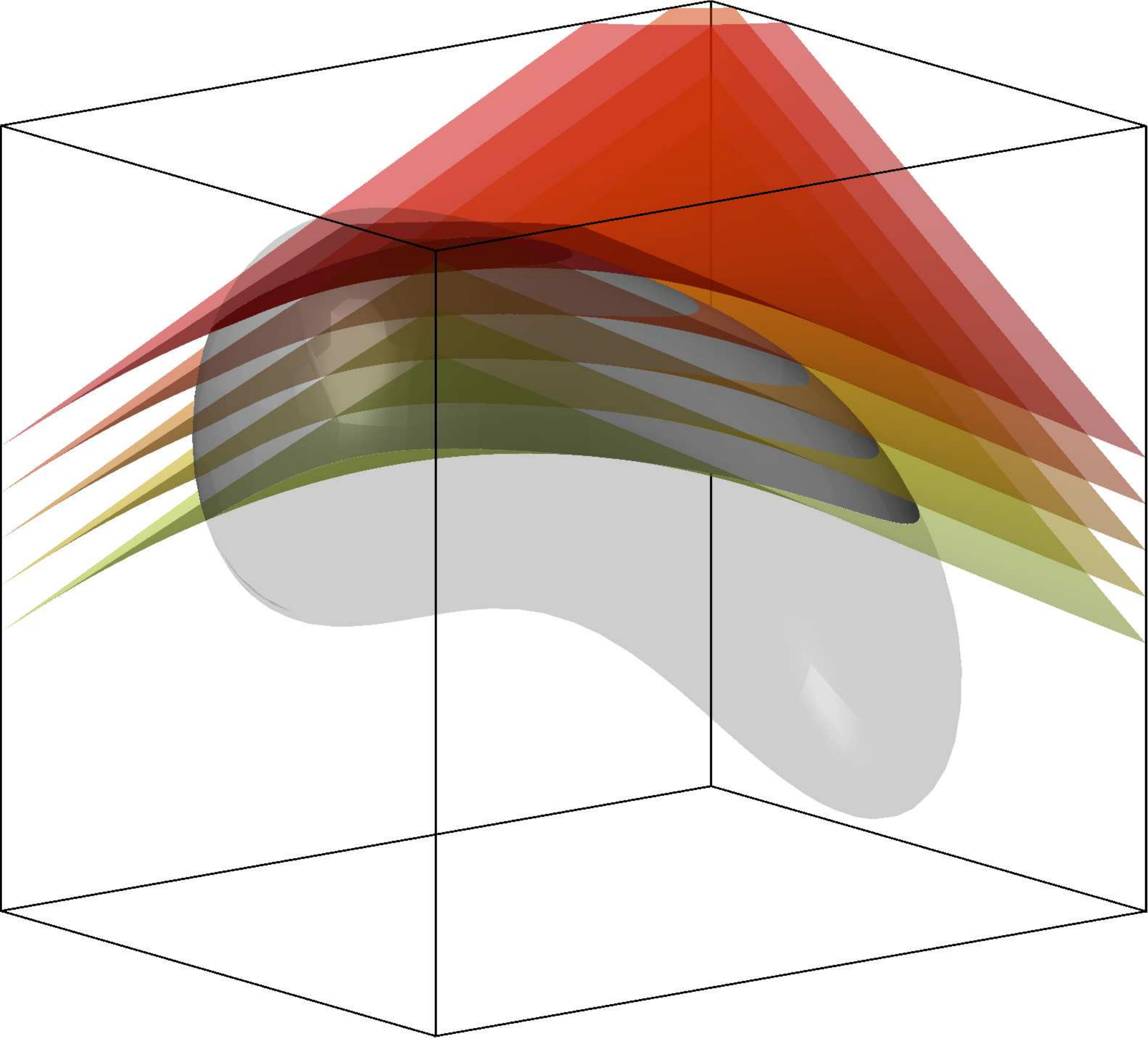}}\hfill\subfigure[$\Omega$ and level sets $\Gamma^{c}$]{\includegraphics[width=0.24\textwidth]{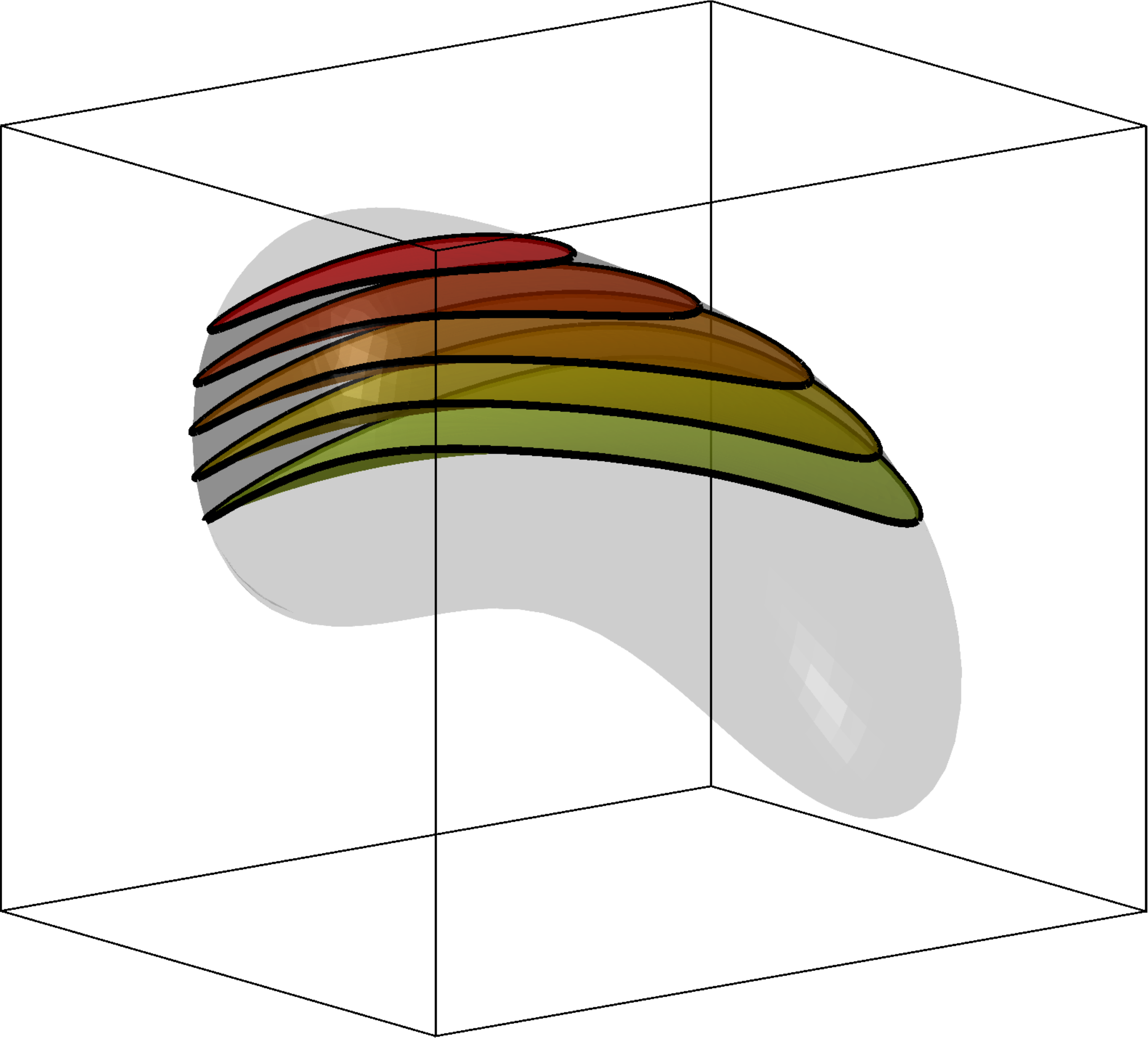}}\hfill\subfigure[level sets $\Gamma^{c}$]{\includegraphics[width=0.24\textwidth]{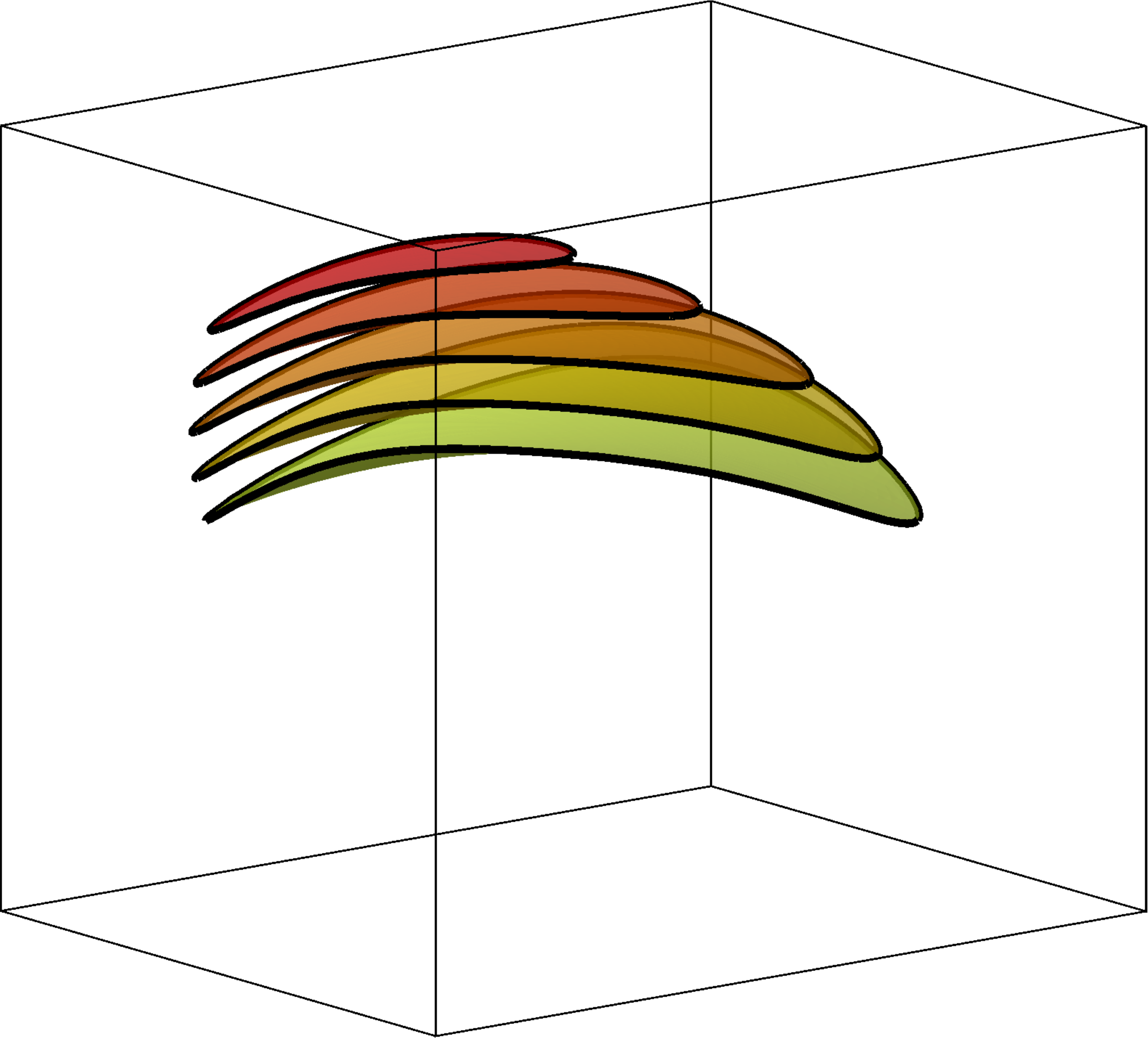}}

\caption{\label{fig:BulkDomainWithLevelSetInterval}The bulk domains $\Omega$
resulting from some larger $\Omega^{\text{sup}}$ in two and three
dimensions and a prescribed level-set interval $\left[\phi^{\min},\phi^{\max}\right]$.
Some selected level sets $\Gamma^{c}$ in this interval are also shown.}
\end{figure}

Instead of defining the bulk domain of interest $\Omega$ directly
with resulting $\phi^{\min}$ and $\phi^{\max}$, one may also prescribe
$\Omega$ \emph{indirectly} by first defining a superset of the bulk
domain, say $\Omega^{\sup}\subset\mathbb{R}^{d}$, and then limit
the bulk domain of interest $\Omega$ using specified values for $\phi^{\min}$
and $\phi^{\max}$,
\begin{equation}
\Omega=\left\{ \vek x\in\Omega^{\sup}:\phi^{\min}\leq\phi(\vek x)\leq\phi^{\max}\right\} ,\label{eq:BulkDomainWithPrescrMinMax}
\end{equation}
see Fig.~\ref{fig:BulkDomainWithLevelSetInterval}. In this case,
the level sets $\Gamma^{c}$ and their boundaries $\partial\Gamma^{c}$
are defined as before, however, we shall restrict the boundary of
the bulk domain $\partial\Omega$ to those parts of the boundary where
$\phi\left(\vek x\right)\neq\phi^{\min}$ and $\phi\left(\vek x\right)\neq\phi^{\max}$.

\subsection{Normal and conormal vectors\label{XX_NormalAndConormalVectors}}

The boundary of the bulk domain $\partial\Omega$ features a unit
normal vector (field) $\vek m\left(\vek x\right)$, $\vek x\in\partial\Omega$.
Depending on whether the bulk domain is defined through a parametrization
or implicitly, these normal vectors may be obtained through different
definitions. However, because the bulk domain is later discretized
by higher-order elements for the numerical analysis, see Section \ref{X_NumericalResults},
the generation of $\vek m$ on element boundaries is a standard task
in the FEM and not further outlined here.

\begin{figure}
\centering

\subfigure[vectors $\vek{n}$, $\vek{m}$, $\vek{q}$]{\includegraphics[width=0.35\textwidth]{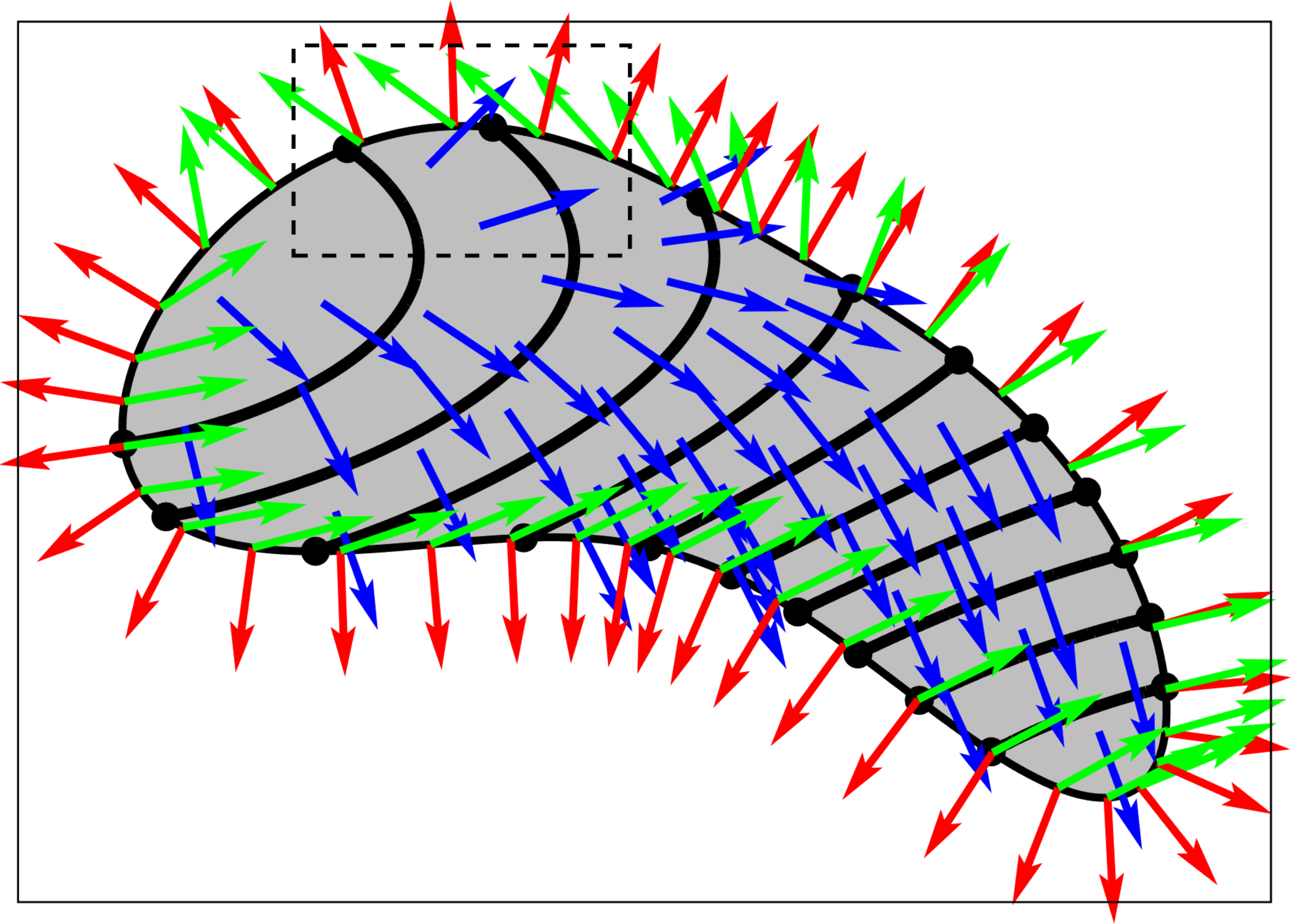}}\qquad\subfigure[zoom]{\includegraphics[width=0.35\textwidth]{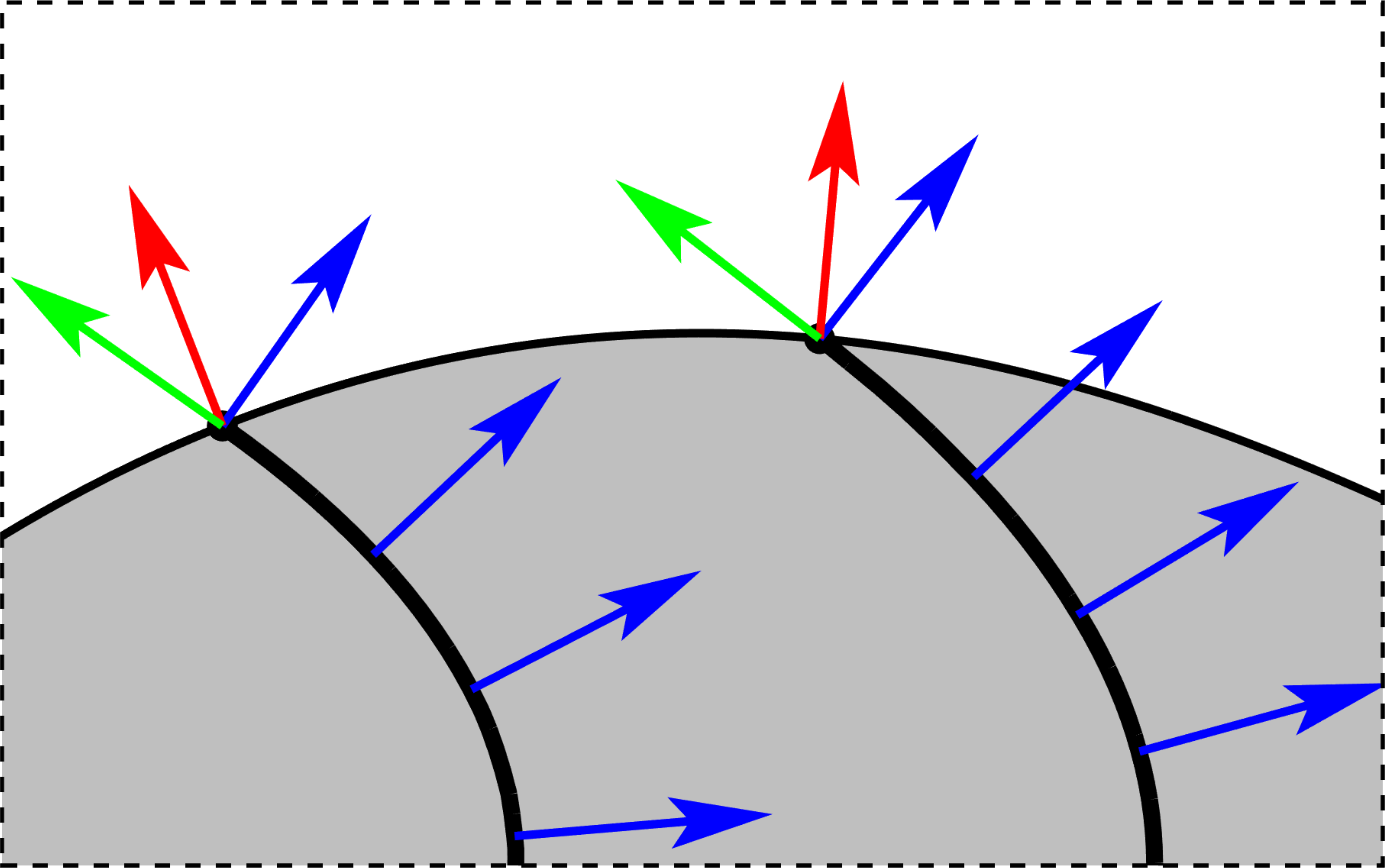}}

\subfigure[vectors $\vek{n}$, $\vek{m}$, $\vek{t}$, $\vek{q}$]{\includegraphics[width=0.35\textwidth]{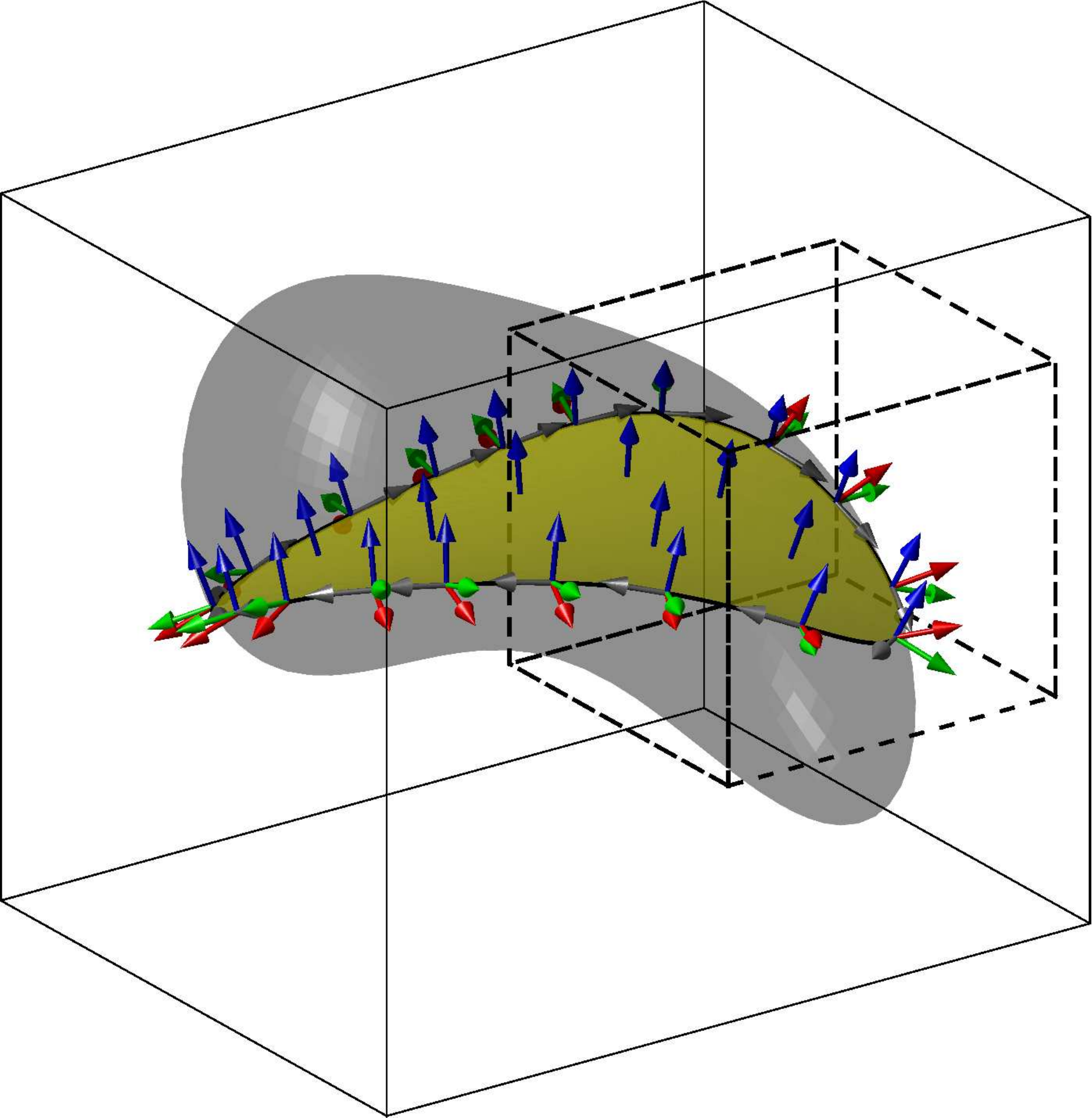}}\qquad\subfigure[zoom]{\includegraphics[width=0.35\textwidth]{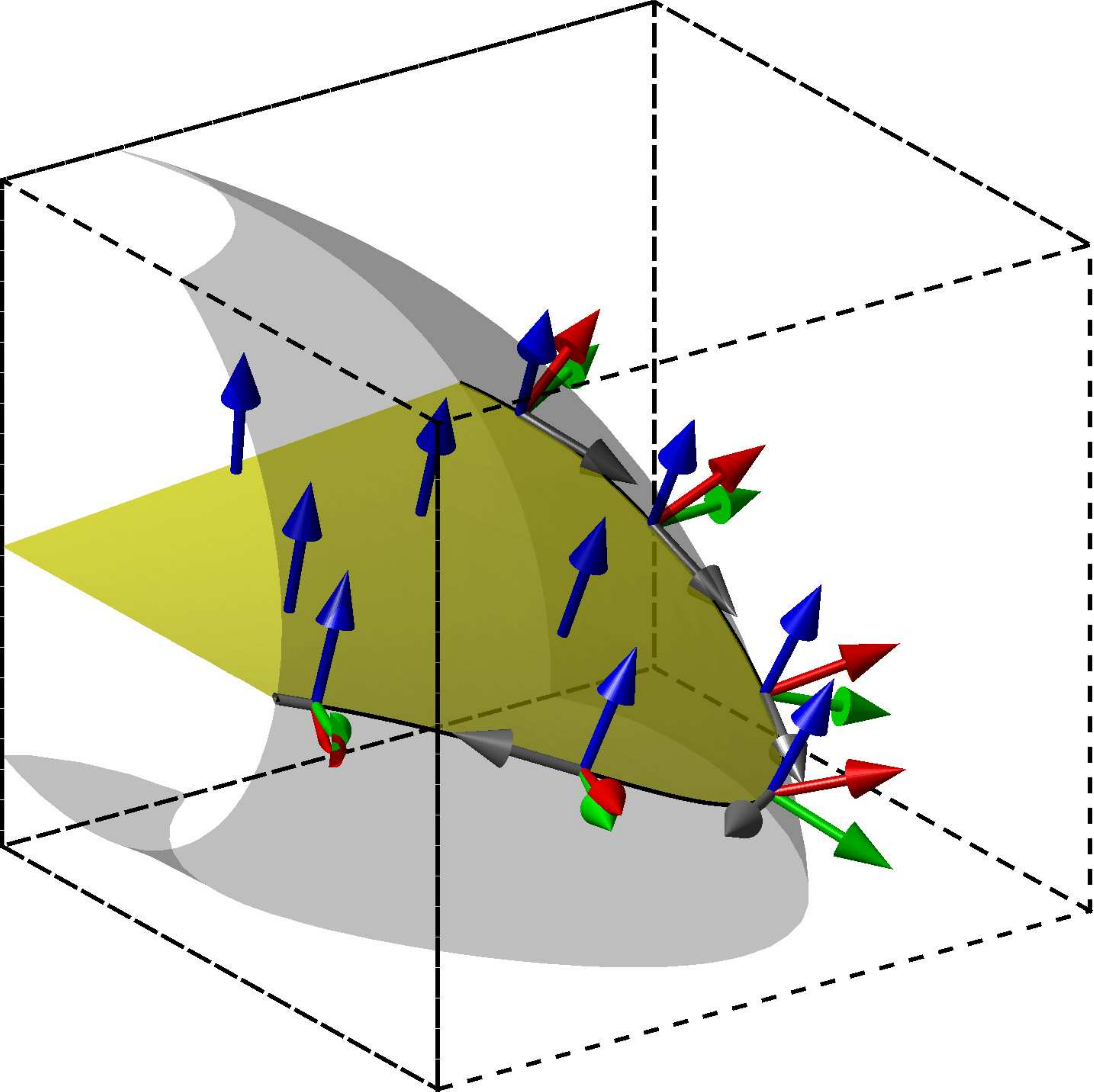}}

\caption{\label{fig:NormalVectors}Vector fields in the domain $\Omega$ and
on the boundary $\partial\Omega$ in two and three dimensions, the
right figures show zooms of the left ones. Normal vectors $\vek n$
with respect to the level sets $\Gamma^{c}$ in $\Omega$ are shown
in blue. Normal vectors $\vek m$ with respect to $\partial\Omega$
are red, tangential vectors $\vek t$ are gray and conormal vectors
$\vek q$ are green.}
\end{figure}

The unit normal vector (field) $\vek n\left(\vek x\right)$ on the
level sets $\Gamma^{c}$ in the whole bulk domain $\Omega$ is obtained
by the gradient of the level-set function,
\begin{equation}
\vek n\left(\vek x\right)=\frac{\vek n^{\star}}{\left\Vert \vek n^{\star}\right\Vert }\quad\textrm{with}\quad\vek n^{\star}=\nabla\phi\left(\vek x\right),\;\vek x\in\Omega.\label{eq:NormalVector_n}
\end{equation}
One may then also construct the projector field $\mat p\left(\vek x\right)\in\mathbb{R}^{d\times d}$,
$\vek x\in\Omega$, 
\begin{equation}
\mat p\left(\vek x\right)=\mat I-\vek n\otimes\vek n.\label{eq:Projector}
\end{equation}
The unit \emph{conormal} vectors $\vek q\left(\vek x\right)$ on $\partial\Omega$
are in the tangent plane of the corresponding level sets at $\partial\Gamma^{c}$
and yet normal to $\vek n\left(\vek x\right)$ from above. In case
of two-dimensional bulk domains, $d=2$, see Figs.~\ref{fig:NormalVectors}(a)
and (b), these vectors are defined without the normal vectors of the
bulk domains $\vek m$ and are simply obtained from $\vek n=[n_{x},\;n_{y}]^{\mathrm{T}}$
as 
\[
\vek q\left(\vek x\right)=\left[\begin{array}{c}
q_{x}\\
q_{y}
\end{array}\right]=\left[\begin{array}{c}
-n_{y}\\
n_{x}
\end{array}\right].
\]
For three-dimensional bulk domains, $d=3$, see Figs.~\ref{fig:NormalVectors}(c)
and (d), one needs to first generate tangent vector fields $\vek t\left(\vek x\right)$
on $\partial\Omega$ using cross products of the normal vector fields
from above, $\vek t=\vek m\times\vek n$, and then 
\[
\vek q=\frac{\vek q^{\star}}{\left\Vert \vek q^{\star}\right\Vert }\,\,\textrm{with}\,\,\vek q^{\star}=\vek n\times\vek t.
\]
These conormal vector fields will later play an important role in
the weak formulation of the boundary value problem and the definition
of boundary conditions. It is noted that for the indirect definition
of the bulk domain $\Omega$ according to Eq.~(\ref{eq:BulkDomainWithPrescrMinMax}),
conormal vectors may only be defined on those parts of the whole boundary
of $\Omega$ which do not coincide with level sets, because there
$\vek m=\vek n$ and tangential vectors $\vek t$ may not be computed
through cross products. This is why in Section \ref{XX_BulkDomainsAndLevelSets},
those parts of the boundary of the bulk domain were excluded from
$\partial\Omega$. In other words, $\partial\Omega$ is the boundary
of the bulk domain $\Omega$ where $\vek m\neq\vek n$ and, hence,
conormal vectors $\vek q$ exist, see, e.g., the black line in Fig.~\ref{fig:BulkDomainWithLevelSetInterval}(c).

\subsection{Differential operators with respect to level sets\label{XX_DifferentialOperators}}

For the definition of BVPs on the level sets, it is important to distinguish
(classical) differential operators acting in the bulk space (such
as the gradient $\nabla$ in Eq.~(\ref{eq:NormalVector_n})) from
differential operators acting on the level sets which may be called
\emph{tangential} or \emph{surface} operators (although, for two-dimensional
bulk domains, the level sets are rather curved \emph{lines}). The
\emph{surface} \emph{gradient} of a scalar function $f\!\left(\vek x\right):\Omega\to\mathbb{R}$
results as \cite{Delfour_2011a,Dziuk_2010a,Jankuhn_2017a,Fries_2019b}
\begin{equation}
\nabla^{\Gamma}f=\mat p\cdot\nabla f,\label{eq:SurfGradScalarImplicit}
\end{equation}
where $\nabla f$ is the classical gradient in the $d$-dimensional
space. It is noted that $\nabla^{\Gamma}\phi=\vek0$. The situation
is analogous for each component of a vector function $\vek u\left(\vek x\right):\Omega\to\mathbb{R}^{d}$,
so that one obtains for the \emph{directional} surface gradient
\begin{eqnarray}
\nabla^{\Gamma,\mathrm{dir}}\vek u & = & \nabla\vek u\cdot\mat p,\label{eq:SurfGradVectorImplicit}\\
\text{for }\vek u=\left[\!\!\begin{array}{c}
u\\
v\\
w
\end{array}\!\!\right]\!\in\mathbb{R}^{3}:\left[\begin{array}{ccc}
\partial_{x}^{\Gamma}u & \partial_{y}^{\Gamma}u & \partial_{z}^{\Gamma}u\\
\partial_{x}^{\Gamma}v & \partial_{y}^{\Gamma}v & \partial_{z}^{\Gamma}v\\
\partial_{x}^{\Gamma}w & \partial_{y}^{\Gamma}w & \partial_{z}^{\Gamma}w
\end{array}\right]\! & = & \!\left[\begin{array}{ccc}
\partial_{x}u & \partial_{y}u & \partial_{z}u\\
\partial_{x}v & \partial_{y}v & \partial_{z}v\\
\partial_{x}w & \partial_{y}w & \partial_{z}w
\end{array}\right]\!\cdot\!\left[\begin{array}{ccc}
p_{11} & p_{12} & p_{13}\\
p_{12} & p_{22} & p_{23}\\
p_{13} & p_{23} & p_{33}
\end{array}\right].\nonumber 
\end{eqnarray}
The \emph{covariant }surface gradient of a vector function $\vek u\left(\vek x\right)$
is based on the projection of the directional one onto the tangent
space, 
\begin{equation}
\nabla^{\Gamma,\mathrm{cov}}\vek u\;=\;\mat p\cdot\nabla^{\Gamma,\mathrm{dir}}\vek u\;=\;\mat p\cdot\nabla\vek u\cdot\mat p.\label{eq:CovariantSurfaceGradient}
\end{equation}
Concerning the \emph{surface divergence }of vector functions $\vek u\left(\vek x\right)$
and tensor functions $\mat A\negmedspace\left(\vek x\right):\Omega\to\mathbb{R}^{d\times d}$,
there holds 
\begin{eqnarray}
\mathrm{div}_{\Gamma}\,\vek u\left(\vek x\right) & = & \mathrm{tr}\left(\nabla^{\Gamma,\mathrm{dir}}\vek u\right)=\mathrm{tr}\left(\nabla^{\Gamma,\mathrm{cov}}\vek u\right)\eqqcolon\nabla^{\Gamma}\cdot\vek u,\label{eq:DivergenceVector}\\
\mathrm{div}_{\Gamma}\,\mat A\negmedspace\left(\vek x\right) & = & \left[\begin{array}{c}
\mathrm{div}_{\Gamma}\left(A_{11},A_{12},A_{13}\right)\\
\mathrm{div}_{\Gamma}\left(A_{21},A_{22},A_{23}\right)\\
\mathrm{div}_{\Gamma}\left(A_{31},A_{32},A_{33}\right)
\end{array}\right]\eqqcolon\nabla^{\Gamma}\cdot\mat A.\label{eq:DivergenceTensor}
\end{eqnarray}

\subsection{Integral theorems on level sets\label{XX_IntegralTheorems}}

\begin{figure}
\centering

\subfigure[21 Gauss points]{\includegraphics[width=0.24\textwidth]{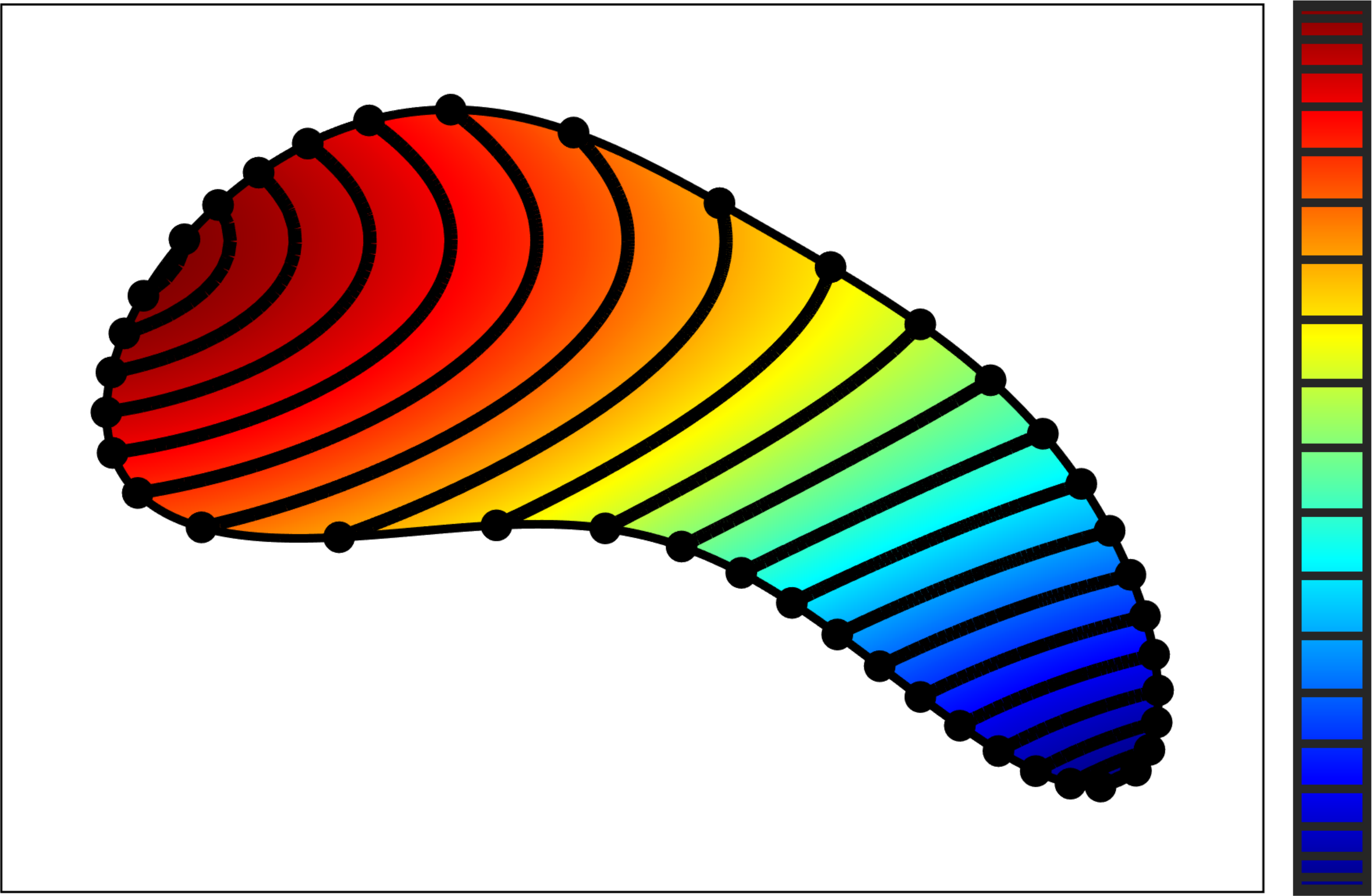}}\hfill\subfigure[11 Gauss points]{\includegraphics[width=0.24\textwidth]{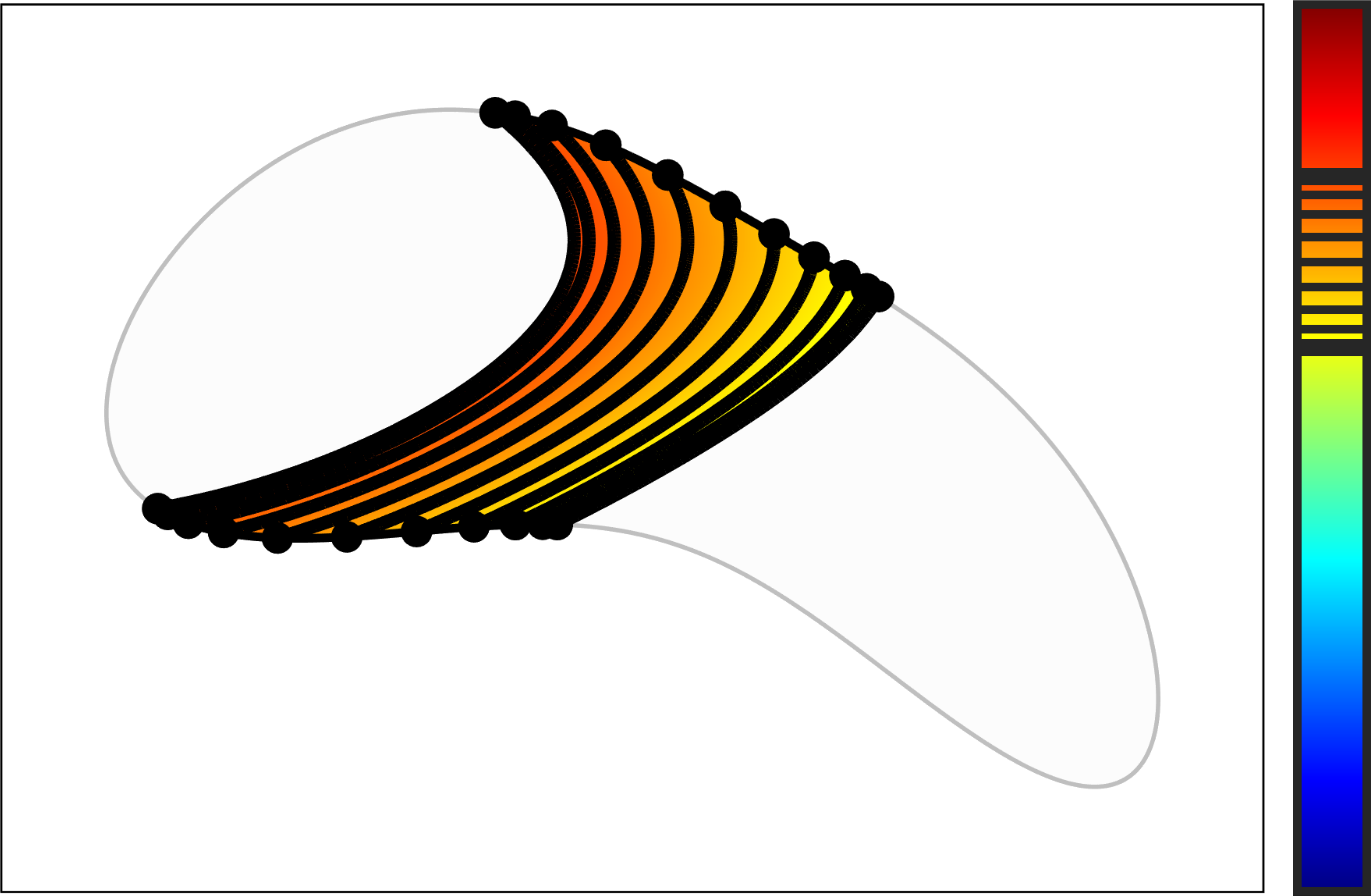}}\hfill\subfigure[21 Gauss points]{\includegraphics[width=0.24\textwidth]{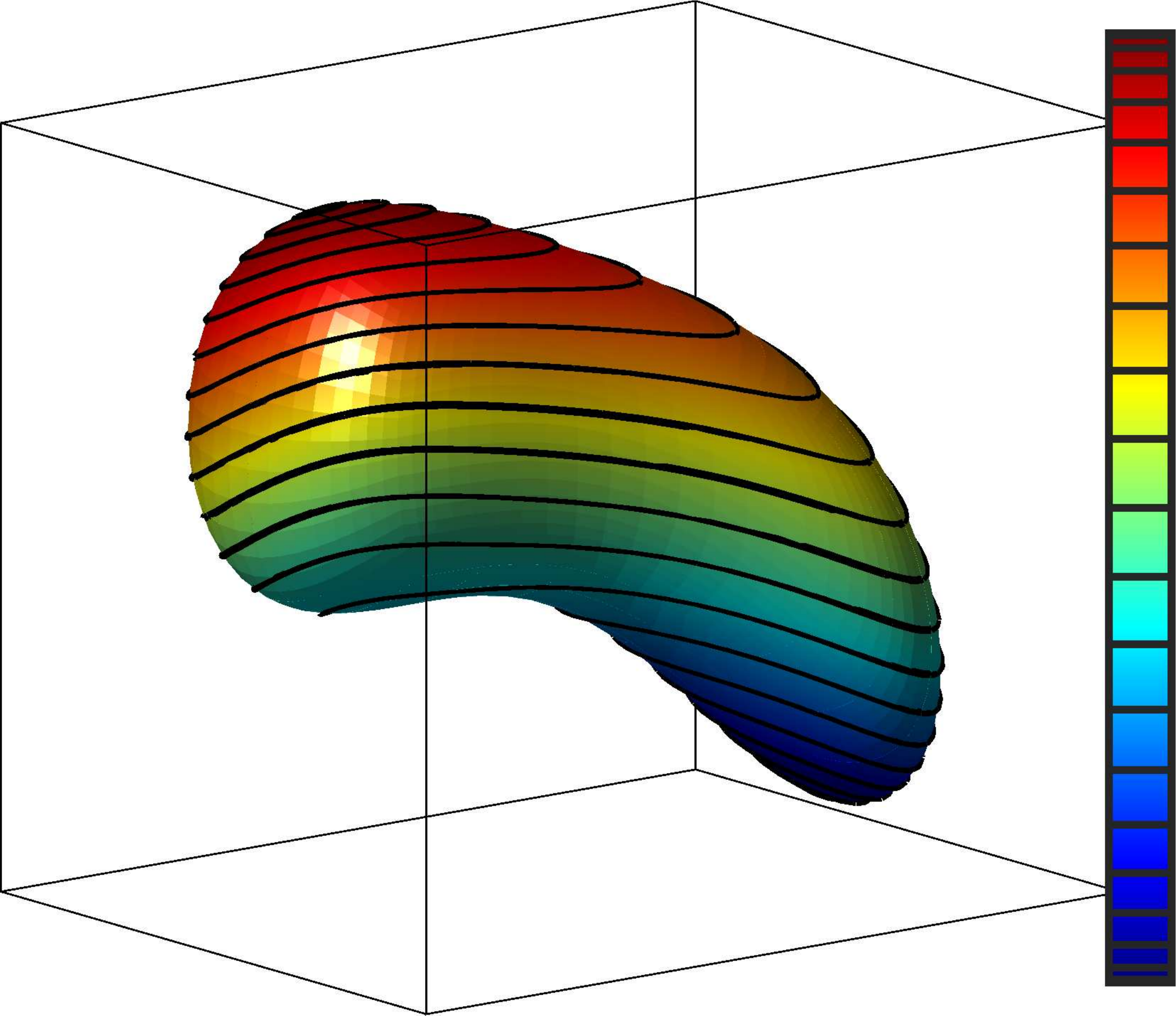}}\hfill\subfigure[11 Gauss points]{\includegraphics[width=0.24\textwidth]{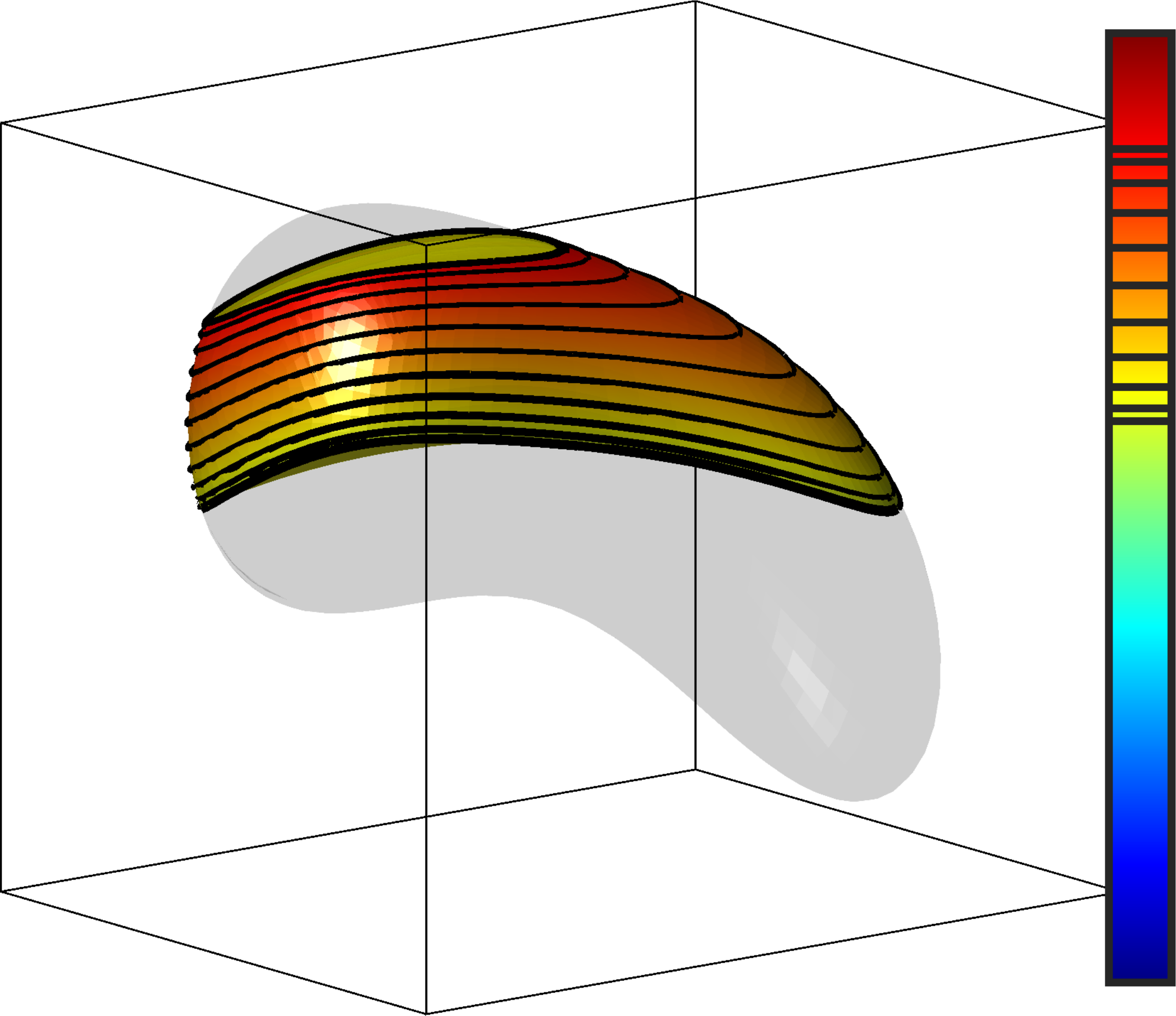}}

\caption{\label{fig:GaussLevelSets}Level sets in the interval $\left[\phi^{\min},\phi^{\max}\right]$
with respect to Gauss integration points, e.g., to numerically confirm
Eq.~(\ref{eq:NumericalIntOverManifolds}). In (a) and (c), $\phi^{\min}=\inf\phi\left(\vek x\right)$
and $\phi^{\max}=\sup\phi\left(\vek x\right)$ with $\vek x\in\Omega$,
in (b) and (d) $\phi^{\min}$ and $\phi^{\max}$ are user-defined
values, see also Section \ref{XX_BulkDomainsAndLevelSets}.}
\end{figure}

The first important integral theorem is given by the \emph{co-area
formula} \cite{Dziuk_2008a,Federer_1969a,Morgan_1988a},
\begin{equation}
\int_{\phi^{\min}}^{\phi^{\max}}\int_{\Gamma^{c}}f\left(\vek x\right)\;\ensuremath{\mathrm{d\ensuremath{\Gamma}}}\;\ensuremath{\mathrm{d}c}=\int_{\Omega}f\left(\vek x\right)\cdot\left\Vert \nabla\phi\right\Vert \;\ensuremath{\mathrm{d\ensuremath{\Omega}}}.\label{eq:CoareaFormulaDomain}
\end{equation}
When integrating over the boundary $\partial\Gamma^{c}$ in the level-set
interval $\left(\phi^{\min},\;\phi^{\max}\right)$, we find
\begin{equation}
\int_{\phi^{\min}}^{\phi^{\max}}\int_{\partial\Gamma^{c}}f\left(\vek x\right)\;\ensuremath{\mathrm{d\ensuremath{\partial\Gamma}}}\;\ensuremath{\mathrm{d}c}=\int_{\partial\Omega}f\left(\vek x\right)\cdot\left(\vek q\cdot\vek m\right)\cdot\left\Vert \nabla\phi\right\Vert \;\ensuremath{\mathrm{d\ensuremath{\partial\Omega}}},\label{eq:CorareaFormulaBoundary}
\end{equation}
which is extended from \cite{Dziuk_2013a,Dziuk_2008a}. Note that
on the right hand side, the conormal vectors $\vek q$ with respect
to $\Gamma^{c}$ as well as the normal vectors $\vek m$ on $\partial\Omega$
are involved. The integral over the level-set interval on the left
hand side of Eq.~(\ref{eq:CoareaFormulaDomain}) may also be evaluated
numerically, e.g., using Gauss quadrature,
\begin{equation}
\int_{\phi^{\min}}^{\phi^{\max}}\int_{\Gamma^{c}}f\left(\vek x\right)\;\ensuremath{\mathrm{d\ensuremath{\Gamma}}}\;\ensuremath{\mathrm{d}c}\approx\sum_{i}w_{i}\cdot\int_{\Gamma^{c_{i}}}f\left(\vek x\right)\;\ensuremath{\mathrm{d\ensuremath{\Gamma}}},\label{eq:NumericalIntOverManifolds}
\end{equation}
where $w_{i}$ are integration weights and $c_{i}$ are selected level-set
values according to the employed Gauss rule, see Fig.~\ref{fig:GaussLevelSets}
for an illustrative setup. The analogy of the two situations of either
integrating over $\Omega$ on the right hand side of Eq.~(\ref{eq:CoareaFormulaDomain})
or numerically on selected level sets according to the right hand
side of Eq.~(\ref{eq:NumericalIntOverManifolds}) carries over to
either formulating BVPs simultaneously for all level sets in a bulk
domain or considering individual BVPs on selected level sets. In this
sense, it is later possible to confirm numerical results (i.e., to
compare mechanical quantities) obtained by the proposed Bulk Trace
FEM for all level sets in a bulk domain with Surface FEM results obtained
on selected level sets (according to some Gauss rule as in Eq.~(\ref{eq:NumericalIntOverManifolds})).

For \emph{one} selected level set $\Gamma^{c}$ related to the constant
value $c$, a scalar function $w\left(\vek x\right)$ and a vector
function $\vek u\left(\vek x\right)$, the following \emph{divergence
theorem on manifolds} is well-known \cite{Delfour_1996a,Delfour_2011a},
\begin{equation}
\int_{\Gamma^{c}}w\cdot\mathrm{div}_{\Gamma}\vek u\,\mathrm{d}\Gamma=-\int_{\Gamma^{c}}\nabla^{\Gamma}w\cdot\vek u\,\mathrm{d}\Gamma+\int_{\Gamma^{c}}\varkappa\cdot w\cdot\left(\vek u\cdot\vek n\right)\,\mathrm{d}\Gamma+\int_{\partial\Gamma^{c}}w\cdot\left(\vek u\cdot\vek q\right)\,\mathrm{d}\partial\Gamma,\label{eq:DivTheoremVector}
\end{equation}
where $\varkappa=\mathrm{div}\vek n=\mathrm{div}_{\Gamma}\vek n$
is the mean curvature. Consequently, when integrating over all level
sets and using the co-area formulas from above,
\begin{align}
\int_{\Omega}w\cdot\mathrm{div}_{\Gamma}\vek u\cdot\left\Vert \nabla\phi\right\Vert \,\mathrm{d}\Omega= & -\int_{\Omega}\nabla^{\Gamma}w\cdot\vek u\cdot\left\Vert \nabla\phi\right\Vert \,\mathrm{d}\Omega+\int_{\Omega}\varkappa\cdot w\cdot\left(\vek u\cdot\vek n\right)\cdot\left\Vert \nabla\phi\right\Vert \,\mathrm{d}\Omega\label{eq:DivTheoremVectorBTF}\\
 & +\int_{\partial\Omega}w\cdot\left(\vek u\cdot\vek q\right)\cdot\left(\vek q\cdot\vek m\right)\cdot\left\Vert \nabla\phi\right\Vert \,\mathrm{d}\partial\Omega.\nonumber 
\end{align}
It is noted that $\left(\vek u\cdot\vek q\right)\cdot\left(\vek q\cdot\vek m\right)=\left(\vek u\cdot\left(-\vek q\right)\right)\cdot\left(\left(-\vek q\right)\cdot\vek m\right)$
so that in Eq.~(\ref{eq:DivTheoremVectorBTF}) the sign of $\vek q$,
hence, the fact whether the conormal vector points inside or outside
of the level sets $\Gamma^{c}$, does not matter. Based on this, one
may immediately state the divergence theorem for a tensor function
$\mat A\!\left(\vek x\right)$ as
\begin{align}
\int_{\Omega}\vek u\cdot\mathrm{div}_{\Gamma}\mat A\cdot\left\Vert \nabla\phi\right\Vert \,\mathrm{d}\Omega= & -\int_{\Omega}\left(\nabla^{\Gamma,\mathrm{dir}}\vek u:\mat A\right)\cdot\left\Vert \nabla\phi\right\Vert \,\mathrm{d}\Omega+\int_{\Omega}\varkappa\cdot\vek u\cdot\left(\mat A\cdot\vek n\right)\cdot\left\Vert \nabla\phi\right\Vert \,\mathrm{d}\Omega\label{eq:DivTheoremTensorBTF}\\
 & +\int_{\partial\Omega}\vek u\cdot\left(\mat A\cdot\vek q\right)\cdot\left(\vek q\cdot\vek m\right)\cdot\left\Vert \nabla\phi\right\Vert \,\mathrm{d}\partial\Omega,\nonumber 
\end{align}
which is later needed for the weak formulation of the mechanical equilibrium
in finite strain theory. Note that for \emph{in-plane} tensor functions
with $\mat A=\mat p\cdot\mat A\cdot\mat p$, the term involving the
curvature $\varkappa$ vanishes due to $\mat A\cdot\vek n=\vek0$.
Furthermore, one finds for in-plane tensors that $\nabla^{\Gamma,\mathrm{dir}}\vek u:\mat A=\nabla^{\Gamma,\mathrm{cov}}\vek u:\mat A$.

\subsection{Invalid combinations of level sets and bulk domains\label{XX_InvalidLevelSets}}

\begin{figure}
\centering

\subfigure[valid]{\includegraphics[width=0.24\textwidth]{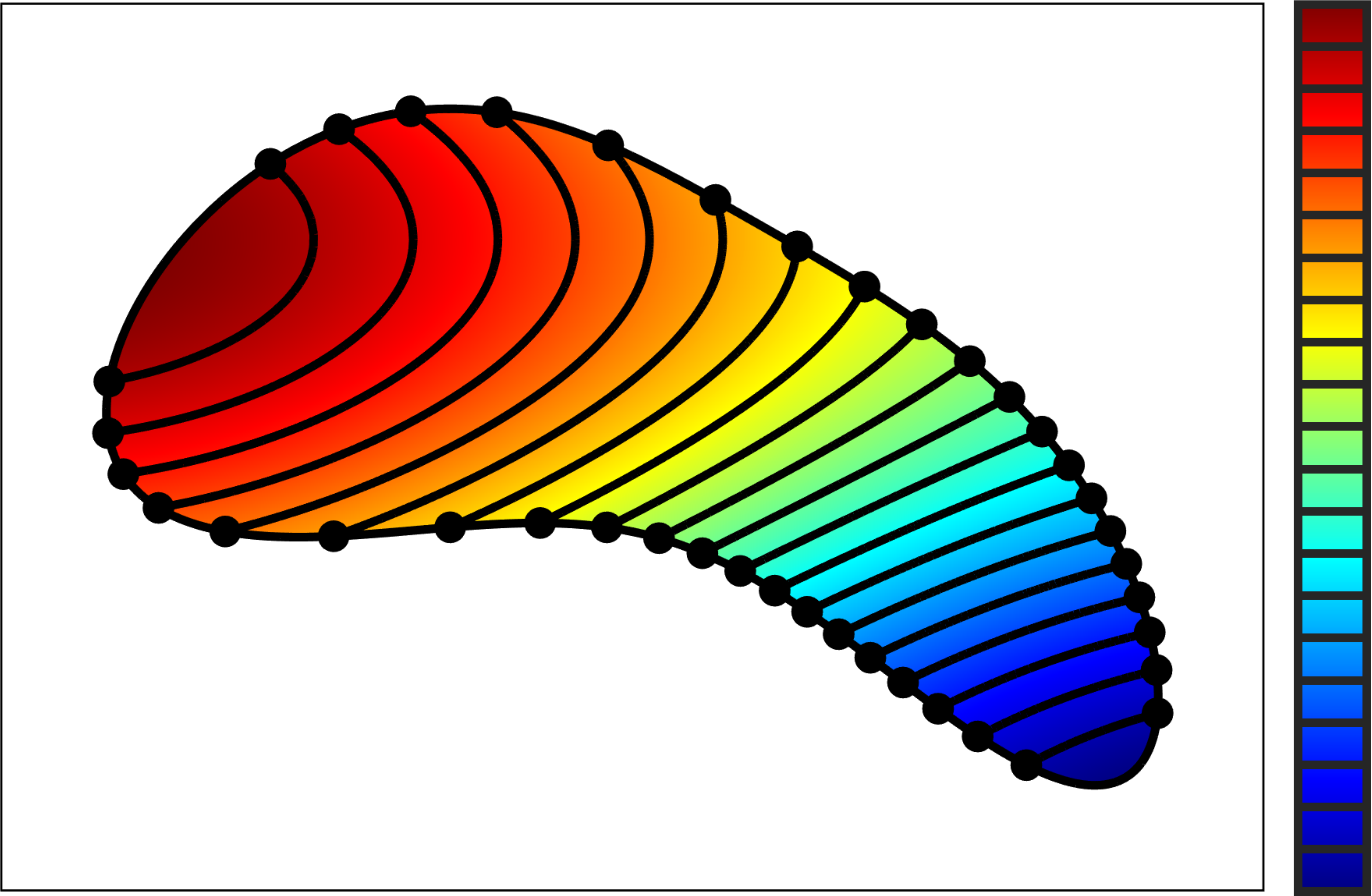}}\hfill\subfigure[invalid]{\includegraphics[width=0.24\textwidth]{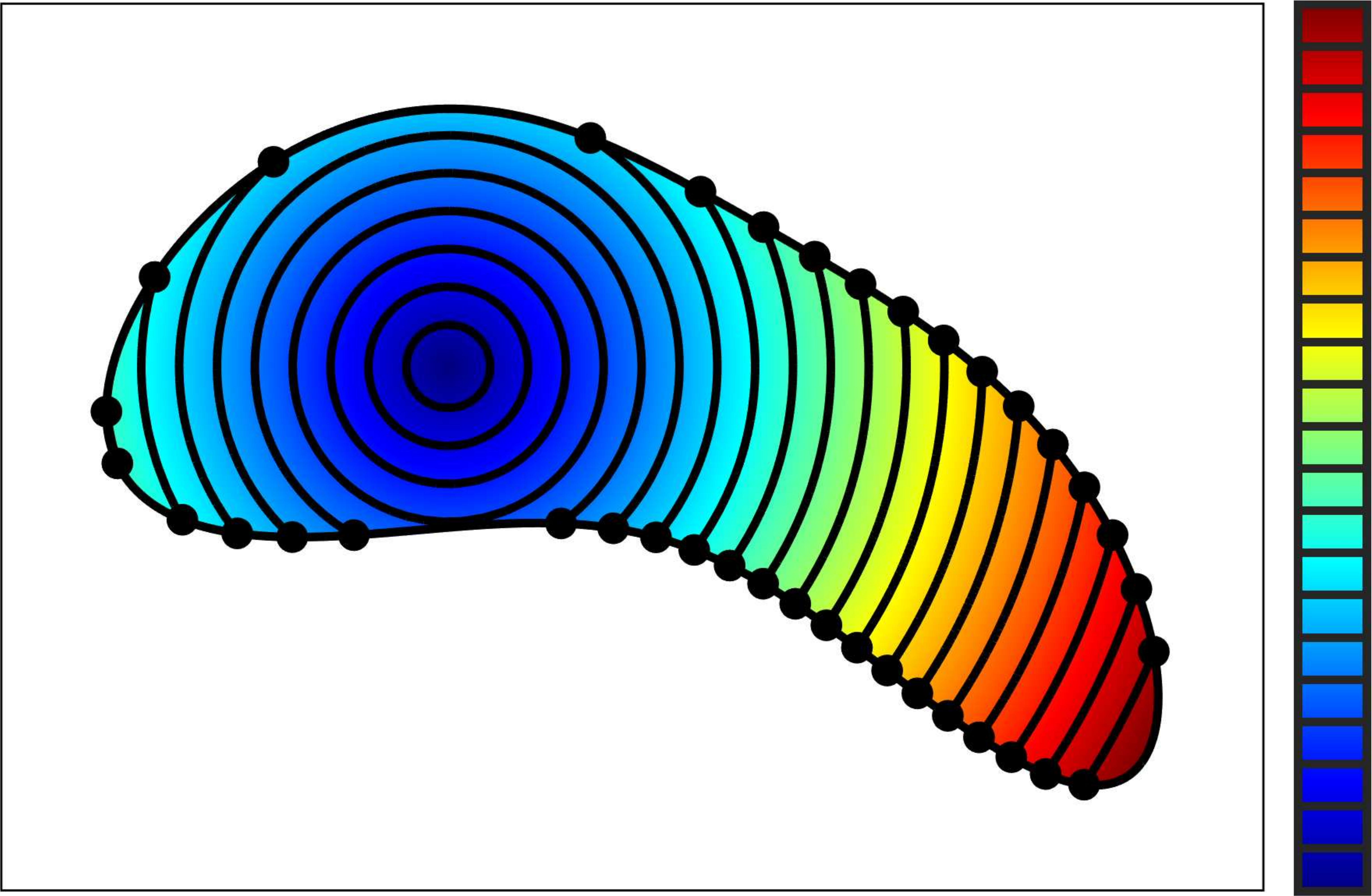}}\hfill\subfigure[invalid]{\includegraphics[width=0.24\textwidth]{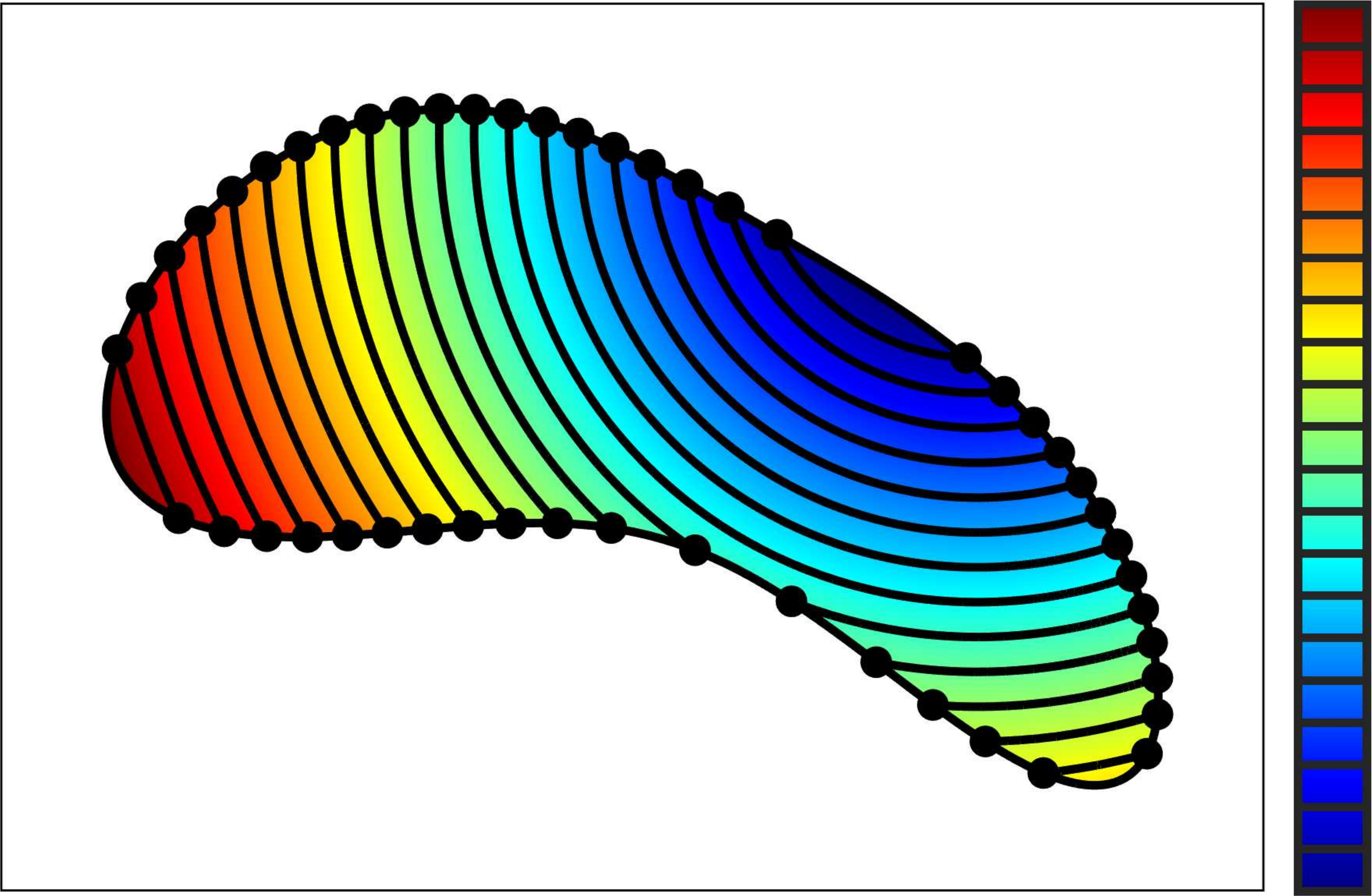}}\hfill\subfigure[invalid]{\includegraphics[width=0.24\textwidth]{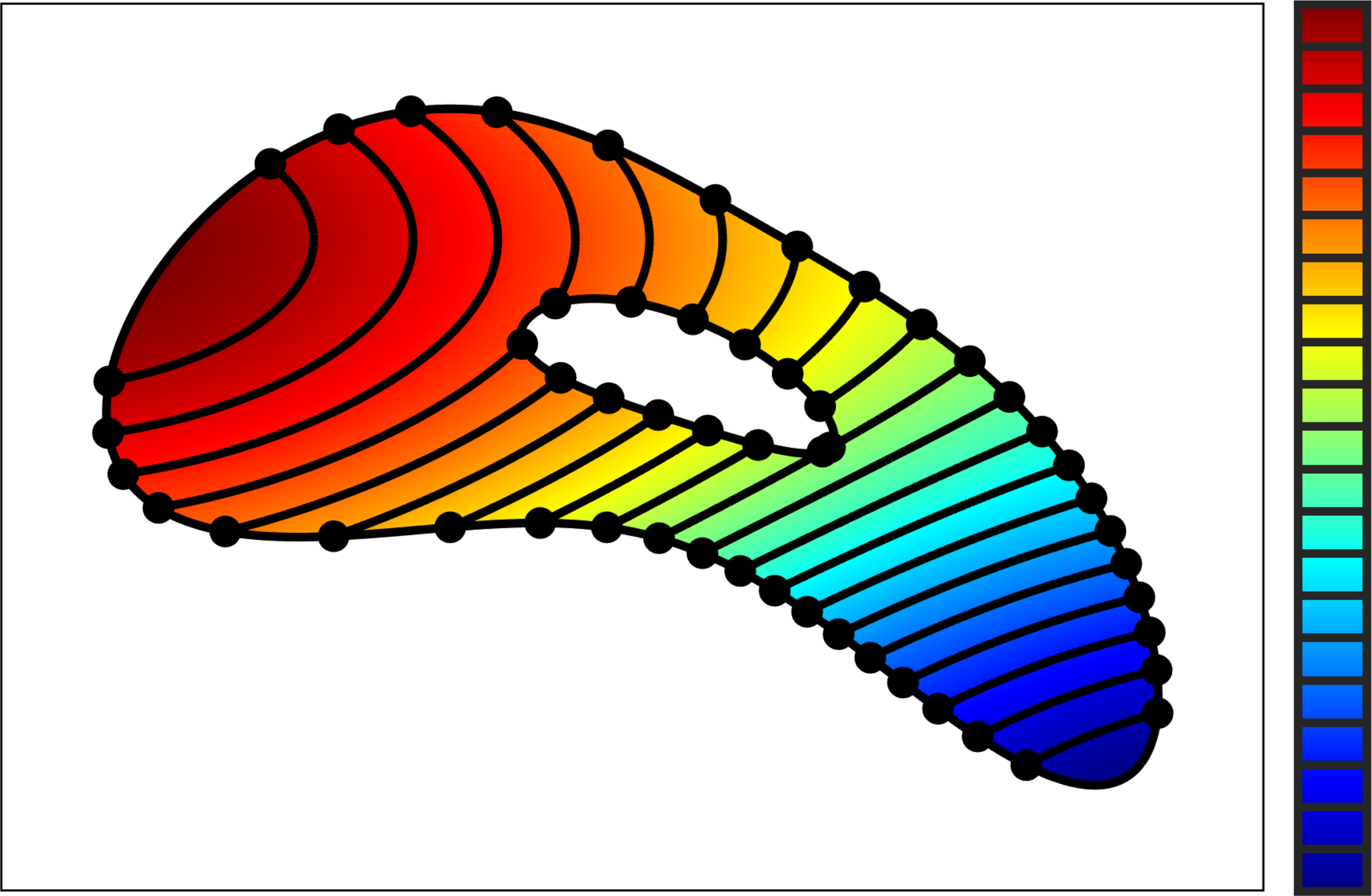}}

\caption{\label{fig:InvalidLevelSets}Valid and invalid combinations of level-set
functions $\phi$ and bulk domains $\Omega$.}
\end{figure}

The simultaneous mechanical modeling and simulation of curved structures
on all level sets over a bulk domain requires everywhere a smooth
transition of the displacement fields between neighbouring level sets.
Most importantly, the topology of neighboring level sets must not
change discontinuously in $\Omega$. Combinations which do not lead
to smooth variations of the implied level-set geometries $\Gamma^{c}$
in some given bulk domain $\Omega$ may be called \emph{invalid}.
Examples are seen in Fig.~\ref{fig:InvalidLevelSets} where (a) shows
a valid situation but (b) to (d) are invalid. In Fig.~\ref{fig:InvalidLevelSets}(b),
a situation is seen where some level sets feature two end points,
however, others are closed and do not feature boundaries $\partial\Gamma^{c}$.
This is the result of a local maximum of the level-set function within
the domain $\Omega$ which, hence, must be excluded. Fig.~\ref{fig:InvalidLevelSets}(c)
shows a situation where some level sets feature two end points, however,
others three or four which results from the opposite curvature of
$\partial\Omega$ and $\Gamma^{c}$ on the lower side of the bulk
domain. In Fig.~\ref{fig:InvalidLevelSets}(d), the level-set function
is identical to the valid situation in Fig.~\ref{fig:InvalidLevelSets}(a),
however, there is now a hole inside the bulk domain $\Omega$, so
that the number of end points of the level sets also varies inside
$\Omega$ being invalid.

Note that in all invalid cases, it is the interplay between the (smooth)
geometry of the bulk domain $\Omega$ and the (smooth) level-set function
$\phi$ which generated these invalid scenarios. Obviously, local
extreme values of $\phi$ inside $\Omega$ have to be excluded (hence,
$\inf\phi$ and $\sup\phi$ must be on the boundary $\partial\Omega$).
Additional requirements are related to the curvature of $\phi$ near
the boundary $\partial\Omega$. The validity of level sets with respect
to bulk domains has also been discussed in \cite{Dziuk_2008a}. Herein,
rather than (re-)formulating mathematically sound requirements for
$\phi$ and $\Omega$ we rather state that only combinations of $\phi$
and $\Omega$ are allowed where all level sets vary smoothly inside
$\Omega$, enabling smooth mechanical fields.

\begin{figure}
\centering

\subfigure[]{\includegraphics[width=0.24\textwidth]{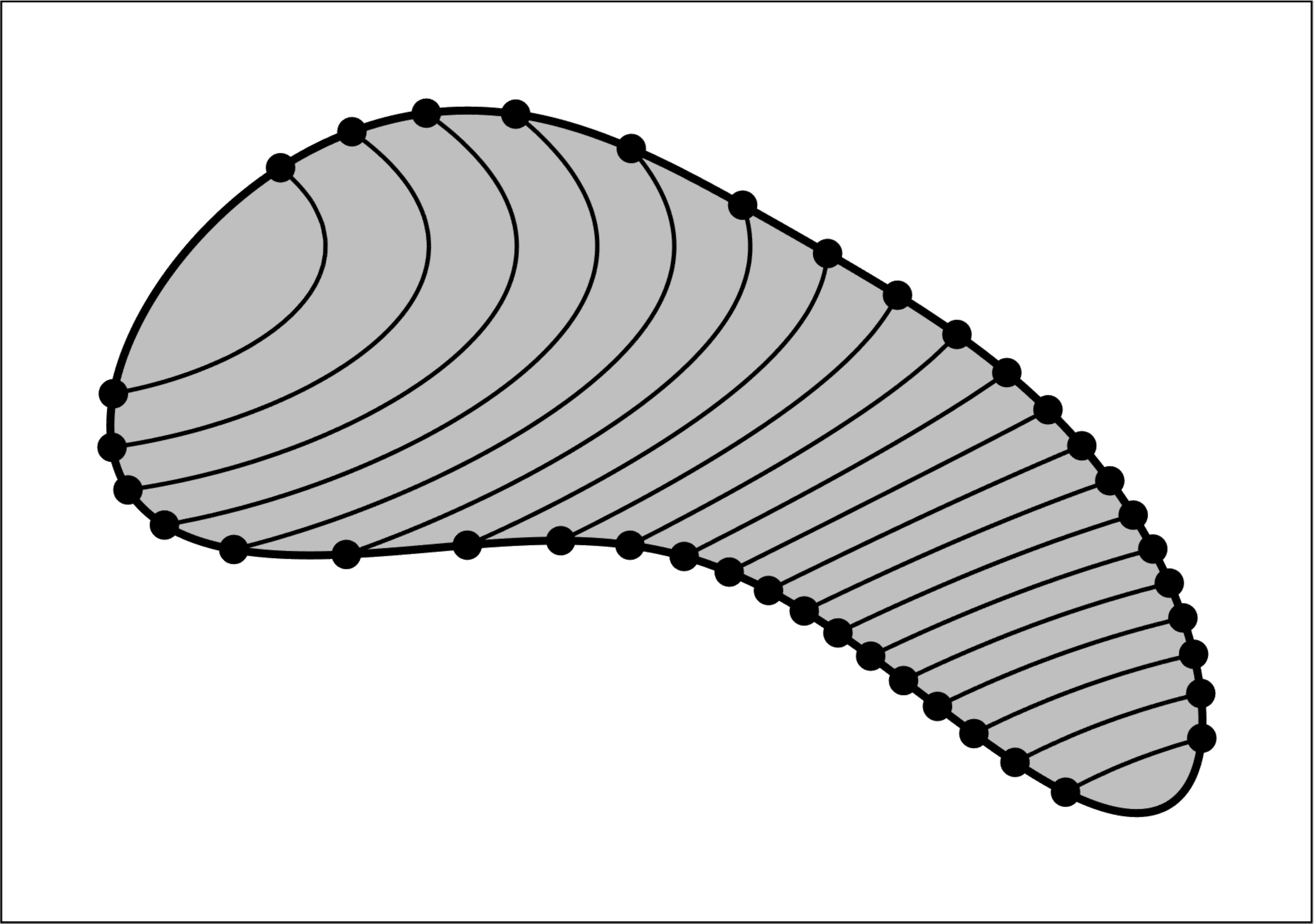}}\hfill\subfigure[]{\includegraphics[width=0.24\textwidth]{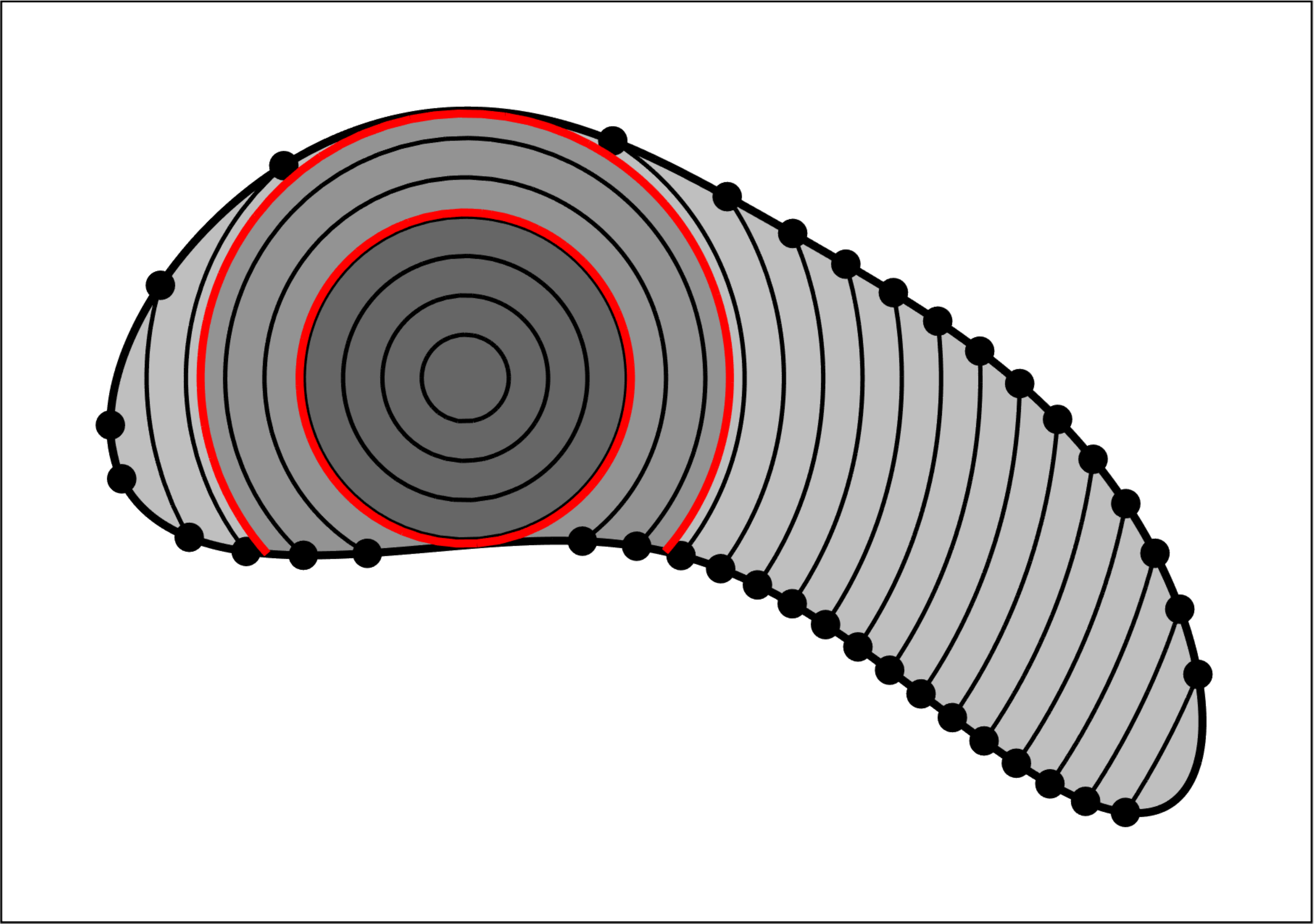}}\hfill\subfigure[]{\includegraphics[width=0.24\textwidth]{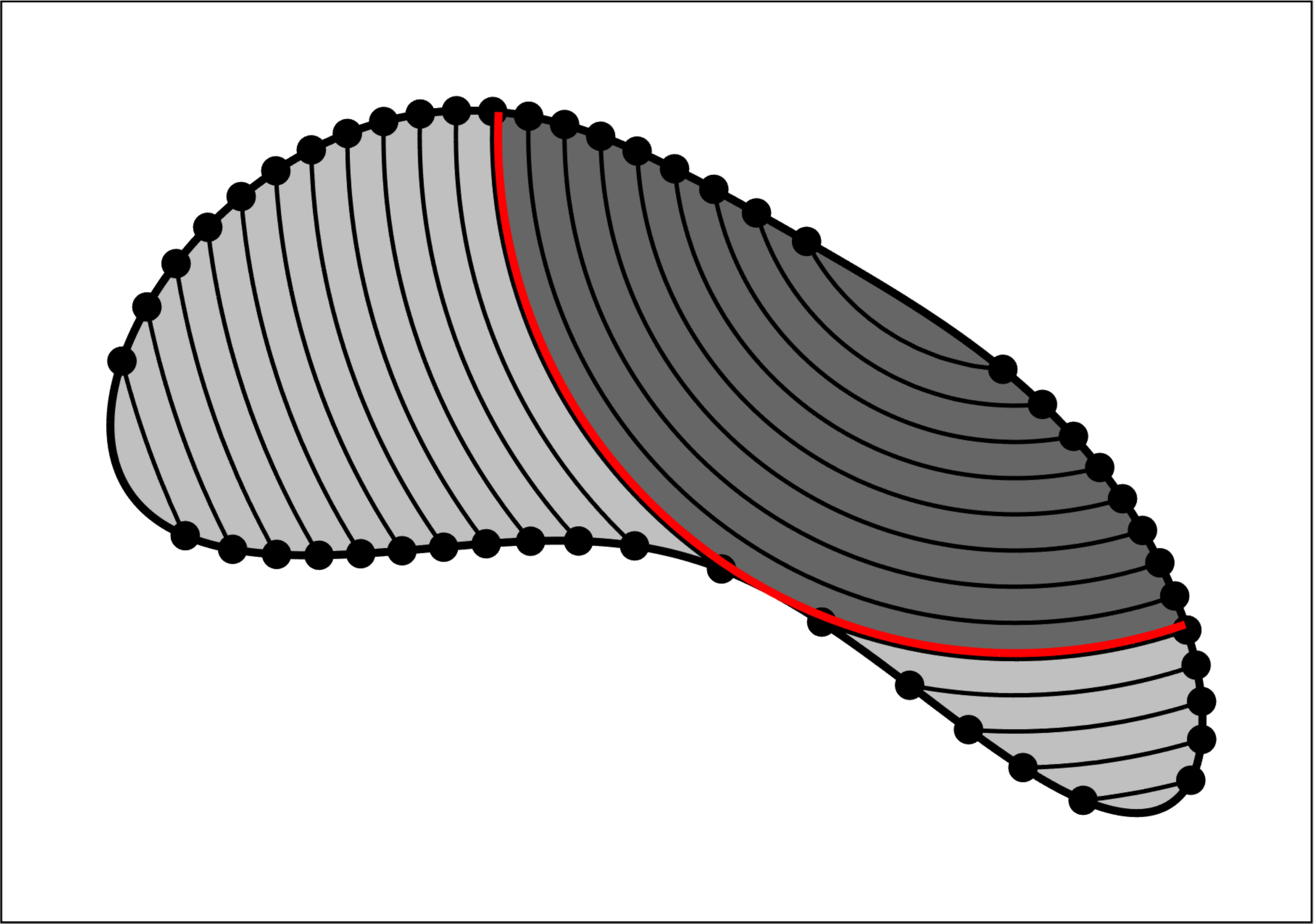}}\hfill\subfigure[]{\includegraphics[width=0.24\textwidth]{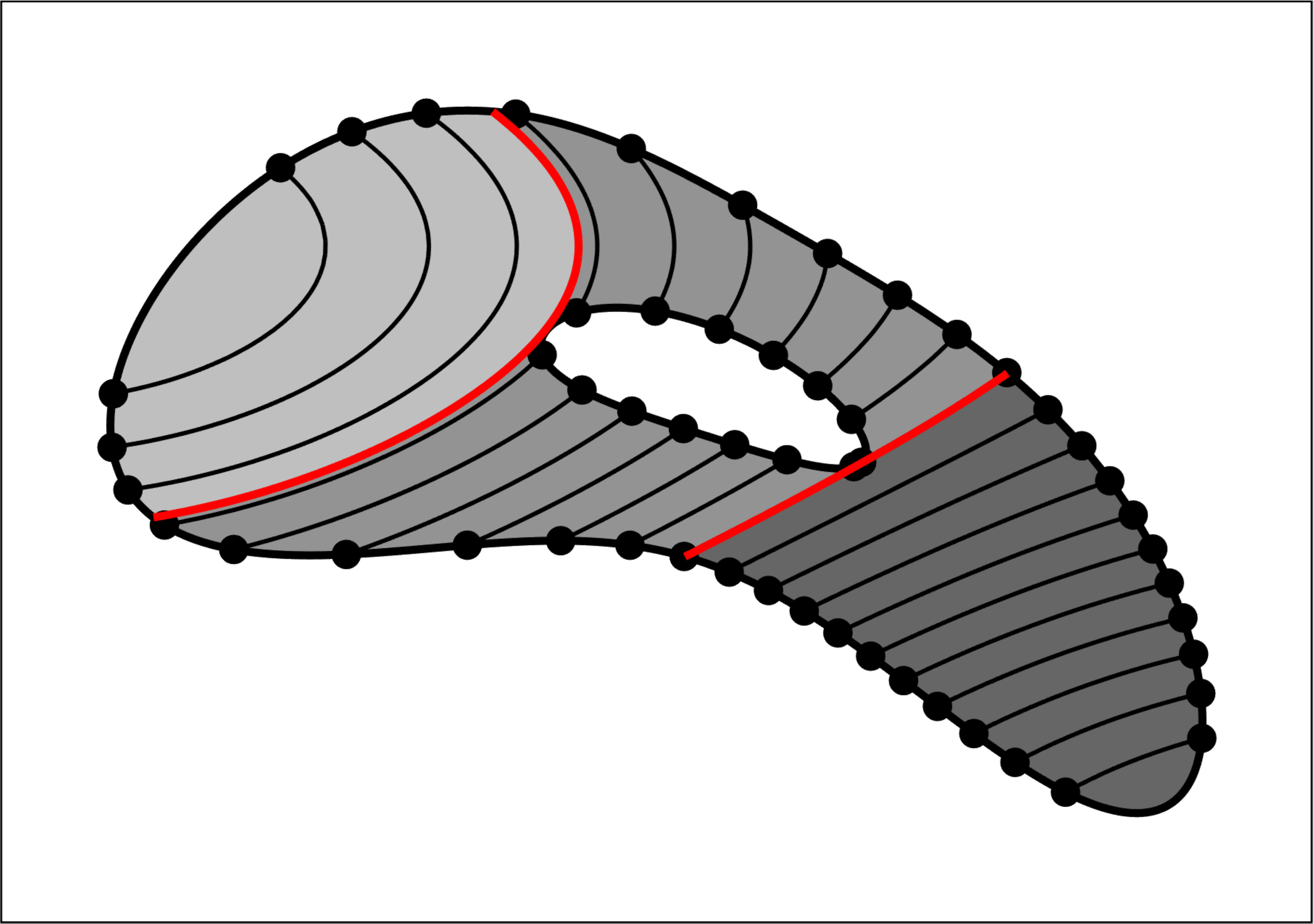}}

\caption{\label{fig:InvalidLevelSetsIntervals}Invalid combinations of $\phi$
and $\Omega$ may still be considered if the bulk domain is reduced
to certain sub-domains (related to certain intervals of the level-set
function) in which the topology does not change, i.e., where the geometries
of $\Gamma^{c}$ vary smoothly (regions with individual gray scales).}
\end{figure}

It is also noteworthy that in invalid cases, one may still pose mechanically
meaningful BVPs over \emph{sub-domains} of $\Omega$ related to individual
intervals of the level-set function $\phi$ where the level sets vary
smoothly and then assess all sub-domains successively. Of course,
in this case, one needs to generate individual meshes in each of the
sub-domains to carry out the numerical analyses (according to Section
\ref{X_BulkTraceFEM}). In Fig.~\ref{fig:InvalidLevelSets}, such
sub-domains (related to certain intervals of the level-set function)
are highlighted, related to the situations shown in Fig.~\ref{fig:InvalidLevelSets},
using different gray scales. This is related to the alternative definition
of bulk domains via prescribed values for $\phi^{\min}$ and $\phi^{\max}$
as described in Section \ref{XX_BulkDomainsAndLevelSets}. It is noted
that the rather strong assumptions on the validity of level sets in
bulk domains are largely alleviated when the bulk domain itself is
equipped with mechanical properties as discussed in Section \ref{XX_ElasticBulkDomains},
leading to a new class of anisotropic material models for continuously
embedded sub-structures in bulk domains.

\section{Mechanical setup in finite strain theory\label{X_MechanicalModel}}

\subsection{Undeformed and deformed configurations\label{XX_Configurations}}

In finite strain or large displacement theory, we distinguish an undeformed
material configuration and a deformed spatial configuration. With
\emph{configuration} we refer to \emph{all} level sets of a level-set
function over some $d$-dimensional bulk domain. As usual in finite
strain theory, we use upper case letters for quantities in the undeformed
configuration and lower case letters for the deformed configuration.

\begin{figure}
\centering

\subfigure[situation in 2D]{\includegraphics[width=0.4\textwidth]{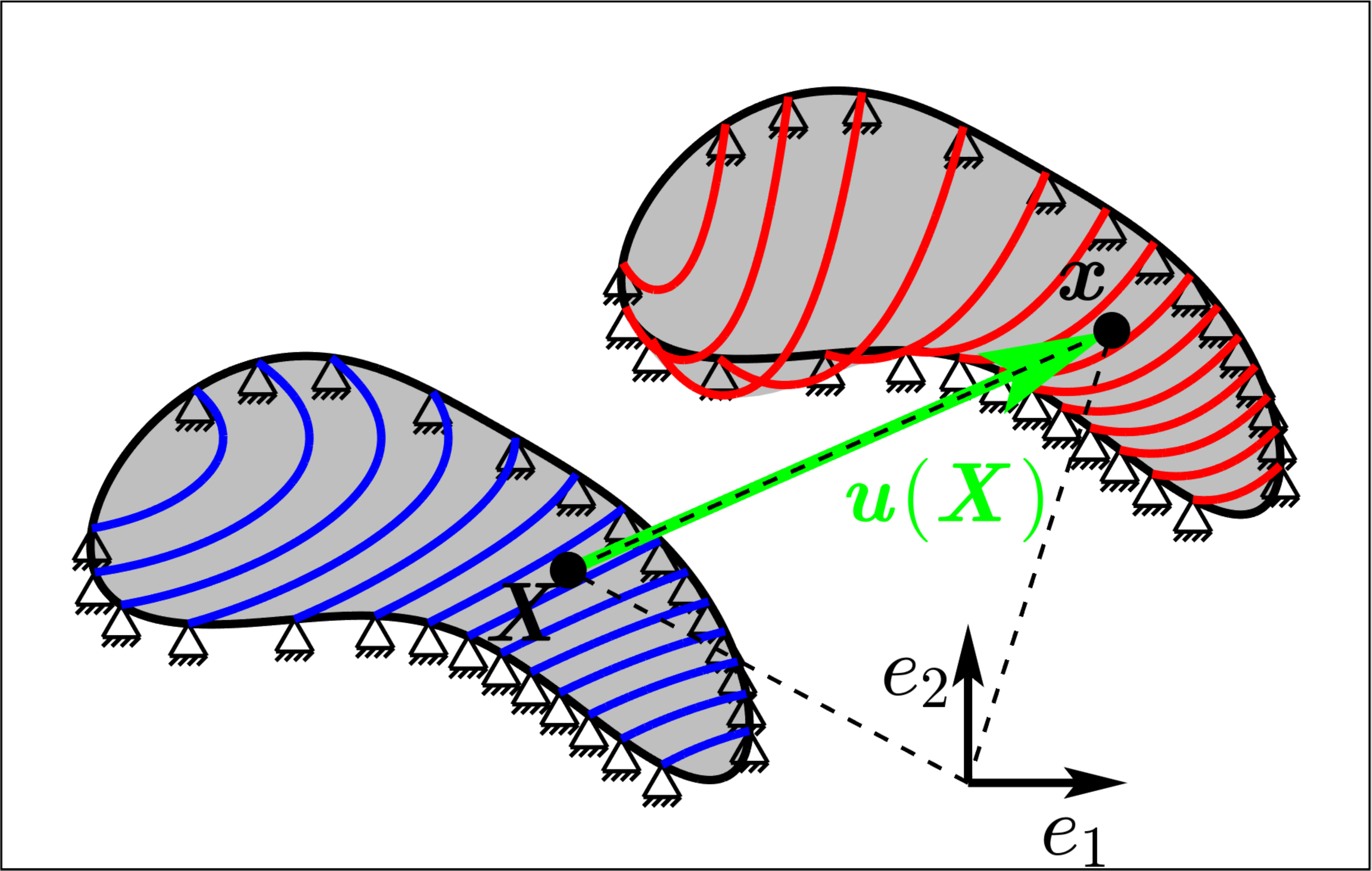}}\qquad\subfigure[mesh in 2D]{\includegraphics[width=0.4\textwidth]{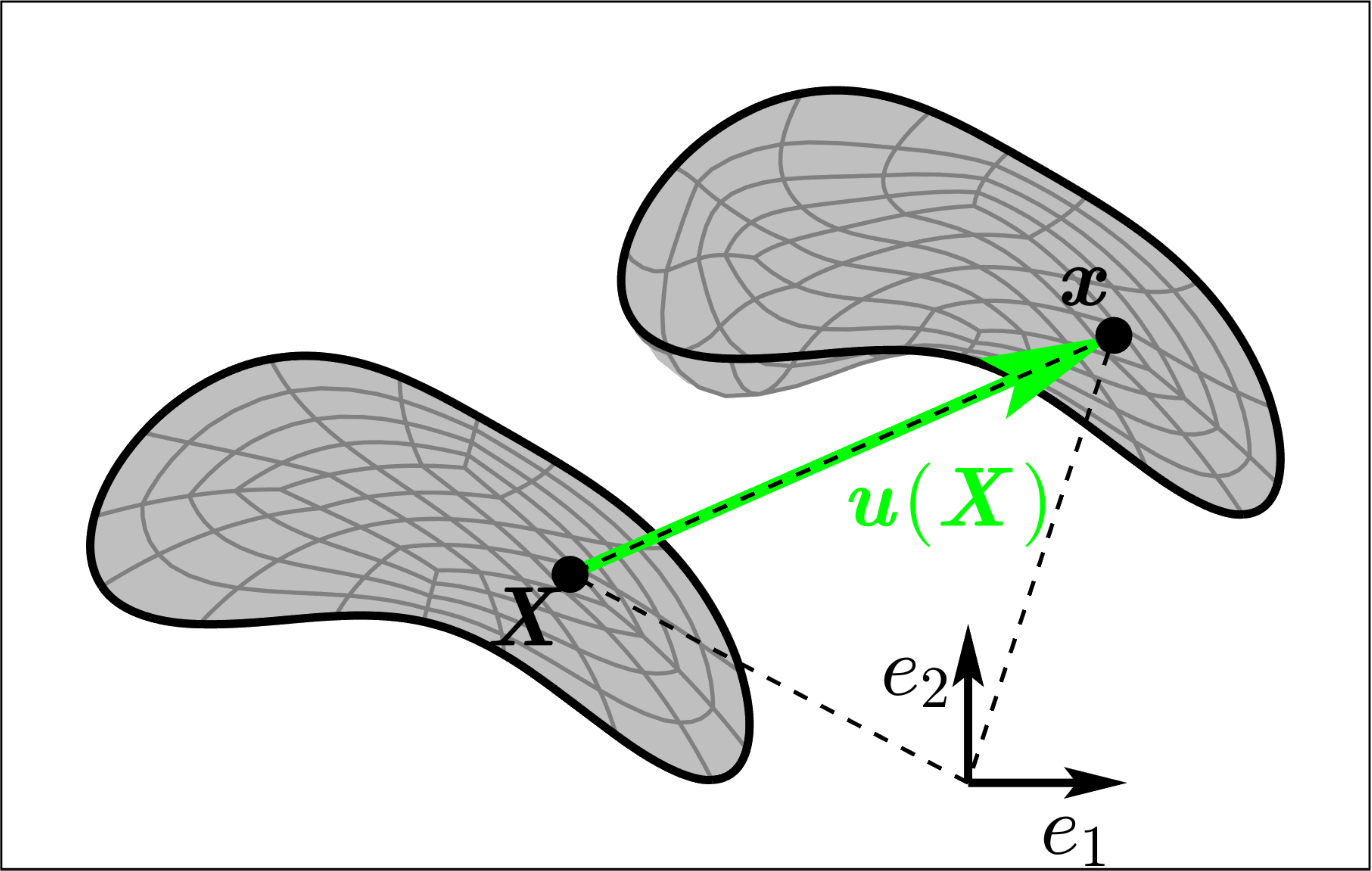}}

\subfigure[situation in 3D]{\includegraphics[width=0.5\textwidth]{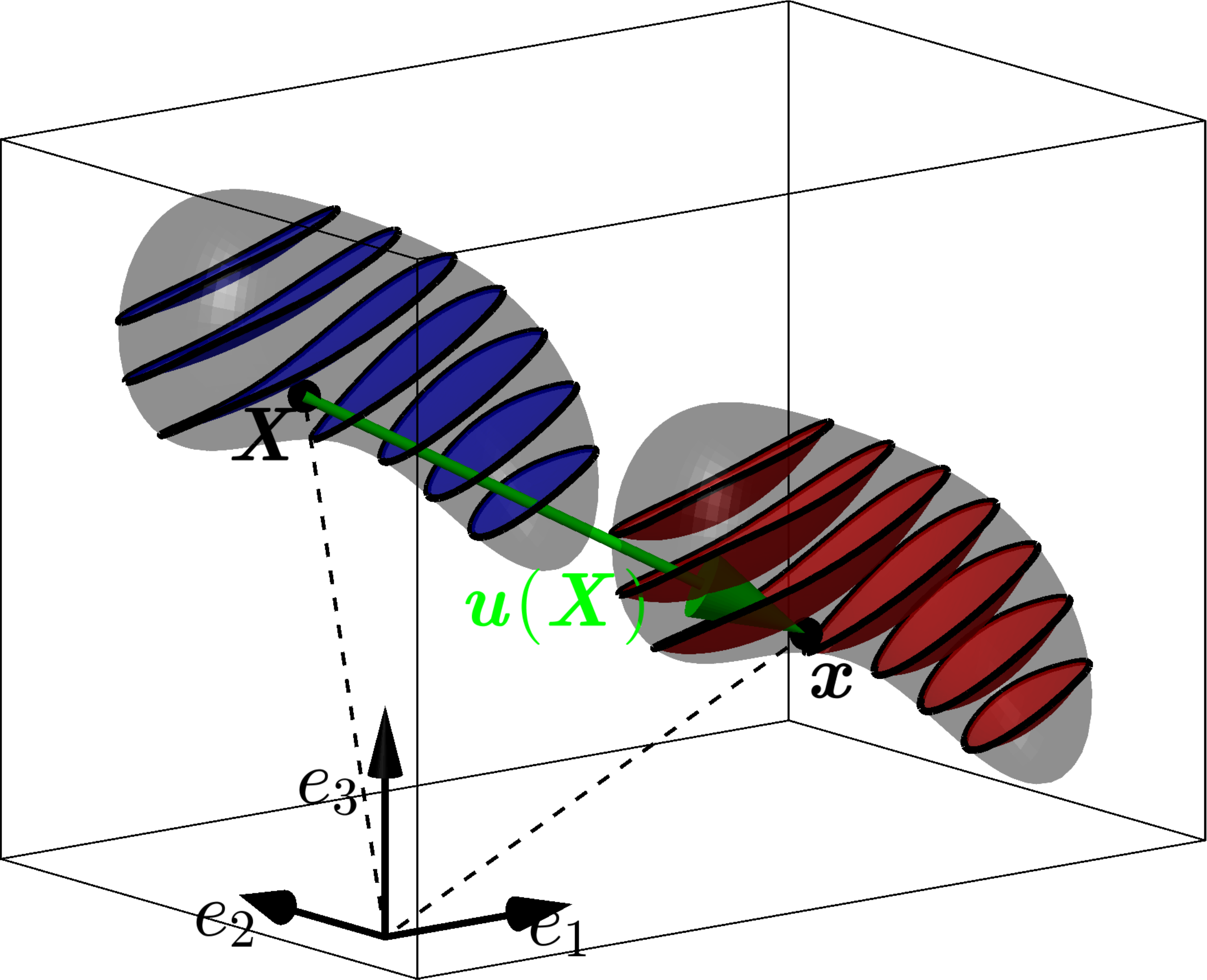}}

\caption{\label{fig:Mechanics}Undeformed \emph{material} configuration and
deformed \emph{spatial} configuration in two and three dimensions.
The bulk domains $\Omega_{\vek X}$ and $\Omega_{\vek x}$ are shown
in gray. The undeformed level sets $\ManUndef$ are blue and the deformed
level sets $\ManDef$ are red. In (b), the corresponding meshes in
the bulk domains $\Omega_{\vek X}$ and $\Omega_{\vek x}$, as used
for the approximation of the displacement field $\vek u\left(\vek X\right)$,
are depicted.}
\end{figure}

Let the undeformed bulk domain be $\Omega_{\vek X}$ and the level-set
function $\phi\left(\vek X\right):\Omega_{\vek X}\rightarrow\mathbb{R}$.
Then, the individual undeformed domains of interest $\ManUndef$ (being
the set of membranes or ropes simultaneously considered) are each
related to constant level-set values,
\begin{equation}
\ManUndef=\left\{ \vek X\in\Omega_{\vek X}:\,\phi(\vek X)=c\right\} ,\,c\in\left(\phi^{\min},\;\phi^{\max}\right),\label{eq:LevelSetsUndef}
\end{equation}
analogously to Eq.~(\ref{eq:LevelSets}). One may now define an individual
boundary value problem for every $\ManUndef$ and determine the individual
displacement field $\vek u\left(\vek X\right)$ of every level set
\cite{Fries_2020a}. However, the displacements between neighboring
level sets (usually) vary smoothly, cf.~Section \ref{XX_InvalidLevelSets},
so that \emph{we rather seek the displacement field of all level sets
at once}. That is, the deformed bulk domain 
\[
\Omega_{\vek x}=\left\{ \vek x=\vek X+\vek u\left(\vek X\right),\vek X\in\Omega_{\vek X}\right\} 
\]
is sought such that
\[
\ManDef=\left\{ \vek x\in\Omega_{\vek x}:\,\phi(\vek x)=c\right\} ,\,c\in\left(\phi^{\min},\;\phi^{\max}\right)
\]
are the deformed structures. Illustratively, plotting selected level
sets of $\phi\left(\vek X\right)$ in $\Omega_{\vek X}$ shows some
undeformed structures whereas plotting the same level sets of $\phi\left(\vek x\right)$
in $\Omega_{\vek x}$ represents the resulting deformed structures,
see Fig.~\ref{fig:Mechanics}.

The geometric quantities and differential operators formulated in
Section \ref{X_Preliminaries} extend straightforwardly to the mechanical
setup outlined here with only small distinctions in the notation to
specify whether these quantities relate to the undeformed or deformed
situation. They are shortly summarized in Table \ref{tab:ImplicitManifolds}.
The bulk domains $\Omega_{\vek X}$ and $\Omega_{\vek x}$ feature
boundaries $\partial\Omega_{\vek X}$ and $\partial\Omega_{\vek x}$
with unit normal vectors $\vek M$ and $\vek m$, respectively. $\vek N$
and $\vek n$ are the respective unit normal vectors on the level
sets $\ManUndef$ and $\ManDef$ in the bulk domains, and $\vek Q$
and $\vek q$ the conormal vectors on the boundaries.

\begin{table}
\begin{tabular}{|c|c|c|}
\hline 
 & undef. material configuration & def. spatial configuration\tabularnewline
\hline 
\hline 
bulk domain & $\begin{array}{l}
\Omega_{\vek X}\subset\mathbb{R}^{d}\\
\text{with boundary }\partial\Omega_{\vek X}
\end{array}$ & $\begin{array}{l}
\Omega_{\vek x}=\left\{ \vek x=\vek X+\vek u\left(\vek X\right),\vek X\in\Omega_{\vek X}\right\} \\
\text{with boundary }\partial\Omega_{\vek x}
\end{array}$\tabularnewline
\hline 
level-set function & $\phi\left(\vek X\right),\;\vek X\in\Omega_{\vek X}$ & $\phi\left(\vek x\right),\;\vek x\in\Omega_{\vek x}$\tabularnewline
\hline 
$\begin{array}{l}
\text{level sets}\\
\text{(=curved structures)}
\end{array}$ & $\begin{array}{l}
\ManUndef=\{\vek X\in\Omega_{\vek X}:\phi(\vek X)=c\},\\
c\in\left(\phi^{\min},\;\phi^{\max}\right)
\end{array}$ & $\begin{array}{l}
\ManDef=\{\vek x\in\Omega_{\vek x}:\phi(\vek x)=c\},\\
c\in\left(\phi^{\min},\;\phi^{\max}\right)
\end{array}$\tabularnewline
\hline 
$\begin{array}{l}
\text{unit normal vectors}\\
\text{w.r.t. bulk domain}
\end{array}$ & $\vek M\;\text{on}\;\partial\Omega_{\vek X}$ & $\vek m\;\text{on}\;\partial\Omega_{\vek x}$\tabularnewline
\hline 
bulk def.~gradient & \multicolumn{2}{c|}{$\mat F_{\Omega}=\nabla_{\vek X}\vek x\left(\vek X\right)=\mat I+\nabla_{\vek X}\vek u\left(\vek X\right)$}\tabularnewline
\hline 
$\begin{array}{l}
\text{unit normal vectors}\\
\text{w.r.t. level sets}
\end{array}$ & $\vek N=\frac{\vek N^{\star}}{\left\Vert \vek N^{\star}\right\Vert },\;\vek N^{\star}=\nabla_{\vek X}\phi$ & $\vek n=\frac{\vek n^{\star}}{\left\Vert \vek n^{\star}\right\Vert },\;\vek n^{\star}=\mat F_{\Omega}^{-\mathrm{T}}\cdot\vek N^{\star}$\tabularnewline
\hline 
projectors & $\mat P=\mat I-\vek N\otimes\vek N$ & $\mat p=\mat I-\vek n\otimes\vek n$\tabularnewline
\hline 
$\begin{array}{l}
\text{unit conormal}\\
\text{vectors in }\mathbb{R}^{2}
\end{array}$ & $\vek Q=\left[\begin{array}{c}
Q_{x}\\
Q_{y}
\end{array}\right]=\left[\begin{array}{c}
-N_{y}\\
N_{x}
\end{array}\right]$ & $\vek q=\left[\begin{array}{c}
q_{x}\\
q_{y}
\end{array}\right]=\left[\begin{array}{c}
-n_{y}\\
n_{x}
\end{array}\right]$\tabularnewline
\hline 
$\begin{array}{l}
\text{unit conormal}\\
\text{vectors in }\mathbb{R}^{3}
\end{array}$ & $\vek Q=\frac{\vek Q^{\star}}{\left\Vert \vek Q^{\star}\right\Vert },\;\vek Q^{\star}=\vek N\times\vek M\times\vek N$ & $\vek q=\frac{\vek q^{\star}}{\left\Vert \vek q^{\star}\right\Vert },\;\vek q^{\star}=\vek n\times\vek m\times\vek n$\tabularnewline
\hline 
line/area stretch & \multicolumn{2}{c|}{$\Lambda=\frac{\left\Vert \vek n^{\star}\right\Vert }{\left\Vert \vek N^{\star}\right\Vert }\cdot\det\mat F_{\Omega}$}\tabularnewline
\hline 
$\begin{array}{l}
\text{surface gradient}\\
\text{of scalar function}
\end{array}$ & $\nabla_{\vek X}^{\Gamma}f=\mat P\cdot\nabla_{\vek X}f$ & $\nabla_{\vek x}^{\Gamma}f=\mat p\cdot\nabla_{\vek x}f$\tabularnewline
\hline 
$\begin{array}{l}
\text{dir. surface gradient}\\
\text{of vector function}
\end{array}$ & $\nabla_{\vek X}^{\Gamma,\mathrm{dir}}\vek u=\nabla_{\vek X}\vek u\cdot\mat P$ & $\nabla_{\vek x}^{\Gamma,\mathrm{dir}}\vek u=\nabla_{\vek x}\vek u\cdot\mat p$\tabularnewline
\hline 
$\begin{array}{l}
\text{cov. surface gradient}\\
\text{of vector function}
\end{array}$ & $\nabla_{\vek X}^{\Gamma,\mathrm{cov}}\vek u=\mat P\cdot\nabla_{\vek X}\vek u\cdot\mat P$ & $\nabla_{\vek x}^{\Gamma,\mathrm{cov}}\vek u=\mat p\cdot\nabla_{\vek x}\vek u\cdot\mat p$\tabularnewline
\hline 
$\begin{array}{l}
\text{relation between}\\
\text{surface gradients}
\end{array}$ & $\nabla_{\vek X}^{\Gamma}f=\mat P\cdot\mat F_{\Omega}^{\mathrm{T}}\cdot\nabla_{\vek x}^{\Gamma}f$ & $\nabla_{\vek x}^{\Gamma}f=\mat p\cdot\mat F_{\Omega}^{-\mathrm{T}}\cdot\nabla_{\vek X}^{\Gamma}f$\tabularnewline
\hline 
surface def.~gradient & \multicolumn{2}{c|}{$\mat F_{\Gamma}=\mat I+\nabla_{\vek X}^{\Gamma,\mathrm{dir}}\vek u$}\tabularnewline
\hline 
\end{tabular}\caption{\label{tab:ImplicitManifolds}Geometric quantities and differential
operators in finite strain theory for level-set geometries in bulk
domains.}
\end{table}

\subsection{Deformation gradients\label{XX_DeformationGradients}}

With the displacement field $\vek u\left(\vek X\right):\mathbb{R}^{d}\rightarrow\mathbb{R}^{d}$
, and $\vek x=\vek X+\vek u\left(\vek X\right)$, the resulting \emph{bulk
deformation gradient} is
\begin{equation}
\mat F_{\Omega}=\nabla_{\vek X}\vek x\left(\vek X\right)=\mat I+\nabla_{\vek X}\vek u\left(\vek X\right),\label{eq:ClassicalDeformationGradient}
\end{equation}
where $\mat I$ is a $(d\times d)$ identity matrix. Note that $\nabla_{\vek X}$
is the classical gradient with respect to the undeformed bulk domain
$\Omega_{\vek X}$ and $\nabla_{\vek x}$ with respect to $\Omega_{\vek x}$.
As such, we have for a scalar function $f\!\left(\vek X\right)$,
the classical bulk gradients $\nabla_{\vek X}f$ and $\nabla_{\vek x}f=\mat F_{\Omega}^{-\mathrm{T}}\cdot\nabla_{\vek X}f$.

Similarly, also the tangential or surface differential operators relate
either to level sets $\ManUndef$ or $\ManDef$. For scalar functions,
\[
\nabla_{\vek X}^{\Gamma}f=\mat P\cdot\nabla_{\vek X}f,\quad\nabla_{\vek x}^{\Gamma}f=\mat p\cdot\nabla_{\vek x}f=\mat p\cdot\mat F_{\Omega}^{-\mathrm{T}}\cdot\nabla_{\vek X}f,
\]
where $\mat P$ and $\mat p$ are projectors obtained from $\vek N$
and $\vek n$ as in Eq.~(\ref{eq:Projector}). For directional and
covariant surface gradients of \emph{vector }functions, $\nabla_{\vek X}^{\Gamma,\mathrm{dir}}\vek u$,
$\nabla_{\vek X}^{\Gamma,\mathrm{cov}}\vek u$, $\nabla_{\vek x}^{\Gamma,\mathrm{dir}}\vek u$,
and $\nabla_{\vek x}^{\Gamma,\mathrm{cov}}\vek u$, see Table \ref{tab:ImplicitManifolds}.

Based on these definitions, the \emph{surface deformation gradient
}$\mat F_{\Gamma}$ may now be defined as
\begin{equation}
\mat F_{\Gamma}=\nabla_{\vek X}^{\Gamma,\mathrm{dir}}\vek x\left(\vek X\right)=\mat I+\nabla_{\vek X}^{\Gamma,\mathrm{dir}}\vek u\left(\vek X\right),\label{eq:SurfaceDeformationGradient}
\end{equation}
and is not to be mixed with the \emph{bulk} deformation gradient in
Eq.~(\ref{eq:ClassicalDeformationGradient}). The stretch of a differential
element in the tangent plane of the level sets upon the deformation
is
\begin{equation}
\Lambda=\frac{\left\Vert \nabla_{\vek x}\phi\right\Vert }{\left\Vert \nabla_{\vek X}\phi\right\Vert }\cdot\det\mat F_{\Omega}=\frac{\left\Vert \vek n^{\star}\right\Vert }{\left\Vert \vek N^{\star}\right\Vert }\cdot\det\mat F_{\Omega}.\label{eq:Strech}
\end{equation}

We are now ready to adapt the definition of well-known stress and
strain tensors in finite strain theory to the level sets in the bulk
domain. These definitions are closely related to \cite{Fries_2020a}
where the authors formulate a mechanical model in the frame of the
tangential differential calculus which applies to \emph{one }selected
level set (in fact, the zero-level set). Here, the same definitions
are repeated as they immediately also apply to \emph{all }level sets.

\subsection{Strain tensors\label{XX_StrainTensors}}

Based on the surface deformation gradient, the directional and tangential
Green-Lagrange strain tensors are defined as 
\begin{eqnarray}
\mat E_{\mathrm{dir}} & = & \nicefrac{1}{2}\cdot\left(\mat F_{\Gamma}^{\mathrm{T}}\cdot\mat F_{\Gamma}-\mat I\right),\label{eq:CauchyGreenDir}\\
\mat E_{\mathrm{tang}} & = & \mat P\cdot\mat E_{\mathrm{dir}}\cdot\mat P,\label{eq:CauchyGreenCov}
\end{eqnarray}
respectively. The Euler-Almansi strain tensors are
\begin{eqnarray}
\mat e_{\mathrm{dir}} & = & \nicefrac{1}{2}\cdot\left(\mat I-\left(\mat F_{\Gamma}\cdot\mat F_{\Gamma}^{\mathrm{T}}\right)^{-1}\right),\label{eq:EulerAlmansiDir}\\
\mat e_{\mathrm{tang}} & = & \mat p\cdot\mat e_{\mathrm{dir}}\cdot\mat p,\label{eq:EulerAlmansiCov}
\end{eqnarray}
where $\mat e_{\mathrm{tang}}$ is tangential to the deformed configuration
$\Gamma_{\vek x}$. There holds $\mat e_{\mathrm{dir}}=\mat F_{\Gamma}^{-\mathrm{T}}\cdot\mat E_{\mathrm{dir}}\cdot\mat F_{\Gamma}^{-1}$
which, however, is not true for the tangential versions of these strain
tensors.

\subsection{Stress tensors\label{XX_StressTensors}}

Conjugated stress tensors are introduced next and only the tangential
versions are considered. Generally speaking, we assume some hyper-elastic
material with an elastic energy function $\Psi\left(\mat E_{\mathrm{tang}}\right)$
and obtain the second Piola-Kirchhoff stress tensor as $\mat S=\frac{\partial\Psi}{\partial\mat E_{\mathrm{tang}}}$.
For simplicity, only Saint Venant--Kirchhoff solids are considered
herein and there follows
\begin{eqnarray}
\mat S & = & \lambda\cdot\mathrm{trace}\left(\mat E_{\mathrm{tang}}\right)\cdot\mat P+2\mu\mat E_{\mathrm{tang}},\label{eq:2ndPiolaKirchhoff}\\
 & = & \mat P\cdot\left(\lambda\cdot\mathrm{trace}\left(\mat E_{\mathrm{dir}}\right)\cdot\mat I+2\mu\mat E_{\mathrm{dir}}\right)\cdot\mat P,\nonumber 
\end{eqnarray}
with $\mat S$ being tangential to $\Gamma_{\vek X}$, $\lambda$
and $\mu$ are the Lam\'e constants. For given Young's modulus $E$
and Poisson's ratio $\nu$, there holds $\lambda=\frac{E\cdot\nu}{1-\nu^{2}}$,
$\mu=\frac{E}{2\left(1+\nu\right)}$ for membranes, and $\lambda=0$,
$\mu=\frac{E}{2}$ for cables. The Cauchy stress tensor reads
\begin{equation}
\vek\sigma=\frac{1}{\Lambda}\cdot\mat F_{\Gamma}\cdot\mat S\cdot\mat F_{\Gamma}^{\mathrm{T}},\label{eq:CauchyStress}
\end{equation}
where $\Lambda$ is a line stretch for cables and an area stretch
for membranes when undergoing the displacement, see Eq.~(\ref{eq:Strech}).
The Cauchy stress is tangential to the level sets in the deformed
configuration $\ManDef$ since $\mat F_{\Gamma}\cdot\mat P=\mat p\cdot\mat F_{\Gamma}\cdot\mat P$
and $\mat P\cdot\mat F_{\Gamma}^{\mathrm{T}}=\mat P\cdot\mat F_{\Gamma}^{\mathrm{T}}\cdot\mat p$,
hence $\vek\sigma=\mat p\cdot\vek\sigma\cdot\mat p.$ The first Piola-Kirchhoff
stress tensor is given by
\begin{equation}
\mat K=\mat F_{\Gamma}\cdot\mat S\label{eq:1stPiolaKirchhoff}
\end{equation}
and there holds $\mat K=\mat K\cdot\mat P=\mat p\cdot\mat K$.

\begin{figure}
\centering

\subfigure[$\sigma_{\mathrm{V}}$ on some $\ManDef\quad$]{\includegraphics[width=0.24\textwidth]{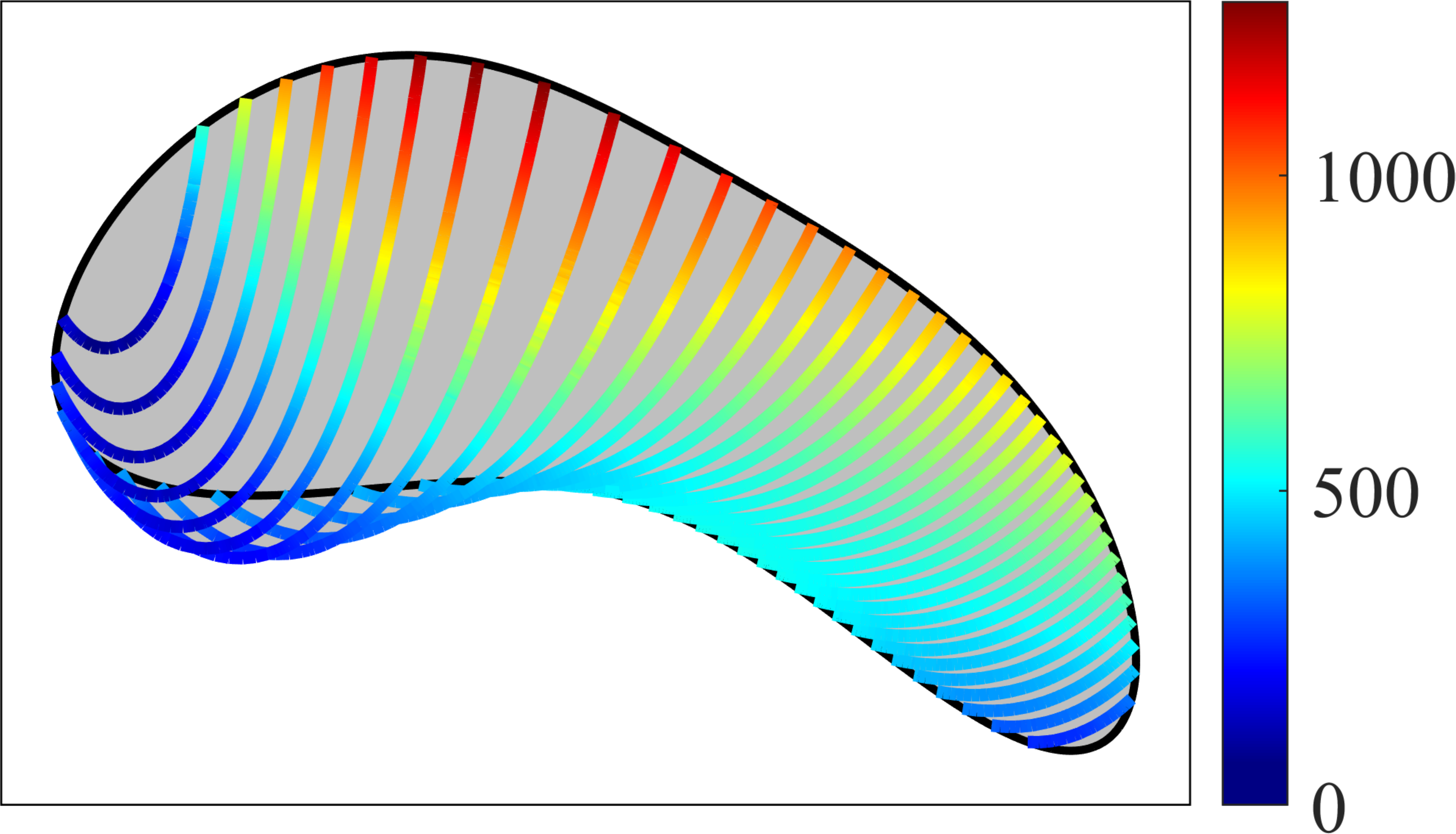}}\hfill\subfigure[$\sigma_{\mathrm{V}}$ in $\Omega_{\vek{x}}\quad$]{\includegraphics[width=0.24\textwidth]{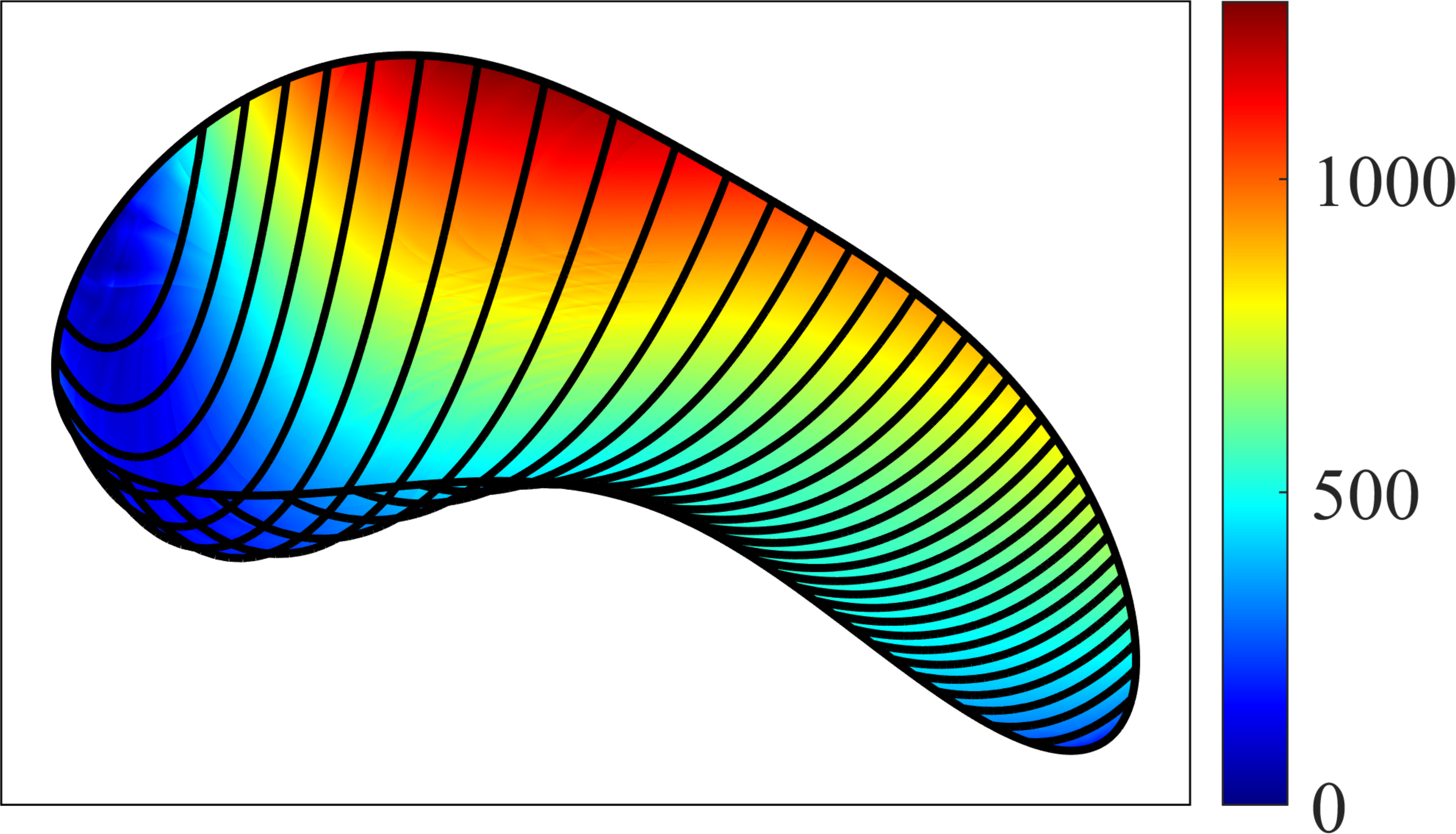}}\hfill\subfigure[$\sigma_{\mathrm{V}}$ on some $\ManDef\quad$]{\includegraphics[width=0.24\textwidth]{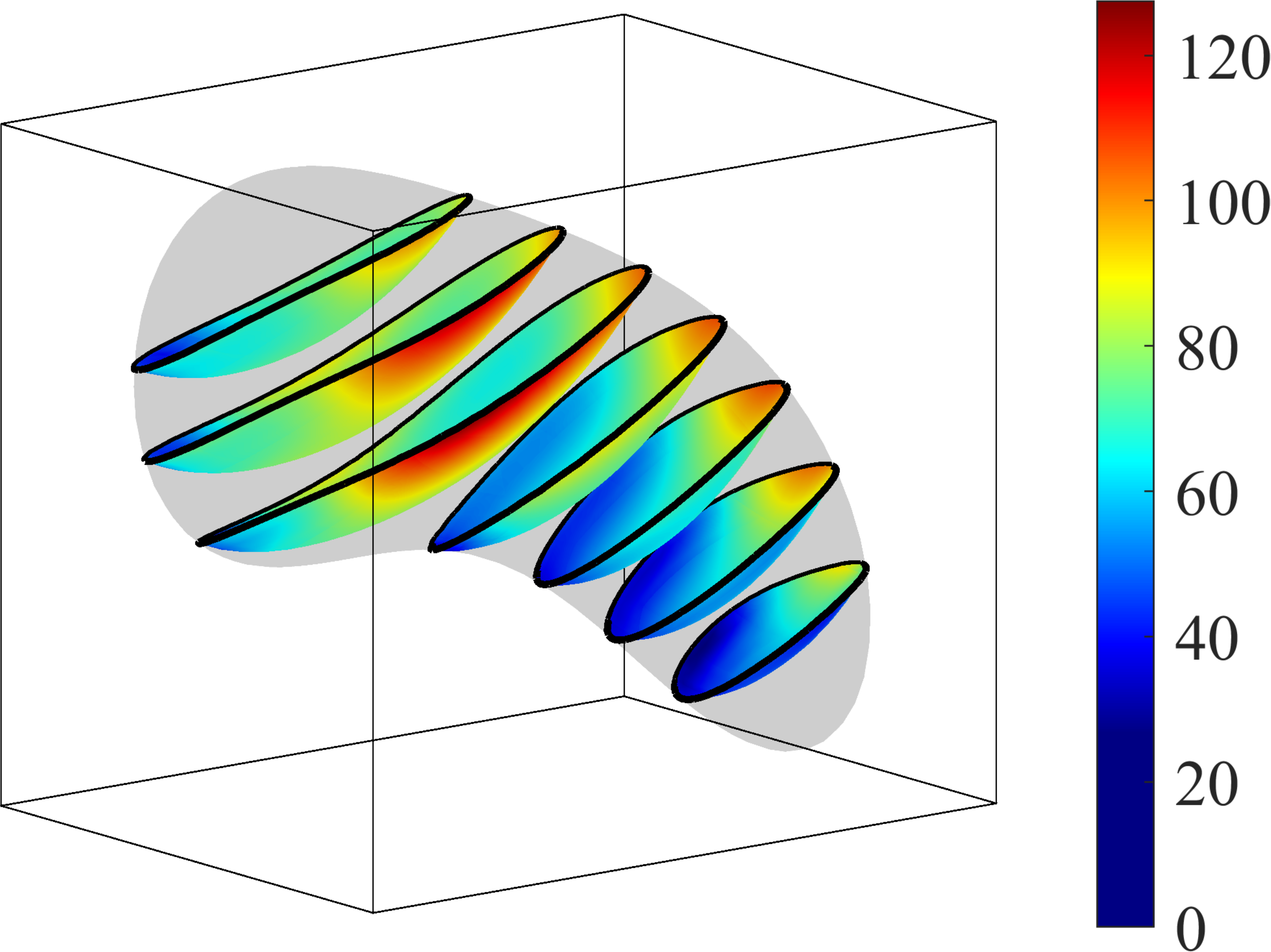}}\hfill\subfigure[$\sigma_{\mathrm{V}}$ on $\partial\Omega_{\vek{x}}\quad$]{\includegraphics[width=0.24\textwidth]{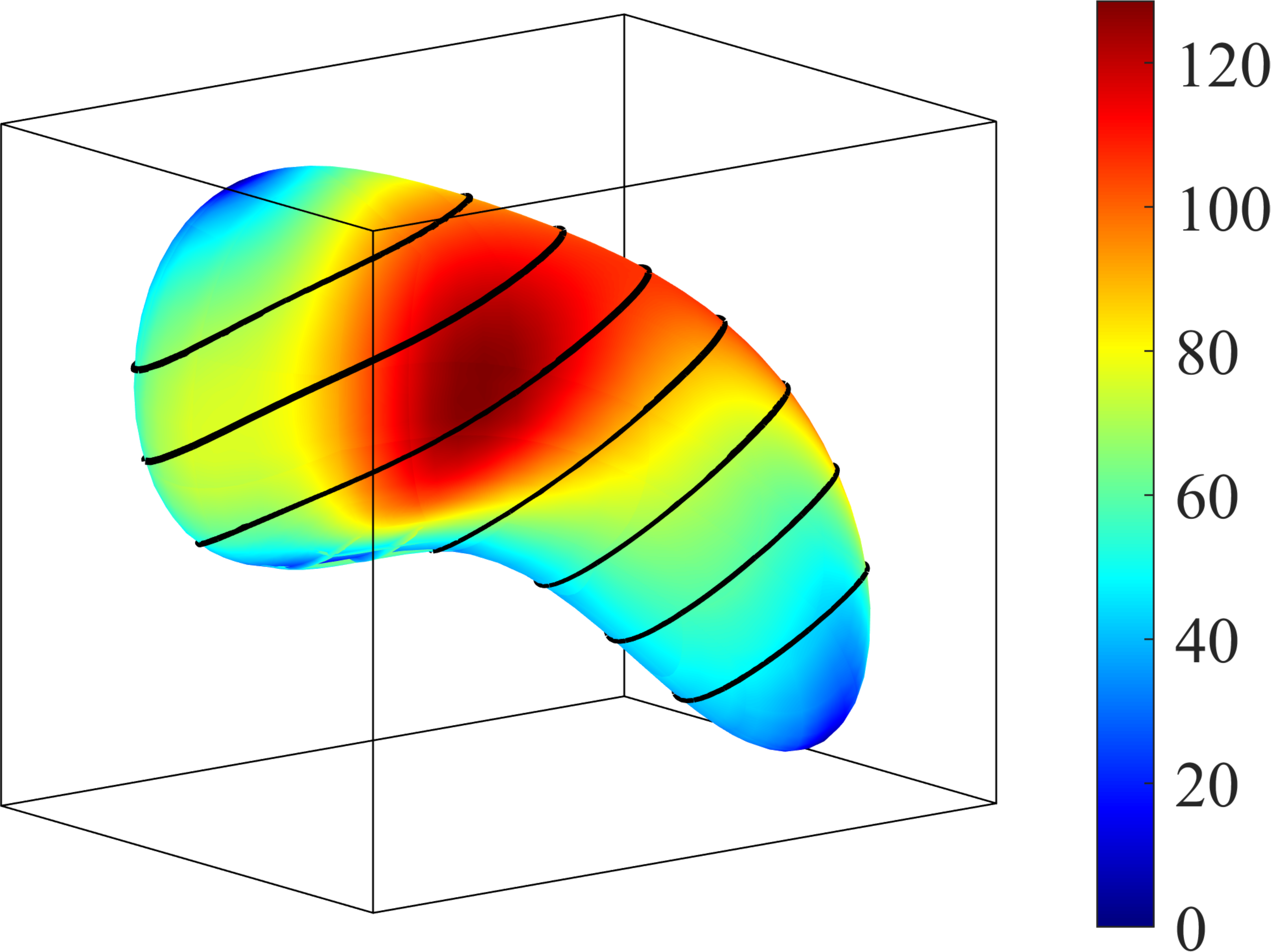}}

\caption{\label{fig:vonMisesStress}Von Mises stress $\sigma_{\mathrm{V}}$
in a two and three-dimensional example related to Fig.~\ref{fig:Mechanics};
(a) and (c) show $\sigma_{\mathrm{V}}$ on many level sets $\ManDef$,
(b) and (d) plot $\sigma_{\mathrm{V}}$ for \emph{all} level sets
in the bulk domain \emph{at once}, resulting in the smooth fields.}
\end{figure}

The usual formulas may be used to compute the (scalar) von Mises stress
$\sigma_{\mathrm{V}}$ from $\vek\sigma$ being of high relevance
in structural design and which is often used in visualizations later
on. It is noted that in the context presented here, one may either
plot results on individual level sets, or as smooth fields over the
bulk domains, see Fig.~\ref{fig:vonMisesStress}.

\subsection{Governing equations\label{XX_GoverningEquations}}

A crucial aspect of finite strain theory is that equilibrium is to
be fulfilled in the \emph{deformed} configuration which is expressed
in strong form as
\begin{equation}
\mathrm{div}_{\Gamma}\,\vek\sigma\!\left(\vek x\right)=-\vek f\!\left(\vek x\right)\quad\forall\vek x\in\Omega_{\vek x},\label{eq:EquilibriumDefConfig}
\end{equation}
where $\vek f$ are body forces. Recall from (\ref{eq:DivergenceTensor})
that $\mathrm{div}_{\Gamma}\,\vek\sigma=\nabla_{\vek x}^{\Gamma,\mathrm{dir}}\cdot\vek\sigma=\nabla_{\vek x}^{\Gamma,\mathrm{cov}}\cdot\vek\sigma$
is the divergence of the Cauchy stress tensor with respect to $\ManDef$.
Furthermore, we have the identity
\begin{equation}
\mathrm{Div}_{\Gamma}\,\mat K\left(\vek X\right)=\mathrm{div_{\Gamma}\,}\vek\sigma\!\left(\vek x\left(\vek X\right)\right)\cdot\Lambda\left(\vek X\right)\label{eq:EquilibriumDivRelation}
\end{equation}
with $\mathrm{Div}_{\Gamma}\,\mat K=\nabla_{\vek X}^{\Gamma,\mathrm{dir}}\cdot\mat K=\nabla_{\vek X}^{\Gamma,\mathrm{cov}}\cdot\mat K$
being the divergence of the first Piola-Kirchhoff stress tensor from
Eq.~(\ref{eq:1stPiolaKirchhoff}) with respect to $\ManUndef$. Due
to $\vek F\!\left(\vek X\right)=\vek f\!\left(\vek x\left(\vek X\right)\right)\cdot\Lambda\left(\vek X\right)$,
the equilibrium in $\ManDef$ can be stated equivalently to Eq.~(\ref{eq:EquilibriumDefConfig})
based on quantities in the undeformed configuration as
\begin{equation}
\mathrm{Div}_{\Gamma}\,\mat K\!\left(\vek X\right)=-\vek F\!\left(\vek X\right)\quad\forall\vek X\in\Omega_{\vek X}.\label{eq:EquilibriumUndefConfig}
\end{equation}

Finally, one may identify the field equations of the BVP modeling
the mechanics of membranes (and ropes) simultaneously on \emph{all}
level sets in some bulk domain through Eq.~(\ref{eq:CauchyGreenCov})
(kinematics), Eq.~(\ref{eq:2ndPiolaKirchhoff}) (constitutive equation),
and Eq.~(\ref{eq:EquilibriumUndefConfig}) (equilibrium). These equations
are formulated with respect to the undeformed configuration which
is often preferred for the numerical treatment (the undeformed domain
is discretized by a mesh once which is then used in the analysis without
any need for mesh updates). One may reformulate these three (vector)
field equations in the usual way to obtain only one (vector) field
equation for the sought displacements. Stress and strain tensors may
then be obtained as a post-processing step. At this stage, the governing
equations are given in strong form and formulated w.r.t.~the undeformed
configuration, to be fulfilled at every point $\vek X\in\Omega_{\vek X}$.
This shall later be converted to the weak form in the usual way to
enable an FEM-based analysis.

\subsection{Boundary conditions\label{XX_BoundaryConditions}}

Recall that $\vek Q$ and $\vek q$ are the conormal vectors on the
boundaries of the bounded level sets $\partial\ManUndef$ and $\partial\ManDef$,
respectively. $\partial\ManUndef$ and $\partial\ManDef$ each fall
into the two non-overlapping Dirichlet and Neumann parts $\left(\partial\ManUndefD,\;\partial\ManUndefN\right)$
and $\left(\partial\ManDefD,\;\partial\ManDefN\right)$. Then, the
boundary conditions in the deformed configuration are 
\begin{align}
\vek u(\vek x) & =\hat{\vek g}(\vek x)\ \text{on}\ \partial\ManDefD,\\
\vek\sigma\!\left(\vek x\right)\cdot\vek q(\vek x) & =\hat{\vek h}(\vek x)\ \text{on}\ \partial\ManDefN,
\end{align}
where $\hat{\vek g}$ are prescribed displacements and $\hat{\vek h}$
are tractions. The equivalent boundary conditions formulated in the
\emph{un}deformed configuration are 
\begin{align}
\vek u(\vek X) & =\hat{\vek G}(\vek X)\ \text{on}\ \partial\ManUndefD,\\
\mat K\!\left(\vek X\right)\cdot\vek Q(\vek X) & =\hat{\vek H}(\vek X)\ \text{on}\ \partial\ManUndefN,
\end{align}
where $\hat{\vek{G}}$ and $\hat{\vek{H}}$ have similar interpretations
as before, see \cite{Calladine_1983a,Fries_2020a} for further information.
We note that because $\partial\ManUndef$ coincides with the boundary
of the bulk domain $\partial\Omega_{\vek X}$, one may also write
$\partial\Omega_{\vek X,\mathrm{D}}$ and $\partial\Omega_{\vek X,\mathrm{N}}$
instead of $\partial\ManUndefD$ and $\partial\ManUndef$, respectively.
With the boundary conditions above, the complete second-order boundary
value problem (BVP) is defined.

\subsection{Optional coupling with elastic bulk domains\label{XX_ElasticBulkDomains}}

So far, the mechanical model outlined above focuses on the mechanics
of (curved) $\left(d-1\right)$-dimensional structures (ropes and
membranes) implied by the level sets in the bulk domain. The bulk
domain geometrically defines the region of interest, hence, it specifies
the considered level-set interval and associates a boundary to the
level sets, thus defining a continuous set of curved, bounded structures.
Later, the bulk domain is discretized by finite elements for the simultaneous
numerical analysis of the embedded curved structures, see Section
\ref{X_BulkTraceFEM}.

A particularly interesting application of the proposed framework is
to additionally equip the bulk domain with mechanical properties.
In the simplest case of a homogenous, isotropic, elastic bulk material,
this may, e.g., be characterized by a bulk Young's modulus $E_{b}$
and Poisson's number $\nu_{b}$. The ropes and membranes may then
be added to the bulk domain as homogenized, continuously embedded
sub-structures. In this case, the bulk domain does not only have a
geometrical but also a mechanical task. For the mechanical model of
the bulk domain in a finite strain context, we refer to existing text
books \cite{Belytschko_2000b,Holzapfel_2000a,Zienkiewicz_2000b}.
It is important to note that an elastic bulk domain alleviates the
rather strong assumptions on the validity of level sets in bulk domains
as outlined in Section \ref{XX_InvalidLevelSets}. Basically any smooth
level-set function may then be employed as the elastic properties
of the bulk domain enforces continuity in normal direction of the
level sets. In this case, the level sets are no longer completely
independent structures but cross-coupled through the bulk behavior.

The focus is now on adding sub-structures to the elastic bulk domain
as assumed in this section. First, consider a finite set of \emph{$n_{\mathrm{discr}}$
discrete} sub-structures implied by the level sets $\Gamma_{\!\!\vek X}^{c_{i}}$
for equally distributed values $c_{i}$ in the level-set interval
$\left[\phi^{\min},\phi^{\max}\right]$ of interest, hence, $c_{i}=\phi^{\min}+\frac{i-1}{n_{\mathrm{discr}}-1}\cdot\left(\phi^{\max}-\phi^{\min}\right)$,
$i=1,\dots,n_{\mathrm{discr}}$. The associated Young's modulus $E_{\mathrm{discr}}$
of each discrete sub-structure shall depend on $n_{\mathrm{discr}}$
so that the integrated stiffness of the bulk domain with embedded
discrete sub-structures yields some chosen target value 
\[
\int_{\Omega_{\vek X}}E_{b}\;\mathrm{d}\Omega+\sum_{i=1}^{n_{\mathrm{discr}}}\int_{\Gamma_{\!\!\vek X}^{c_{i}}}E_{\mathrm{discr}}\;\mathrm{d}\Omega=\mathfrak{E}_{\mathrm{int}},
\]
which is easily solved for $E_{\mathrm{discr}}$ with given $E_{b}$
and $\mathfrak{E}_{\mathrm{int}}$ when assuming constant $E_{b}$
and $E_{\mathrm{discr}}$. A unit analysis, in fact, shows that $E_{\mathrm{discr}}$
is rather the Young's modulus multiplied by an assumed unit thickness
of the membrane, so as to avoid the explicit introduction of the thickness
$t$ in the model. For $n_{\mathrm{discr}}\rightarrow\infty$, the
mechanical quantities such as displacements, stresses, and strains
converge to the same values as obtained with the homogenized, continuously
embedded sub-structures as proposed herein when choosing the Young's
modulus $E_{\mathrm{cont}}$ to fulfill
\[
\int_{\Omega_{\vek X}}E_{b}\;\mathrm{d}\Omega+\int_{\Omega_{\vek X}}E_{\mathrm{cont}}\cdot\left\Vert \nabla_{\!\!\vek X}\phi\right\Vert \;\mathrm{d}\Omega=\mathfrak{E}_{\mathrm{int}}.
\]
This is later confirmed in Section \ref{XX_Testcase5}. It is noted
that embedding individual discrete sub-structures in continua is described,
e.g., in \cite{Burman_2019b,Formaggia_2014a,Hansbo_2022a}. This often
introduces discontinuities in the physical fields, resulting in the
need for mesh refinements, which is not the case in the homogenized
model proposed here. We believe that this application of the proposed
framework has large potential in defining advanced material models
with embedded sub-structures with possible applications, e.g., in
textile, biomechanical or fiber-reinforced structures and composite
laminates. One may label related mechanical models as continuously
embedded sub-structure models or embedded, layered manifold models.

\section{The Bulk Trace FEM\label{X_BulkTraceFEM}}

\begin{figure}
\centering

\subfigure[Surface FEM]{\includegraphics[width=0.3\textwidth]{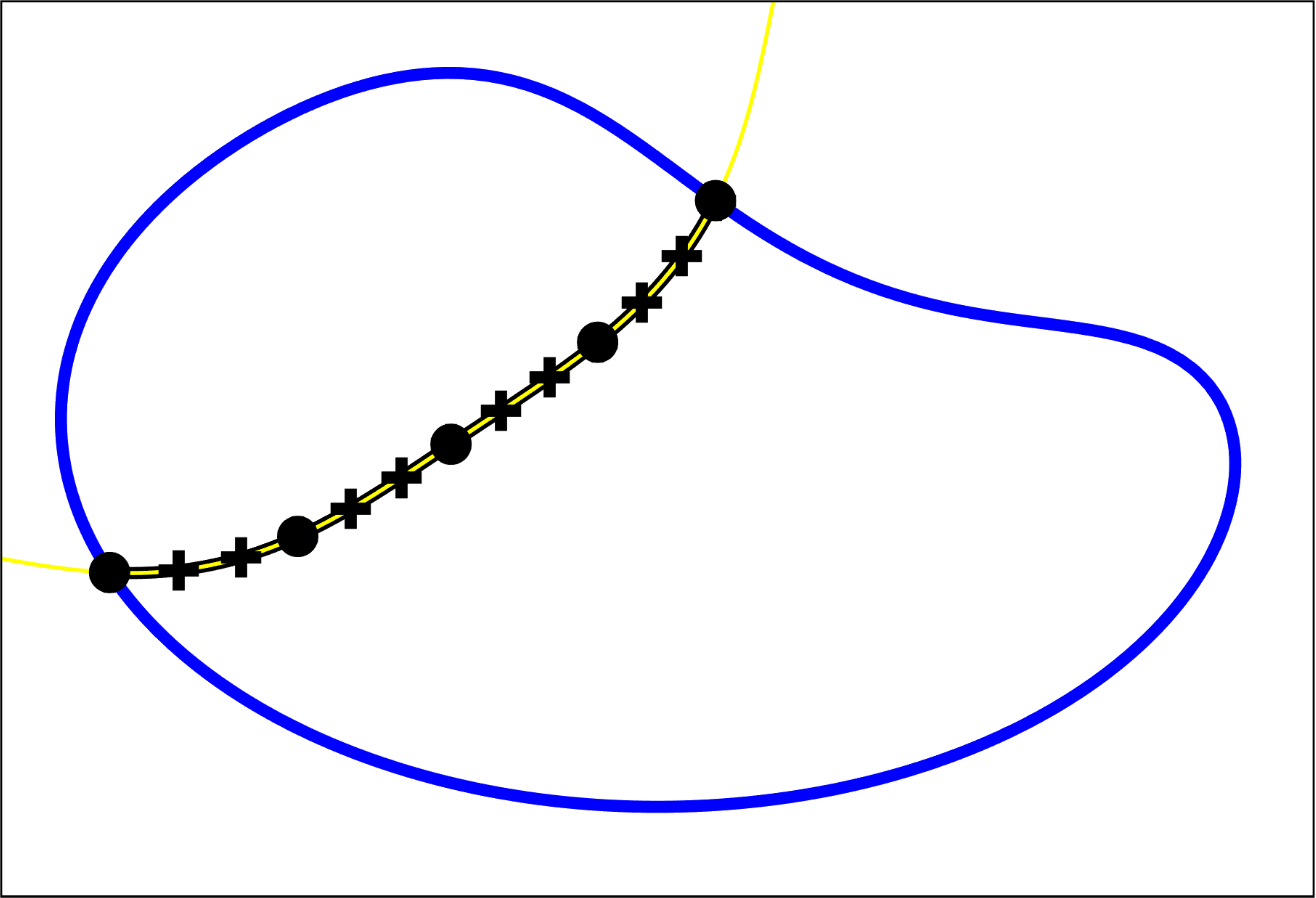}}\hfill\subfigure[Trace FEM]{\includegraphics[width=0.3\textwidth]{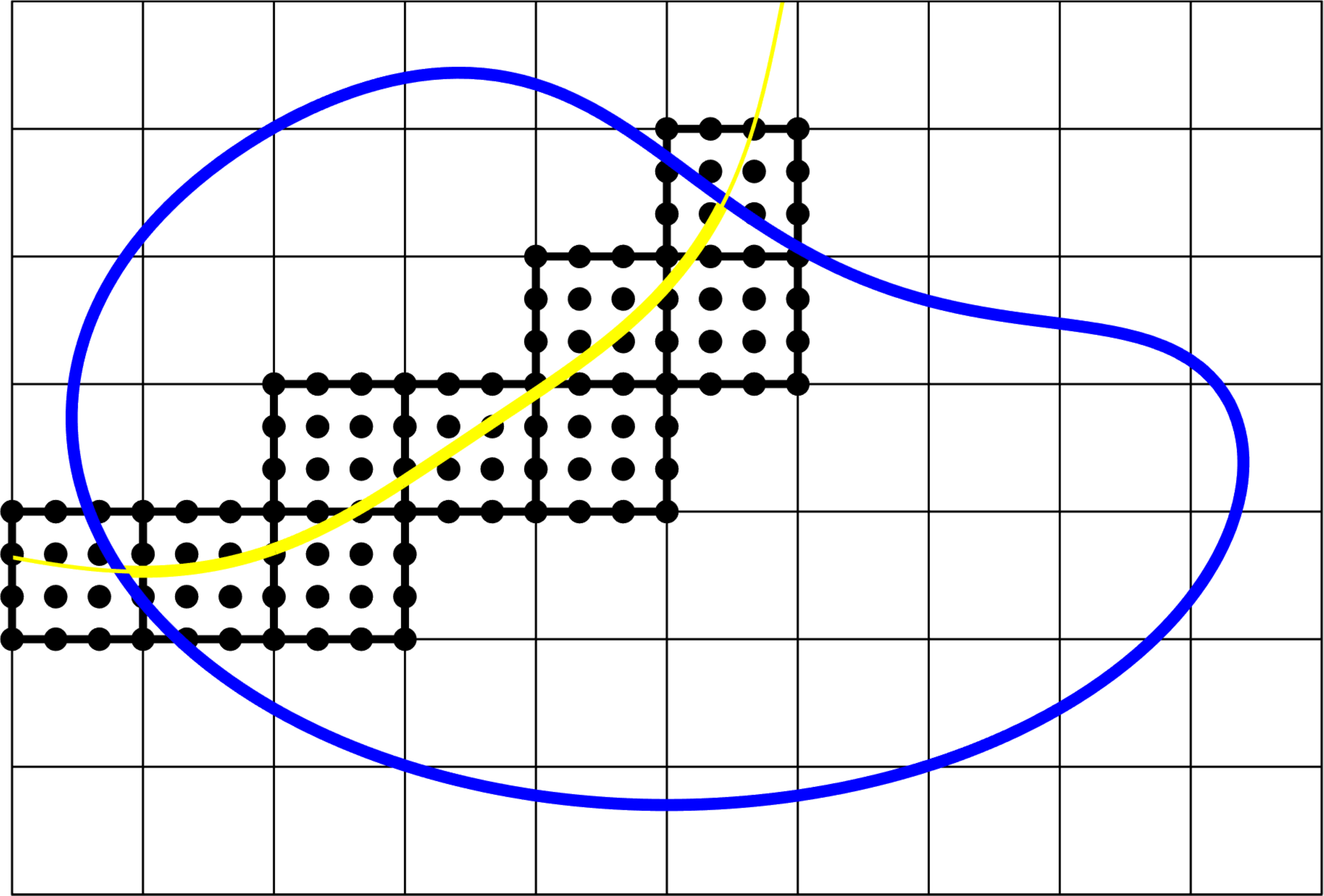}}\hfill\subfigure[Bulk Trace FEM]{\includegraphics[width=0.3\textwidth]{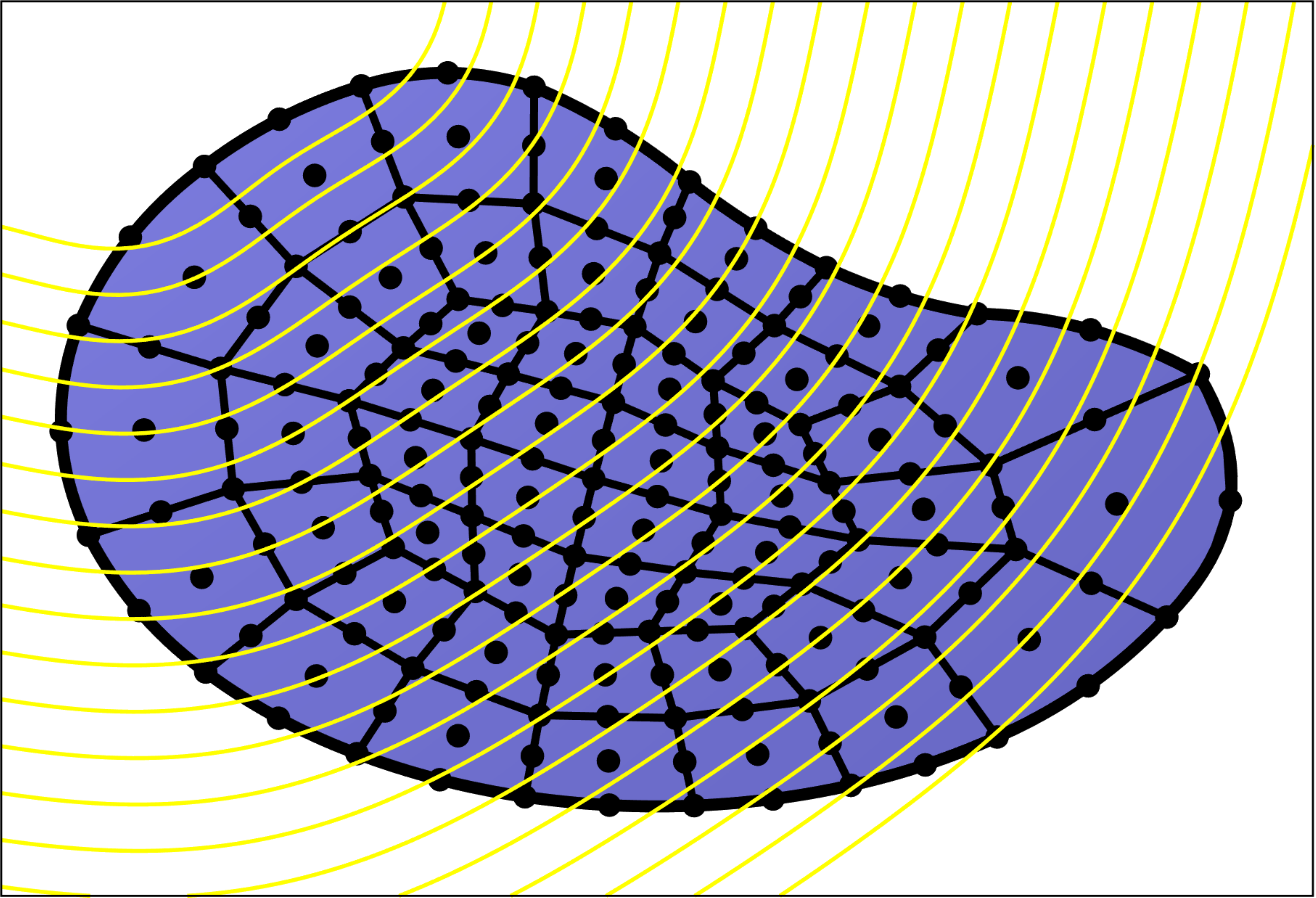}}

\subfigure[Surface FEM]{\includegraphics[width=0.3\textwidth]{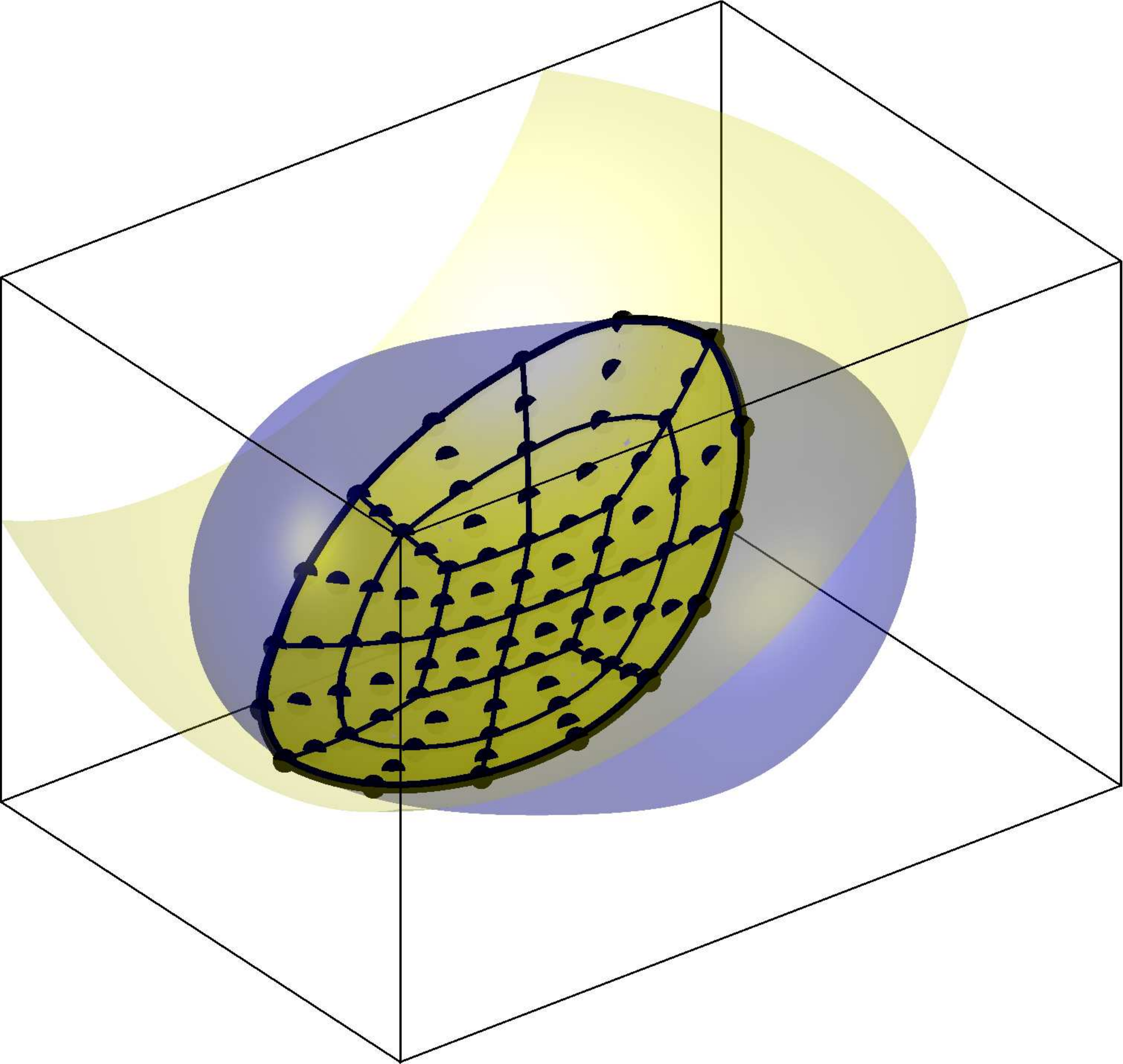}}\hfill\subfigure[Trace FEM]{\includegraphics[width=0.3\textwidth]{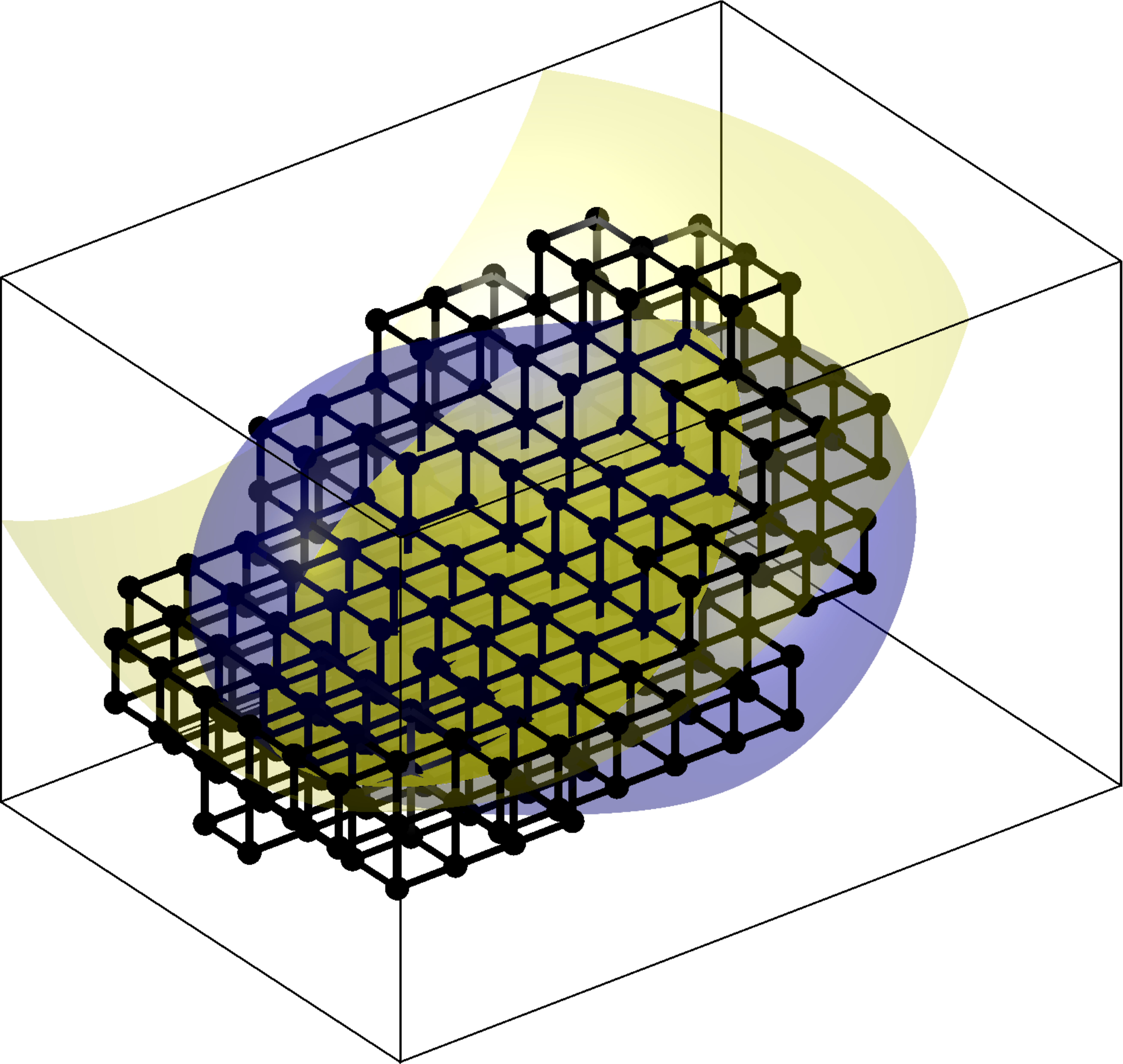}}\hfill\subfigure[Bulk Trace FEM]{\includegraphics[width=0.3\textwidth]{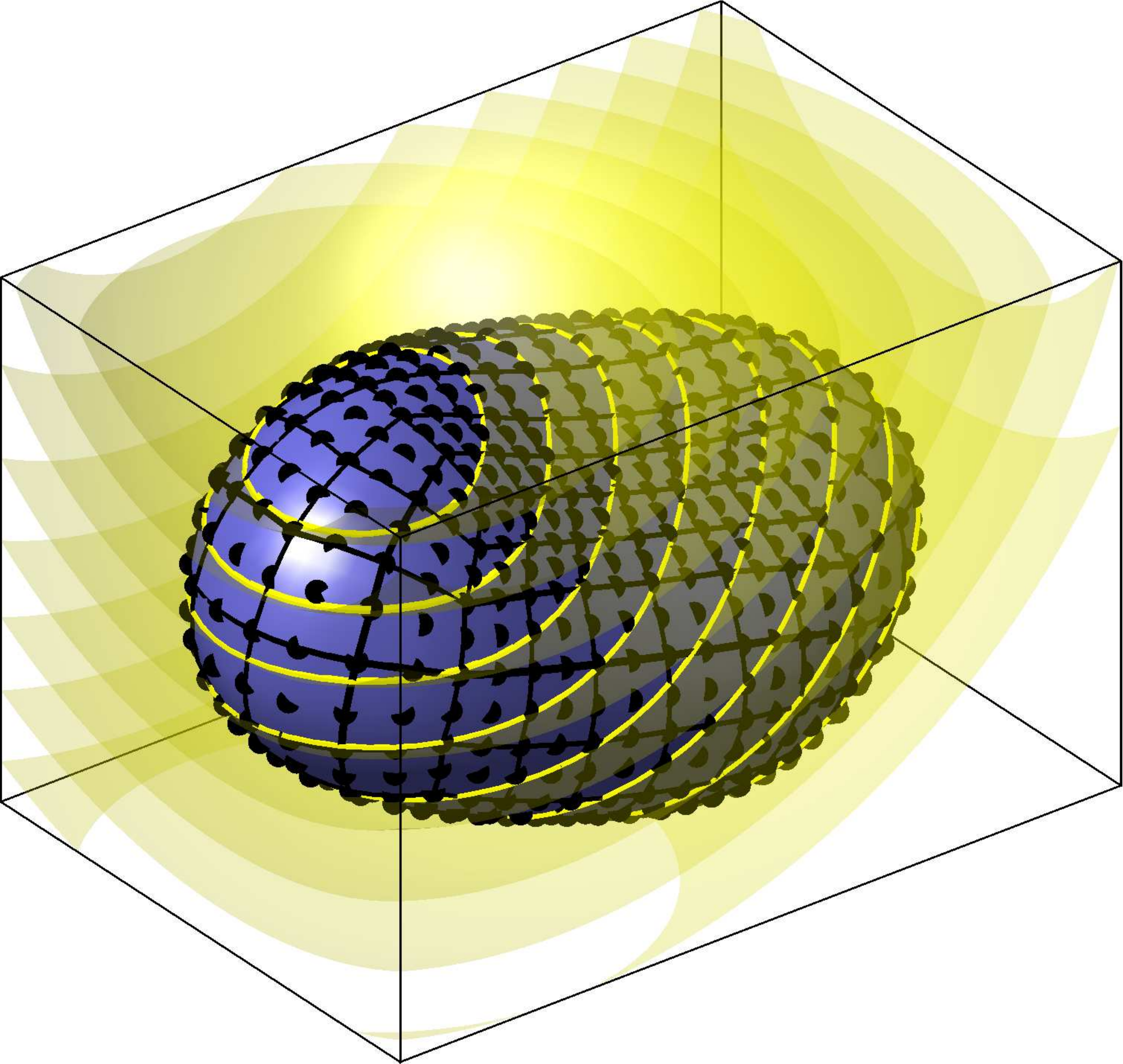}}

\caption{\label{fig:VariantsFEM}Variants of the FEM for BVPs on curved lines
((a) to (c)) and surfaces ((d) to (f)) with focus on the employed
meshes; note that the number and order of the elements is only examplarily.
The classical Surface FEM applies to (a) and (d) where the manifold
is discretized by a conforming mesh, (b) and (e) is the Trace FEM
where the zero-level set is the domain of interest and the cut elements
of a background mesh are used for the discretization, (c) and (f)
is the proposed Bulk Trace FEM where the bulk domain is discretized
and BVPs on all level sets are considered simultaneously.}
\end{figure}

For the numerical analysis of lower-dimensional structures embedded
in some exterior space, the classical approach is to discretize the
(curved) structure using some conforming surface mesh and approximate
the BVP posed with respect to \emph{one} geometry of interest, see
Figs.~\ref{fig:VariantsFEM}(a) and (d). Then, the BVP is typically
defined using curvilinear coordinates as they immediately result from
using the surface elements in the mesh \cite{Bischoff_2017a,Calladine_1983a,Ciarlet_1997a,Chapelle_2011a},
implying a parametrization of the surface. This approach may be called
the Surface FEM. An alternative numerical treatment would be to define
\emph{one} geometry of interest implicitly through the zero-level
set of a scalar function and use an immersing, non-conforming, background
mesh with the dimensionality $d$ of the embedding space. Only the
elements cut by the zero-level set are considered in the analysis,
see Figs.~\ref{fig:VariantsFEM}(b) and (e). The resulting method
was labeled Trace FEM, see, e.g., \cite{Olshanskii_2009b,Olshanskii_2017a,Gross_2018a,Grande_2018a},
and is a fictitious domain method where the numerical integration
in the cut elements \cite{Fries_2015a,Fries_2017c,Cheng_2009a,Abedian_2013a,Moumnassi_2011a,Mueller_2013a},
stabilization \cite{Grande_2016a,Olshanskii_2017a,Gross_2018a}, and
the enforcement of boundary conditions \cite{Burman_2012a,Burman_2012b,Schillinger_2016a,Mendez_2004a,Ruess_2013a}
are crucial aspects for the success of the approximations.

For the simultaneous analysis of structures implied by \emph{all}
level sets over a bulk domain as proposed herein, the bulk domain
$\Omega_{\vek X}$ is discretized by a background mesh which is conforming
to the boundary $\partial\Omega_{\vek X}$ (and, hence, also $\partial\ManUndef$),
yet typically \emph{not} conforming to the individual level sets $\ManUndef$
of $\phi$, see Figs.~\ref{fig:VariantsFEM}(c) and (f). The method
is similar to the Trace FEM in that it uses test and trial functions
for the numerical analysis which are implied by the $d$-dimensional
background mesh. However, no special numerical integration and stabilization
is needed (as there are no cut elements), and boundary conditions
are enforced as in the Surface FEM. Therefore, the numerical method
employed here, and closely related to \cite{Dziuk_2013a,Dziuk_2008a,Dziuk_2008b,Burger_2009a}
for transport problems, may be seen as a hybrid between the Surface
and Trace FEM. Hence, we label this approach the Bulk Trace FEM.

It is noted that formulating the mechanical models for ropes and membranes
based on the tangential differential calculus using surface operators
as in the previous section results in a unified and general description
which may be used in all the resulting variants of the FEM with only
minor adaptions. This was already emphasized in \cite{Schoellhammer_2019a,Schoellhammer_2019b,Fries_2020a}
in the context of the Surface and Trace FEM and is herein confirmed
for the proposed Bulk Trace FEM.

\subsection{Governing equations in weak form\label{XX_WeakForm}}

For any FEM simulations, it is necessary to state the governing equations
in weak form. Therefore, the following test and trial function spaces
are introduced, 
\begin{align}
\mathcal{S}_{\vek u} & =\left\lbrace \vek v\in\left[\mathcal{H}^{1}(\Omega_{\vek X})\right]^{d}:\ \vek v=\hat{\vek G}\ \text{on}\ \partial\Omega_{\vek X,\mathrm{D}}\right\rbrace ,\label{eq:TrialFctSpaceCont}\\
\mathcal{V}_{\vek u} & =\left\lbrace \vek v\in\left[\mathcal{H}^{1}(\Omega_{\vek X})\right]^{d}:\ \vek v=\vek0\ \text{on}\ \partial\Omega_{\vek X,\mathrm{D}}\right\rbrace ,\label{eq:TestFctSpaceCont}
\end{align}
where $\mathcal{H}^{1}$ is the Sobolev space of functions with square
integrable first derivatives. The task is to find $\vek u\in\mathcal{S}_{\vek u}$
such that for all $\vek w\in\mathcal{V}_{\vek u}$, there holds
\begin{align}
\int_{\Omega_{\vek X}}\nabla_{\vek X}^{\Gamma,\mathrm{dir}}\vek w:\mat K\left(\vek u\right)\cdot\left\Vert \nabla_{\!\!\vek X}\phi\right\Vert \,\mathrm{d}\Omega & =\int_{\Omega_{\vek X}}\!\!\!\vek w\cdot\vek F\cdot\left\Vert \nabla_{\!\!\vek X}\phi\right\Vert \,\mathrm{d}\Omega\label{eq:WeakFormCont}\\
 & +\int_{\partial\Omega_{\vek X,\text{N}}}\!\!\!\!\!\!\vek w\cdot\hat{\vek H}\cdot\left(\vek Q\cdot\vek M\right)\cdot\left\Vert \nabla_{\!\!\vek X}\phi\right\Vert \;\mathrm{d}\partial\Omega
\end{align}

This weak form is obtained after the following sequence of steps:
Multiply Eq.~(\ref{eq:EquilibriumUndefConfig}) with test functions
$\vek w$$\left(\vek X\right)$ and integrate over the level sets
in some interval $\int_{\phi^{\min}}^{\phi^{\max}}\int_{\ManUndef}\square\,\ensuremath{\mathrm{d\ensuremath{\Gamma}}}\,\ensuremath{\mathrm{d}c}$
which, through the co-area formula (\ref{eq:CoareaFormulaDomain})
is converted into an integral over the bulk domain $\int_{\Omega_{\vek X}}\square\cdot\left\Vert \nabla_{\!\!\vek X}\phi\right\Vert \,\ensuremath{\mathrm{d\ensuremath{\Omega}}}$.
Then, the divergence theorem for tensors (\ref{eq:DivTheoremTensorBTF})
is applied to obtain (\ref{eq:WeakFormCont}), noting that the curvature
term vanishes due to $\mat K\cdot\vek N=\vek0$. A similar weak form
could be posed over the deformed configuration which, however, would
be less desirable for the numerical analysis (as it would require
mesh updates during the iterative procedure to solve this non-linear
problem).

\subsection{Discretized weak form\label{XX_DiscreteWeakForm}}

Let there be a discretization of the bulk domain, $\Omega_{\vek X}^{h}$,
in the form of a conforming, $d$-dimensional mesh, herein composed
by higher-order elements of the Lagrange type with equally spaced
nodes in the reference element (with arbitrary element types such
as triangular/quadrilateral in 2D, or tetrahedral/hexahedral in 3D).
The nodal coordinates in the undeformed configuration are labeled
$\vek X_{i}$ with $i=1,\dots,n_{q}$ and $n_{q}$ being the number
of nodes in the mesh, see Figs.~\ref{fig:VariantsFEM}(c) and (f)
for generic examples. The resulting nodal basis functions $\ShapeFcts\!\left(\vek X\right)$,
used as test and trial functions, span a $C^{0}$-continuous finite
element space as 
\begin{align}
\mathcal{Q}_{\Omega_{\vek X}}^{h}:=\left\lbrace v_{h}\in C^{0}(\Omega_{\vek X}^{h}):\ v_{h}=\sum_{i=1}^{n_{q}}\ShapeFcts(\vek X)\cdot\hat{v}_{i}\text{ with }\hat{v}_{i}\in\mathbb{R}\right\rbrace \subset\mathcal{H}^{1}(\Omega_{\vek X}^{h})\ .\label{eq:SurfaceFEMFctSpace}
\end{align}
$\ShapeFcts(\vek X)$ are obtained from isoparametric mappings from
the $d$-dimensional reference element to the physical elements. Furthermore,
the level-set function $\phi$ is replaced by its interpolation $\phi_{h}\left(\vek X\right)\in\mathcal{Q}_{\Omega_{\vek X}}^{h}$
with prescribed nodal values $\hat{\phi}_{i}=\phi\left(\vek X_{i}\right)$.
Based on Eq.~(\ref{eq:SurfaceFEMFctSpace}), the following discrete
test and trial function spaces are introduced 
\begin{align}
\mathcal{S}_{\Omega_{\vek X}}^{h} & =\left\lbrace \vek v_{h}\in\left[\mathcal{Q}_{\Omega_{\vek X}}^{h}\right]^{d}:\ \vek v_{h}=\hat{\vek G}\ \text{on}\ \partial\Omega_{\vek X,\mathrm{D}}^{h}\right\rbrace ,\label{eq:TrialFctSpaceDiscr}\\
\mathcal{V}_{\Omega_{\vek X}}^{h} & =\left\lbrace \vek v_{h}\in\left[\mathcal{Q}_{\Omega_{\vek X}}^{h}\right]^{d}:\ \vek v_{h}=\vek0\ \text{on}\ \partial\Omega_{\vek X,\mathrm{D}}^{h}\right\rbrace .\label{eq:TestFctSpaceDiscr}
\end{align}
The discrete weak form of Eq.~(\ref{eq:WeakFormCont}) reads as follows:
Given the level-set function $\phi_{h}$, Lam\'e constants $(\lambda,\mu)\in\mathbb{R}^{+}$,
body forces $\vek F\in\mathbb{R}^{d}$ on $\Omega_{\vek X}^{h}$,
tractions $\hat{\vek H}\in\mathbb{R}^{d}$ on $\partial\Omega_{\vek X,\text{N}}^{h}$,
find the displacement field $\vek u_{h}\in\mathcal{S}_{\Omega_{\vek X}}^{h}$
such that for all test functions $\vek w_{h}\in\mathcal{V}_{\Omega_{\vek X}}^{h}$
there holds in $\Omega_{\vek X}^{h}$:
\begin{align}
\int_{\Omega_{\vek X}^{h}}\nabla_{\vek X}^{\Gamma,\mathrm{dir}}\vek w_{h}:\mat K\left(\vek u_{h}\right)\cdot\left\Vert \nabla_{\!\!\vek X}\phi_{h}\right\Vert \,\mathrm{d}\Omega= & \int_{\Omega_{\vek X}^{h}}\!\!\!\vek w_{h}\cdot\vek F\cdot\left\Vert \nabla_{\!\!\vek X}\phi_{h}\right\Vert \,\mathrm{d}\Omega\label{eq:WeakFormDiscr}\\
 & +\int_{\partial\Omega_{\vek X,\text{N}}^{h}}\!\!\!\!\!\!\vek w_{h}\cdot\hat{\vek H}\cdot\left(\vek Q_{h}\cdot\vek M_{h}\right)\cdot\left\Vert \nabla_{\!\!\vek X}\phi_{h}\right\Vert \;\mathrm{d}\partial\Omega.
\end{align}
For brevity, we avoid adding an $h$ also to the discretized differential
operators, e.g., we do not write $\nabla_{\!\!\vek X}^{h}$. The sought
discrete displacement field $\vek u_{h}(\vek X)$ is obtained solving
a non-linear system of equations for the $n_{\text{DOF}}=d\cdot n_{q}$
nodal values (degrees of freedom) as usual in the context of finite
strain theory.

\subsection{Technical aspects\label{XX_TechnicalAspects}}

For the implementation of the Bulk Trace FEM, some useful hints are
given next. It is found beneficial to split the implementation into
an application-independent part related to finite element technology
and the part which is related to the concrete BVP of interest, i.e.,
the application. The first part may be reused in any other Bulk Trace
FEM simulation, e.g., in the context of solving transport problems
on all level sets in a bulk domain. Of course, the situation in the
reference elements is completely identical to any other FE application
and not further detailed here. As such, a standard set of integration
points and element functions, being related to the type and order
of the \emph{reference} element, is given as a starting point. Let
the reference element feature the nodal coordinates $\vek r_{i}$
and element functions $\ShapeFcts\left(\vek r\right)$ with $i=1,\dots n^{\mathrm{npe}}$
and $n^{\mathrm{npe}}$ being the number of nodes per element. A proper
Gauss quadrature rule, related to the order of the reference element,
is defined by a set of $n^{\mathrm{int}}$ integration points $\vek r_{i}^{\mathrm{int}}$
and weights $w_{i}^{\mathrm{int,ref}}$ with $i=1,\dots n^{\mathrm{int}}$.

The mapping from the reference element to any of the \emph{physical}
elements in the undeformed configuration is given by the isoparametric
concept, $\vek X\!\left(\vek r\right)=\sum_{i=1}^{n^{\mathrm{npe}}}B_{i}\!\left(\vek r\right)\cdot\vek X_{i}^{\mathrm{el}}$
where $\vek X_{i}^{\mathrm{el}}$ are the respective nodal element
coordinates. In every element, this results in mapped integration
points $\vek X_{i}^{\mathrm{int}}$ with modified integration weights
$w_{i}^{\mathrm{int,phys}}$. In a standard FE context, $w_{i}^{\mathrm{int,phys}}=w_{i}^{\mathrm{int,ref}}\cdot\det\mat J\left(\vek r_{i}^{\mathrm{int}}\right)$
where $\mat J=\nicefrac{\partial\vek X}{\partial\vek r}$ is the Jacobi
matrix of the finite element map. However, because all domain integrals
in Eq.~(\ref{eq:WeakFormDiscr}) involve the term $\left\Vert \nabla_{\!\!\vek X}\phi_{h}\right\Vert $,
it is useful to also modify the weights according to this term as
well, hence,
\begin{align*}
\int_{\Omega_{\vek X}^{\mathrm{el}}}f\left(\vek X\right)\cdot\left\Vert \nabla_{\!\!\vek X}\phi_{h}\left(\vek X\right)\right\Vert \,\mathrm{d}\Omega\approx & \;\sum_{i=1}^{n^{\mathrm{int}}}f\left(\vek X_{i}^{\mathrm{int}}\right)\cdot w_{i}^{\mathrm{int,phys}},\\
\text{with }w_{i}^{\mathrm{int,phys}}= & \;w_{i}^{\mathrm{int,ref}}\cdot\left\Vert \nabla_{\!\!\vek X}\phi_{h}\left(\vek X_{i}^{\mathrm{int}}\right)\right\Vert \cdot\det\left(\mat J\left(\vek r_{i}^{\mathrm{int}}\right)\right),\\
\text{and \ensuremath{\nabla_{\!\!\vek X}\phi_{h}\left(\vek X(\vek r_{i}^{\mathrm{int}})\right)}}= & \;\mat J^{-\mathrm{T}}\!\left(\vek r_{i}^{\mathrm{int}}\right)\cdot\nabla_{\!\vek r}\phi_{h}\left(\vek r_{i}^{\mathrm{int}}\right).
\end{align*}

It is also found useful to apply some standard differential operators
such as the gradient operator to the test and trial functions in the
FE analysis which may also be recycled in various applications of
the Bulk Trace FEM. For example, when the gradients $\nabla_{\!\vek X}\ShapeFcts$
are precomputed, it is trivial to specify the gradient of some (scalar)
approximation as $\nabla_{\!\vek X}u_{h}=\sum_{i}\nabla_{\!\vek X}\ShapeFcts\cdot\hat{u}_{i}$.
Of course, in the reference element, differential operators with respect
to the coordinates $\vek r$ may easily be applied to the element
functions, e.g., to generate $\nabla_{\vek r}\ShapeFcts\left(\vek r\right)$.
Upon mapping these functions to the (undeformed) physical elements
and seeking the standard (bulk) gradient with respect to the coordinates
$\vek X$, we have $\nabla_{\vek X}\ShapeFcts\left(\vek X\right)=\mat J^{-\mathrm{T}}\left(\vek r\right)\cdot\nabla_{\vek r}\ShapeFcts\left(\vek r\right)$
as usual. As discussed in Section \ref{X_Preliminaries}, in the Bulk
Trace FEM, differential operators with respect to the level sets of
$\phi$ play a crucial role. For the surface gradient of the element
functions $\ShapeFcts$, we have
\begin{align*}
\nabla_{\!\vek X}^{\Gamma}\ShapeFcts\left(\vek X\right)= & \;\mat P\left(\vek X\right)\cdot\nabla_{\!\vek X}\ShapeFcts\left(\vek X\right),\\
= & \;\mat P\left(\vek X\right)\cdot\mat J^{-\mathrm{T}}\left(\vek r\right)\cdot\nabla_{\vek r}\ShapeFcts\left(\vek r\right),\\
\text{with }\mat P= & \;\mat I-\vek N\otimes\vek N,\;\vek N=\frac{\vek N^{\star}}{\left\Vert \vek N^{\star}\right\Vert },\;\vek N^{\star}=\nabla_{\!\vek X}\phi_{h},
\end{align*}
cf.~Table \ref{tab:ImplicitManifolds}. With these special integration
points and differential operators according to the proposed Bulk Trace
FEM, it is noteworthy that for the application-dependent (element-by-element)
integration of the weak form (\ref{eq:WeakFormDiscr}), one may employ
the same routines as for a classical implementation of finite strain
theory in $d$ dimensions. This largely simplifies the implementation
and enables a good software design.

\section{Numerical results\label{X_NumericalResults}}

Test cases for the simultaneous analysis of ropes and membranes in
bulk domains are presented in this section. In order to confirm the
success of the proposed method, different error measures are evaluated
in the deformed bulk domains, namely measuring (i) the integrated
length/area of the level sets, (ii) the error in the stored elastic
energy, and (iii) the residual error. 

For the error in the integrated level sets $\varepsilon_{\phi}$,
one integrates the length/area of the deformed structures (i.e., of
the level sets in the deformed bulk domain $\Omega_{\vek x}$), directly
based on the co-area formula in Eq.~(\ref{eq:CoareaFormulaDomain})
and compares it with the analytical one:
\begin{equation}
\varepsilon_{\phi}=\left|\mathfrak{D}\left(\vek u\right)-\mathfrak{D}\left(\vek u_{h}\right)\right|,\;\text{with }\;\mathfrak{D}\left(\vek u\right)=\int_{\Omega_{\vek x}}\left\Vert \nabla_{\!\!\vek x}\phi\right\Vert \,\mathrm{d}\Omega.\label{eq:LevelSetError}
\end{equation}
The ``stored energy error'' $\varepsilon_{\mathfrak{e}}$, see,
e.g., \cite[p.~229]{Zienkiewicz_2013a}, compares the approximated
stored elastic energy of all deformed structures in $\Omega_{\vek x}$
with the analytical one,
\begin{equation}
\varepsilon_{\mathfrak{e}}=\left|\mathfrak{e}\left(\vek u\right)-\mathfrak{e}\left(\vek u_{h}\right)\right|,\label{eq:EnergyError}
\end{equation}
with 
\begin{eqnarray}
\mathfrak{e}\left(\vek u\right) & = & \frac{1}{2}\int_{\Omega_{\vek x}}\mat e_{\mathrm{tang}}\left(\vek u\right):\vek\sigma\left(\vek u\right)\cdot\left\Vert \nabla_{\!\!\vek x}\phi\right\Vert \,\mathrm{d}\Omega,\label{eq:EnergyDef}\\
 & = & \frac{1}{2}\int_{\Omega_{\vek X}}\mat E_{\mathrm{tang}}\left(\vek u\right):\mat S\left(\vek u\right)\cdot\left\Vert \nabla_{\!\!\vek X}\phi\right\Vert \,\mathrm{d}\Omega.\label{eq:EnergyUndef}
\end{eqnarray}

The ``stored energy error'' used here is not to be mixed with the
classical energy error norm, see, e.g., \cite[p.~494]{Zienkiewicz_2013a}.
The two error measures (\ref{eq:LevelSetError}) and (\ref{eq:EnergyError})
require the knowledge of the analytical displacement field $\vek u\left(\vek X\right)$.
One common approach would be to generate manufactured solutions. However,
$\mathfrak{D}\left(\vek u\right)$ and $\mathfrak{e}\left(\vek u\right)$
may also be obtained by overkill approximations based on extremely
fine meshes with elements of very high order, which is the path chosen
herein. The respective numbers are given with large accuracy in the
respective test cases below, so that they may also serve as benchmark
values. Provided that geometry and boundary conditions allow for sufficiently
smooth solutions, the expected convergence rates in these two error
norms are $p+1$ with $p$ being the order of the elements. Note that
this order does not only depend on the integrand but also on the accuracy
of the numerical integration and the ability of the finite elements
to represent arbitrary yet smooth geometries. When using the two error
measures (\ref{eq:LevelSetError}) and (\ref{eq:EnergyError}) adapted
to a standard two- or three-dimensional finite strain context \cite{Belytschko_2000b,Holzapfel_2000a,Zienkiewicz_2000b},
it is actually found that odd element orders converge with $p+1$
and even orders with $p+2$. These orders are also confirmed in the
present mechanical context in the numerical test cases below.

The ``residual error'' $\varepsilon_{\mathrm{res}}$ integrates
the error in the equilibrium as stated in Eq.~(\ref{eq:EquilibriumDefConfig}),
that is,
\begin{align}
\varepsilon_{\mathrm{res}}=\; & \sqrt{\sum_{i=1}^{n_{\mathrm{el}}}\int_{\Omega_{\vek x}^{\mathrm{el},\,i}}\mathfrak{r}\left(\vek u_{h}\right)\cdot\mathfrak{r}\left(\vek u_{h}\right)\cdot\left\Vert \nabla_{\!\!\vek x}\phi\right\Vert \,\mathrm{d}\Omega},\label{eq:ResidualError}\\
\text{with }\;\mathfrak{r}\left(\vek u_{h}\right)=\; & \mathrm{div}_{\Gamma}\,\vek\sigma\!\left(\vek u_{h}\right)+\vek f\!\left(\vek x\right).
\end{align}
This error obviously vanishes for the analytical solution. Note that
the integrand in (\ref{eq:ResidualError}) involves second-order derivatives,
therefore, the integral must not be carried out over the whole (discretized)
domain $\Omega_{\vek x}$ but integrated element by element as indicated
by the summation. That is, element boundaries, where already the first
derivatives of the $C^{0}$-continuous shape functions feature jumps,
are neglected in computing $\varepsilon_{\mathrm{res}}$. Due to the
presence of second-order derivatives, the expected convergence rates
are $p-1$ which also indicates that higher-order elements are crucial
for convergence in $\varepsilon_{\mathrm{res}}$. Similar studies
on energy and residual errors have been conducted by the authors in
\cite{Fries_2020a,Schoellhammer_2019a,Schoellhammer_2019b} in the
context of the Surface and Trace FEM.

It is also noted that reference values for the stored energy and integrated
level sets may be computed using the classical Surface FEM and numerically
integrating in the interval $\left[\phi_{\min},\phi_{\max}\right]$
according to the analogy shown in Eq.~(\ref{eq:NumericalIntOverManifolds}).
Thereby, one may confirm with high reliability the success of the
proposed Bulk Trace FEM.

\subsection{Test case 1 in 2D: Circular bulk domain\label{XX_Testcase1}}

\begin{figure}
\centering

\subfigure[setup]{\includegraphics[width=0.2\textwidth]{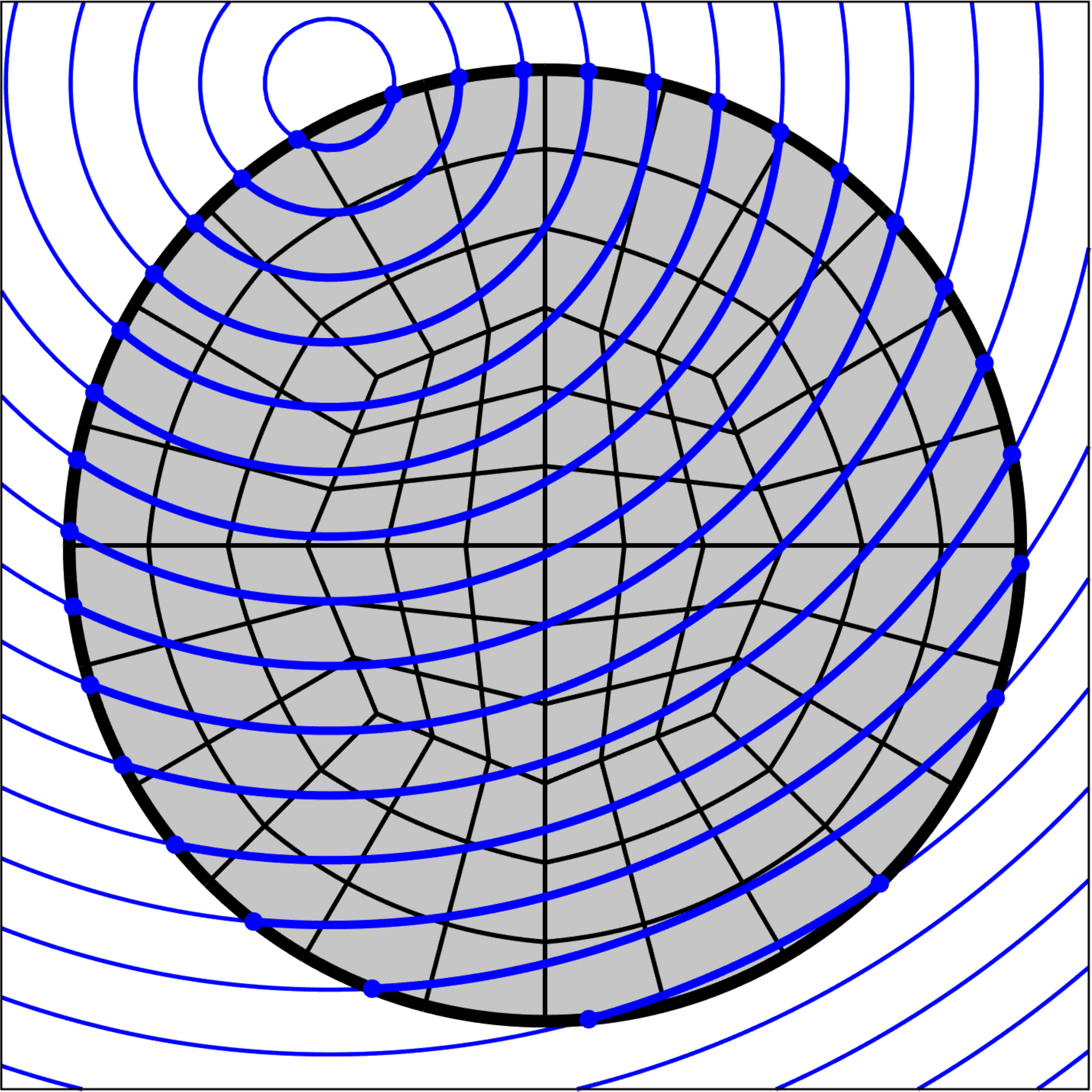}}\hfill\subfigure[result]{\includegraphics[width=0.23\textwidth]{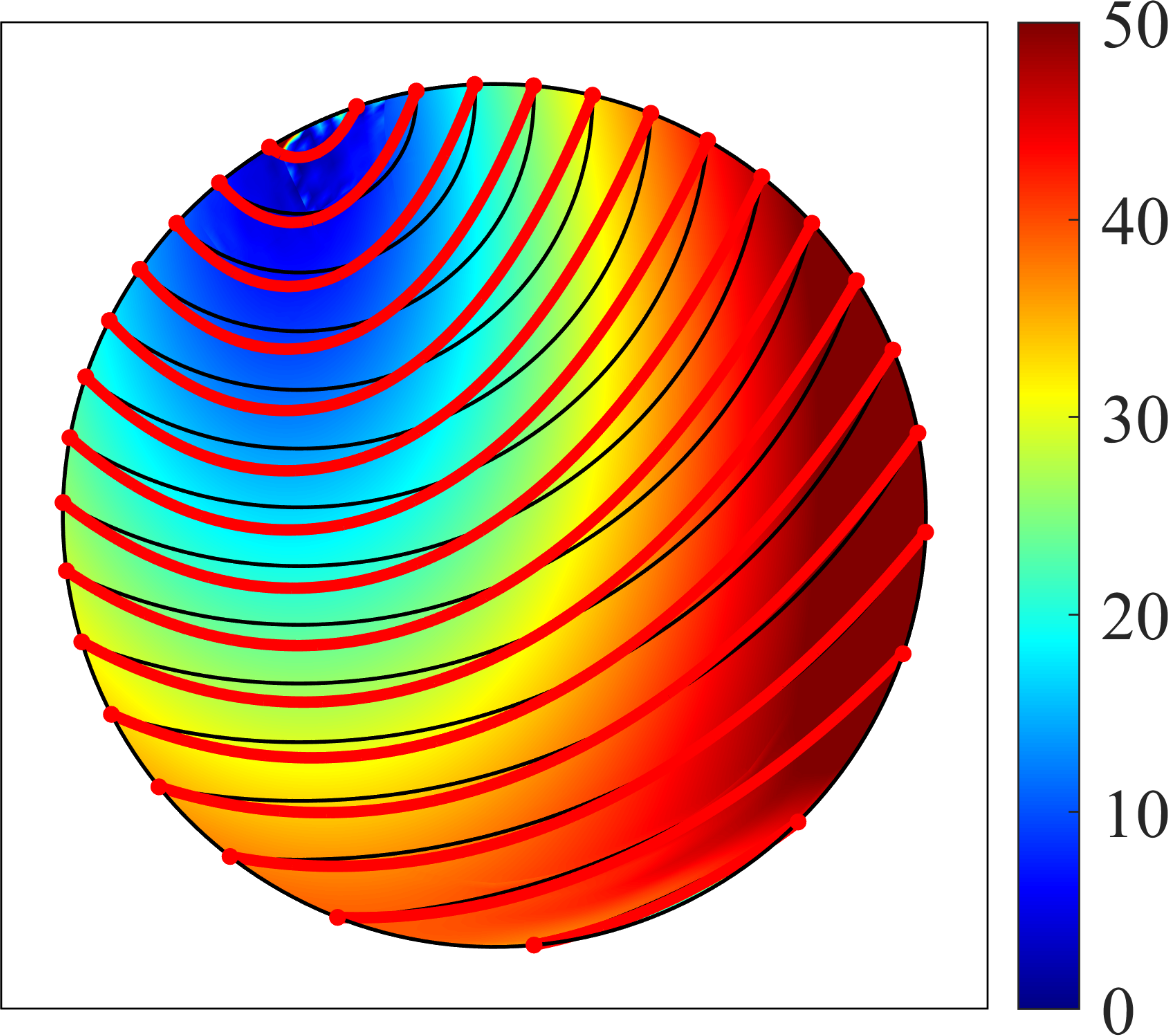}}\hfill\subfigure[result]{\includegraphics[width=0.23\textwidth]{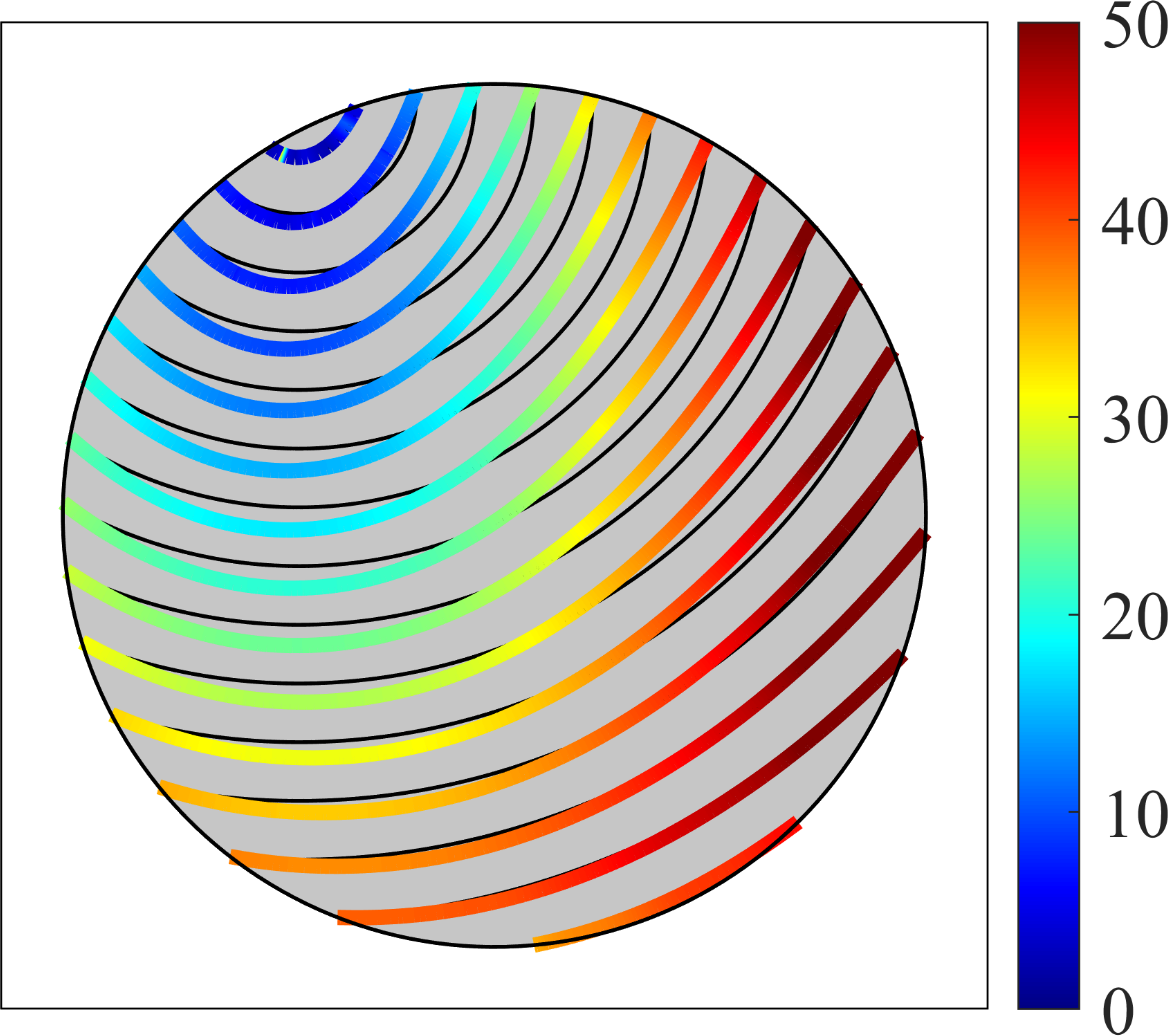}}\hfill\subfigure[result]{\includegraphics[width=0.2\textwidth]{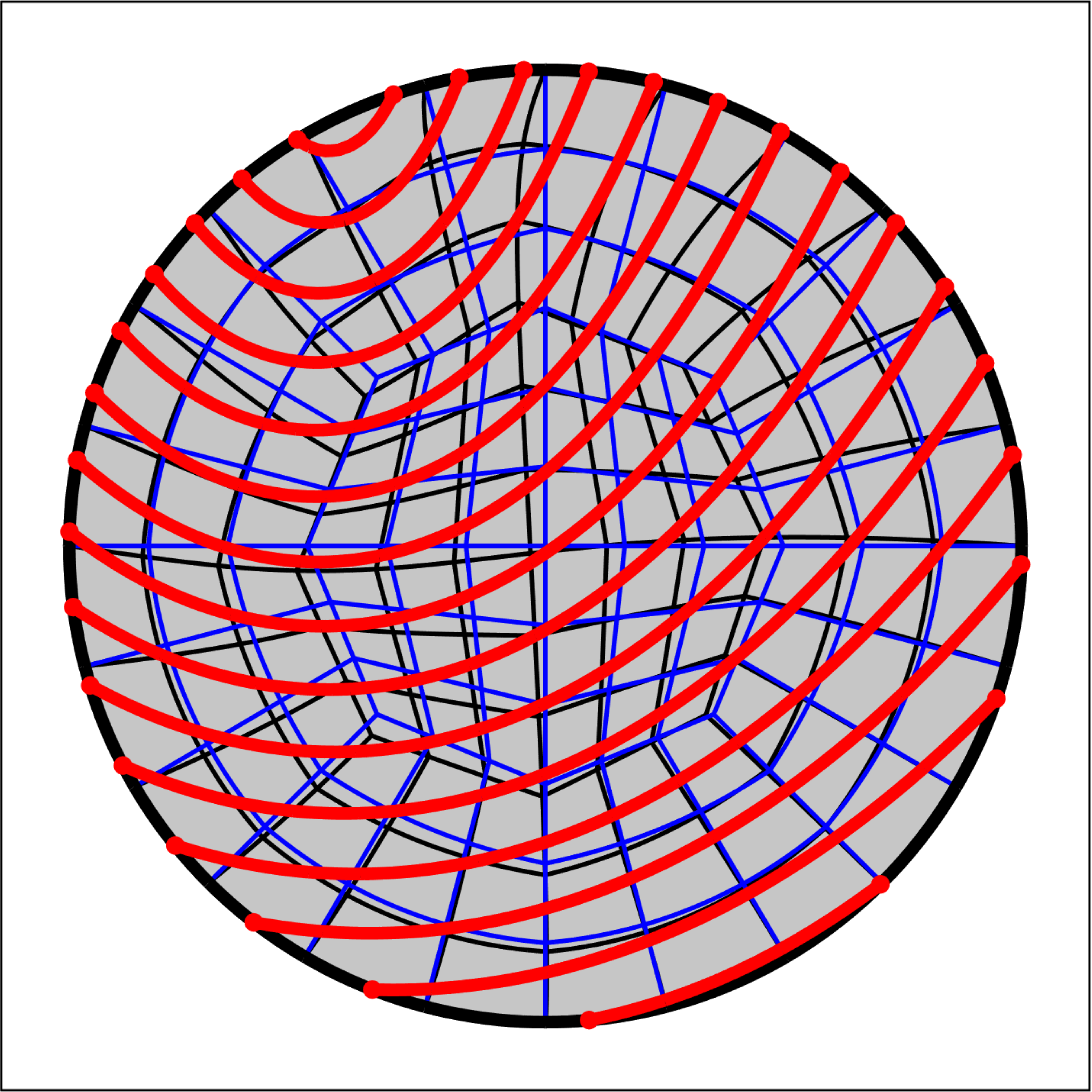}}

\caption{\label{fig:TC2d1}Setup and results for test case 1 in 2D: (a) shows
the bulk domain $\Omega_{\vek X}$ with selected level sets of $\phi\left(\vek X\right)$
and an example mesh used for the analysis, (b) shows the von Mises
stress in $\Omega_{\vek x}$ and the deformed level sets in red, (c)
is an alternative visualization with the same content where the von
Mises stress is shown on the deformed level sets, (d) highlights the
deformation of the bulk domain by showing elements in the undeformed
and deformed meshes.}
\end{figure}

The undeformed bulk domain of interest $\Omega_{\vek X}$ is a circle
with radius $r=0.28$ centered at the origin. The level-set function
$\phi\left(\vek X\right)$ implying the undeformed ropes is
\begin{align}
\phi\left(\vek X\right) & =\left\Vert \vek X-\vek X_{C}\right\Vert -R_{c},\label{eq:LevelSetCircle}\\
 & =\sqrt{\left(X-X_{c}\right)^{2}+\left(Y-Y_{c}\right)^{2}}-R_{c}
\end{align}
with $X_{C}=-0.3\cdot\sin25^{{^\circ}}$, $Y_{C}=0.3\cdot\cos25^{{^\circ}}$,
and $R_{C}=0.3$. See Fig.~\ref{fig:TC2d1}(a) for a sketch of the
setup and an example mesh. Zero displacements are prescribed on the
boundary $\partial\Omega_{\vek X}$. Young's modulus $E$ is set to
$10\,000$ and, as there is no Poisson's ratio for ropes, $\lambda=0$
and $\mu=E/2$. For the loading, we consider dead load with $\vek F\!\left(\vek X\right)=\left[0,-100\right]^{\mathrm{T}}$.
In the equations above, we avoided to explicitly mention the cross
section $A$ for ropes (and thickness $t$ for membranes) and assume
that they are $1$ in all test cases discussed herein. Otherwise,
they could easily be considered by properly manipulating the material
parameters above.

Figs.~\ref{fig:TC2d1}(b) to (d) show different visualizations obtained
by a single analysis with the Bulk Trace FEM: (b) shows selected deformed
level sets in red and von Mises stress obtained from the Cauchy stress
tensor $\vek\sigma$ in the deformed bulk domain as a color field,
(c) is an alternative visualization where the von Mises stress is
plotted on the deformed level sets themselves; doing this for all
level sets would result in the color field shown in (b). Finally,
Fig.~\ref{fig:TC2d1}(d) shows the elements of an example mesh in
the undeformed and deformed bulk domain and the deformed level sets
in red.

\begin{figure}
\centering

\subfigure[convergence in $\varepsilon_{\mathfrak{e}}$]{\includegraphics[width=0.32\textwidth]{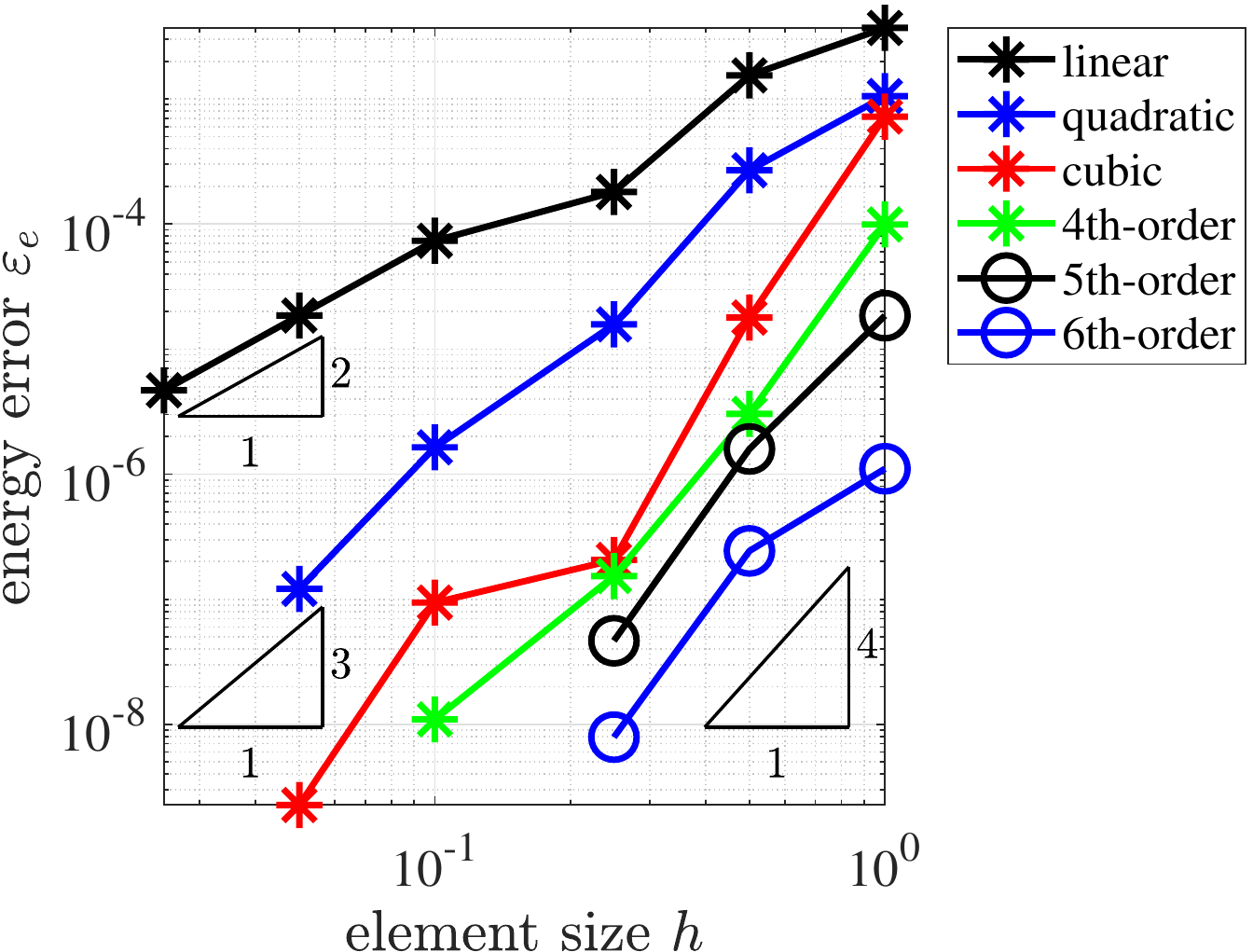}}\hfill\subfigure[convergence in $\varepsilon_{\mathrm{res}}$]{\includegraphics[width=0.32\textwidth]{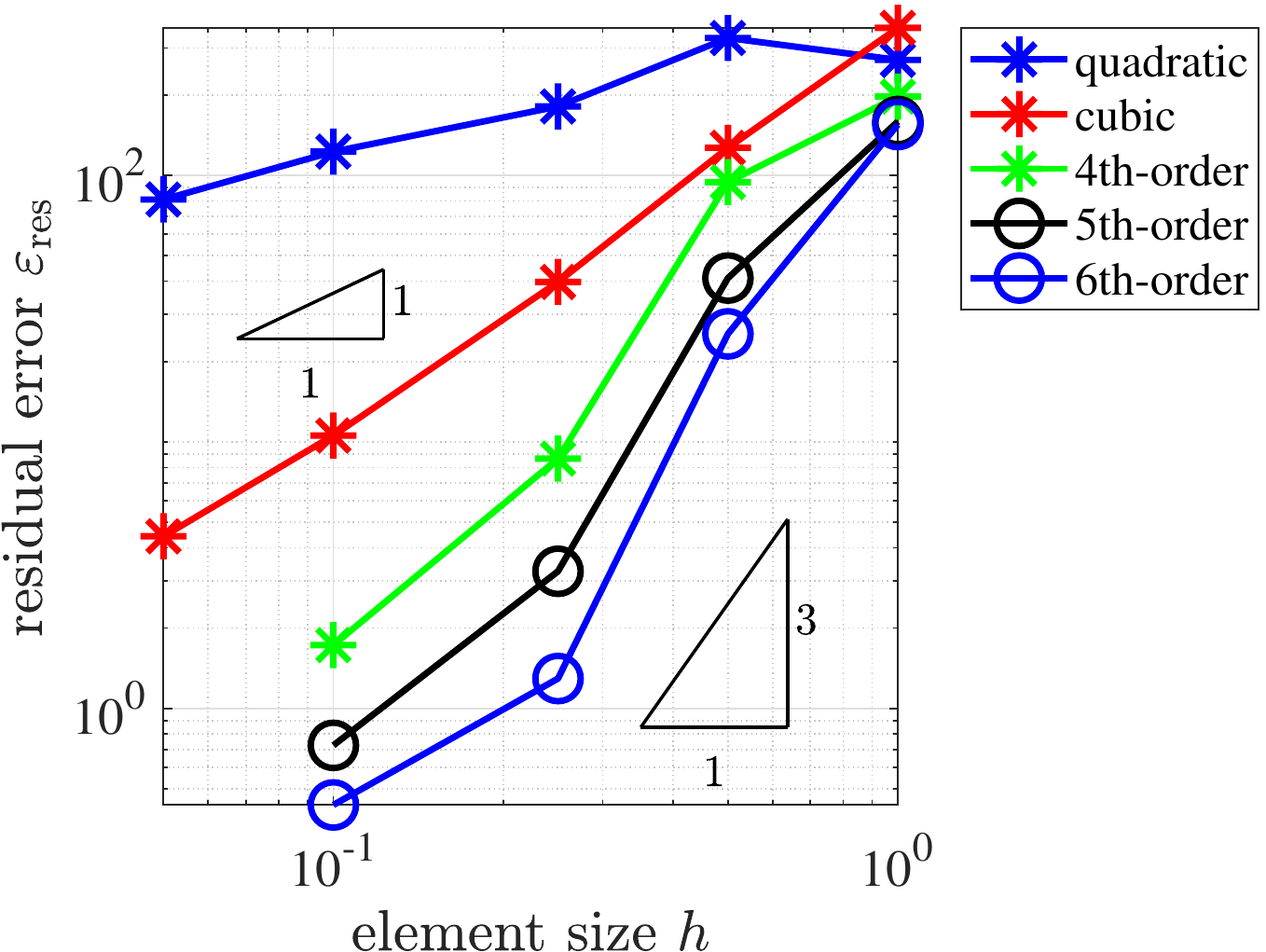}}\hfill\subfigure[von Mises stress]{\includegraphics[width=0.3\textwidth]{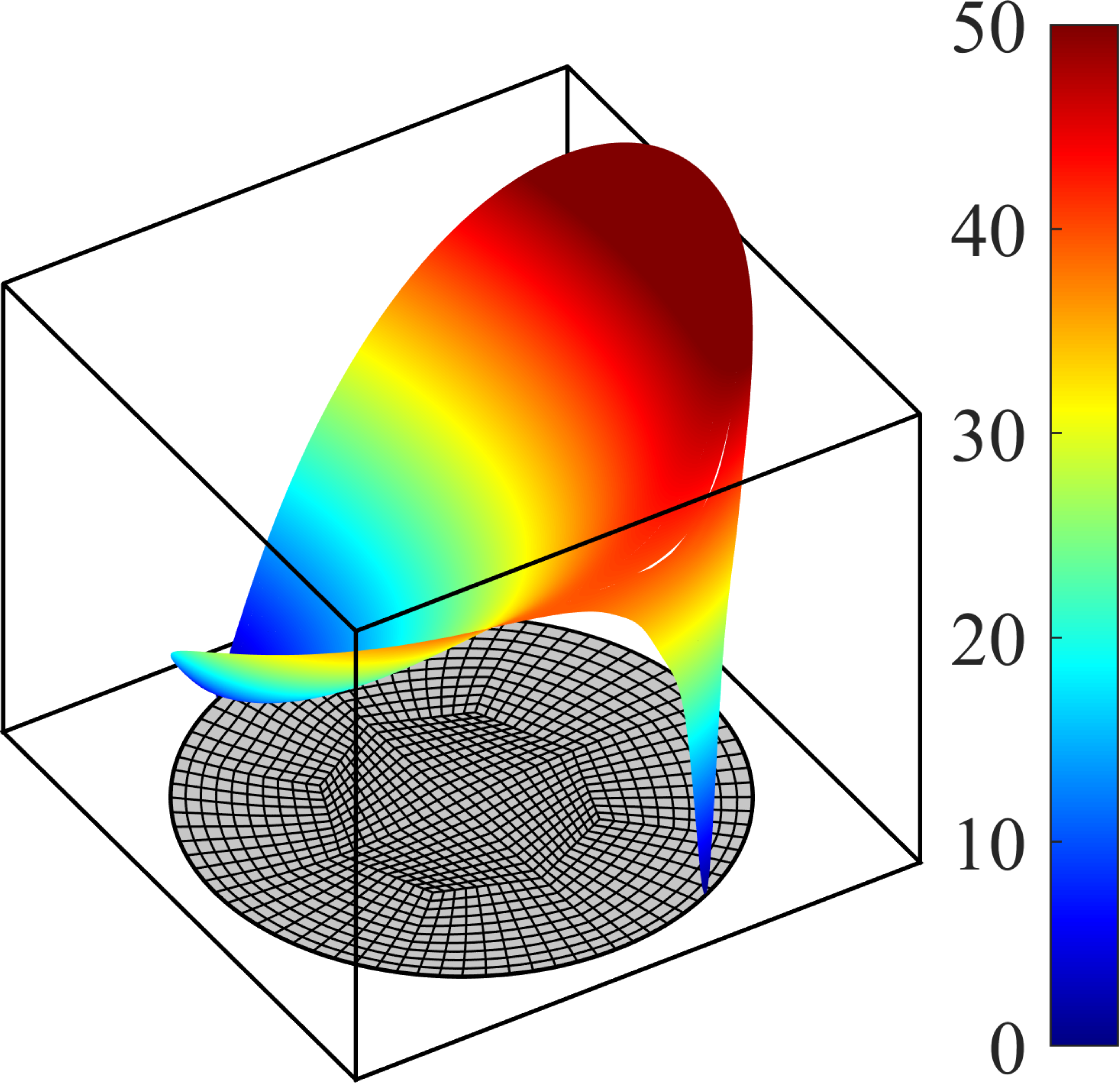}}

\caption{\label{fig:TC2d1Conv}Convergence results for test case 1 in 2D: (a)
and (b) show sub-optimal convergence rates in $\varepsilon_{\mathfrak{e}}$
and $\varepsilon_{\mathrm{res}}$, (c) shows the von Mises stress
in the domain confirming locally steep gradients causing the sub-optimal
results.}
\end{figure}

In the numerical analyses based on the Bulk Trace FEM, we have used
sequences of meshes with different resolution and order (from 1 to
6), enabling systematic convergence studies in the error measures
introduced above. Convergence results are seen in Fig.~\ref{fig:TC2d1Conv}(a)
for the stored energy error $\varepsilon_{\mathfrak{e}}$ and (b)
for the residual error $\varepsilon_{\mathrm{res}}$. It is seen that
for this test case, the convergence is sub-optimal because of locally
steep gradients in the mechanical fields as shown for the example
of the resulting von Mises stress field in Fig.~\ref{fig:TC2d1Conv}(c).

\begin{figure}
\centering

\subfigure[setup]{\includegraphics[width=0.2\textwidth]{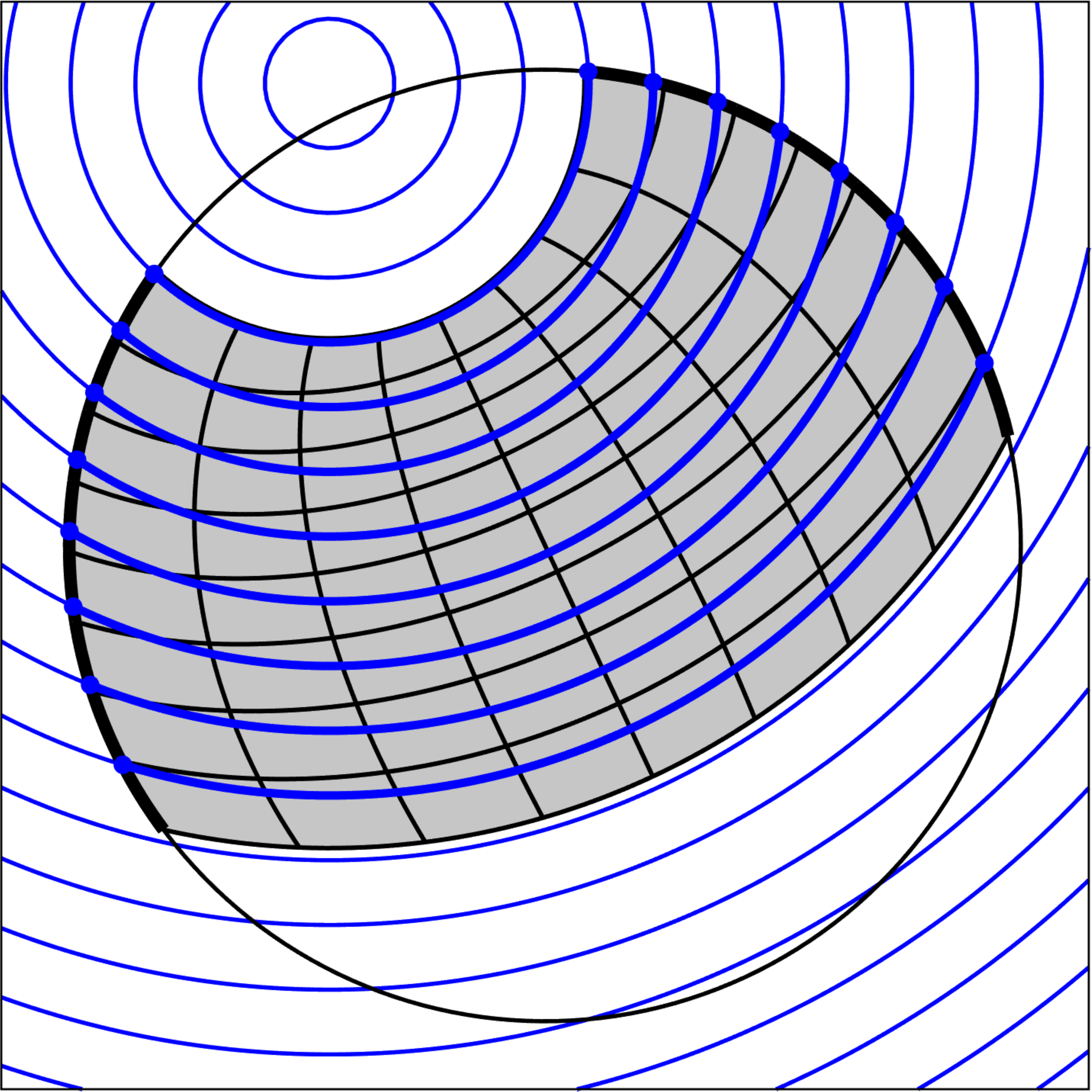}}\hfill\subfigure[result]{\includegraphics[width=0.23\textwidth]{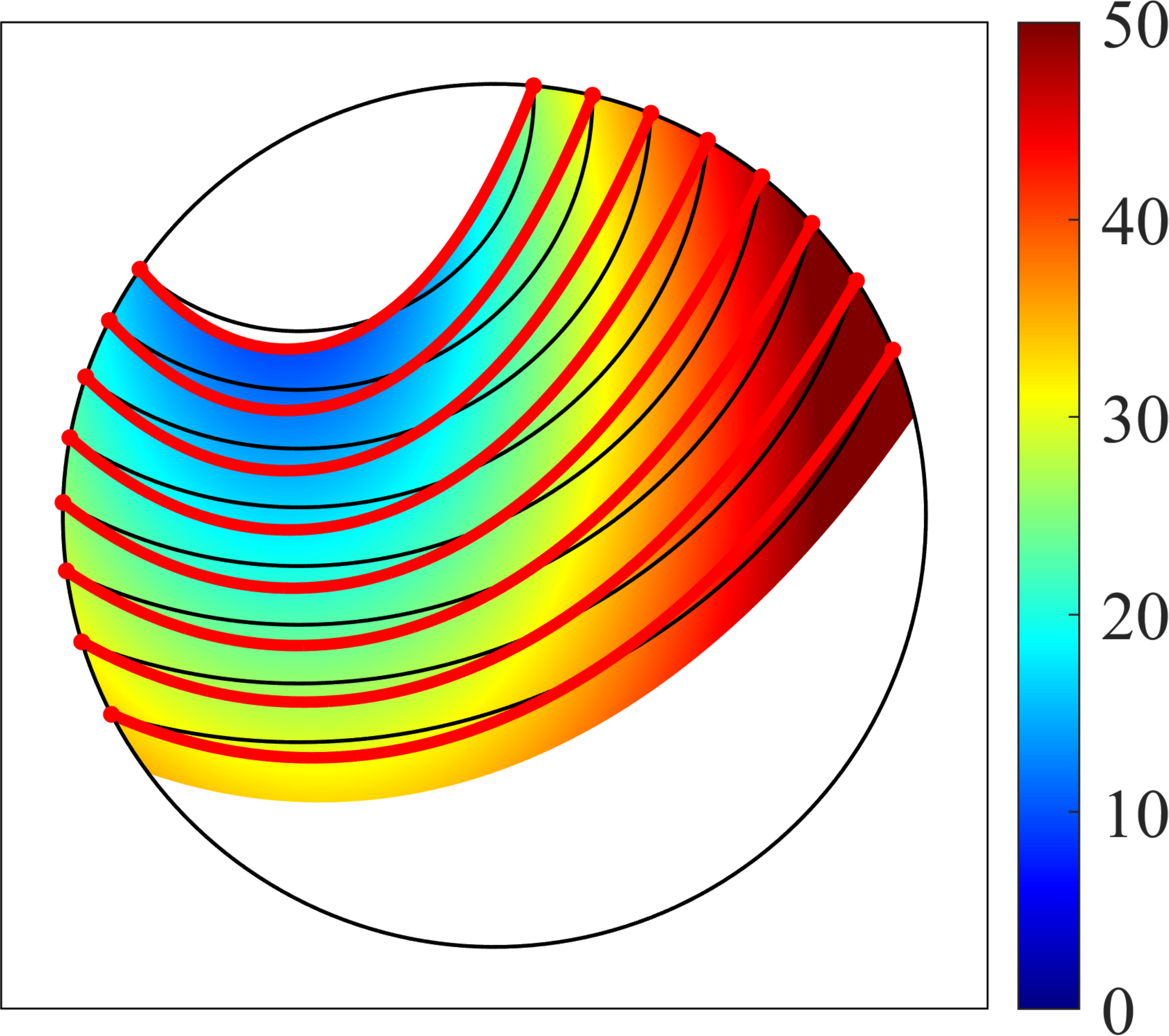}}\hfill\subfigure[result]{\includegraphics[width=0.23\textwidth]{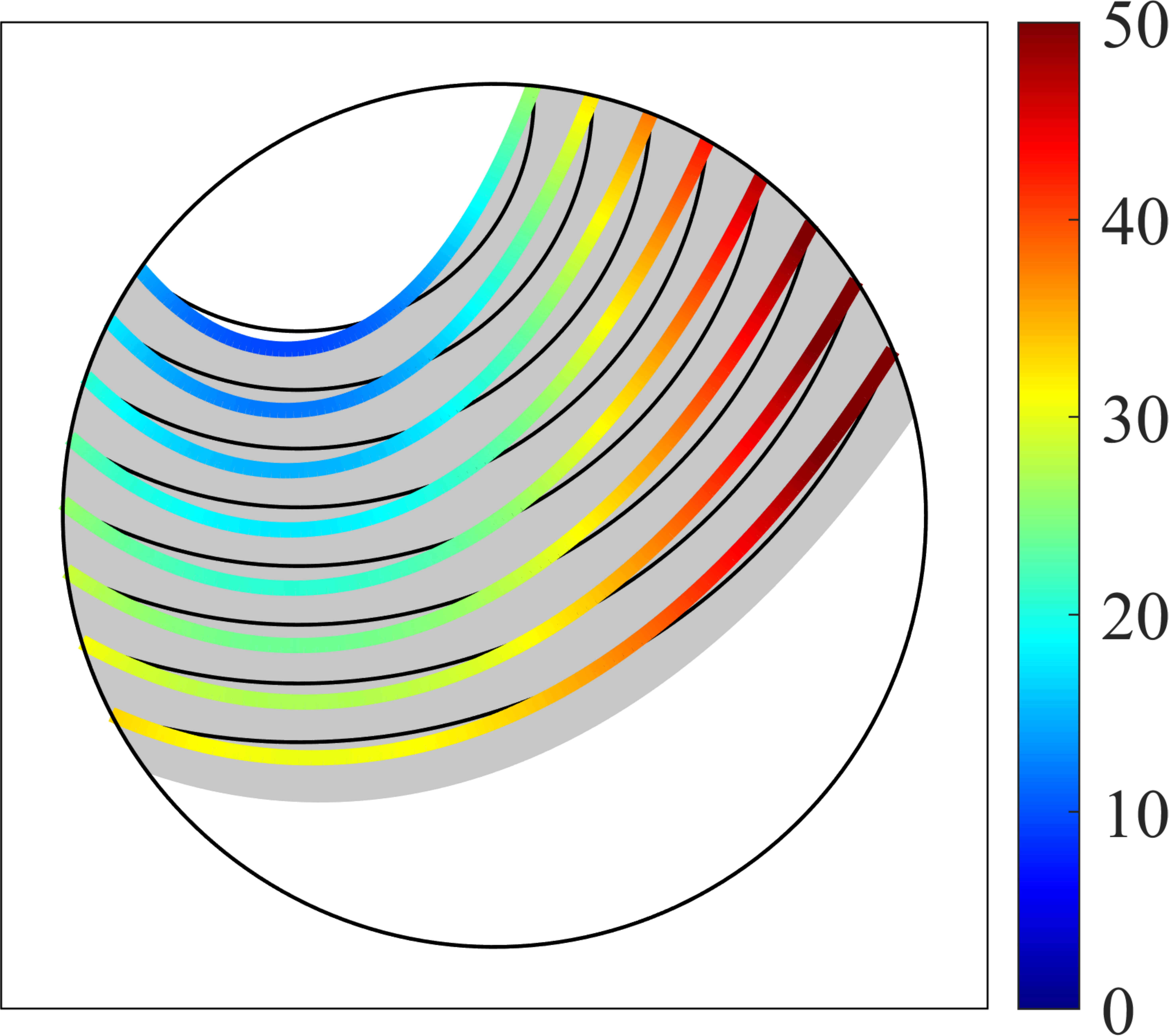}}\hfill\subfigure[result]{\includegraphics[width=0.2\textwidth]{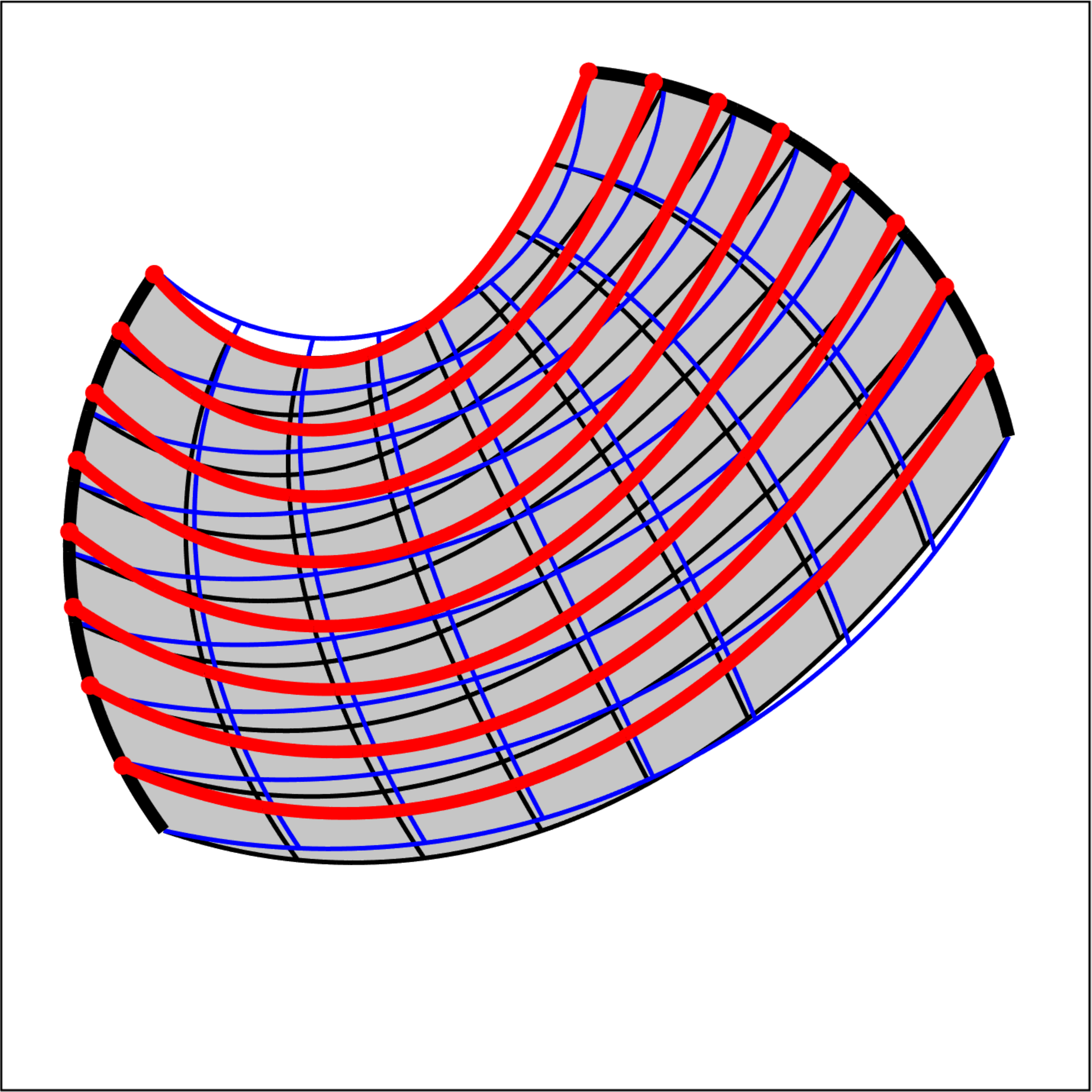}}

\caption{\label{fig:TC2d1Interval}Setup and results for test case 1 in 2D
in the prescribed interval $\phi\in\left[-0.15,0.15\right]$.}
\end{figure}

\begin{figure}
\centering

\subfigure[convergence in $\varepsilon_{\phi}$]{\includegraphics[width=0.32\textwidth]{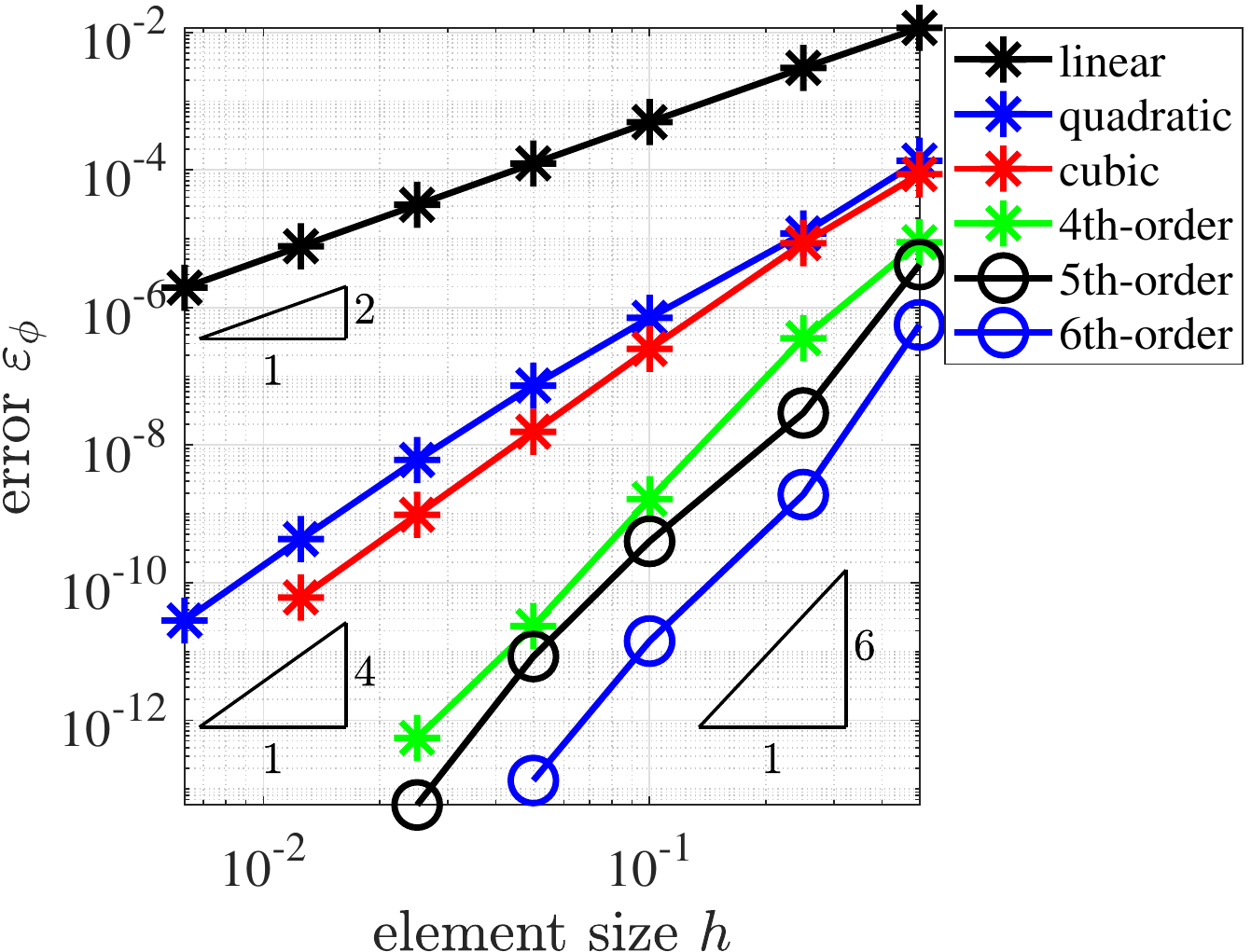}}\hfill\subfigure[convergence in $\varepsilon_{\mathfrak{e}}$]{\includegraphics[width=0.32\textwidth]{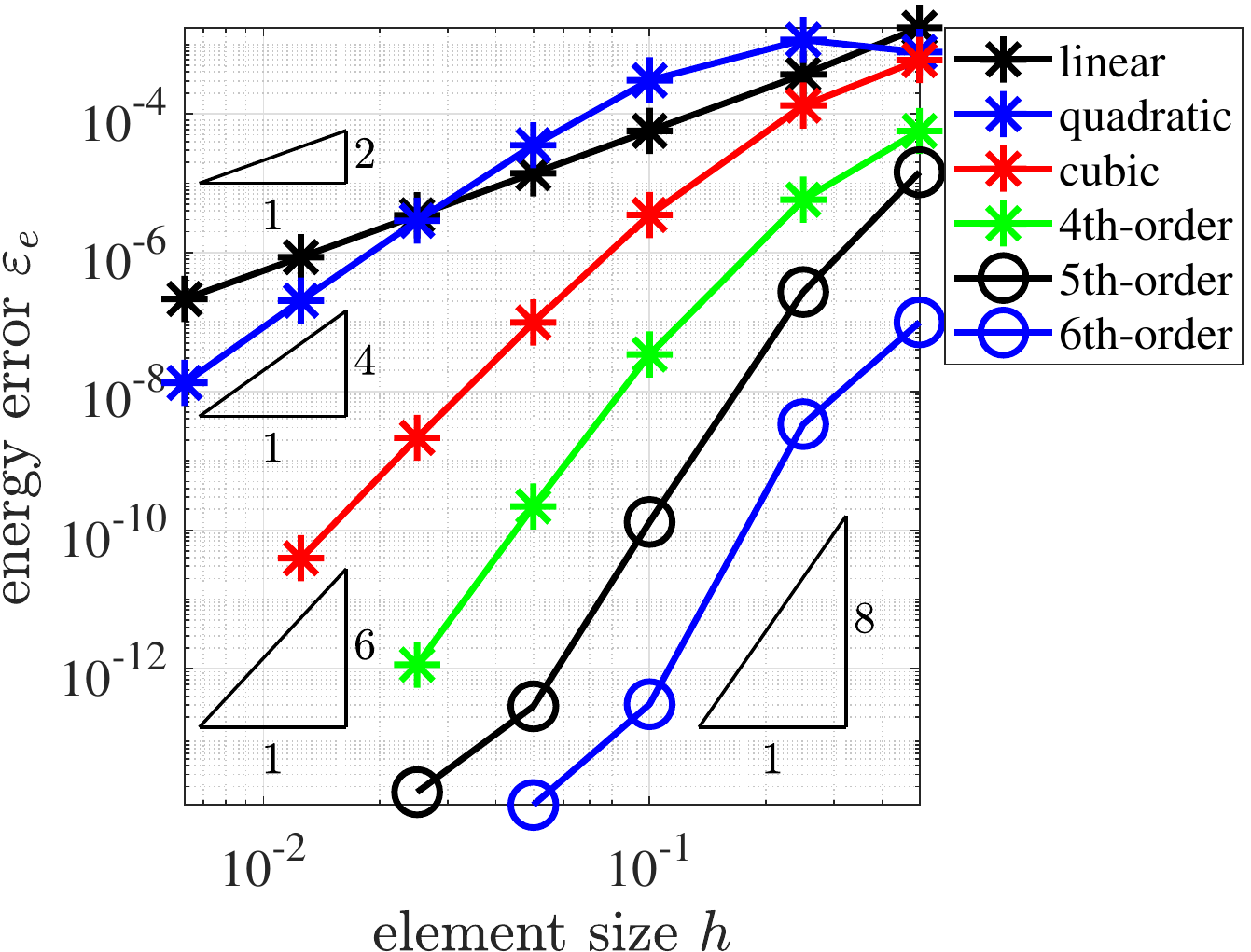}}\hfill\subfigure[convergence in $\varepsilon_{\mathrm{res}}$]{\includegraphics[width=0.32\textwidth]{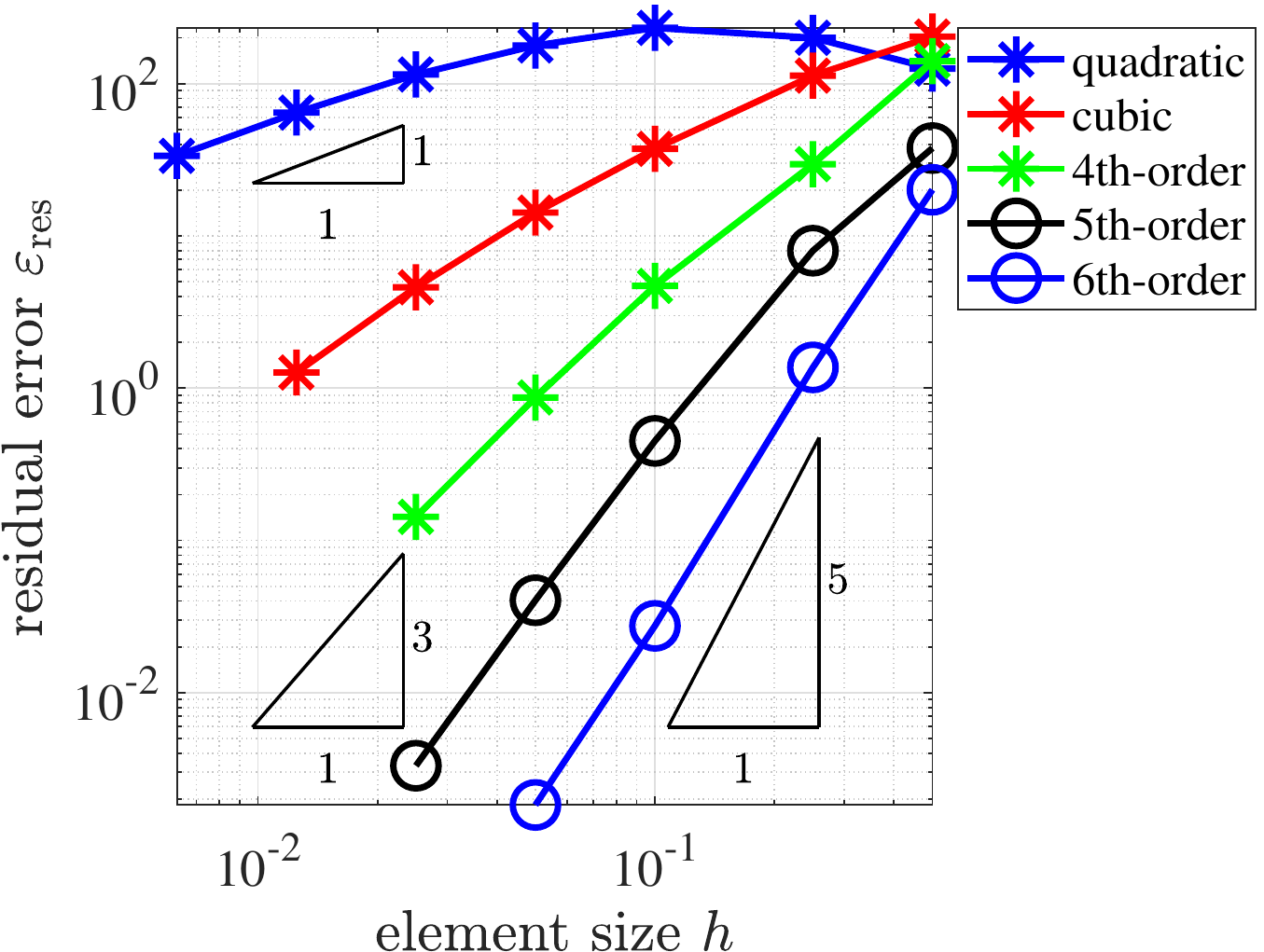}}

\caption{\label{fig:TC2d1ConvInterval}Convergence results for test case 1
in 2D in the interval $\phi\in\left[-0.15,0.15\right]$ confirming
optimal convergence rates in $\varepsilon_{\phi}$, $\varepsilon_{\mathfrak{e}}$
, and $\varepsilon_{\mathrm{res}}$.}
\end{figure}

Therefore, the analysis is now restricted to the level-set interval
$\phi\in\left[-0.15,0.15\right]$ resulting in a subset of the previous
domain as shown in Fig.~\ref{fig:TC2d1Interval}(a) including an
example mesh. Some results obtained with this mesh are visualized
in Figs.~\ref{fig:TC2d1Interval}(b) to (d) following the style of
Fig.~\ref{fig:TC2d1}; all mechanical fields are smooth in this interval.
The reference values for the integrated level sets are $\mathfrak{D}\left(\vek u\right)=0.1644415441226$
and for the stored elastic energy $\mathfrak{e}\left(\vek u\right)=7.792649686407\cdot10^{-3}$.
Convergence results are seen in Fig.~\ref{fig:TC2d1ConvInterval}
in all three error measures defined above. As can be seen, we obtain
at least $p+1$ in $\varepsilon_{\phi}$ and $\varepsilon_{\mathfrak{e}}$,
and $p-1$ in the residual error $\varepsilon_{\mathrm{res}}$ as
expected. In Fig.~\ref{fig:TC2d1ConvInterval}(a), one can see that
\emph{even} element orders achieve $p+2$ in $\varepsilon_{\phi}$,
whereas \emph{odd} orders achieve $p+1$, which is also confirmed
in the next test cases. In Fig.~\ref{fig:TC2d1ConvInterval}(b) one
can see an even better convergence in the stored energy error $\varepsilon_{\mathfrak{e}}$,
however, this can be traced back to the special case of using \emph{circular
}undeformed ropes as implied by Eq.~(\ref{eq:LevelSetCircle}). Using
more general definitions rather leads to the expected rates of $p+1$,
which is also confirmed in the next studies. Note that for the residual
error shown in Fig.~\ref{fig:TC2d1ConvInterval}(c), the results
obtained with \emph{linear} meshes are omitted because they cannot
be expected to converge as the expected rate is $p-1$ due to the
involved second-order derivatives in $\varepsilon_{\mathrm{res}}$.

\begin{figure}
\centering

\subfigure[front view, undef.]{\includegraphics[width=0.24\textwidth]{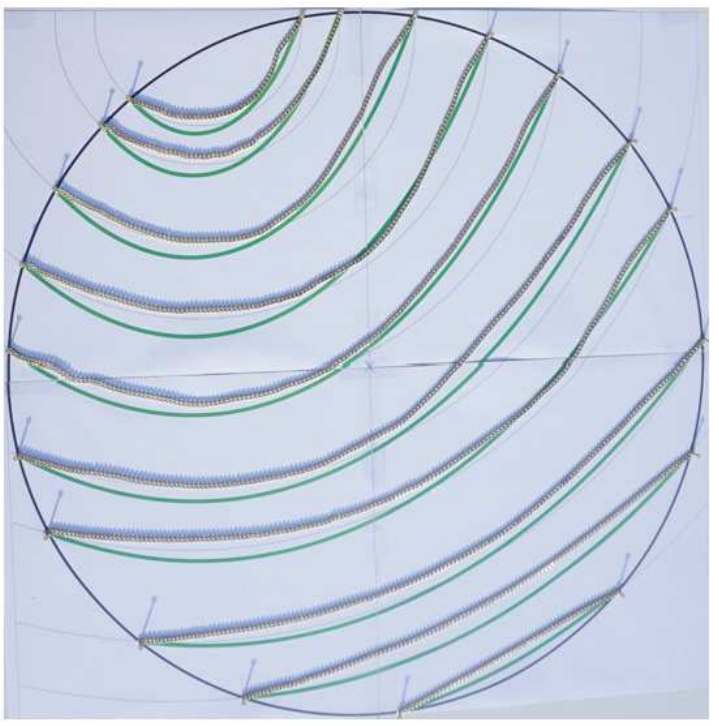}}\hfill\subfigure[zoom]{\includegraphics[width=0.24\textwidth]{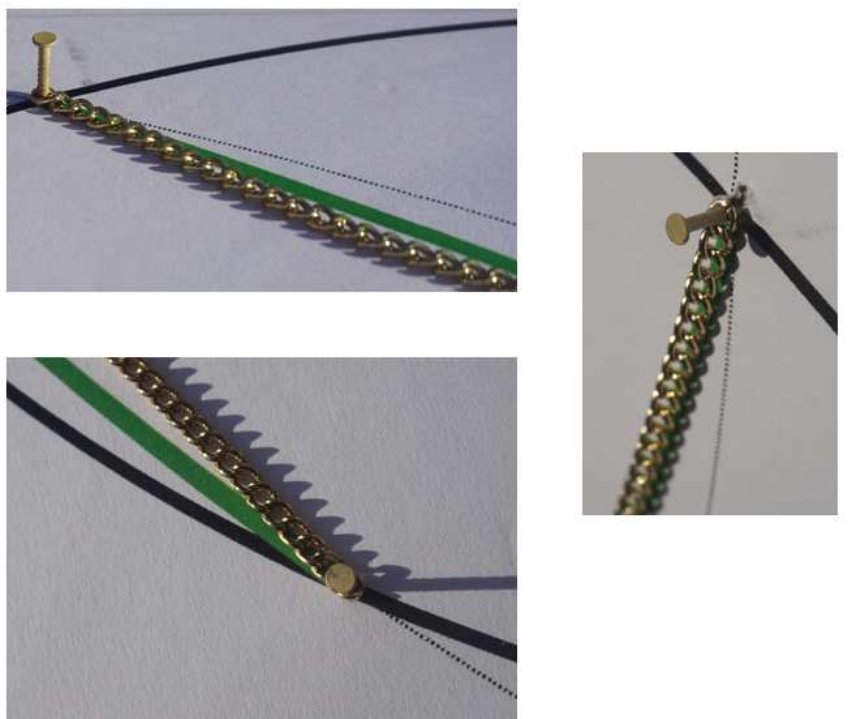}}\hfill\subfigure[front view, def.]{\includegraphics[width=0.24\textwidth]{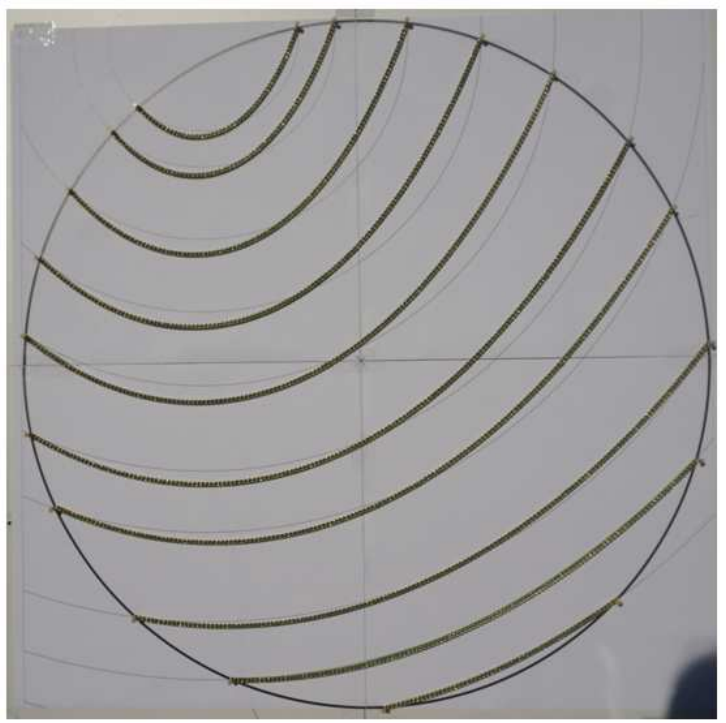}}\hfill\subfigure[side view, def.]{\includegraphics[width=0.23\textwidth]{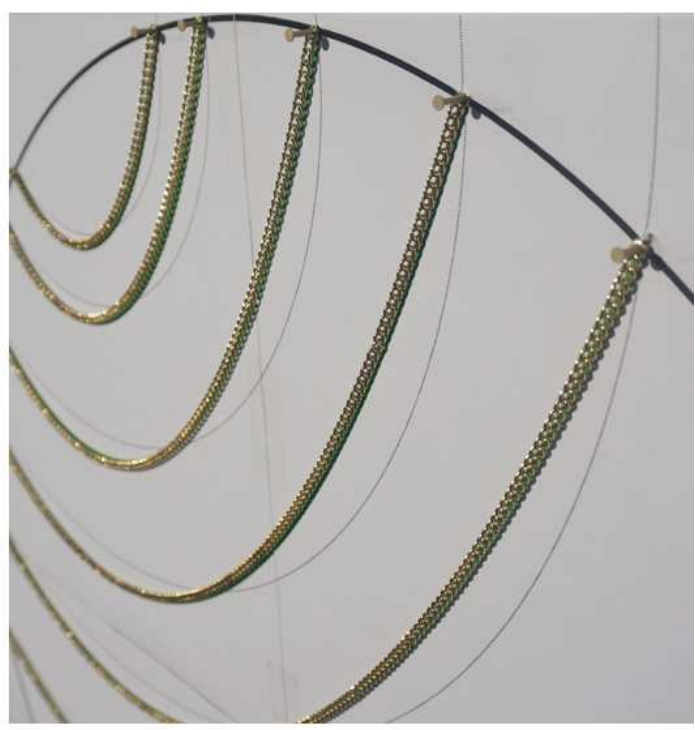}}

\caption{\label{fig:TC2d1Experiment}Small-scale experiment to confirm that
the Bulk Trace FEM is able to simultaneously capture the expected
catenaries in the geometrical setup of test case 1 in 2D.}
\end{figure}

Finally, a small experiment is described to confirm that the computed
results are mechanically meaningful. Therefore, we select 10 level
sets $\ManUndef$ of Eq.~(\ref{eq:LevelSetCircle}) with
\[
c=\left\{ -0.22,-0.19,-0.13,-0.07,-0.01,0.06,0.11,0.183,0.23,0.27\right\} ,
\]
compute the corresponding arc lengths in the undeformed configuration
and cut 10 chains with the respective lengths. The resulting displacement
field is computed by means of the Bulk Trace FEM with largely increased
$E$ (because the elastic strains in the chains under dead load are
negligible). The computed deformed level sets are printed out true
to scale and fixed to a board, see the green lines in Fig.~\ref{fig:TC2d1Experiment}(a).
The chains are fixed with small nails at the corresponding end points
of the level sets $\partial\ManUndef$, see Figs.~\ref{fig:TC2d1Experiment}(a)
and (b) for the board lying horizontally. Then, the board is set vertically
so that the chains deform under their own weight. The chains hang
in the shape of catenaries as expected and are perfectly on top of
the deformed level sets predicted with the Bulk Trace FEM, see Figs.~\ref{fig:TC2d1Experiment}(c)
and (d). This is just to confirm that although the overall setup discussed
in this work is rather unusual, the results are just as physically
meaningful as any other mechanical model.

\subsection{Test case 2 in 2D: Implicit bulk domain\label{XX_Testcase2}}

\begin{figure}
\centering

\subfigure[level sets]{\includegraphics[width=0.2\textwidth]{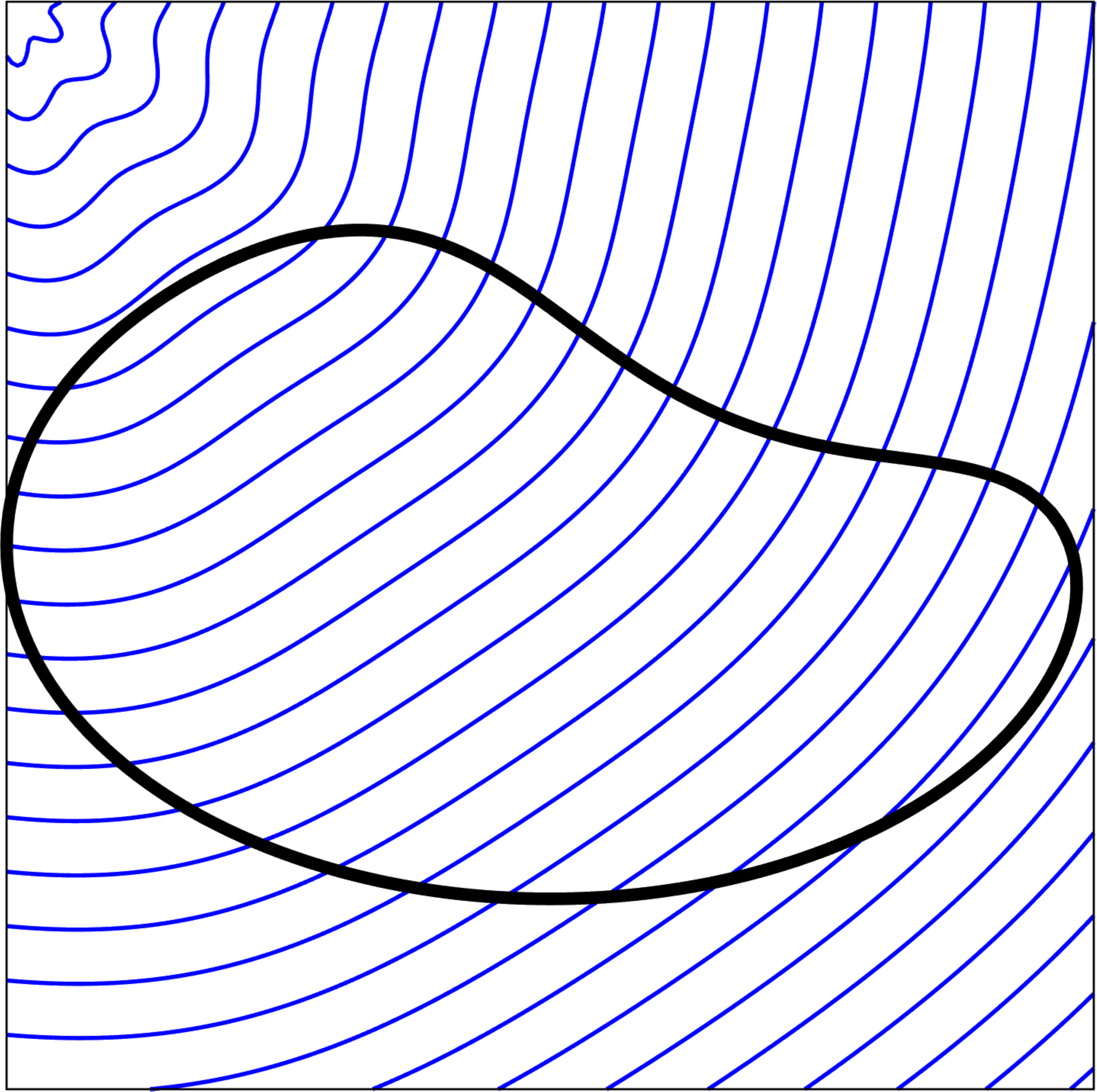}}\hfill\subfigure[$\Omega_{\vek X}$ and mesh]{\includegraphics[width=0.2\textwidth]{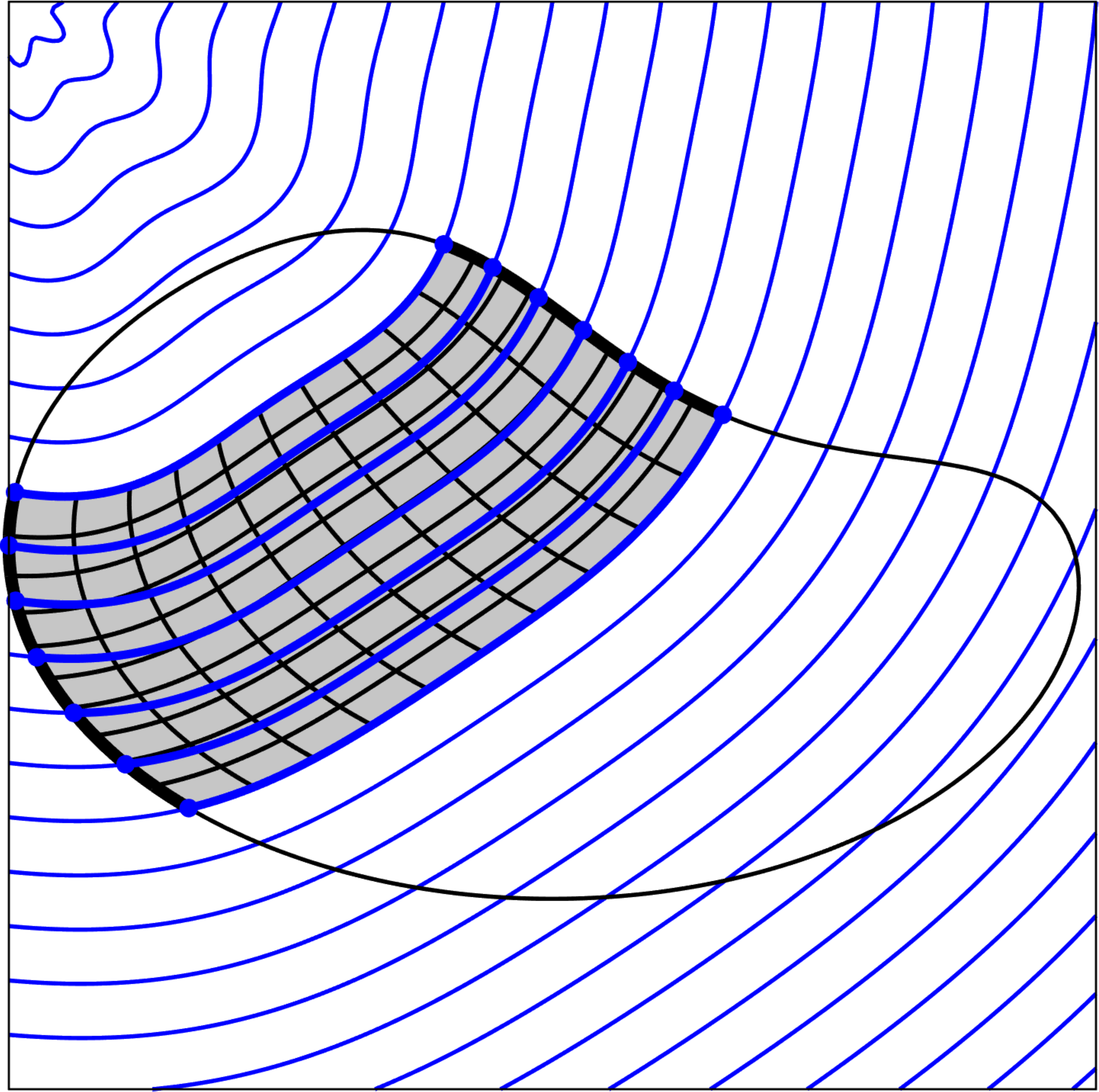}}\hfill\subfigure[result]{\includegraphics[width=0.25\textwidth]{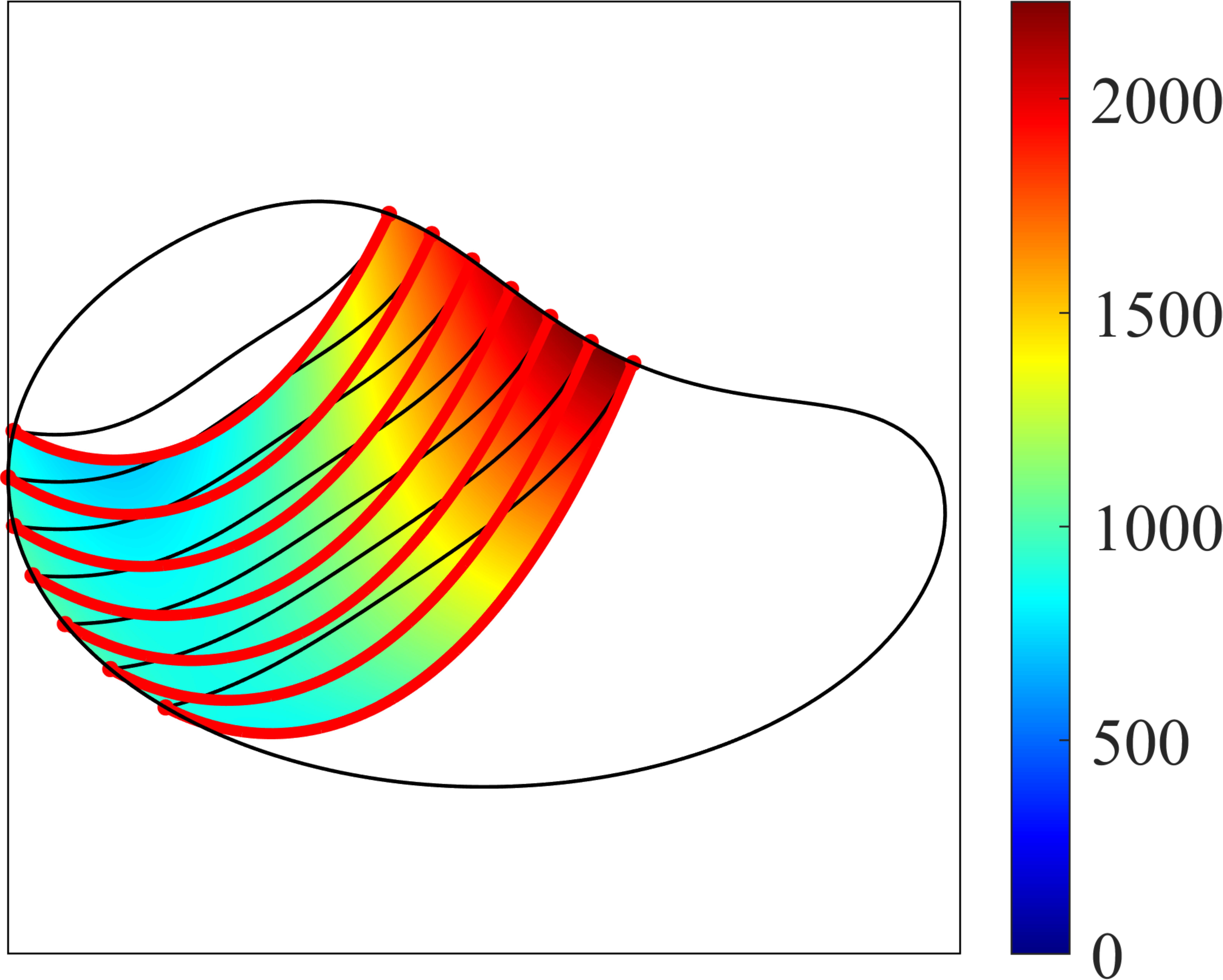}}\hfill\subfigure[result]{\includegraphics[width=0.2\textwidth]{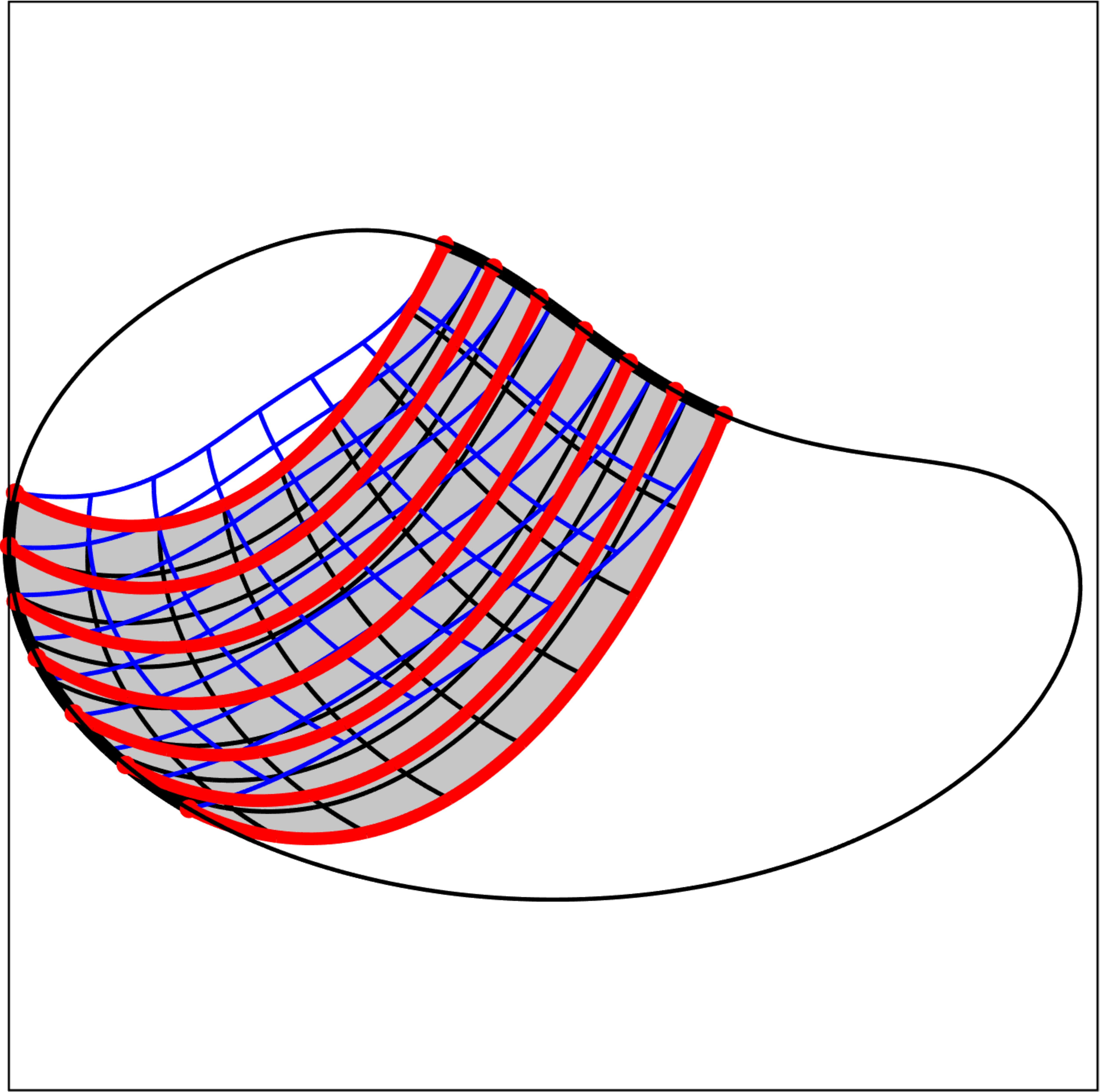}}

\caption{\label{fig:TC2d2Interval}Setup and results for test case 2 in 2D
in the prescribed interval $\phi\in\left[4,7\right]$.}
\end{figure}

Next, a more general shape of the bulk domain and level sets is suggested.
The bulk domain $\Omega_{\vek X}$ is defined implicitly by the zero-level
set of 
\begin{align*}
\psi\left(\vek X\right)= & \psi_{1}\left(\vek X\right)+\psi_{2}\left(\vek X\right)\\
\text{with }\;\psi_{1}\left(\vek X\right)= & \left(\frac{X}{R_{X}}\right)^{2}+\left(\frac{Y}{R_{Y}}\right)^{2}-1,\quad R_{X}=10,\;R_{Y}=6.5,\\
\text{and }\;\psi_{2}\left(\vek X\right)= & f_{\text{Bell}}\left(a\right).
\end{align*}
Therein, $f_{\text{Bell}}$ is a $C_{4}$-continuous bell-shaped function
defined as

\begin{align*}
f_{\text{Bell}}\left(a\right)= & \begin{cases}
\begin{array}{c}
-1024\cdot a^{10}+5120\cdot a^{9}-10240\cdot a^{8}+\\
10240\cdot a^{7}-5120\cdot a^{6}+1024\cdot a^{5}
\end{array} & 0\leq a\leq1,\\
\;0 & \text{else},
\end{cases}\\
\text{with }\;a= & \;\frac{r+R_{B}}{2\cdot R_{B}}\quad\text{and}\quad r=\left\Vert \vek X-\vek X_{B}\right\Vert .
\end{align*}
In this example, $\vek X_{\!B}=\left[4,5\right]^{\mathrm{T}}$ and
$R_{B}=12$. The resulting boundary of the implied bulk domain $\Omega_{\vek X}$,
being the zero-level set of $\psi\left(\vek X\right)$, is the black
line in Fig.~\ref{fig:TC2d2Interval}(a). For the level-set function
$\phi\left(\vek X\right)$ implying the geometries of interest, i.e.,
the undeformed ropes, we define
\begin{align*}
\phi\left(\vek X\right)= & \;\nicefrac{1}{2}\left\Vert \vek X-\vek X_{F}\right\Vert -\nicefrac{1}{10}\cdot\sin\left(8\cdot\theta\left(\vek X\right)\right)-\nicefrac{1}{2},\\
\text{with }\;\theta\left(\vek X\right)= & \;\text{atan}\frac{Y-Y_{F}}{X-X_{F}}\quad\text{and}\quad\vek X_{\!F}=\left[-10,10\right]^{\mathrm{T}},
\end{align*}
see the blue lines in Fig.~\ref{fig:TC2d2Interval}(a). Finally,
the bulk domain of interest is defined as
\[
\Omega_{\vek X}=\left\{ \vek X\in\mathbb{R}^{2}:\psi\left(\vek X\right)\leq0\;\;\text{and}\;\;4\leq\phi(\vek X)\leq7\right\} 
\]
and is shown in Fig.~\ref{fig:TC2d2Interval}(b) with an example
mesh and selected level sets of $\phi\left(\vek X\right)$.

\begin{figure}
\centering

\subfigure[convergence in $\varepsilon_{\phi}$]{\includegraphics[width=0.32\textwidth]{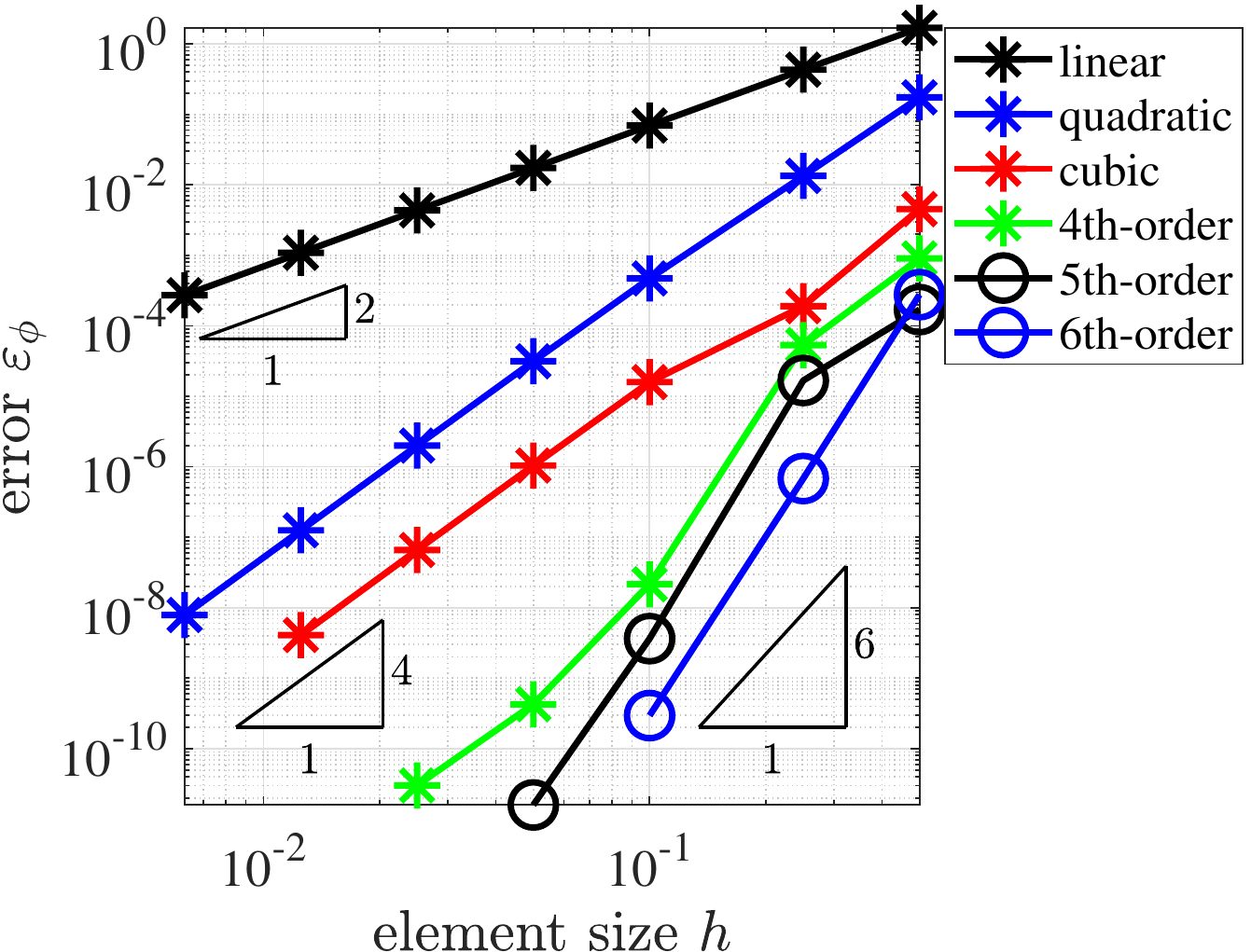}}\hfill\subfigure[convergence in $\varepsilon_{\mathfrak{e}}$]{\includegraphics[width=0.32\textwidth]{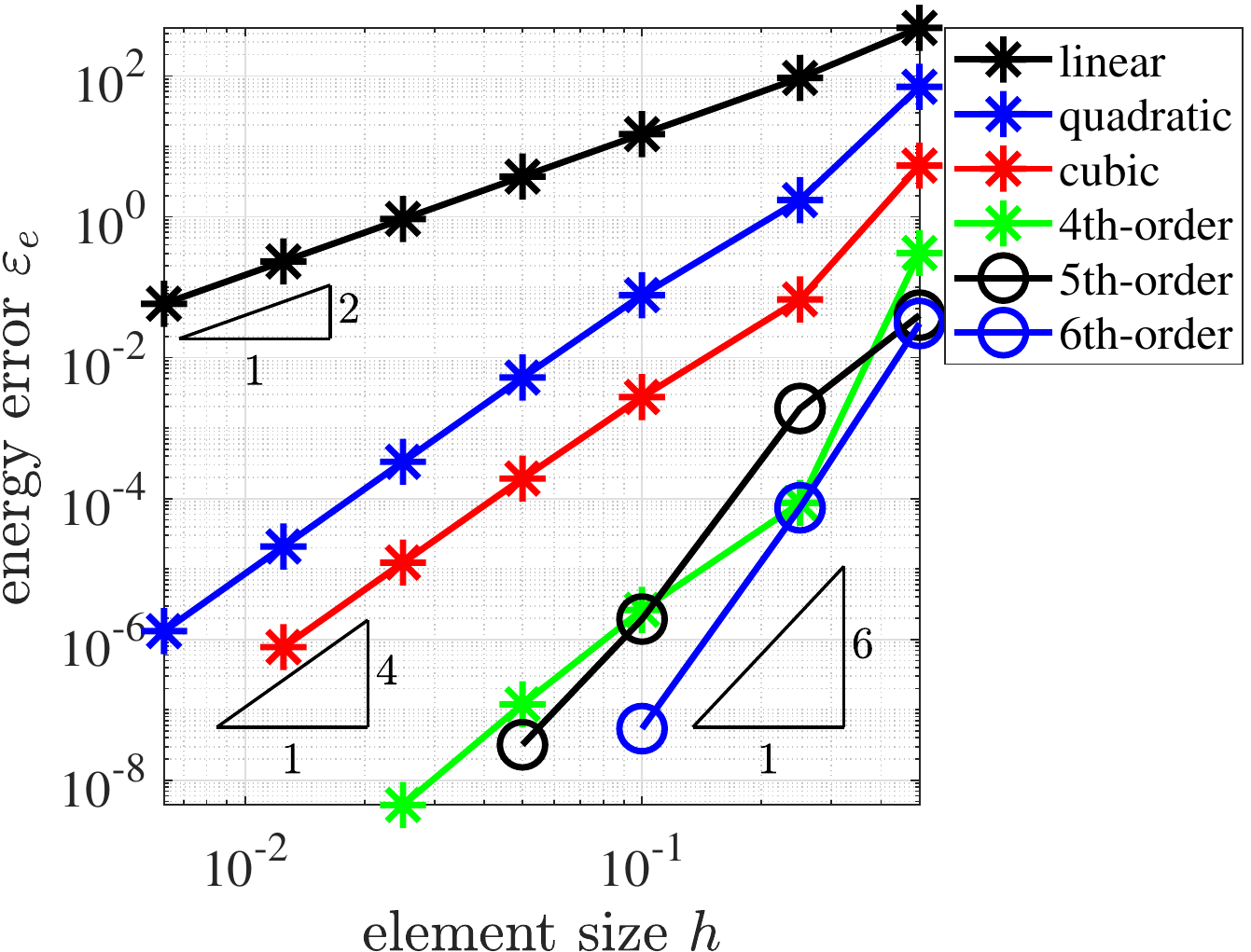}}\hfill\subfigure[convergence in $\varepsilon_{\mathrm{res}}$]{\includegraphics[width=0.32\textwidth]{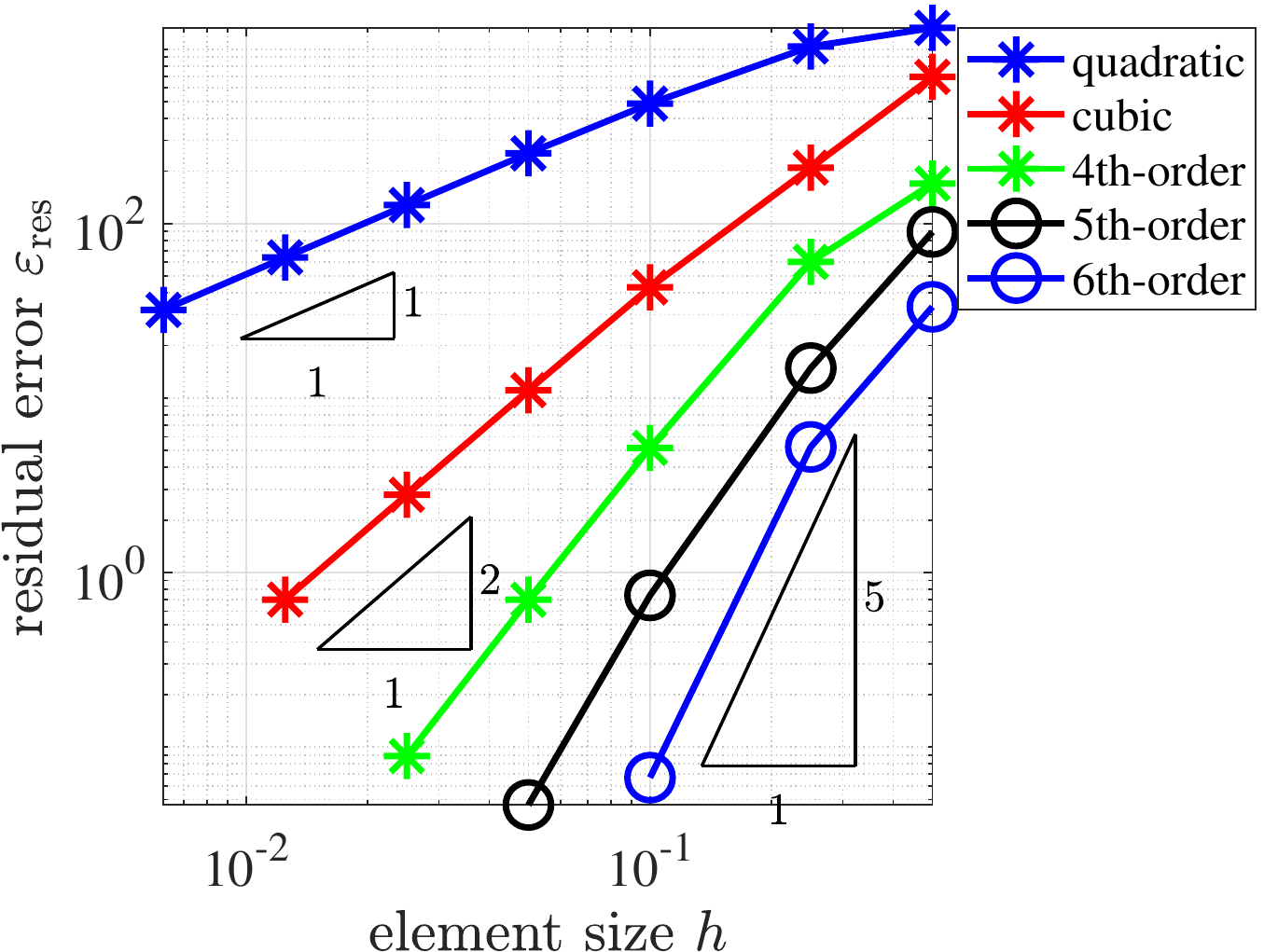}}

\caption{\label{fig:TC2d2ConvInterval}Convergence results for test case 2
in 2D confirming optimal convergence rates in $\varepsilon_{\phi}$,
$\varepsilon_{\mathfrak{e}}$ , and $\varepsilon_{\mathrm{res}}$.}
\end{figure}

Material parameters are again set to $E=10\,000$ and the bulk forces
are $\vek F\!\left(\vek X\right)=\left[0,-200\right]^{\mathrm{T}}$.
Zero displacements are prescribed on $\partial\Omega_{\vek X}$. Benchmark
values for the integrated level sets are $\mathfrak{D}\left(\vek u\right)=39.05000466379$
and for the stored energy $\mathfrak{e}\left(\vek u\right)=2317.129363166$.
The deformed bulk domain $\Omega_{\vek x}$ is seen in Fig.~\ref{fig:TC2d2Interval}(c)
and (d). Convergence results are seen in Fig.~\ref{fig:TC2d2ConvInterval}
in all three error measures. The convergence in $\varepsilon_{\phi}$
and $\varepsilon_{\mathfrak{e}}$ is $p+1$ for odd element orders
and $p+2$ for even orders. The convergence in $\varepsilon_{\mathrm{res}}$
is $p-1$ as expected. The same rates of convergence have been obtained
using triangular and quadrilateral elements in the Bulk Trace FEM
analyses, respectively.

\subsection{Test case 3 in 3D: Spherical bulk domain\label{XX_Testcase3}}

\begin{figure}
\centering

\subfigure[level sets]{\includegraphics[width=0.3\textwidth]{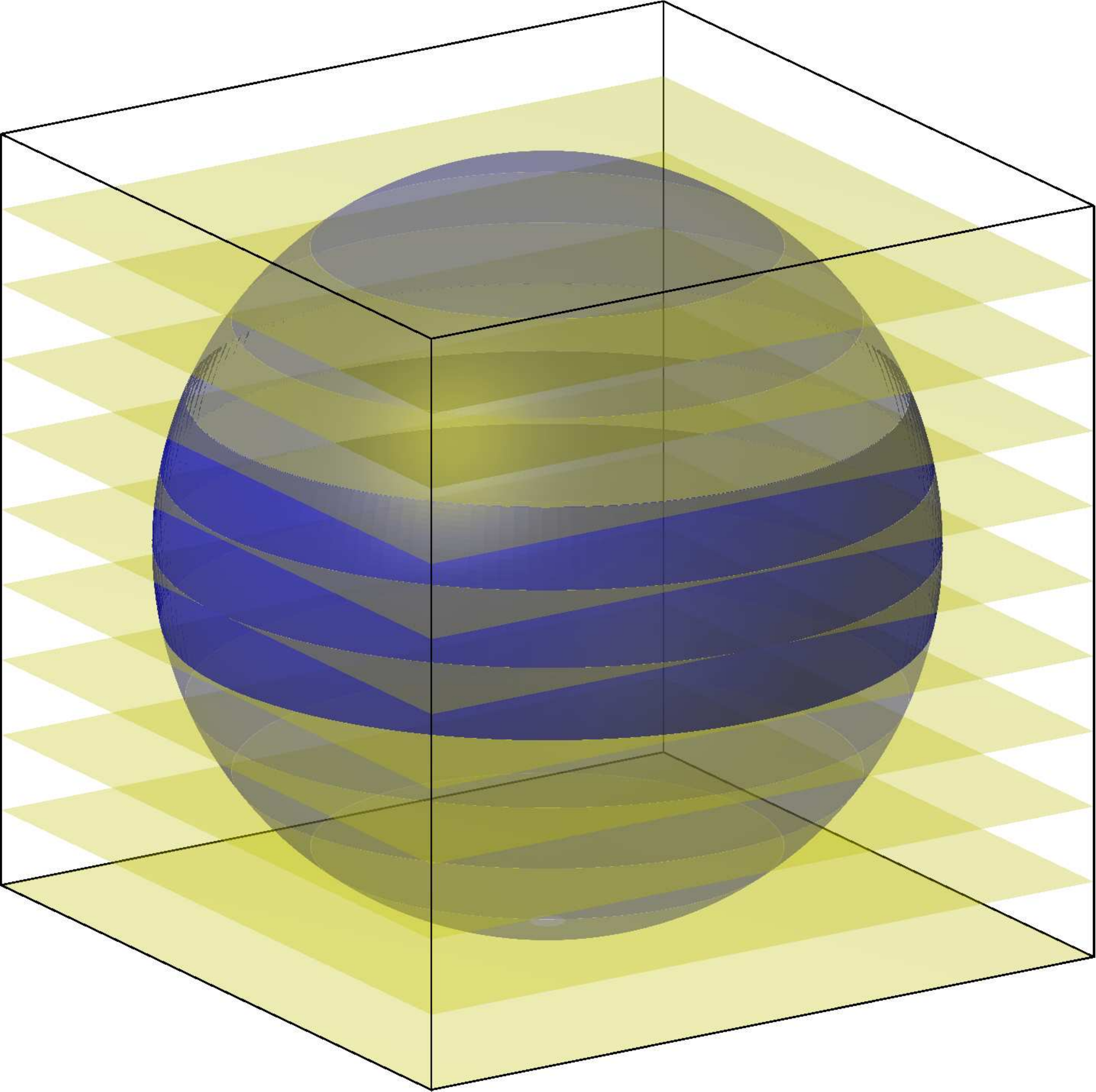}}$\qquad$\subfigure[$\Omega_{\vek X}$ and mesh]{\includegraphics[width=0.3\textwidth]{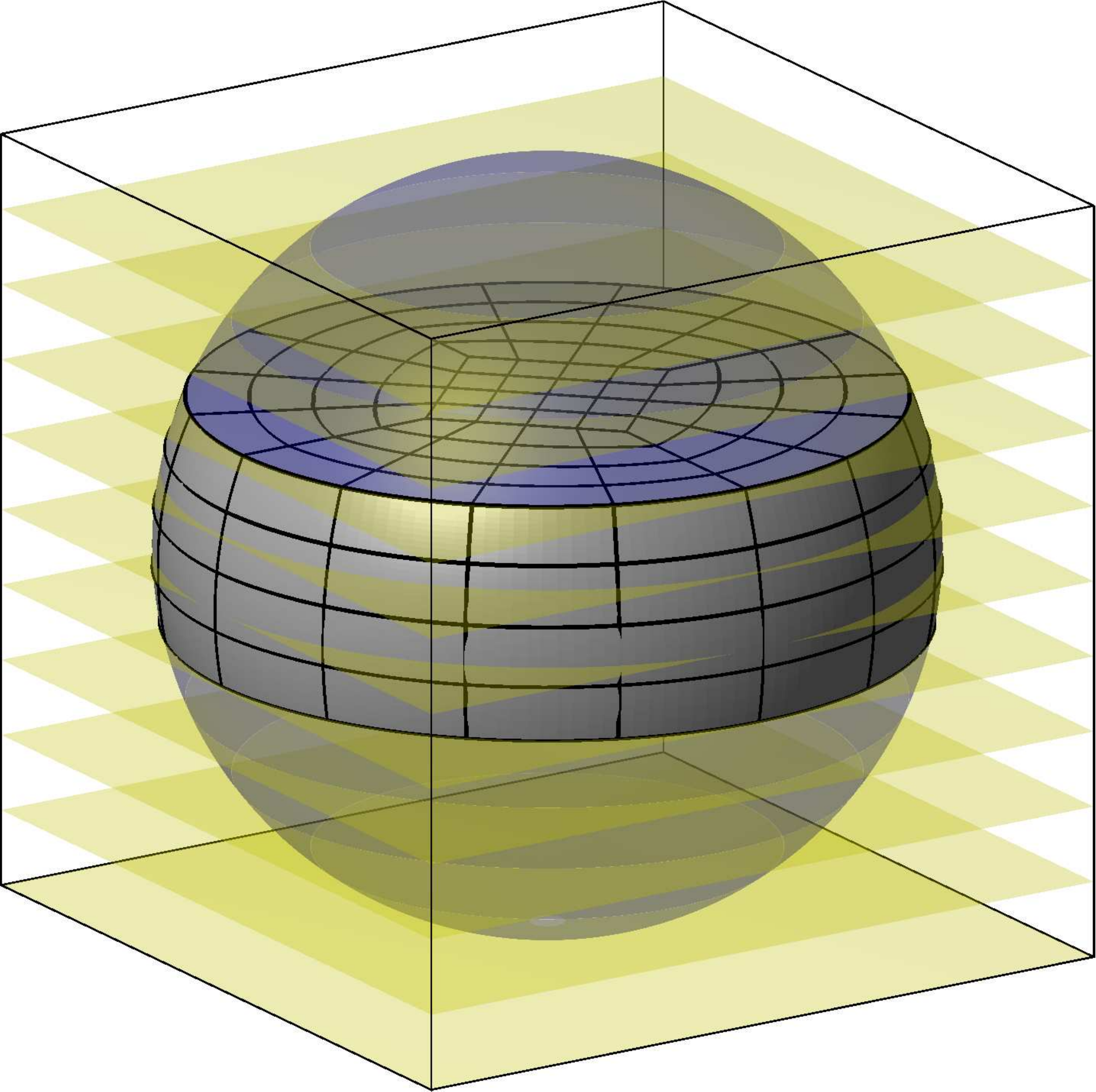}}

\subfigure[result]{\includegraphics[width=0.3\textwidth]{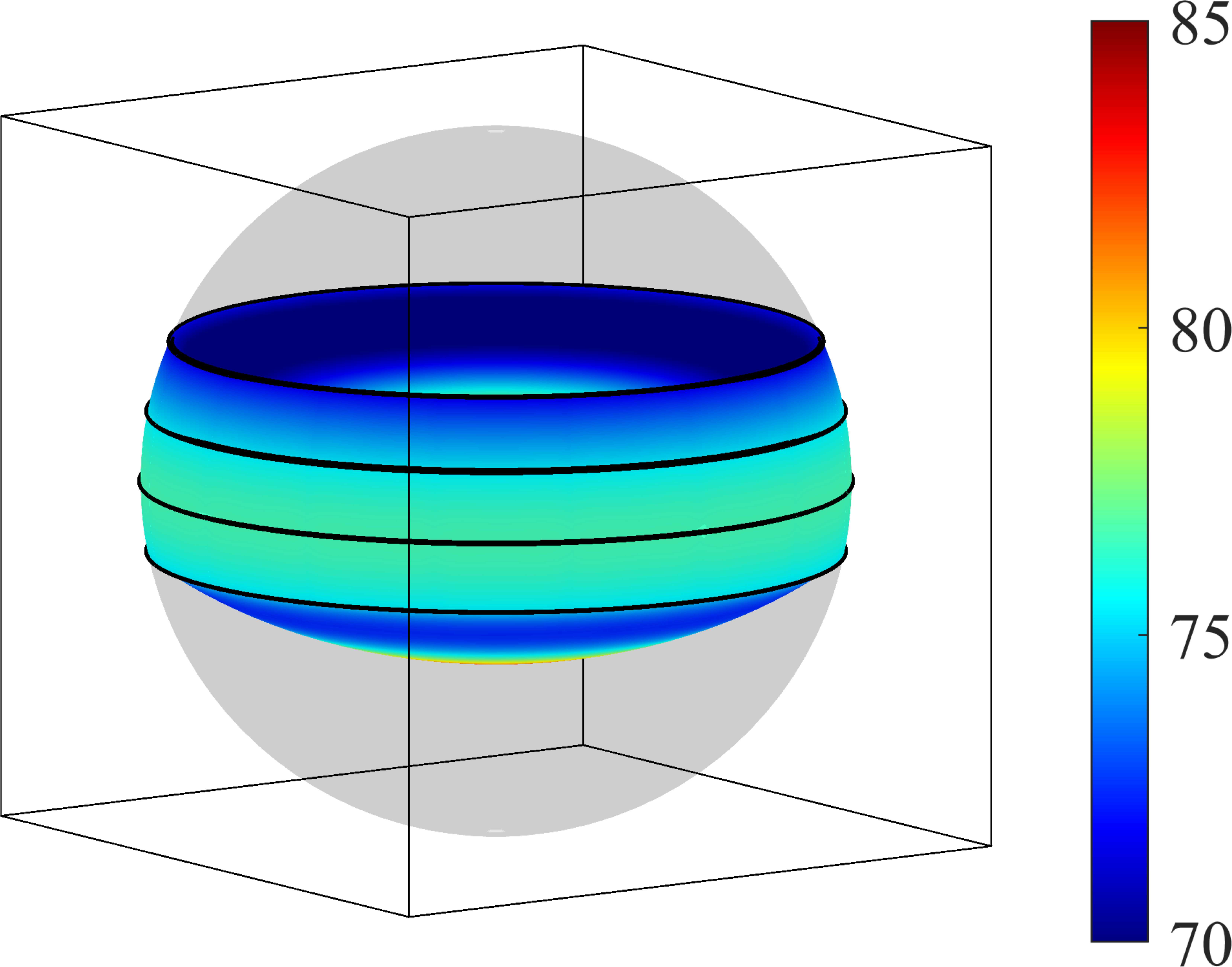}}\hfill\subfigure[result]{\includegraphics[width=0.3\textwidth]{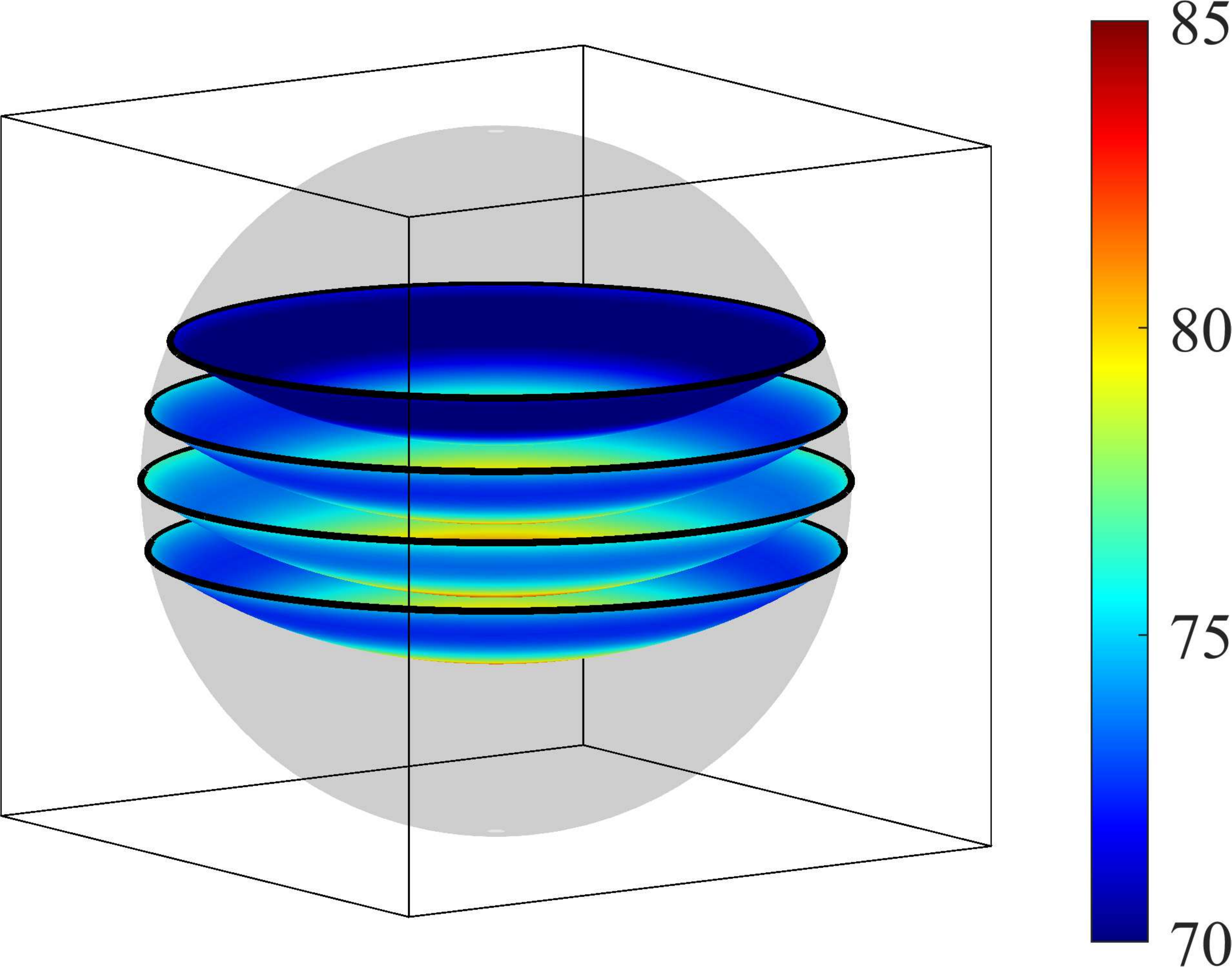}}\hfill\subfigure[result]{\includegraphics[width=0.25\textwidth]{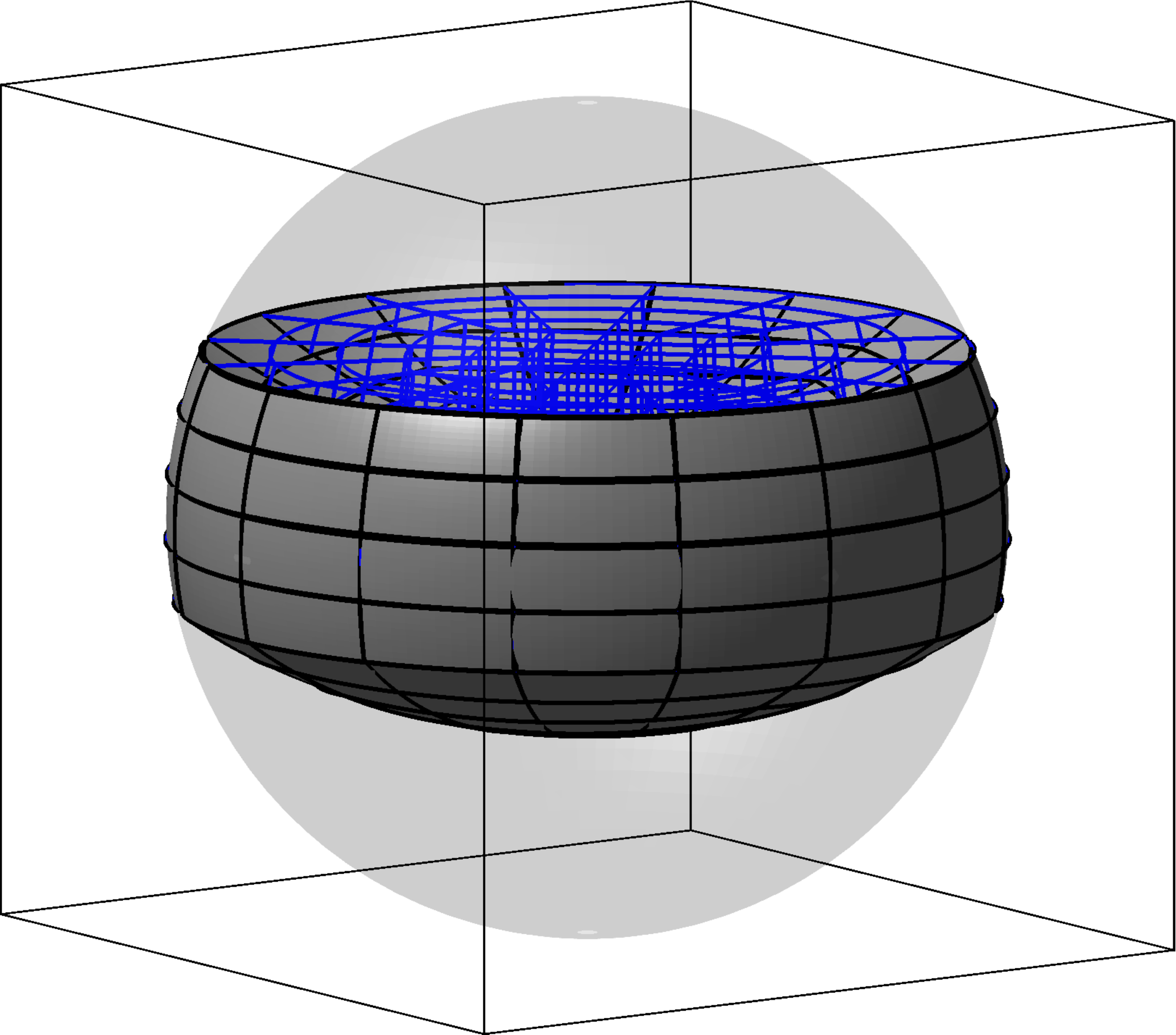}}

\caption{\label{fig:TC3d1Interval}Setup and results for test case 3 in 3D
in the prescribed interval $\phi\in\left[-\nicefrac{1}{5},\nicefrac{2}{5}\right]$.}
\end{figure}

The bulk domain is a subset of a sphere with radius $1$ centered
at the origin. Let $\psi\left(\vek X\right)=\left\Vert \vek X\right\Vert -1$
and $\phi\left(\vek X\right)=Z$, then 
\[
\Omega_{\vek X}=\left\{ \vek X\in\mathbb{R}^{3}:\psi\left(\vek X\right)\leq0\;\;\text{and}\;\;-\nicefrac{1}{5}\leq\phi(\vek X)\leq\nicefrac{2}{5}\right\} .
\]
The resulting undeformed bulk domain is seen in Fig.~\ref{fig:TC3d1Interval}(a)
and an example mesh in (b). It can also be seen that a set of planar
and circular membranes is implied in the undeformed configuration.
The thickness of the membranes is $t=1$, Young's modulus is $E=1\,000$
and Poisson ratio $\nu=0.3$, which is easily converted into the Lam\'e
parameters. The loading is gravity acting on the membrane surfaces
with $\vek F\!\left(\vek X\right)=\left[0,0,-100\right]^{\mathrm{T}}$
for all $\vek X\in\Omega_{\!\vek X}$. The whole boundary is treated
as a Dirichlet boundary with prescribed zero-displacements.

\begin{figure}
\centering

\subfigure[convergence in $\varepsilon_{\phi}$]{\includegraphics[width=0.32\textwidth]{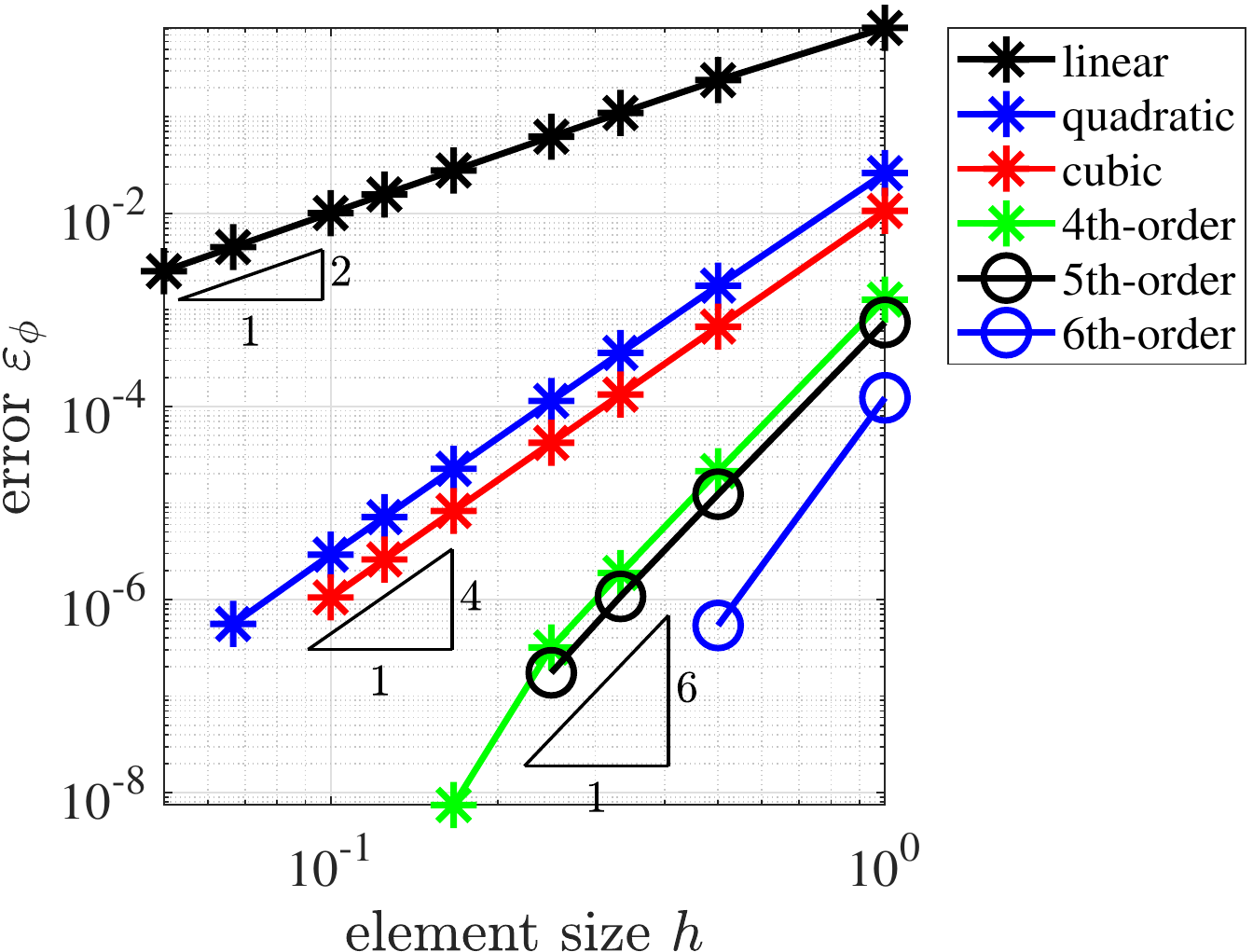}}\hfill\subfigure[convergence in $\varepsilon_{\mathfrak{e}}$]{\includegraphics[width=0.32\textwidth]{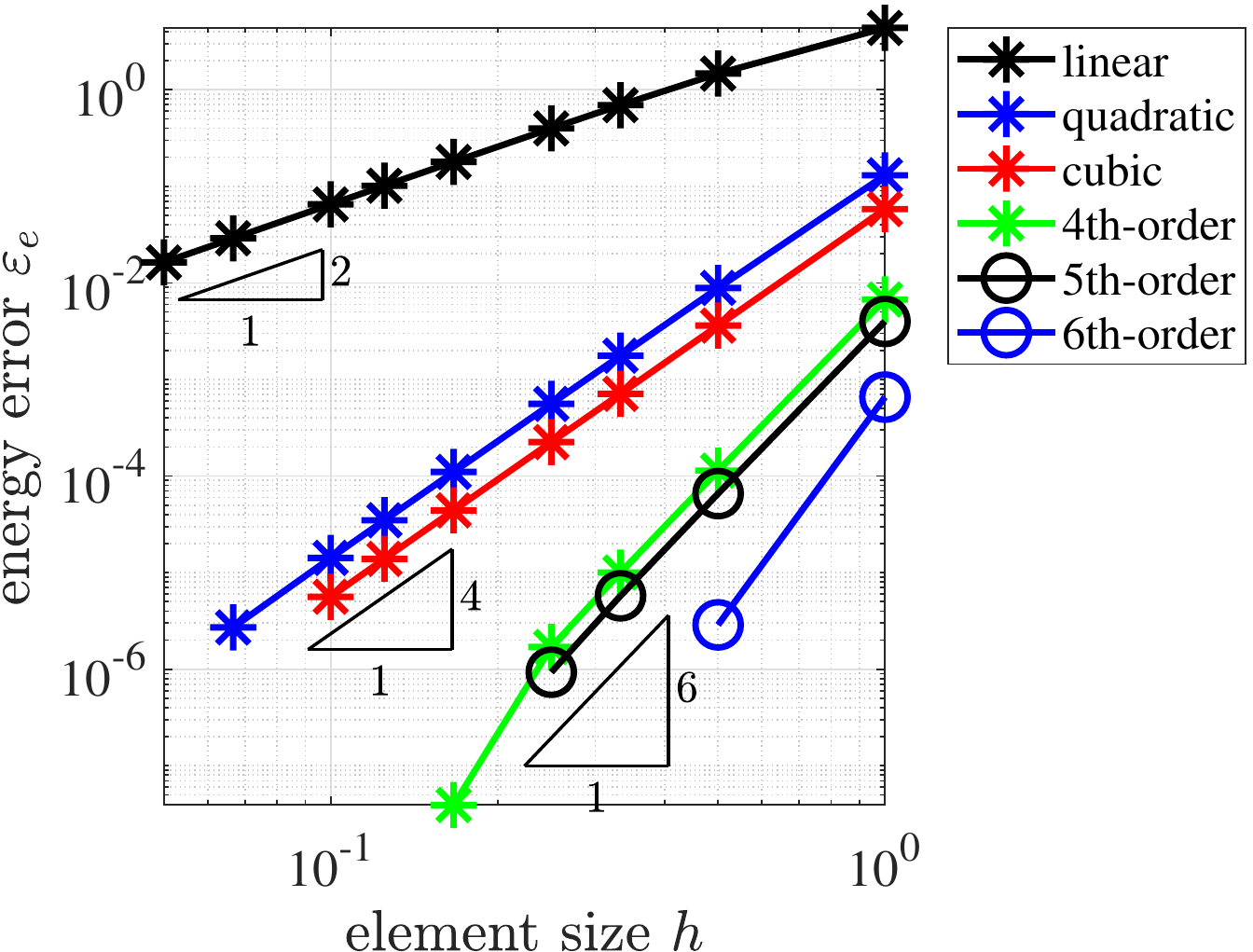}}\hfill\subfigure[convergence in $\varepsilon_{\mathrm{res}}$]{\includegraphics[width=0.32\textwidth]{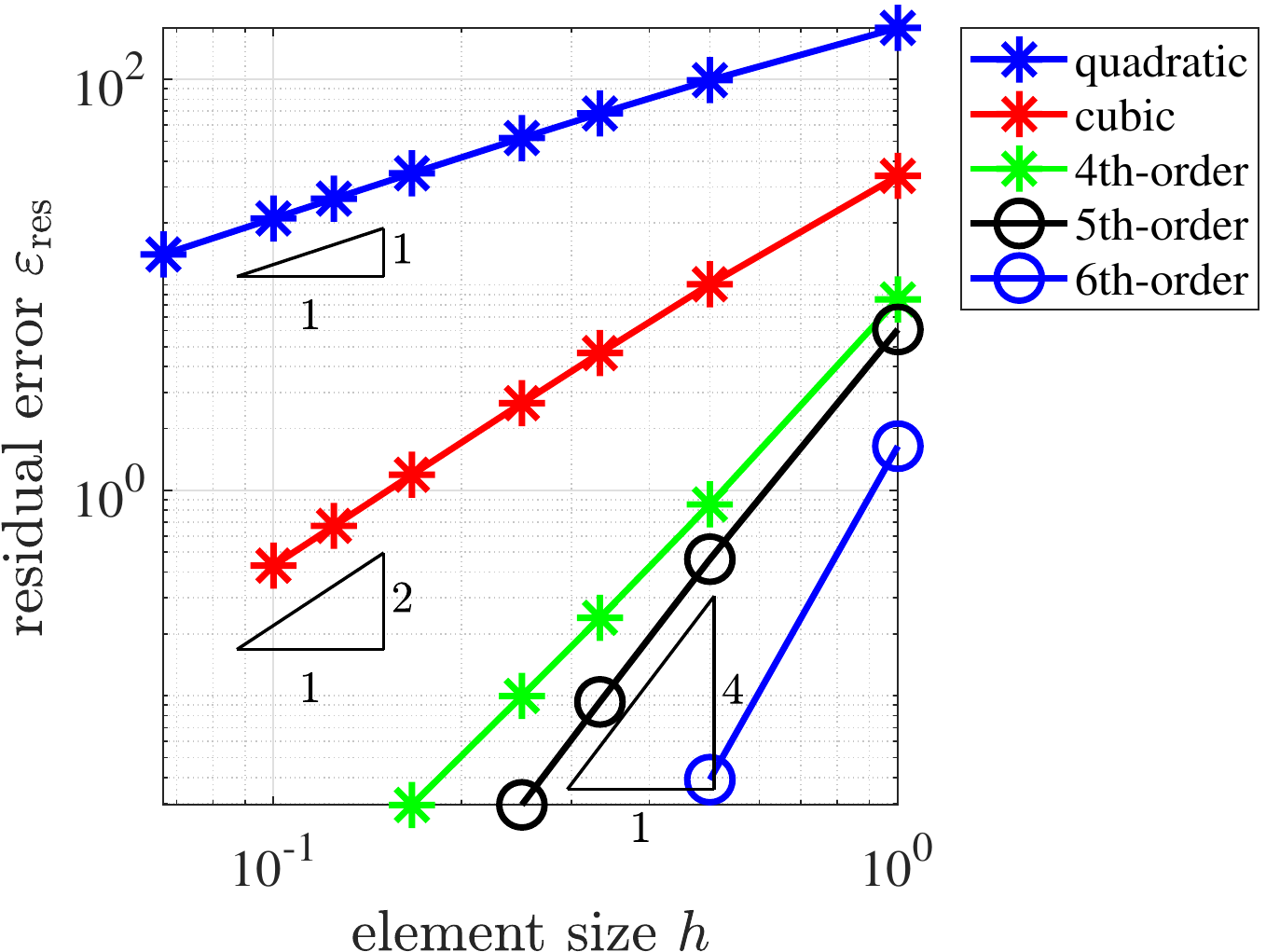}}

\caption{\label{fig:TC3d1ConvInterval}Convergence results for test case 1
in 3D confirming optimal convergence rates in $\varepsilon_{\phi}$,
$\varepsilon_{\mathfrak{e}}$ , and $\varepsilon_{\mathrm{res}}$.}
\end{figure}

Results of a single analysis are seen in Figs.~\ref{fig:TC3d1Interval}(c)
to (e); (c) and (d) are alternative visualizations of the von Mises
stress based on $\vek\sigma$, once shown on the boundary $\partial\Omega_{\vek x}$
and once on selected level sets $\ManDef$ with $c\in\left\{ -\nicefrac{1}{5},0,\nicefrac{1}{5},\nicefrac{2}{5}\right\} $.
Convergence results are seen in Fig.~\ref{fig:TC3d1ConvInterval}.
Therefore, benchmark values of $\mathfrak{D}\left(\vek u\right)=1.981355380281$
and $\mathfrak{e}\left(\vek u\right)=6.588725461796$ have been used
to compute $\varepsilon_{\phi}$ and $\varepsilon_{\mathfrak{e}}$;
in three dimensions, $\varepsilon_{\phi}$ is the error in the integrated
areas of the level sets over the deformed bulk domain. As can be seen,
optimal results are again achieved in all three error measures. For
$\varepsilon_{\phi}$ and $\varepsilon_{\mathfrak{e}}$, we again
confirm $p+1$ for odd element orders and $p+2$ for even orders.

\subsection{Test case 4 in 3D: Ellipsoidal bulk domain\label{XX_Testcase4}}

\begin{figure}
\centering

\subfigure[level sets]{\includegraphics[width=0.3\textwidth]{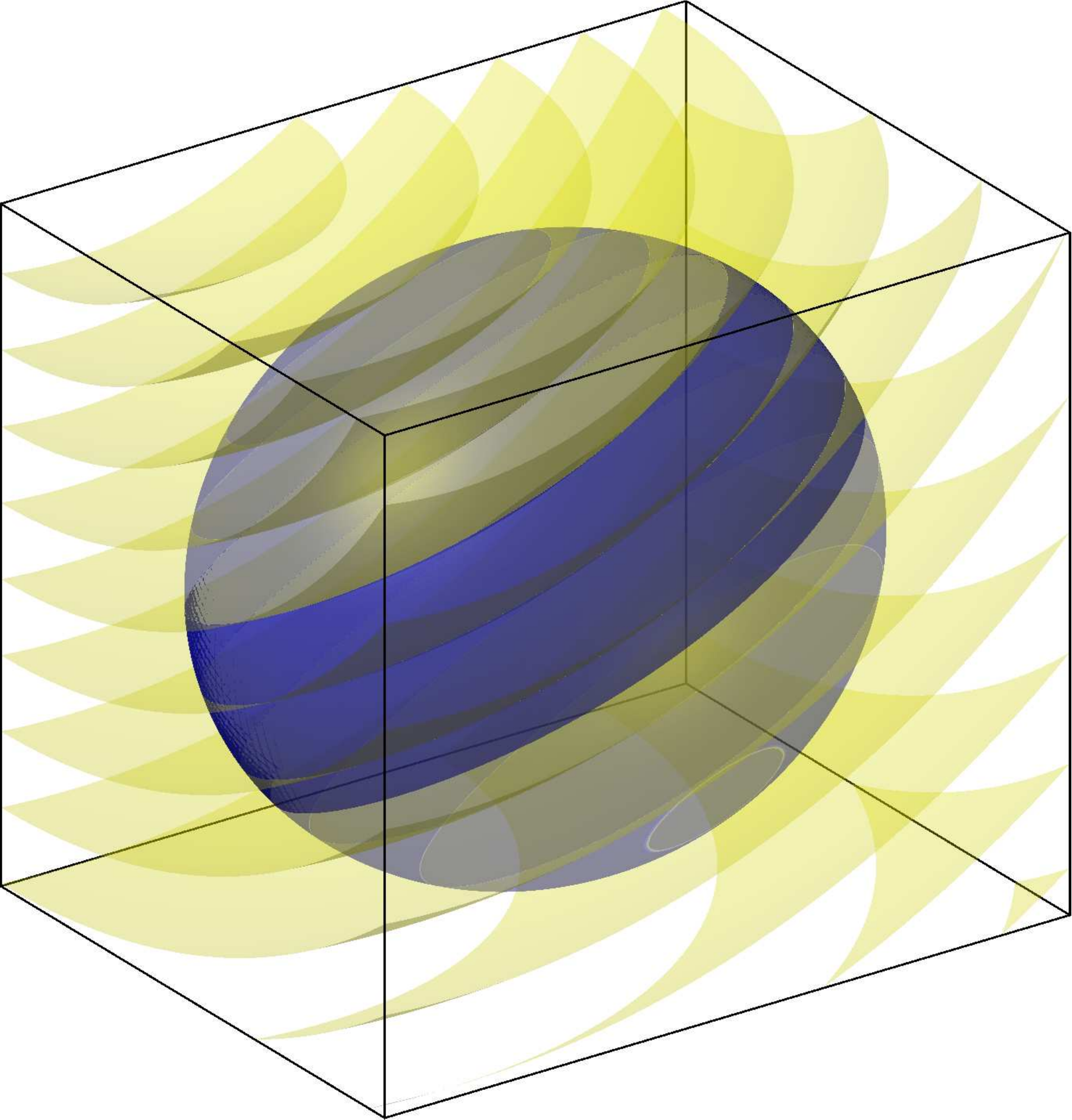}}$\qquad$\subfigure[$\Omega_{\vek X}$ and mesh]{\includegraphics[width=0.3\textwidth]{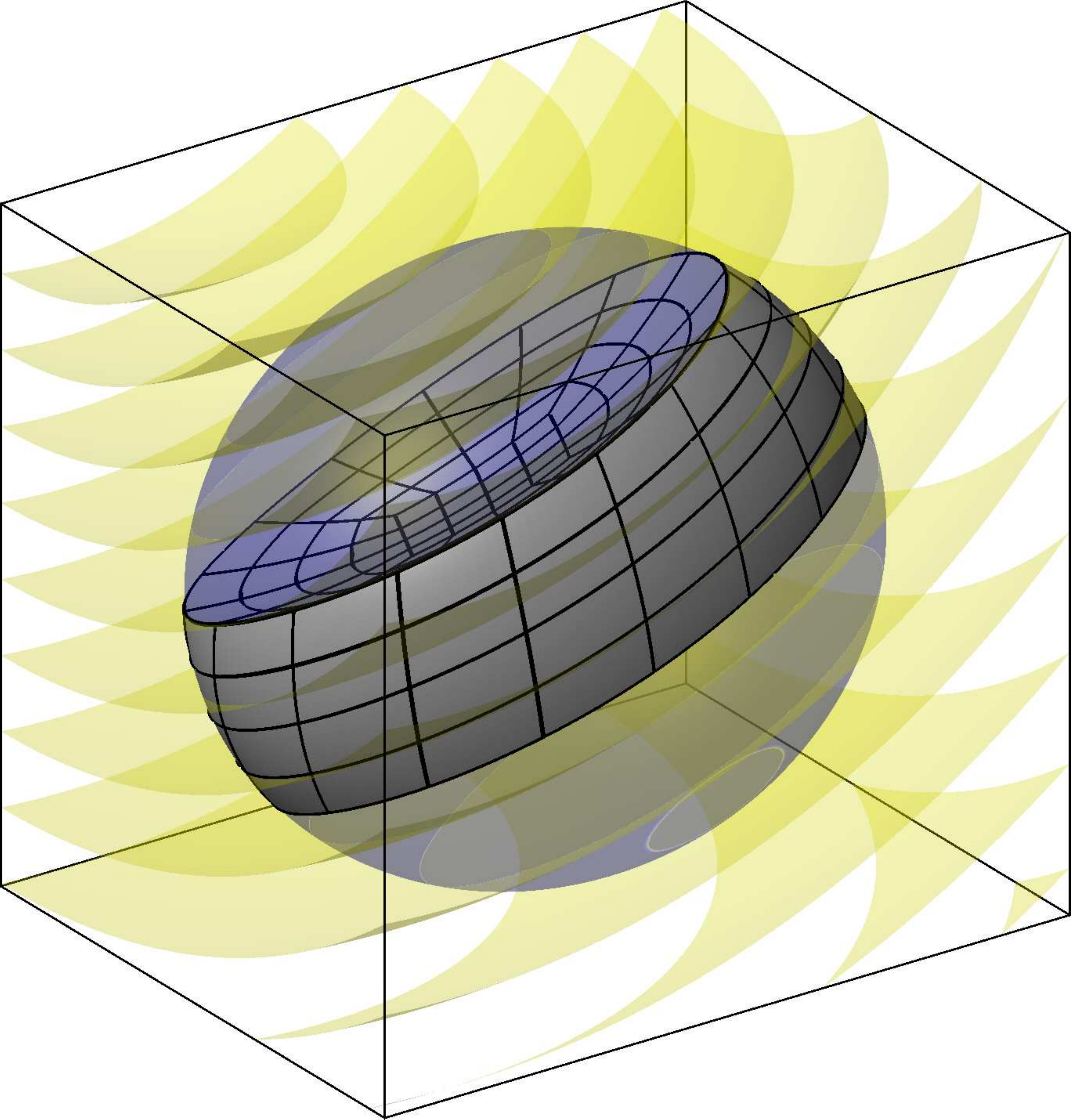}}

\subfigure[result]{\includegraphics[width=0.3\textwidth]{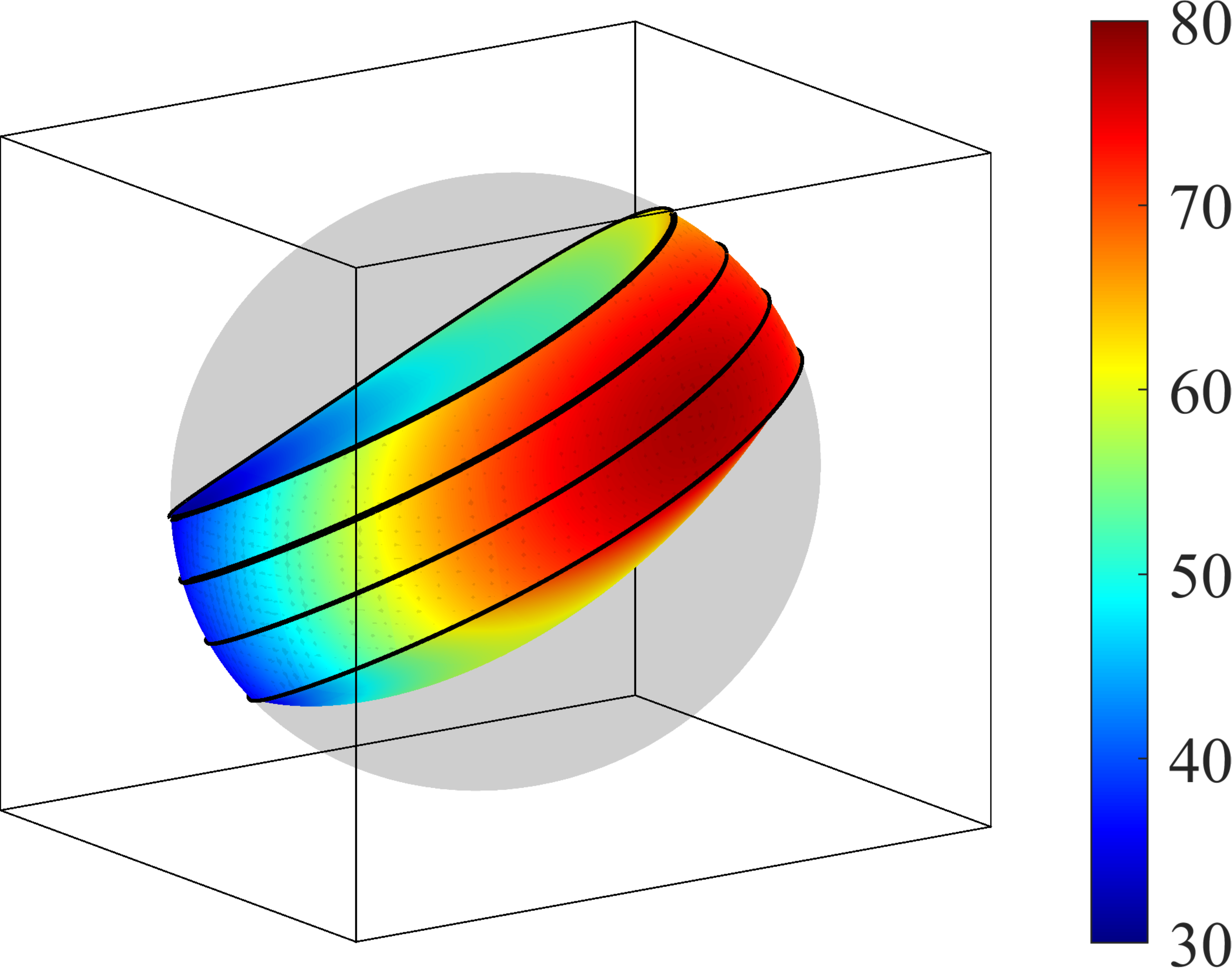}}\hfill\subfigure[result]{\includegraphics[width=0.3\textwidth]{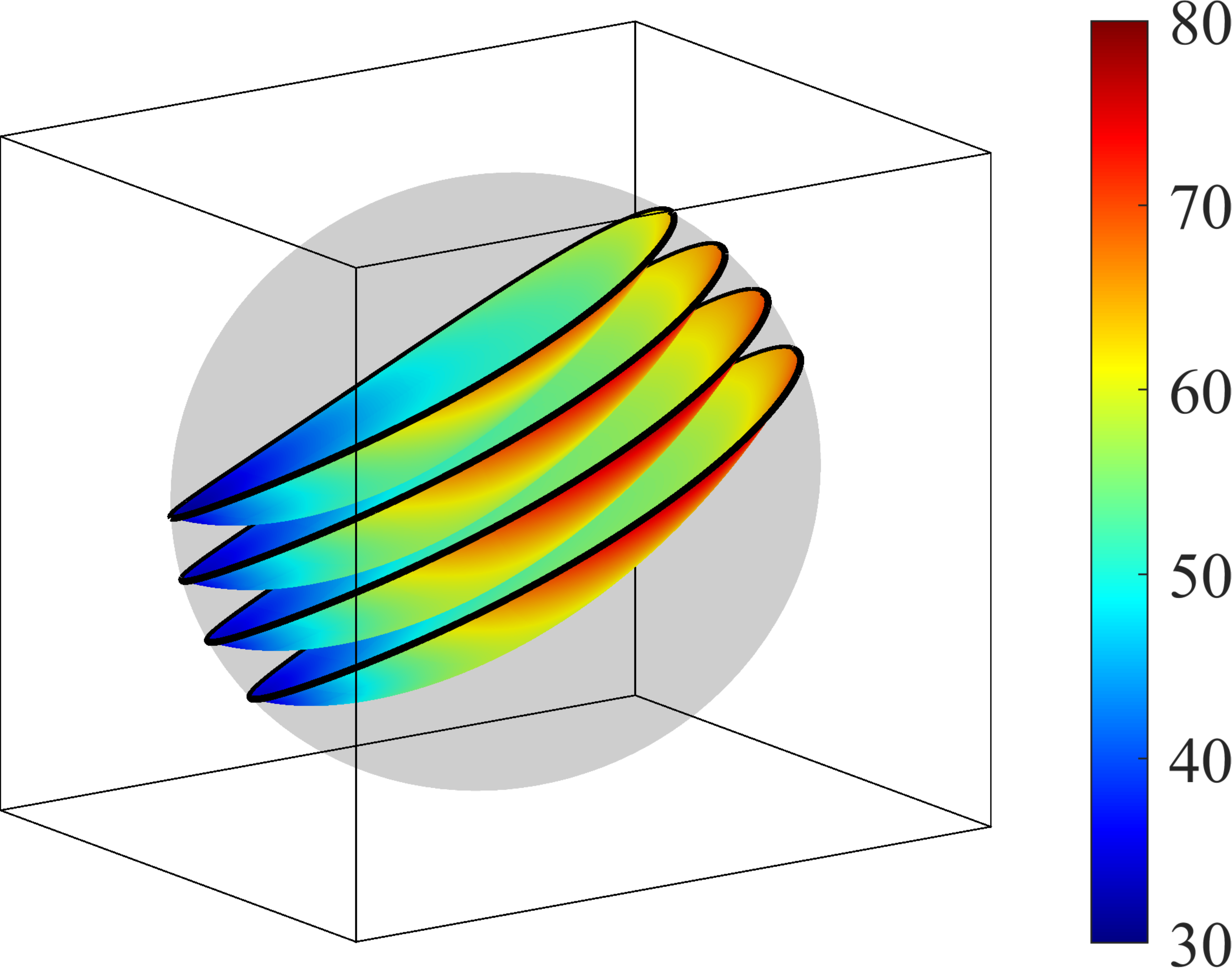}}\hfill\subfigure[convergence]{\includegraphics[width=0.32\textwidth]{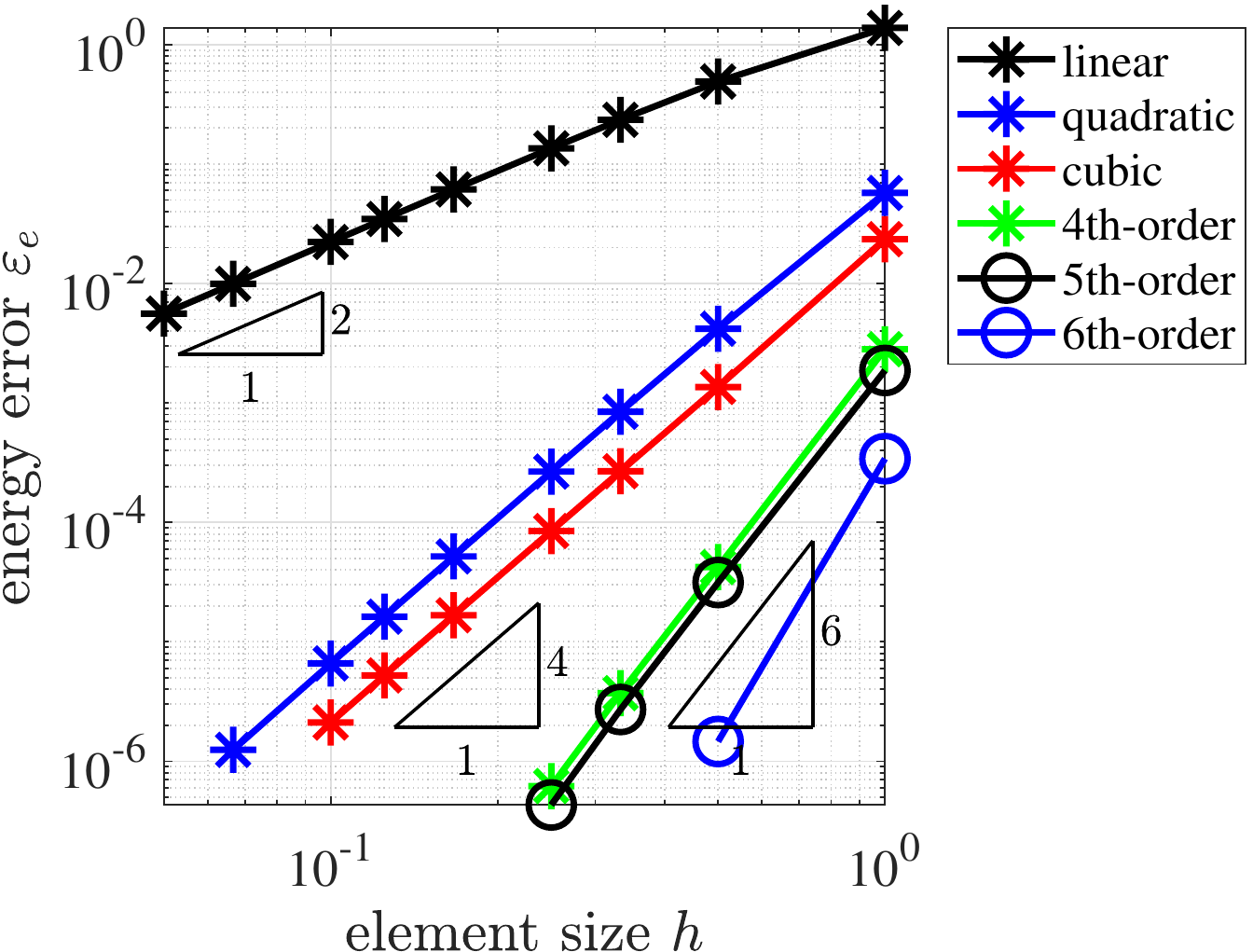}}

\caption{\label{fig:TC3d2Interval}Setup and results for test case 4 in 3D
in the prescribed interval $\phi\in\left[0,\nicefrac{1}{2}\right]$.}
\end{figure}

We define the two level-set functions
\[
\psi\left(\vek X\right)=\left(\frac{X-X_{E}}{R_{X}}\right)^{2}+\left(\frac{Y-Y_{E}}{R_{Y}}\right)^{2}+\left(\frac{Z-Z_{E}}{R_{Z}}\right)^{2}-1,
\]
with $\vek X_{\!E}=\left[-0.2,0.2,0.1\right]^{\mathrm{T}}$ and $R_{X}=0.9$,
$R_{Y}=0.7$, $R_{Z}=0.8$ and 
\[
\phi\left(\vek X\right)=\left\Vert \vek X-\vek X_{\!S}\right\Vert -R_{S}
\]
with $\vek X_{\!S}=\left[-1,1,2\right]^{\mathrm{T}}$ and $R_{S}=2$,
see Fig.~\ref{fig:TC3d2Interval}(a). The bulk domain is defined
as 
\[
\Omega_{\vek X}=\left\{ \vek X\in\mathbb{R}^{3}:\psi\left(\vek X\right)\leq0\;\;\text{and}\;\;0\leq\phi(\vek X)\leq\nicefrac{1}{2}\right\} .
\]
The resulting undeformed bulk domain with an example mesh is seen
in Fig.~\ref{fig:TC3d2Interval}(b). All material parameters, loading
and boundary conditions are given as in the previous test case in
Section \ref{XX_Testcase3}. Results are seen in Figs.~\ref{fig:TC3d2Interval}(c)
and (d) following previous visualizations. Benchmark values for the
convergence studies are $\mathfrak{D}\left(\vek u\right)=1.032907088507$
and $\mathfrak{e}\left(\vek u\right)=1.863258461070$. However, because
the resulting convergence plots are virtually identical to previous
test cases, only the convergence in the stored energy error $\varepsilon_{\mathfrak{e}}$
is shown in Fig.~\ref{fig:TC3d2Interval}(e).

\subsection{Test case 5 in 2D: Bulk material with embedded fibres\label{XX_Testcase5}}

\begin{figure}
\centering

\subfigure[$n_{\mathrm{rad}}=2$]{\includegraphics[width=0.23\textwidth]{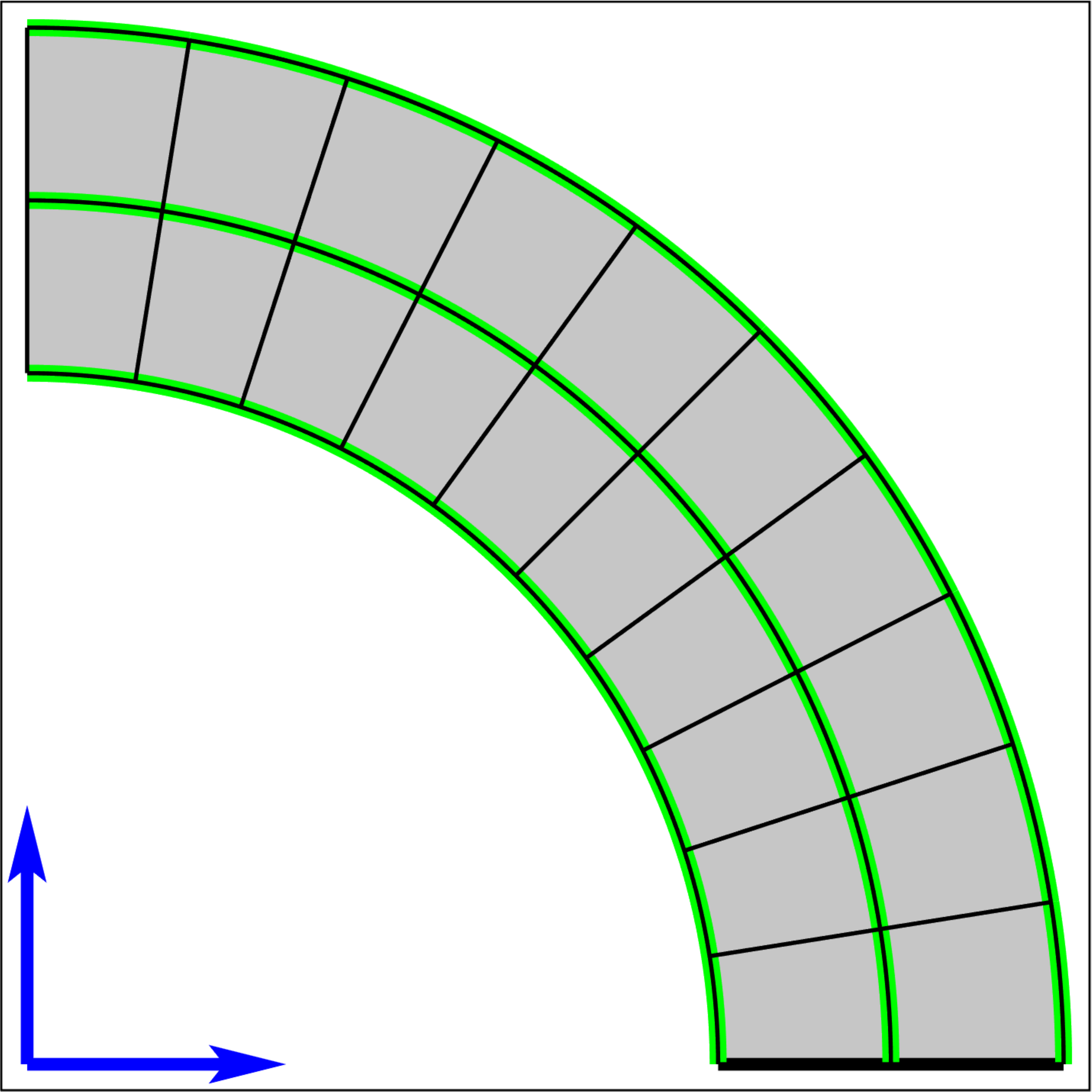}}\hfill\subfigure[$n_{\mathrm{rad}}=3$]{\includegraphics[width=0.23\textwidth]{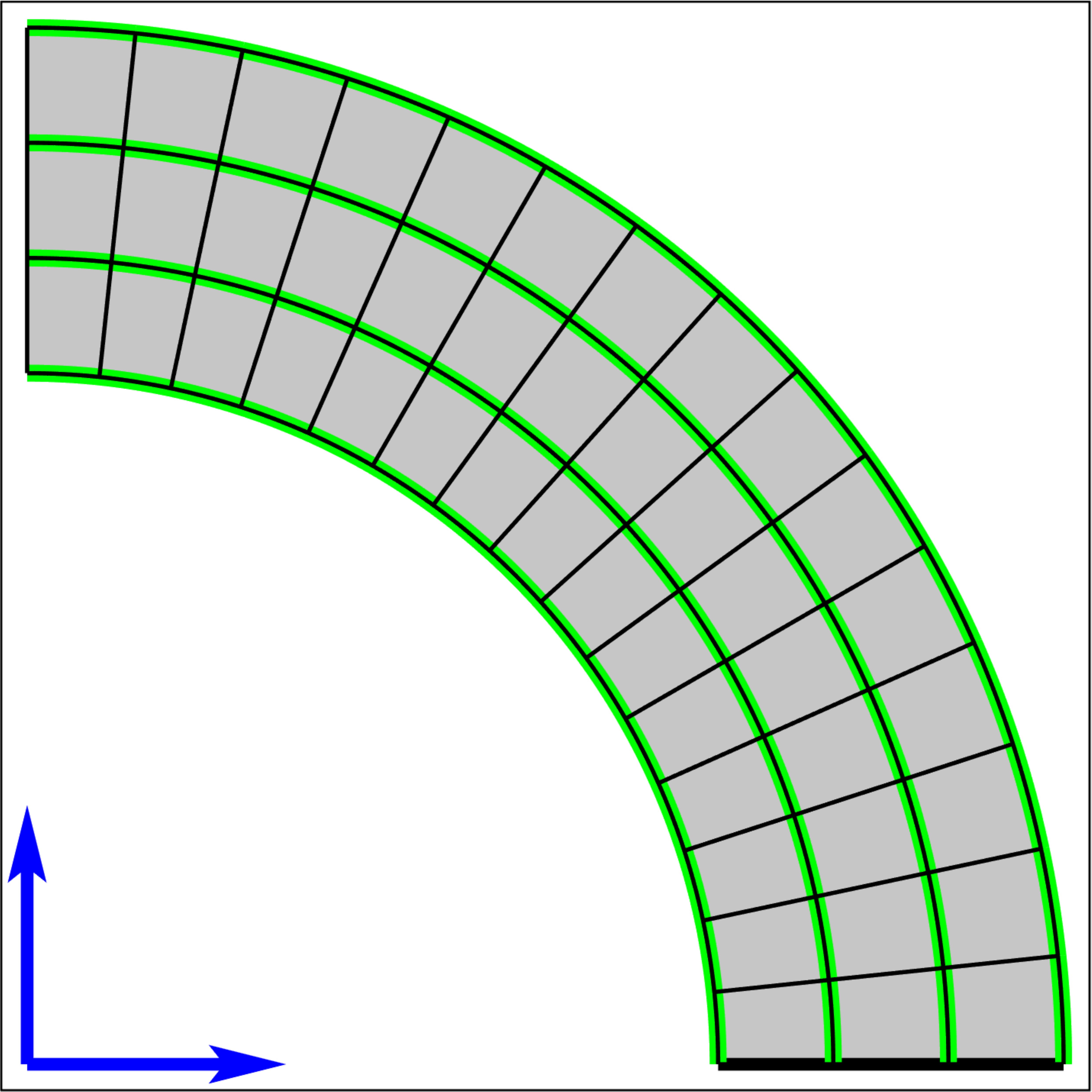}}\hfill\subfigure[$n_{\mathrm{rad}}=5$]{\includegraphics[width=0.23\textwidth]{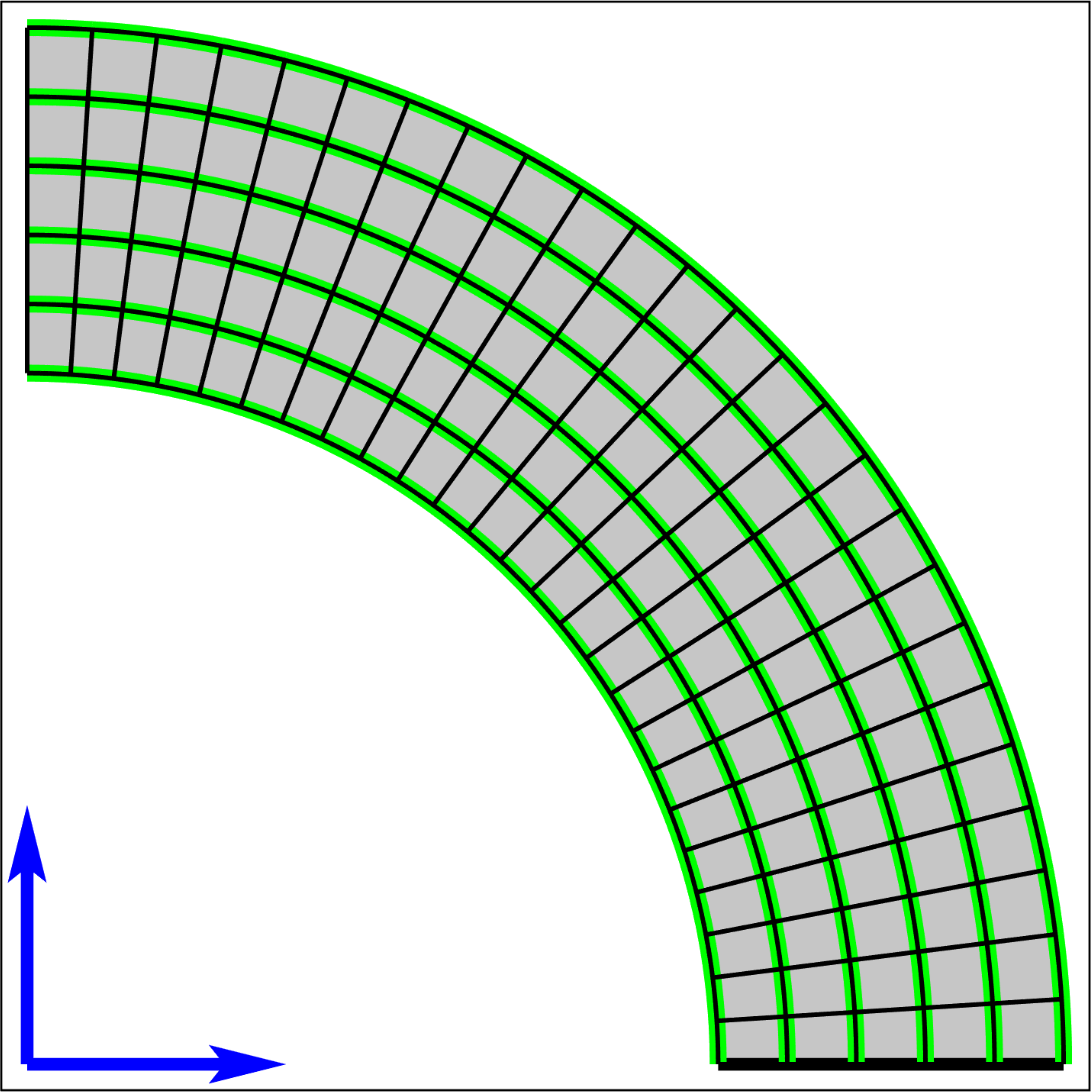}}\hfill\subfigure[$n_{\mathrm{rad}}=10$]{\includegraphics[width=0.23\textwidth]{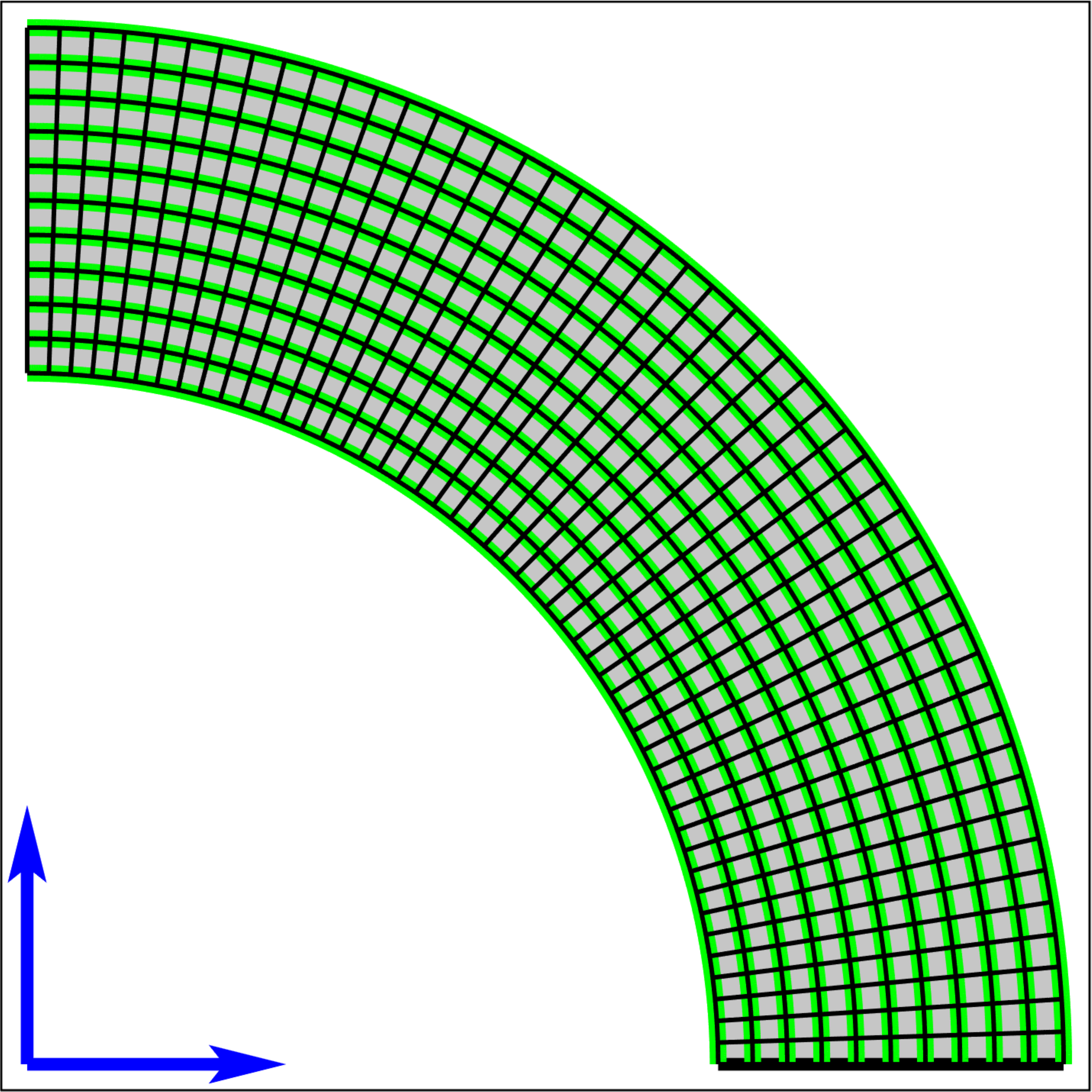}}

\caption{\label{fig:TC2d3DiscreteFibreModel}\emph{Discrete} fibre model in
an annulus where the number of fibres (and their stiffness) is associated
to the number of elements in radial direction $n_{\mathrm{rad}}$.
The embedded fibres are meshed conformingly by the element edges shown
as green lines.}
\end{figure}

The final example considers a coupled problem where some isotropic
and homogeneous bulk material is reinforced with curved, one-dimensional
fibres. This application relates to the discussion of continuously
embedded sub-structure models in Section \ref{XX_ElasticBulkDomains}.
The bulk domain is a quarter annulus centered at the origin, an inner
radius of $R_{i}=8$ and an outer radius of $R_{o}=12$, see Fig.~\ref{fig:TC2d3DiscreteFibreModel}(a).
The resulting area is $A=\nicefrac{1}{4}\cdot\pi\cdot\left(R_{o}^{2}-R_{i}^{2}\right)=20\pi$.
In contrast to the previous examples where the focus was exclusively
on the behaviour of ropes and membranes within bulk domains, in this
study, there is also an isotropic \emph{bulk material} characterized
by a Young's modulus of $E_{b}=10\,000$ and a Poisson's number of
$\nu_{b}=0.3$; the Lam\'e constants are then computed according
to plane strain conditions. The behaviour of the bulk material is
modeled according to standard finite strain theory in two dimensions.
There is a body force of $\vek F\!\left(\vek X\right)=\left[0,-20\right]^{\mathrm{T}}$
acting on the bulk material only (i.e., there are no additional body
forces on the fibres). Zero-displacements are prescribed on the lower
side of the bulk domain where $Y\!=0$.

\begin{figure}
\centering

\subfigure[deformation]{\includegraphics[width=0.27\textwidth]{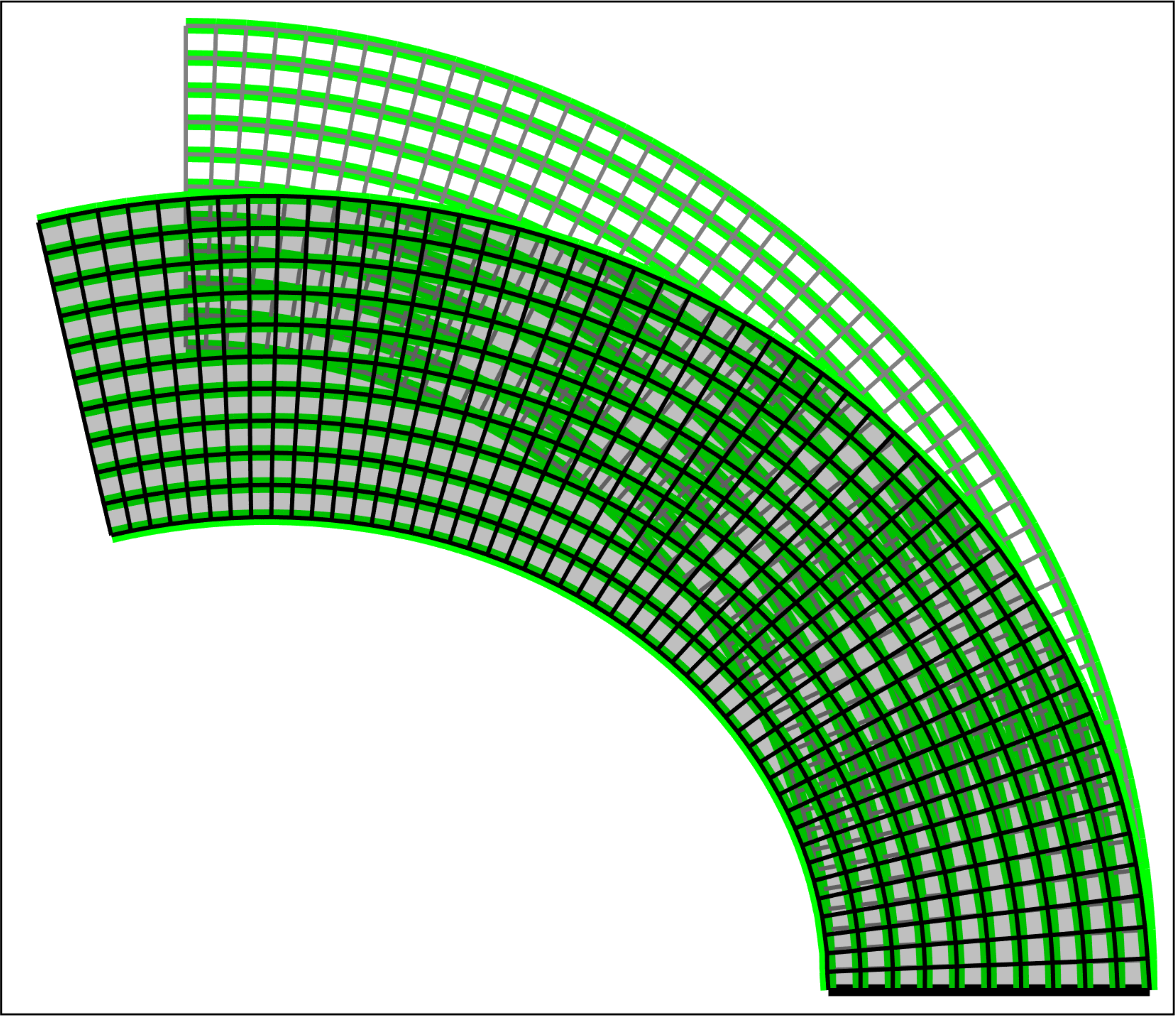}}$\qquad$\subfigure[cont.~fibre model]{\includegraphics[width=0.23\textwidth]{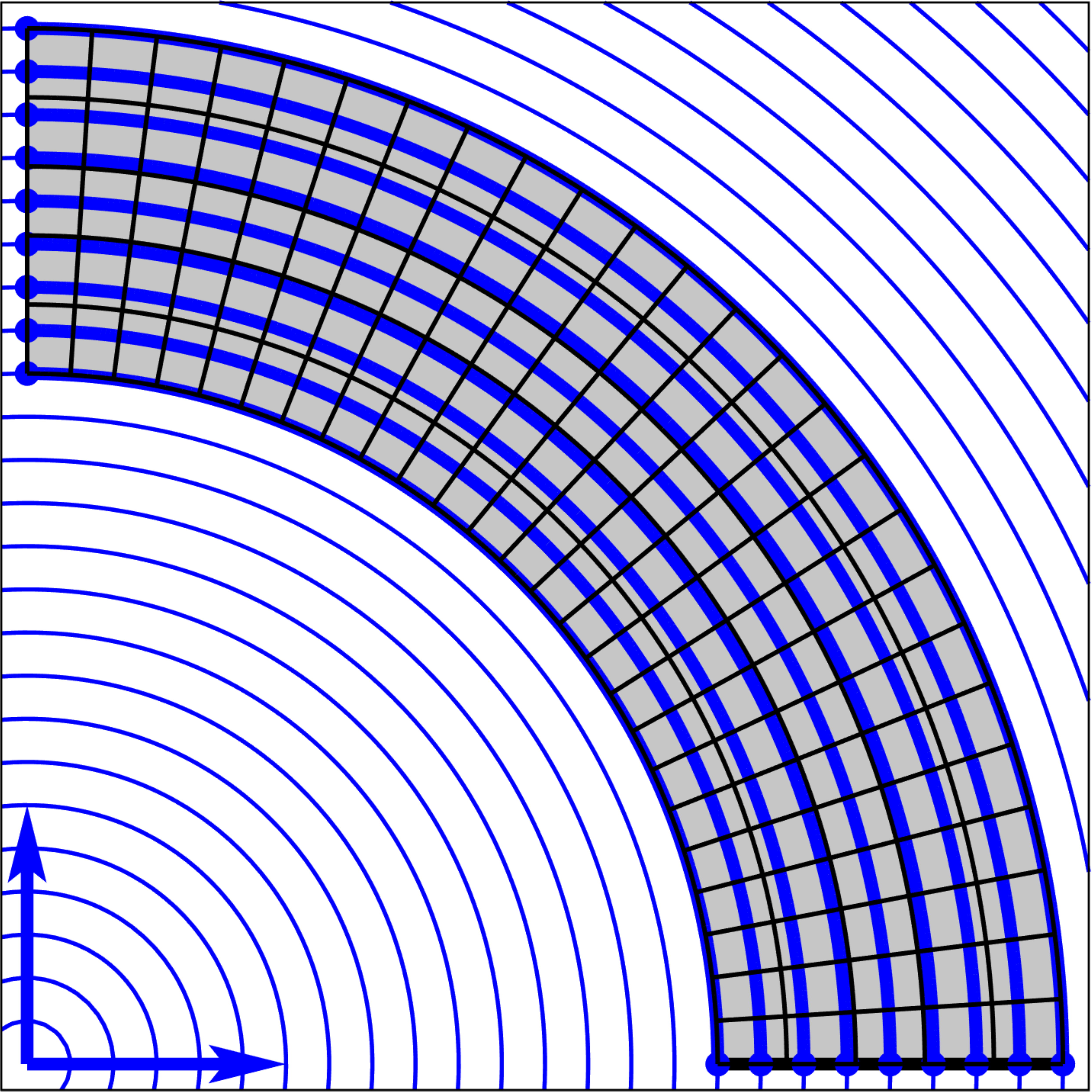}}$\qquad$\subfigure[deformation]{\includegraphics[width=0.27\textwidth]{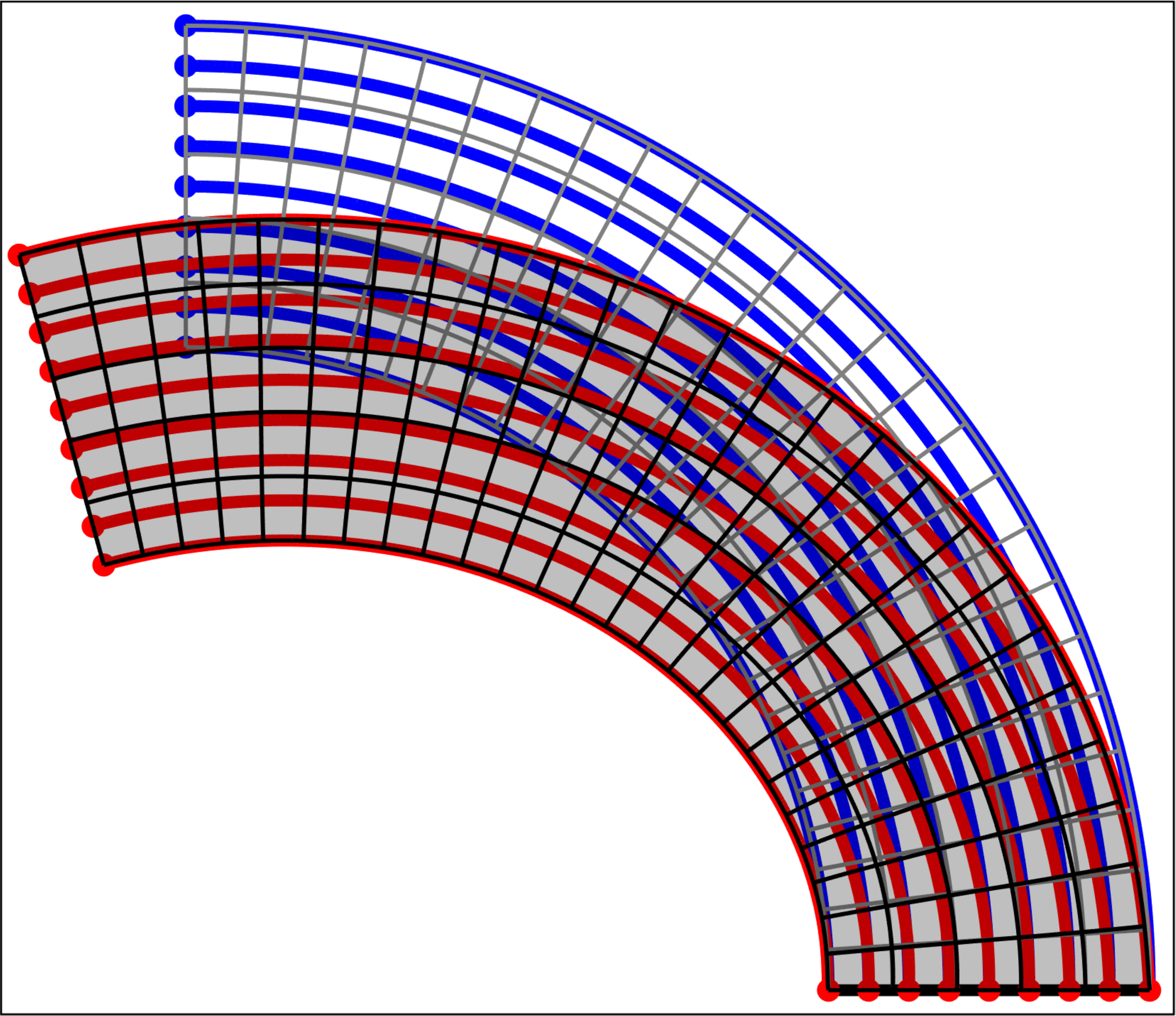}}

\caption{\label{fig:TC2d3ContinuousFibreModel}(a) shows the deformed configuration
according to the discrete fibre model with $n_{\mathrm{rad}}=10$,
(b) shows the \emph{continuous} fibre model in an annulus where the
fibres are implied by all level sets in the interval $\phi\in\left[8,12\right]$,
(c) is the deformed configuration for the continuous fibre model as
obtained with the Bulk Trace FEM.}
\end{figure}

For the modeling of the fibres, two strategies are compared: In the
first approach, we consider a \emph{discrete} set of fibres where
the number of fibres is coupled to the element number used in the
analysis, see Fig.~\ref{fig:TC2d3DiscreteFibreModel}. As can be
seen, the discrete fibres are meshed conformingly by curved, one-dimensional
elements and one may use standard approaches (i.e., classical Surface
FEM) for the modeling and analysis, see, e.g., \cite{Bischoff_2017a,Calladine_1983a,Ciarlet_1997a,Chapelle_2011a,Fries_2020a};
of course, coupled to the analysis of the bulk material. Let $n_{\mathrm{rad}}$
be the number of elements in radial direction of the annulus, then
there are $n_{\mathrm{rad}}+1$ fibres with geometries $\Gamma_{\!\vek X}^{i}$,
$i=1,\dots n_{\mathrm{rad}}+1$ in the analysis. In order not to make
the fibre-reinforced annulus stiffer upon increasing the element number,
we reduce the stiffness of the fibres with each refinement step. Therefore,
it is useful to define a target value for the \emph{integrated} Young's
modulus over the bulk domain \emph{and} the fibres as
\begin{align*}
\int_{\Omega_{\vek X}}E_{b}\;\mathrm{d}\Omega+\sum_{i=1}^{n_{\mathrm{rad}}+1}\int_{\Gamma_{\!\vek X}^{i}}E_{\mathrm{discr}}\;\mathrm{d}\Omega & =580\,000\pi,\\
\Rightarrow\sum_{i=1}^{n_{\mathrm{rad}}+1}\int_{\Gamma_{\!\vek X}^{i}}E_{\mathrm{discr}}\;\mathrm{d}\Omega & =580\,000\pi-20\pi\cdot10\,000,\\
\Rightarrow E_{\mathrm{discr}} & =76\,000\,/\,\left(n_{\mathrm{rad}}+1\right).
\end{align*}
This enables us to keep the cross section of the fibres at $1$. Of
course, one could also define a constant Young's modulus for the discrete
fibres and adjust the cross section with respect to $n_{\mathrm{rad}}$.
Note that the Poisson number plays no role for fibres, hence, $\nu=0$.
A sketch of the deformed domain may be seen in Fig.~\ref{fig:TC2d3ContinuousFibreModel}(a).

The other strategy for the fibres is to associate a \emph{continuous}
set of fibres to the level sets of $\phi\left(\vek X\right)=\left\Vert \vek X\right\Vert $
in the interval $\phi_{\min}=8$ and $\phi_{\max}=12$, see Fig.~\ref{fig:TC2d3ContinuousFibreModel}(b),
as proposed in this work. Then, the Young's modulus is computed as
\begin{align*}
\int_{\Omega_{\vek X}}E_{b}\;\mathrm{d}\Omega+\int_{\Omega_{\vek X}}E_{\mathrm{cont}}\cdot\left\Vert \nabla_{\!\!\vek X}\phi\right\Vert \;\mathrm{d}\Omega & =580\,000\pi,\\
\Rightarrow E_{\mathrm{cont}}\cdot20\pi & =580\,000\pi-20\pi\cdot10\,000,\\
\Rightarrow E_{\mathrm{cont}} & =19\,000,
\end{align*}
keeping the cross section constant at $1$. Together with $\nu=0$,
these material parameters are used for the analysis based on the Bulk
Trace FEM as outlined above, this time coupled with the classical
2D analysis of the bulk material. The deformed domain, obtained with
the Bulk Trace FEM, is shown in Fig.~\ref{fig:TC2d3ContinuousFibreModel}(c).
We observe a stored elastic energy of $\mathfrak{e}\left(\vek u\right)=674.363$
and a displacement of the lower left node as $\vek u\left(0,8\right)=\left[-1.0194534,-2.710712\right]^{\mathrm{T}}$.
Most importantly, we confirm that the strategy based on the discrete
fibres converges to the case of continuous fibres as proposed herein.
However, it is also important to note that due to the boundary conditions,
there is a singularity at the corner points at $\vek X=\left[8,0\right]^{\mathrm{T}}$
and $\left[12,0\right]^{\mathrm{T}}$ which hinders optimal convergence
rates. Therefore, these results are given with reduced accuracy compared
to the previous test cases.

\begin{figure}
\centering

\subfigure[prescribed displ.]{\includegraphics[width=0.27\textwidth]{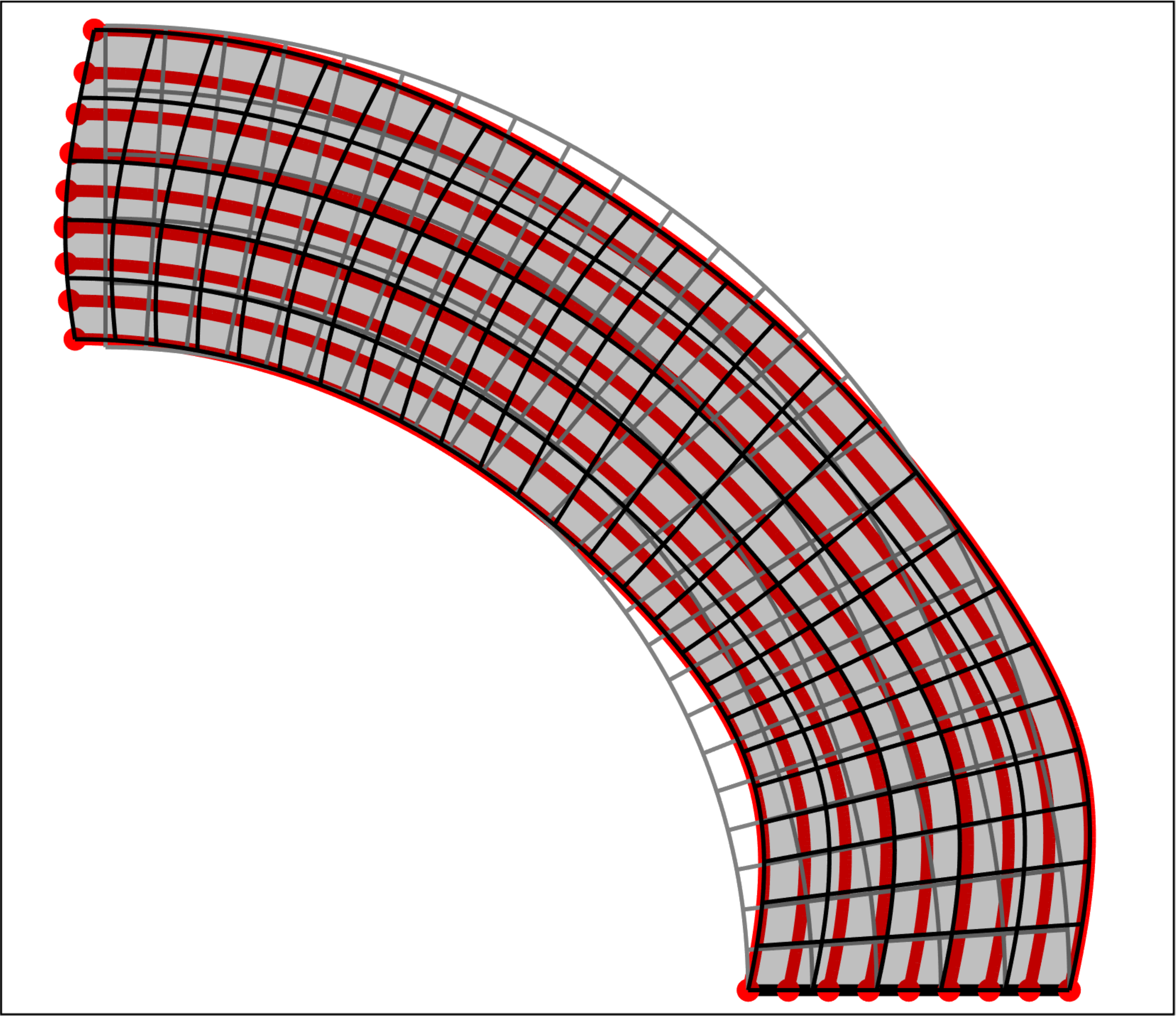}}$\qquad$\subfigure[cont.~fibre model]{\includegraphics[width=0.27\textwidth]{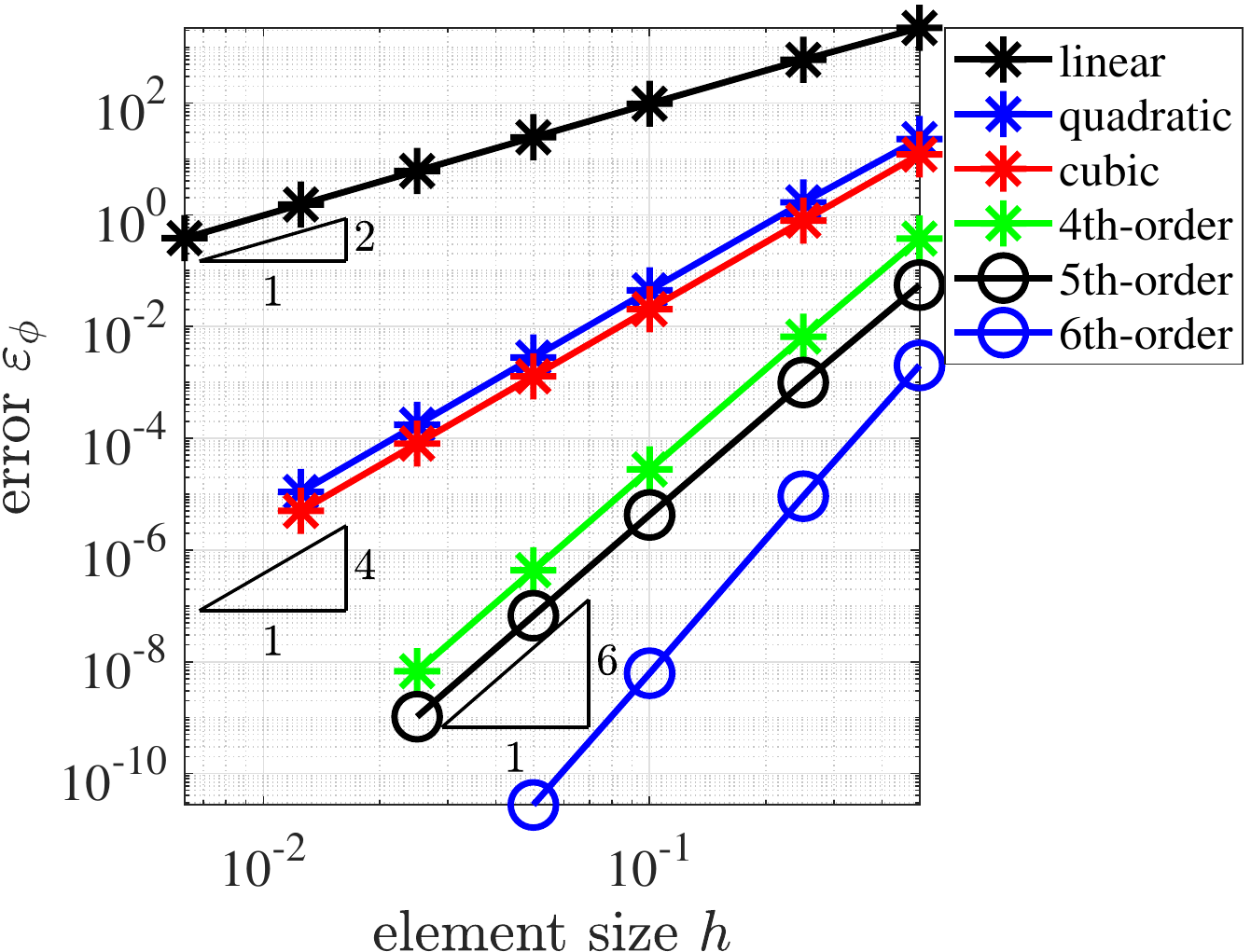}}$\qquad$\subfigure[discrete fibre model]{\includegraphics[width=0.27\textwidth]{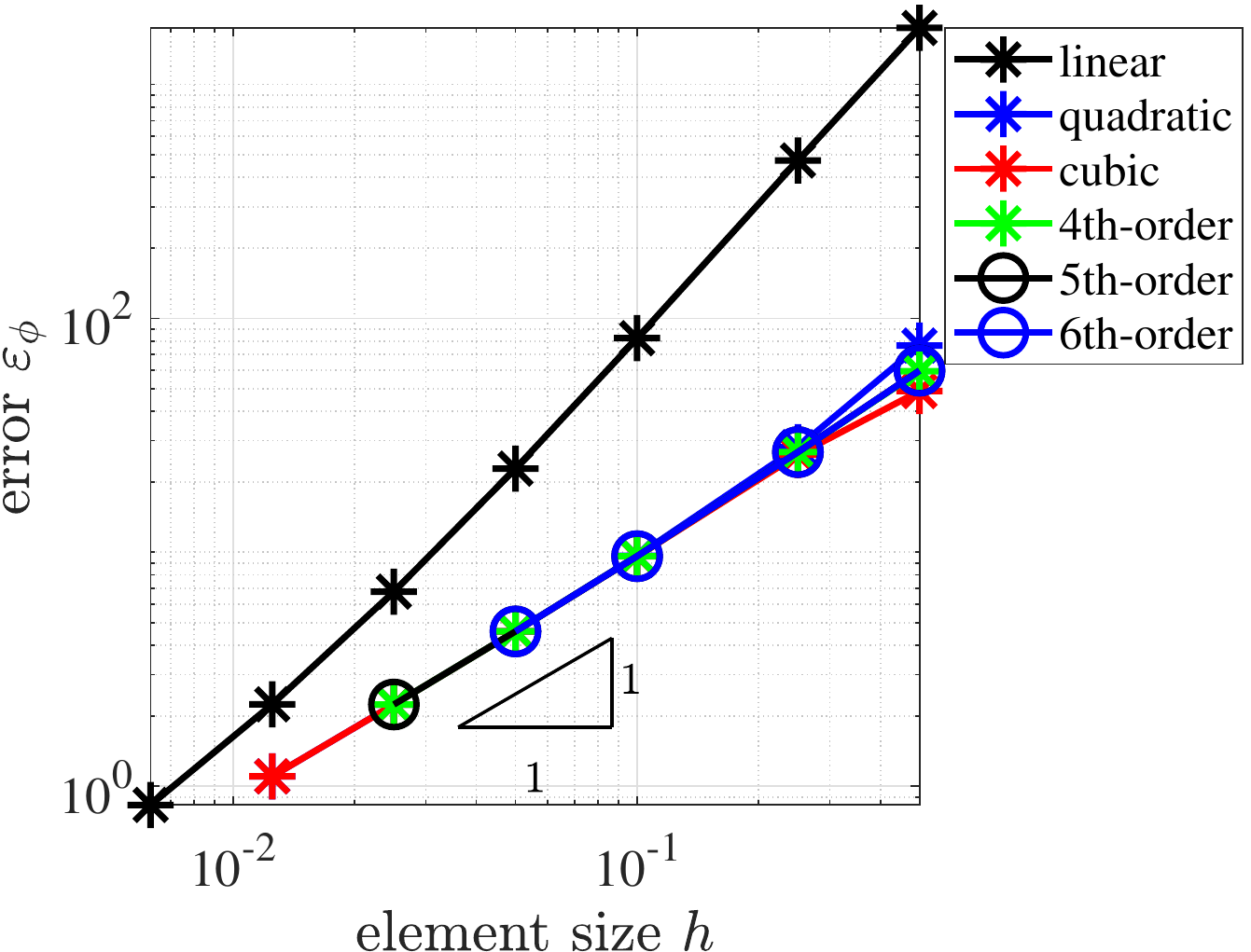}}

\caption{\label{fig:TC2d3FibreModelsPrescrSol}(a) shows the deformed configuration
according to the prescribed displacement field in Eq.~(\ref{eq:PrescrDisplField}),
(b) shows convergence rates for the continuous fibre model and (c)
for the discrete fibre model towards the reference energy of the continuous
model.}
\end{figure}

In order to make the results more quantifiable, it was found useful
to provide energy results for a prescribed displacement field of 
\begin{equation}
\vek u\left(\vek X\right)=\left[\begin{array}{c}
u\left(X,Y\right)\\
v\left(X,Y\right)
\end{array}\right]=\left[\begin{array}{c}
\nicefrac{1}{2}\sin\left(\nicefrac{1}{2}Y\right)\\
\nicefrac{1}{10}\sin\left(Y\right)
\end{array}\right],\label{eq:PrescrDisplField}
\end{equation}
see Fig.~\ref{fig:TC2d3FibreModelsPrescrSol}(a). Then, a benchmark
energy for the scenario with continuously embedded fibres may be provided
with $\mathfrak{e}\left(\vek u\right)=11499.322459892$ with high
accuracy. The Bulk Trace FEM converges with optimal accuracy to this
benchmark value, see Fig.~\ref{fig:TC2d3FibreModelsPrescrSol}(b).
It can also be confirmed that the discrete-fibre setting converges
to this energy as seen in Fig.~\ref{fig:TC2d3FibreModelsPrescrSol}(c).
However, one may then only expect a first-order convergence (independently
of the element orders) due to the different geometric representation
of the fibres. It is thus confirmed that the proposed modeling and
analysis of structures implied by all level sets within a bulk domain
provides the basis for continuously embedded sub-structure models
within (isotropic) bulk materials which will be further investigated
in future works. 

\section{Conclusions\label{X_Conclusions}}

A mechanical model for structural ropes and membranes is formulated
which simultaneously applies to \emph{all} level sets in a bulk domain.
Interpreting the level sets in the undeformed and deformed bulk domains
yields the material and spatial configurations of the (curved) structures.
For the modeling, it is crucial to employ classical differential operators
with respect to the bulk domain as well as surface differential operators
with respect to the level sets. Then, all mechanical quantities such
as displacements, stresses and strains may be specified accordingly.
The resulting BVP in strong form may be reformulated in weak form
using the co-area formula and the divergence theorem on manifolds,
resulting in integral terms over the bulk domain. 

For the numerical analysis, one may then employ a (higher-order) mesh
in the bulk domain which conforms to the boundaries of the bulk domain,
however, it does (typically) not conform to the geometry of the level-set
structures. This method is called the Bulk Trace FEM as it features
similarities to the classical FEM with conforming meshes and the Trace
FEM using non-conforming meshes. Most importantly, the Bulk Trace
FEM does not feature cut elements and, therefore, standard methods
for the numerical integration and enforcement of boundary conditions
may be employed without any need for stabilization. Technical aspects
of the implementation are outlined and enable a straightforward implementation
of the method in an existing FE solver.

The numerical results confirm that higher-order convergence rates
are achieved as expected. The potential of the proposed method for
the simultaneous assessment of mechanical properties of structures
implied by all level sets in a bulk domain is shown. Furthermore,
the described model may easily be combined with conventional (mechanical)
bulk models for $d$-dimensional structures, introducing a new concept
for advanced material models, possibly labelled continuously embedded
sub-structure models or embedded, layered manifold models. In future
works, we shall extend the present approach for membranes also to
classical theories of shells with the same motivation to enable the
simultaneous analysis of shells on all level sets in a bulk domain
or to add sub-structures to existing bulk materials.

\bibliographystyle{/home/tpfries/Publications/schanz}
\addcontentsline{toc}{section}{\refname}\bibliography{FriesRefsNew}

\end{document}